\documentclass[11pt]{article}
\usepackage[utf8]{inputenc}
\usepackage{custom}
\usepackage{tabularx}

\usepackage{subcaption}
\usepackage{todonotes}
\setuptodonotes{fancyline, color=blue!10, size=\small}

\title{{Technical Proposal for the \\
Atom Interferometer CERN Experiment (AICE) Facility}}

\author[1]{Gianluigi~Arduini,}
\affiliation[1]{CERN, 1211 Geneva 23, Switzerland}

\author[2]{{Nadja~Augst},}
\affiliation[2]{German Aerospace Center (DLR), Institute of Quantum Technologies, Wilhelm-Runge-Stra\ss e 10, 89081 Ulm, Germany}

\author[3]{Mark~G.~Bason,}
\affiliation[3]{Rutherford Appleton Laboratory, UKRI-STFC, Harwell Campus, Didcot, OX11 0QX, United Kingdom}

\author[4]{Charles~Baynham,}
\affiliation[4]{High Energy Physics Group, Blackett Laboratory, Imperial College, Prince Consort Road, London, SW7 2AZ, UK}

\author[5]{Andrea~Bertoldi,}
\affiliation[5]{IOGS, LP2N, Université Bordeaux, CNRS, UMR 5298, F-33400 Talence, France}

\author[6]{Gianfranco Bertone,}
\affiliation[6]{Gravitation Astroparticle Physics Amsterdam (GRAPPA),\\ University of Amsterdam, 1098 XH Amsterdam, The Netherlands}

\author[7,8]{Diego~Blas,}
\affiliation[7]{Institut de F\'{i}sica d'Altes Energies (IFAE), The Barcelona Institute of Science and Technology, Campus UAB, 08193 Bellaterra (Barcelona), Spain}
\affiliation[8]{Instituci\'{o} Catalana de Recerca i Estudis Avan\c{c}ats (ICREA), Passeig Llu\'{i}s Companys 23, 08010 Barcelona, Spain}

\author[9]{Daniela~Bortoletto,}
\affiliation[9]{Department of Physics, University of Oxford, Parks Road, Oxford OX1 3PU, UK}

\author[10]{Sougato~Bose,}
\affiliation[10]{Department of Physics \& Astronomy, University College London, Gower Street, London, WC1E 6BT, UK}

\author[1]{Roberto~Ales~Bozzi,}
\author[4,9]{Oliver~Buchmueller$^*$,}
\author[1]{Tamara~Alice~Bud,}
\author[11]{Clare~Burrage,}
\affiliation[11]{School of Physics and Astronomy, University of Nottingham, University Park, Nottingham NG7 2RD, UK}

\author[1]{Sergio~Calatroni$^*$,}
\author[12]{{John~Carlton},}
\affiliation[12]{DESY, Notkestrasse 85, 22607 Hamburg, Germany}

\author[13]{Vassilis~Charmandaris,}
\affiliation[13]{Chairman, Board of Directors, Foundation for Research and Technology -- Hellas (FORTH), Vassilika Vouton, 70013 Heraklion, Creta, Greece}

\author[14]{Maria~Luisa~(Maril\`u)~Chiofalo,}
\affiliation[14]{Department of Physics, University of Pisa, Largo Bruno Pontecorvo 3 56126 Pisa, Italy; INFN-Pisa, Largo Bruno Pontecorvo 3 56126 Pisa, Italy}

\author[15]{Pierre Clad\'e,}
\affiliation[15]{Laboratoire Kastler Brossel, Sorbonne Universit{\'{e}} PSL, Coll{\`{e}}ge de France, 75005 Paris, France}

\author[16]{Jonathon~Coleman,}
\affiliation[16]{Department of Physics, University of Liverpool, Merseyside, L69 7ZE, UK}

\author[1]{Fabio~Corsanego,}
\author[4,17]{Albert~De~Roeck,}
\affiliation[17]{Department of Physics, Faculty of Science, Chulalongkorn University, Bangkok, 10330, Thailand}

\author[1]{Arnaud~Devienne,}
\author[18]{Fabio~Di~Pumpo,}
\affiliation[18]{Institut f{\"u}r Quantenphysik and Center for Integrated Quantum Science and Technology (IQST), Universit{\"a}t Ulm, Albert-Einstein-Allee 11, 89081 Ulm, Germany}

\author[1,19]{John~Ellis$^*$,}
\affiliation[19]{Physics Department, King's College London, Strand, London, WC2R 2LS, UK}

\author[20]{Pierre~Fayet,}
\affiliation[20]{Laboratoire de physique de l’\'Ecole normale sup\'erieure, ENS-PSL, CNRS, Sorbonne Univ., Univ. Paris Cit\'e, 24 rue Lhomond, 75231 Paris Cedex 05, France}

\author[9]{Chris~Foot,}
\author[12,21]{{Elina~Fuchs},}
\affiliation[21]{Institut f{\" u}r Theoretische Physik, Leibniz Universit{\" a}t Hannover, Appelstrasse 2, 30167 Hannover, Germany}

\author[22]{Naceur~Gaaloul,}
\affiliation[22]{Leibniz Universit\"at Hannover, Institut f\"ur Quantenoptik, Welfengarten 1, 30167 Hannover, Germany}

\author[23]{Susan~Gardner,}
\affiliation[23]{Department of Physics and Astronomy, University of Kentucky, Lexington, KY 40506-0055, USA}

\author[24]{Enno~Giese,}
\affiliation[24]{Technische Universit{\"a}t Darmstadt, Fachbereich Physik, Institut f{\"u}r Angewandte Physik, 64289 Darmstadt, Germany}

\author[1]{Eduardo~Granados,}
\author[15]{Sa\"{\i}da~Guellati-Khelifa,}
\author[1]{Michael~Guinchard,}
\author[1]{Timo~Hakulinen,}
\author[25,26]{Tiffany~Harte,}
\affiliation[25]{Cavendish Laboratory, University of Cambridge, J J Thomson Avenue, Cambridge, CB3 0US, UK}
\affiliation[26]{Department of Physics and Astronomy, University of Manchester, Manchester M13 9PL, UK}

\author[4]{Richard~Hobson$^*$,}
\author[27]{Michael~Holynski,}
\affiliation[27]{School of Physics and Astronomy, University of Birmingham, B152TT Edgbaston, UK}

\author[1]{Angelo~Infantino,}
\author[28]{Alex~Kehagias,}
\affiliation[28]{School of Applied Mathematical and Physical Sciences, National Technical University of Athens, Hroon Polytechniou 9, Athens, 15780, Greece}

\author[10]{Eva~Kilian-Rademacher,}
\author[23]{Wolfgang~Korsch,}
\author[1]{Damien~Lafarge,}
\author[27,29]{Samuel~Lellouch,}
\affiliation[29]{School of Engineering, University of Birmingham, B152TT Edgbaston, UK}

\author[30,31]{Lucas~Lombriser,}
\affiliation[30]{Department of Applied Future Technologies, University of Applied Sciences of the Grisons, Pulverm\"uhlestrasse 57, 7000 Chur, Switzerland}
\affiliation[31]{D\'epartement de Physique Th\'eorique, Universit\'e de Gen\`eve, 24 quai Ernest Ansermet, 1211 Gen\`eve 4, Switzerland}

\author[32]{Elias~Lopez~Asamar,}
\affiliation[32]{Departamento de F{\'i}sica Te{\'o}rica, Universidad Autonoma de Madrid, 28049 Madrid, Spain; Instituto de F{\'i}sica Te{\'o}rica UAM-CSIC, 28049 Madrid, Spain}

\author[33]{J.~Luis~Lopez-Gonzalez,}
\affiliation[33]{Department of Mathematics and Physics, Autonomous University of Aguascalientes, Av. Universidad 940, Aguascalientes 20100, Mexico}

\author[31,34]{Michele~Maggiore,}
\affiliation[34]{Gravitational Wave Science Centre, Universit\'e de Gen\`eve, Gen\`eve, Switzerland}

\author[1]{Charline~Marcel,}
\author[3]{Anna~Marchant,}
\author[19]{Christopher~McCabe,}
\author[35]{Gaetano~Mileti,}
\affiliation[35]{University of Neuch{\^ a}tel, Avenue de Bellevaux 51, CH-2009 Neuch{\^ a}tel, Neuch{\^ a}tel, Switzerland}

\author[26]{Peter~Millington,}
\author[25]{Jeremiah~Mitchell,}
\author[25]{Noam~Mouelle,}
\author[1]{{Angel~Navascues~Cornago},}
\author[1]{{François~Pillon},}
\author[14]{Rosa~Poggiani,}
\author[36]{Johann~Rafelski,}
\affiliation[36]{Department of Physics, The University of Arizona, Tucson, AZ 85721-0081, USA}

\author[22]{Ernst~M.~Rasel,}
\author[21]{{Maike~Reckermann},}
\author[2]{{Albert~Roura},}
\author[1]{{Rui~Samoes},}
\author[34,37]{Federico~Sanchez~Nieto,}
\affiliation[37]{D\'epartement de Physique Nucl\'eaire et Corpusculaire, Université de Genève, 24 quai Ernest Ansermet, 1211 Genève 4, Switzerland}

\author[9]{Jack~Sander,}
\author[1]{Carlo~Scarcia,}
\author[9]{Jesse~Schelfhout,}
\author[18,38]{Wolfgang~Schleich,}
\affiliation[38]{Institute for Quantum Science and Engineering (IQSE), and Texas A\&M AgriLife Research and Hagler Institute for Advanced Study, Texas A\&M University, College Station, TX 77843-4242, USA}

\author[22]{Dennis~Schlippert,}
\author[25]{Ulrich~Schneider,}
\author[34,37]{Steven~Schramm,}
\author[39,40]{Florian~Schreck,}
\affiliation[39]{Van der Waals-Zeeman Institute, Institute of Physics, University of Amsterdam, Science Park 904, 1098 XH Amsterdam, The Netherlands}
\affiliation[40]{QuSoft, Science Park 123, 1098XG Amsterdam, The Netherlands}

\author[41]{Olga~Sergijenko,}
\affiliation[41]{Faculty of Space Technologies, AGH University of Krakow, Aleja Mickiewicza 30, Krakòw 30-059, Poland}

\author[42]{Marcelle~Soares-Santos,}
\affiliation[42]{University of Zurich, Winterthurerstrasse 190, 8057 Zurich, Switzerland}

\author[43]{Guglielmo~M.~Tino,}
\affiliation[43]{Dipartimento di Fisica e Astronomia and LENS, Universit\`{a} di Firenze, INFN Sezione di Firenze, CNR-INO, via Sansone 1, I-50019 Sesto Fiorentino, Italy}

\author[16]{Jonathan~N.~Tinsley,}
\author[3]{Tristan~Valenzuela,}
\author[1]{Daniel~Valuch,}
\author[3]{Maurits~van~der~Grinten,}
\author[39]{Klaasjan~van~Druten,}
\author[44,45,46]{{Ville~Vaskonen},}
\affiliation[44]{Laboratory of High Energy and Computational Physics, KBFI, R{\"a}vala 10, Tallinn, 10143, Estonia}
\affiliation[45]{Dipartimento di Fisica e Astronomia, Universit\`a degli Studi di Padova, Via Marzolo 8, 35131 Padova, Italy}
\affiliation[46]{Istituto Nazionale di Fisica Nucleare, Sezione di Padova, Via Marzolo 8, 35131 Padova, Italy}

\author[47]{Wolf~von~Klitzing,}
\affiliation[47]{Foundation for Research and Technology -- Hellas (FORTH), Institute of Electronic Structure and Lasers (IESL), Vassilika Vouton, 70013 Heraklion, Crete, Greece}

\author[22]{Michael~Werner}
\date{\today}

\begin{document}
%\linenumbers

\graphicspath{{figures/}}

\abstract{%
We present the technical proposal for the Atom Interferometer CERN Experiment (AICE), a $\mathcal{O}(100)$\,m vertical atom interferometer to be installed against the wall of the PX46 access shaft to the LHC.
AICE is conceived as a {versatile and flexible} long-baseline atom-interferometry facility whose primary scientific goal is probing for bosonic ultralight dark matter (ULDM) in a mass range inaccessible to other experiments, with a secondary goal of pioneering the exploration of gravitational waves (GWs) with frequencies in the range ${\sim}$0.03--3\,Hz as a pathfinder for future longer-baseline detectors.
{The Initial configuration of the experiment employs ultracold $^{87}$Sr atoms in a single-photon 698-nm interferometer with three shaft-based atom sources in a multi-source gradiometer geometry, supported by one surface reference source for laser stabilisation and diagnostics, to target scalar ULDM. Operation with $^{88}$Sr will give sensitivity to axion-like particles (ALPs), vector ULDM with $B{-}L$ couplings and violation of the principle of equivalence, while a $^{171}$Yb upgrade will improve the sensitivity to $B{-}L$ couplings and equivalence violations. Probing the Einstein equivalence principle (EP) and measuring $\alpha$ will proceed in parallel with the ULDM searches.} 
A conceptual feasibility study~\cite{Arduini:2851946} and a detailed technical implementation study~\cite{Arduini:2025jhe} have established that PX46 is a uniquely mature and implementation-ready site, with no technical showstoppers.
{Completing site preparation works during LS3 would enable the subsequent installation and operation of AICE without impacting HL-LHC operations.}
The detector design builds on {experience with the design and construction of the VLBAI and MAGIS experiments~\cite{schlippert2020matter,MAGIS-100:2021etm} as well as the AION-10 Technical Design Report~\cite{Bongs:2025rqe}}, scaling the  strontium gradiometer architecture validated in~\cite{baynham_prototype_2026} to the $\sim$100\,m baseline.
{The AICE experimental programme of dark matter searches, precision  measurements and probes of gravity is well aligned with CERN's mission and priorities.}
AICE is {endorsed} by the TVLBAI proto-collaboration, comprising 57~institutions in 22~countries.
\\[5pt]
%\noindent CERN-LHCC-20XX-XXX / LHCC-P-XXX / CERN-PBC-REPORT-2026-001 \\%[10pt]\
\noindent CERN-PBC-REPORT-2026-001 \\%[10pt]\
$^*$Contact person}

\maketitle

% ---------------------------------------------------------------
% Main sections
% ---------------------------------------------------------------
%\input{sections/sec_intro}
%\input{sections/sec_concept}
%\input{sections/sec_physics}
%\input{sections/sec_detector}
%\input{sections/sec_site}
%\input{sections/sec_civil}
%\input{sections/sec_noise}
%\input{sections/sec_ops}

\section{Introduction}
\label{sec:intro}

Over the past two decades, atom interferometry has developed into a powerful quantum
sensor {for fundamental physics and other precision measurements}.
In more recent years, several programmes directed towards long-baseline atom
interferometry {have been launched}, including
{VLBAI (Germany)~\cite{schlippert2020matter},
MIGA (France)~\cite{Canuel:2017rrp},
MAGIS-100 (US)~\cite{MAGIS-100:2021etm},
AION (UK)~\cite{Badurina:2019hst,Bongs:2025rqe,baynham_prototype_2026} and
ZAIGA (China)~\cite{Zhan:2019quq}}, and
are {establishing an} experimental basis for long-baseline operation in
fundamental physics.
These programmes, together with pioneering demonstrations at LENS in Florence and at
Stanford, have demonstrated single-photon clock-transition
interferometry~\cite{hu2017atom,rudolf2020_689LMT,AION:2025igp},
% NOTE (references stage): twins to deduplicate: hu2017atom = AIClock2017;
% LMTClockHogan2022 = wilkasonFloquet.
large-momentum transfer beyond $100\,\hbar k$~\cite{Asenbaum:2017},
multi-interferometer gradiometry~\cite{AION:2025igp,Canuel:2017rrp},
and dual-species differential acceleration measurements at the
1--10\,m scale~\cite{Asenbaum2020}.
They are now {linked through} the Terrestrial Very-Long-Baseline Atom
Interferometry (TVLBAI) proto-collaboration~\cite{TVLBAISummary,abdalla_terrestrial_2025,TVLBAI3,TVLBAIESPP},
which {coordinates routes} towards instruments at the 100\,m scale and beyond located, e.g., in the UK~\cite{abdalla_terrestrial_2025}, Switzerland~\cite{Guinchard:2026cen} and the US~\cite{Heise:2026obv}.

\begin{figure}[htbp]
    \centering
    \includegraphics[width=0.45\linewidth]{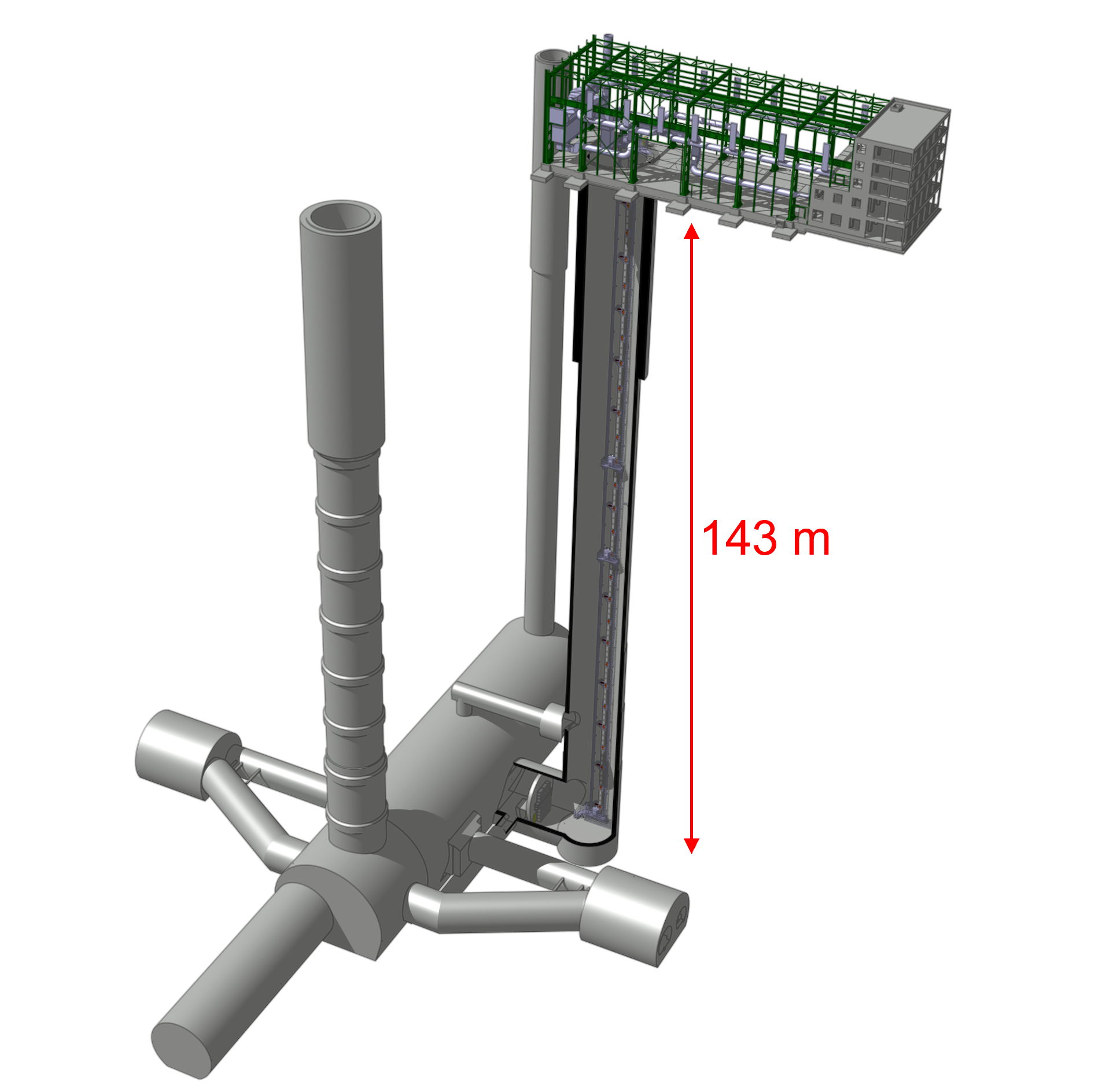}
    \includegraphics[width=0.45\linewidth]{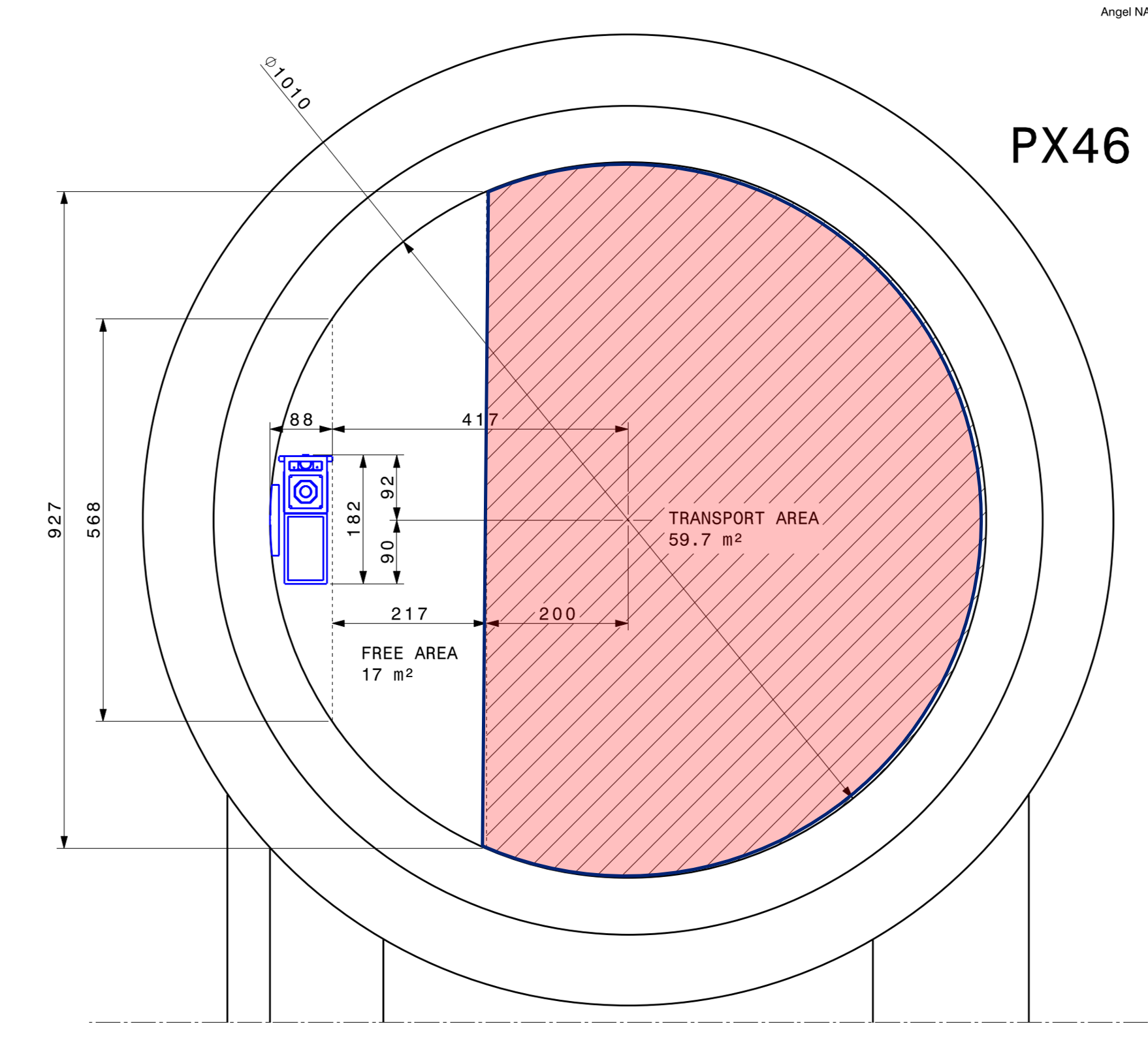}
    \caption{{\it Left panel}: General schematic view of the proposed implementation of AICE in the PX46 access shaft~\cite{Arduini:2851946,Arduini:2025jhe}. {\it Right panel}: Horizontal cross section of the PX46 shaft, showing that the proposed location of AICE (blue) is compatible with the space required for transport of LHC components (red hatched area).}
    \label{fig:PX46}
\end{figure}

Within this framework, the possibility of a long-baseline atom interferometer facility at CERN has been
studied in detail.
A conceptual feasibility study~\cite{Arduini:2851946} established the viability of a
$\mathcal{O}(100\,\mathrm{m})$ atom interferometer in the PX46 access shaft to the LHC, {a uniquely favourable European site (see Section~\ref{sec:whycern}).
The left panel of Fig.~\ref{fig:PX46} shows a schematic view of PX46 and the layout of the civil engineering infrastructure at Point 4 on the LHC ring. The vertical height from the SX4 surface building down to the level of the LHC is $\sim 143$\,m and the internal diameter of the shaft is 10.1\,m. PX46 provides access to the main LHC radiofrequency (RF) system and its primary use is for occasionally raising and lowering technical equipment. However, as seen in the right  panel of Fig.~\ref{fig:PX46}, a substantial fraction of the horizontal cross-section of the PX46 shaft is not required for LHC access, and would be large enough to accommodate an atom-interferometer experiment.}
A subsequent implementation study~\cite{Arduini:2025jhe} developed the corresponding
technical concept, including infrastructure integration, civil engineering,
radiation protection, and compatibility with HL-LHC operations.
Together, these studies show that PX46 can host a vertical interferometer at the required
scale without fundamental technical obstacles and {without impacting the LHC}
programme.

The AICE (Atom Interferometer CERN Experiment) proposal is the {fruit} of this effort.
It %is a 140\,m vertical atom interferometer installed in the PX46 shaft, which 
would be a pioneering project at the 100\,m scale within the TVLBAI roadmap.
AICE is conceived as a facility rather than a single experiment.
It is intended to operate in a staged way, with different configurations serving different
physics goals over time.

Its main scientific objective is the search for ultralight dark matter in scalar,
vector $B{-}L$, and {pseudoscalar axion-like particle (ALP)} channels.
The same infrastructure can also be used in parallel to support precision measurements of the fine-structure
constant $\alpha$, tests of the universality of free fall, {and other fundamental physics measurements}. At later stages it also provides a pathfinder for gravitational-wave searches in the
mid-frequency band.

The remainder of this introduction frames the scientific motivation for the dark matter and
precision-physics programmes (Section~\ref{sec:motivation}), describes the staged facility
concept and performance roadmap (Section~\ref{sec:staged}), situates AICE within the wider
atom-interferometry programme on which it is based (Section~\ref{sec:wider}), explains why
CERN is the right host and why now is the right time (Section~\ref{sec:whycern}), and maps the
structure of this technical proposal (Section~\ref{sec:docmap}).

\subsection{Motivation}
\label{sec:motivation}

\paragraph{Ultralight dark matter.}

The identity of dark matter remains unresolved despite extensive experimental effort.
Collider searches and direct-detection experiments have placed stringent bounds on
weakly-interacting particles at and above the electroweak scale, motivating attention to
candidates at much lower masses.
Ultralight bosonic fields with masses in the range $10^{-22}$--$10^{-12}$\,eV arise
naturally in several well-motivated extensions of the Standard Model, including scalar
moduli and dilaton-like states, axion-like particles, and low-mass vector bosons with $B{-}L$
gauge couplings~\cite{Arvanitaki:2014faa,Graham:2013gfa,Graham:2015ifn}.
These fields behave on laboratory scales as coherently oscillating classical backgrounds,
which may induce time-dependent perturbations of fundamental constants and of free-fall
acceleration at frequencies in the range $10^{-7}$--$1$\,Hz.
A unified
treatment of their common response framework is given in Section~\ref{sec:uldm_framework}, and the prospective AICE sensitivities to these and other dark matter channels {using cold $^{87}$Sr, $^{88}$Sr and $^{171}$Yb atoms} are described in Section~\ref{sec:uldm}. 

Detecting these signatures requires phase, frequency, or acceleration sensitivity across
this low-frequency band.
Long-baseline atom interferometers are naturally suited to this requirement.
The phase accumulated over a free-fall time $T$ is sensitive to all these dark matter channels, {large-momentum-transfer
(LMT) atom optics using a narrow clock transition amplify the scalar and vector signals, and the
gradiometer configuration provides intrinsic common-mode rejection of laser
phase noise and shared mechanical vibration, and suppresses spatially common
electromagnetic perturbations. Fluctuating gravity gradient noise acts
directly on the atoms. Its common component cancels, but its differential
component is not rejected, and its impact is assessed in
Section~\ref{sec:noise:GGN:impact}.}
An interferometer at the 100\,m scale, operating at the phase-noise level now being
demonstrated by current prototypes, would have sensitivities to the couplings of these ultralight
dark matter channels that are competitive with or better than the best existing
laboratory bounds, as shown in Section~\ref{sec:uldm}.

\paragraph{Precision physics: the fine-structure constant and the equivalence principle.}

The $^1S_0$--$^3P_0$ strontium clock transition at 698\,nm and the analogous ytterbium
transition at 578\,nm, used in AICE for interferometry, are the same transitions exploited
by state-of-the-art optical clocks.
Their interferometric use enables determinations of the fine-structure constant $\alpha$
via the atom-recoil method.
The two most precise atom-interferometry values of $\alpha$, obtained with
caesium in 2018~\cite{parker2018} and rubidium in 2020~\cite{morel2020}, are
in tension at more than $5\sigma$.
Strontium and ytterbium measurements use single-photon clock transitions rather than the
two-photon methods used for caesium and rubidium.
They therefore constitute an independent determination of $\alpha$ using a
systematically distinct family of measurements, and their consistency with the alkali
values would itself be a probe of the origin of the tension.
AICE delivers these measurements through staged campaigns within the science
programme. An initial $\alpha(\mathrm{Sr})$ result appears during commissioning, a
species cross-check with $\alpha(\mathrm{Yb})$ follows during the vector phase, and
progressively more precise determinations are obtained at later facility stages,
culminating in an Ultimate $\alpha$ campaign at Stretch parameters targeting
$\sigma_\alpha/\alpha \lesssim 10^{-11}$.

AICE also tests the weak equivalence principle {(WEP)} via the universality of
free fall (UFF). Running an atom interferometer with two different atomic species
simultaneously provides a ground-based test of the E\"otv\"os ratio
$\eta_{AB} = 2(a_A-a_B)/(a_A+a_B)$, where $a_A$ and $a_B$ are the
measured free-fall accelerations of the two species, bridging the gap between current
10\,m atom-based tests~\cite{Asenbaum2020} and space-borne UFF missions
including MICROSCOPE~\cite{Touboul:2022yrw}.
The WEP programme is distributed across the same facility stages as the
ULDM and $\alpha$ campaigns. A $^{87}$Sr/$^{88}$Sr isotopic
configuration validates dual-species systematics during early facility
operation, the $^{87}$Sr/$^{171}$Yb dual-wavelength configuration
delivers the flagship WEP result during the vector phase, and a revisit
at Stretch parameters targets $\eta \lesssim$ few$\,\times 10^{-16}$,
competitive with MICROSCOPE and positioning AICE as a terrestrial
bridge to future space missions.
The precision-physics case is described in Section~\ref{sec:alpha}, together with possible probes of gravitational physics and quantum mechanics.

\paragraph{Gravitational waves in the mid-frequency band.}

The same infrastructure is sensitive to gravitational waves in the
${\sim}0.03$--$10$\,Hz frequency band, which lies between the sensitive regions of ground-based laser
interferometers and the future space-based LISA mission.
The long free-fall baseline and low-frequency interrogation window would offer pioneering terrestrial
exploration of this band with an underground atom-interferometry facility.
This band may contain signals from intermediate-mass black hole mergers and
early inspiral phases of LIGO/Virgo/KAGRA sources that could be observable with an ${\cal O}(1)$~km atom interferometer.
AICE operates as a pathfinder in this mid-frequency band; its prospects are discussed in Section~\ref{sec:gw}.

\paragraph{Alignment with the CERN mission.}
The AICE facility offers opportunities for several fundamental measurements in particle physics, such as searches for dark matter that complement the LHC and a precision measurement of the fine-structure constant, and in other aspects of fundamental physics, such as pathfinding searches for gravitational waves in a frequency range inaccessible to other experiments. AICE forms an integral part of the Physics Beyond Colliders programme and has strong synergies with the Quantum Technology Initiative, as discussed further in Section~\ref{sec:whycern}.

\subsection{AICE as a staged facility}
\label{sec:staged}

\paragraph{Infrastructure and configuration layers.}

AICE is conceived as a facility rather than a single-purpose experiment.
A single 140\,m vertical vacuum tower in PX46 serves all three ultralight dark matter
campaigns and the embedded precision-physics programme.
It is useful to distinguish three layers of the facility:
the fixed infrastructure (the PX46 shaft, vacuum tower, support structure, site services,
and CERN integration), which is established during construction and remains unchanged
throughout the programme;
the shared core systems (the 698\,nm strontium clock laser, the control architecture, and
the atom source platform), which are common to all science phases;
and the campaign-dependent configuration (species choice and source count, optical configuration
for gradiometer or full-shaft Mach-Zehnder mode, and, from the vector phase
onwards, the 578\,nm ytterbium laser system), which evolves across campaigns as the
science programme advances.

This architecture allows the same facility to support different physics objectives
through limited additions between campaigns.

\paragraph{Performance roadmap.}

\begin{table}[htbp]
\centering
\caption{%
  Performance parameter ladder for AICE and the TVLBAI roadmap.
  The Initial column corresponds to the AICE commissioning and first-physics regime.
  The Baseline and Stretch columns define the two successive AICE science-phase
  operating levels and drive all dark matter sensitivity projections in this proposal.
  The 1\,km column corresponds to a future TVLBAI facility and is provided for roadmap
  context only.
  $T_{\mathrm{grad}}$, $T_{\mathrm{MZ}}$ and $T_{\mathrm{ax}}$ are the interrogation
  times used in the scalar gradiometer, the vector/equivalence-principle
  full-baseline dual-species Mach-Zehnder, and the axion full-shaft single-species
  Mach-Zehnder modes respectively (see Section~\ref{sec:concept}). The vector/WEP and
  axion modes both interrogate over the full one-way drop and therefore share the
  same time, $T_{\mathrm{MZ}} = T_{\mathrm{ax}} \simeq \sqrt{2L/g}$; only the scalar
  gradiometer, whose clouds split the shaft, runs at the shorter $T_{\mathrm{grad}}$.
  $n$ is the LMT order of the atom optics, corresponding to $n\hbar k_{0}$
  momentum imparted between the two interferometer arms on the 698\,nm
  $^{1}S_{0}$--$^{3}P_{0}$ clock transition. For the Stretch goal $n$, multi-loop pulse sequences must be implemented to fit within the instrument~\cite{Schach_2025} (see Section~\ref{sec:concept:modes}).
  $N_{\mathrm{source}}$ is the number of cold-atom sources, and
  $\delta\phi_{\mathrm{noise}}$ is the assumed atom-shot-noise amplitude spectral density.%
}
\vspace{3mm}
\label{tab:roadmap}
\begin{tabular}{lcccc}
\hline
Parameter & Initial & Baseline & Stretch & 1\,km \\
\hline
Programme phase        & Comm./Fund.\ Phys.
                       & Baseline
                       & Stretch Goal
                       & Future Facility \\
Baseline $L$ [m]       & $140$
                       & $140$
                       & $140$
                       & $1000$ \\
$T_{\mathrm{grad}}$ [s]
                       & $2.5$
                       & $2.5$
                       & $2.5$
                       & $7.0$ \\
$T_{\mathrm{MZ}}$ [s]
                       & $5.0$
                       & $5.0$
                       & $5.0$
                       & $14.0$ \\
$T_{\mathrm{ax}}$ [s]
                       & $5.0$
                       & $5.0$
                       & $5.0$
                       & $14.0$ \\
$n$ (LMT)              & {$\sim 100$}
                       & $10^{3}$
                       & $4\times10^{4}$
                       & $4\times10^{4}$ \\
$N_{\mathrm{source}}$  & $3$
                       & $5$--$10$
                       & $10$--$20$
                       & $\mathcal{O}(100)$ \\
$\delta\phi_{\mathrm{noise}}$ [rad/$\sqrt{\mathrm{Hz}}$]
                       & ${\sim} 10^{-3}$
                       & ${\sim} 10^{-4}$
                       & $10^{-5}$
                       & $10^{-5}$ \\
$t_{\mathrm{int}}$     & ---
                       & 1\,yr
                       & 1\,yr
                       & 1\,yr \\
\hline
\end{tabular}
\end{table}

The performance of AICE evolves through three levels that mirror the broader TVLBAI
roadmap~\cite{TVLBAISummary} and are summarised in Table~\ref{tab:roadmap}.
The Initial level, applicable during commissioning and first-physics running, corresponds
to a momentum transfer of $n \approx 100$ and phase noise of $\delta\phi \approx 10^{-3}$\,rad/$\sqrt{\mathrm{Hz}}$
with approximately three atom sources.
This parameter set has been chosen in line with the performance of recent cold-atom experiments in Tables~\ref{tab:LMT_history} and \ref{tab:atom_production_rates}, and is now actively being reproduced across the national
prototype programmes described in Section~\ref{sec:wider}. The step from the few-meter prototypes to the $L \approx 140\,\mathrm{m}$ scale of AICE is one of integration and scale, not the introduction
of a new experimental principle. The Baseline and Stretch parameters will require R\&D, particularly towards higher-power clock-interferometry lasers to extend large momentum transfer $n$ (Section~\ref{sec:laser-upgrades}) and higher-flux sources of cold atoms to reduce phase noise $\delta\phi_\mathrm{noise}$ (see Section~\ref{sec:atom-sources}).
The Baseline level, which drives all primary dark matter sensitivity projections in this
proposal, reaches $\delta\phi \approx 10^{-4}$\,rad/$\sqrt{\mathrm{Hz}}$ with
$N_{\mathrm{source}} = 5$--$10$ and an LMT order of $n = 10^{3}$ on the 698\,nm clock transition.
The Stretch level, targeted for the final science campaign, achieves
$\delta\phi \approx 10^{-5}$\,rad/$\sqrt{\mathrm{Hz}}$ with
$N_{\mathrm{source}} = 10$--$20$ and an LMT order of $n = 4\times10^{4}$ on the
698\,nm clock transition, extending the axion-nucleon reach to approximately one
order of magnitude below the SN\,1987A astrophysical bound
(which is subject to supernova modelling uncertainties).
For context, the 1\,km column in Table~\ref{tab:roadmap} extrapolates to the scale of the
future TVLBAI facility; those numbers are included for roadmap orientation and do not
represent an AICE deliverable.

\begin{table}[htbp]
  \begin{minipage}{\textwidth}
  \centering
  \caption{Progress in Large Momentum Transfer (LMT) atom interferometry ($n$ in Table~\ref{tab:roadmap}).}
  \label{tab:LMT_history}
    \begin{tabular}{llll}
    \hline
    Momentum transfer method & Atom & LMT ($n$) & Year\\
    \hline
    Bragg diffraction & Rb & $102\,\hbar k$~\cite{chiow_102ensuremathhbark_2011} & 2011\\
    Kapitza--Dirac \& Bragg diffraction & Yb & $112\,\hbar k$~\cite{plotkin-swing_three-path_2018} & 2018 \\
    Symmetric Bloch oscillation & Cs & $240\,\hbar k$~\cite{pagel_symmetric_2020} & 2020 \\
    Single-photon (7~kHz transition) & Sr & $400\,\hbar k$~\cite{LMTClockHogan2022} & 2022 \\
    Double Bragg diffraction \& Bloch oscillation & Rb & $408\,\hbar k$~\cite{gebbe_twin-lattice_2021} & 2021 \\
    Bragg diffraction & Rb & $600\,\hbar k$~\cite{rodzinka_optimal_2024} & 2024 \\
    Single-photon (1~mHz transition)~\footnote{Planned AICE interferometry type; the tabulated $1\,\hbar k$ is the published demonstration, while $61\,\hbar k$ LMT had been observed in an AICE laboratory as of June 2026 (unpublished).} & Sr & $1\,\hbar k$~\cite{baynham_prototype_2026} & 2026 \\
    \hline
    \end{tabular}
    \end{minipage}
\end{table}

\begin{table}[htbp]
\begin{minipage}{\textwidth}
\centering
\caption{Reported production rates for Sr and Yb atoms---crucial for $\delta\phi_\mathrm{noise}$ in the roadmap Table~\ref{tab:roadmap}. The AICE experiment requires cooling close to Bose-Einstein Condensation (BEC) or Degenerate Fermi Gas (DFG) to limit atom cloud expansion. Recent magneto-optical trap (MOT) experiments produce cold atoms quickly---including the three compact sources developed by AICE collaborators~\cite{walker2026high,feng2024,wodey_robust_2021}, discussed in Section~\ref{sec:atom-sources}---but further cooling is needed before their use for atom interferometry. The phase-noise standard quantum limit scales as $\delta\phi_\mathrm{noise} \sim 1/\sqrt{N_\mathrm{1\,s}}.$}
\label{tab:atom_production_rates}
\begin{tabular}{lcccc}
\toprule
Species & Atoms/second ($N_\mathrm{1\,s}$) & Preparation time & Sample type & Reference \\
\midrule
$^{87}$Sr  & $8\times10^3$   & 13 s        & DFG          & \cite{campbell_fermi-degenerate_2017} \\
$^{87}$Sr  & $\sim10^4$      & $\sim20$ s & DFG          & \cite{stellmer_production_2013} \\
$^{174}$Yb & $8\times10^4$   & 1.8--15 s  & BEC          & \cite{roy_rapid_2016} \\
$^{171}$Yb & $1\times10^4$   & 24 s       & DFG          & \cite{vaidya_degenerate_2015} \\
$^{84}$Sr  & $2.4\times10^5$ & Continuous & BEC          & \cite{chen_continuous_2022} \\
$^{87}$Sr  & $1.3\times10^7$ & Continuous & MOT (\qty{12}{\micro\kelvin}) & \cite{escudero_steady-state_2021} \\
$^{174}$Yb & $1\times10^9$ & Continuous & MOT ($\sim$\qty{10}{\milli\kelvin}) & \cite{wodey_robust_2021} \\
$^{88}$Sr  & $5\times10^{9}$ & Continuous & MOT ($\sim$\qty{10}{\milli\kelvin}) & \cite{feng2024} \\
$^{88}$Sr  & $4\times10^{10}$ & Continuous & MOT ($\sim$\qty{10}{\milli\kelvin}) & \cite{walker2026high} \\
\bottomrule
\end{tabular}
\end{minipage}
\end{table}

\paragraph{Science programme and timeline.}

\begin{figure}[htbp]
\centering
\includegraphics[width=\textwidth]{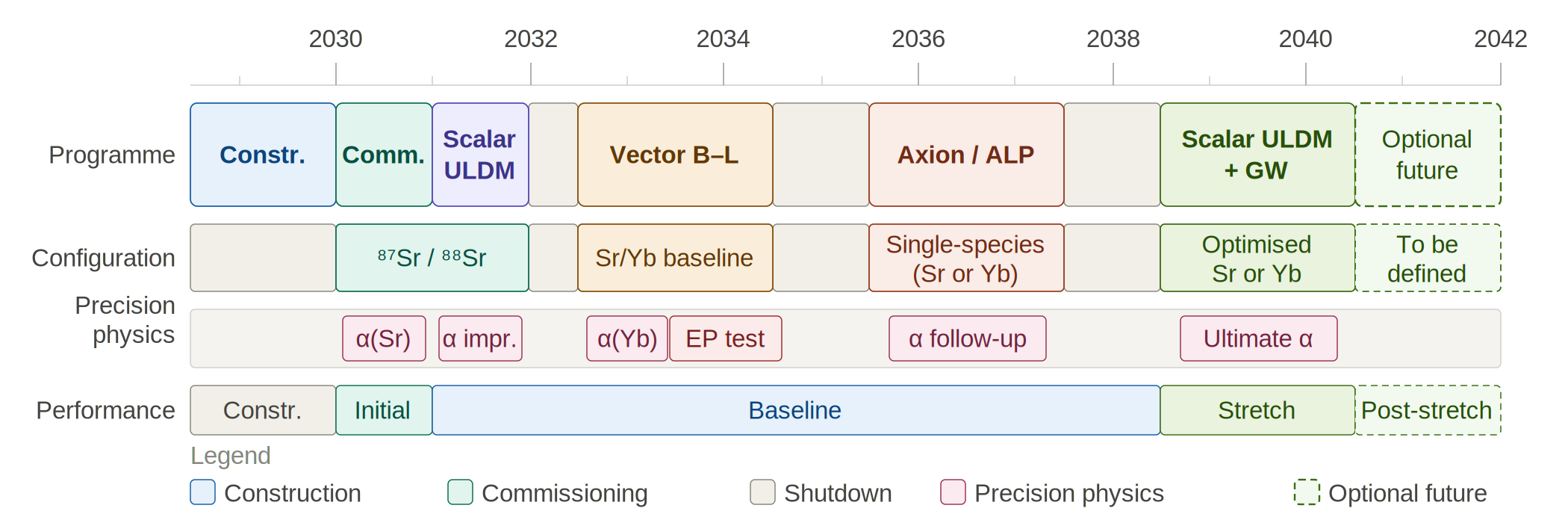}
\caption{%
  Indicative timeline of the science programme for AICE from mid-2028 to the end of 2042, showing   the main science phases, the atom species in use, the short precision physics campaigns embedded within each science phase, and the   projected facility operating levels. Shutdowns for reconfiguration are indicated with grey background, and an optional future campaign is shown with a dashed border. Detailed descriptions of the operational phases are given in Section~\ref{sec:ops}.%
}
\label{fig:timeline}
\end{figure}

An indicative timeline is shown in Figure~\ref{fig:timeline}. The sequence is designed so that each phase builds directly on the hardware and operational experience of the preceding one, with three short shutdowns used for specific upgrades rather than major reconfiguration.

Construction runs from mid-2028 to early 2030, coinciding with LS3. Commissioning begins at the start of Run~4 in 2030 and continues for one year at Initial parameters, without interfering with HL-LHC operations. The first science phase (Scalar ULDM, $\sim$2031--2032) extends this to Baseline parameters in the compact single-species $^{87}$Sr gradiometer configuration, the most straightforward operating mode.
Shutdown~1 ($\sim$2032) will enable installation of the $^{171}$Yb source and 578\,nm clock laser and reconfigures the interferometry from the compact gradiometer to full-baseline dual-species Mach-Zehnder operation, enabling the vector $B{-}L$ and equivalence-principle searches over the full one-way drop ($T_{\mathrm{MZ}} = 5.0$\,s); no change to the shaft or primary laser is required.
The Vector $B{-}L$ phase ($\sim$2032--2034) exploits this to access the $^{87}$Sr/$^{171}$Yb species pair, which provides the best $B{-}L$ sensitivity; $^{87}$Sr/$^{88}$Sr also serves as a validation configuration. Shutdown~2 ($\sim$2034--2035) will enable the reconfiguration of the optics and control systems for the axion-specific single-species clock-gradient mode (single $^{87}$Sr species, LMT disabled so that $k_\mathrm{eff} = k_0$), which is distinct from the dual-species Mach-Zehnder operation already established for the vector phase even though both use the full baseline, and which requires a dedicated reconfiguration period.
The Axion/ALP phase ($\sim$2035--2037) uses this mode, starting at Baseline parameters and advancing towards Stretch performance, with species choice ($^{87}$Sr or $^{171}$Yb) driven by the nuclear coupling factor.
Shutdown~3 ($\sim$2037--2038) will enable the installation of the hardware required for Stretch parameters. The Scalar ULDM + GW phase ($\sim$2038--2040) delivers the full Stretch science output, combining the most sensitive ULDM running with first gravitational-wave pathfinder operation. An optional future campaign after $\sim$2040 remains to be defined and is shown with a dashed border in Figure~\ref{fig:timeline}.

The precision-physics campaigns are embedded throughout. The $\alpha$ programme delivers an initial $\alpha(\mathrm{Sr})$ measurement during commissioning, a species cross-check $\alpha(\mathrm{Yb})$ during the vector phase, and progressively more sensitive determinations culminating in the Ultimate $\alpha$ run at Stretch parameters. The equivalence-principle programme begins with a $^{87}$Sr/$^{88}$Sr isotopic validation during early facility operation, delivers a flagship Sr/Yb dual-wavelength measurement during the vector phase, and culminates in an Ultimate WEP campaign at Stretch parameters targeting $\eta \lesssim$ few$\,\times 10^{-16}$. Section~\ref{sec:ops} gives a detailed account of each science phase.

The timeline in Figure~\ref{fig:timeline} is indicative.
The relative durations and ordering will be adjusted in response to commissioning outcomes and
{the evolution of the experimental and theoretical landscape}.

\subsection{Context for the AICE proposal}
%Relation to the wider programme}
\label{sec:wider}

AICE rests on an already substantial foundation.
The technologies it requires are not speculative: they have been developed and, in most
cases, demonstrated individually or in partial-system form across a set of national and
international programmes (see Tables \ref{tab:LMT_history} and \ref{tab:atom_production_rates}).
The step from the existing 10\,m scale prototypes to AICE is one of integration and
scaling within a well-characterised engineering envelope, not an unvalidated technology leap.

\paragraph{Technology readiness across national programmes.}

The following programmes provide the direct technical basis for AICE.
Each contributes one or more elements of the demonstrated capability on which AICE
is based.

AION (UK)~\cite{Badurina:2019hst,Bongs:2025rqe,AION:2025igp} contributes single-photon strontium
clock-transition interferometry experience and an integrated 10\,m system design.
The AION prototype has demonstrated differential interferometry using the 698\,nm $^{87}$Sr transition~\cite{AION:2025igp},
directly validating the core gradiometry technique to be used by AICE for
dark-matter and gravitational-wave searches.

MAGIS-100 (Fermilab, US)~\cite{MAGIS-100:2021etm} contributes 100\,m-scale implementation and
LMT scaling in a strontium single-photon system. Its Stanford predecessors include the
10\,m rubidium tower, the first long-baseline atom interferometer, which demonstrated LMT
at the $100\,\hbar k$ level~\cite{Asenbaum:2017}, and strontium LMT at the $400\,\hbar k$ level on the 689\,nm
$^{1}S_{0}$--$^{3}P_{1}$ intercombination line~\cite{LMTClockHogan2022}; a 10\,m
strontium prototype is under construction as a direct MAGIS-100 pathfinder. AION--MAGIS cooperation
is covered by a US--UK agreement, and operation as a network similar to LIGO--Virgo--KAGRA
(LVK, currently evolving towards the International Gravitational-Wave Observatory Network, IGWN)
is envisaged.

VLBAI (Leibniz University Hannover, Germany)~\cite{schlippert2020matter} contributes
10\,m instrument construction and operation, and dual-species capability with rubidium and ytterbium, providing initial experience relevant to both dark matter and gravitational-wave searches.

MIGA (University of Bordeaux, France)~\cite{Canuel:2017rrp} contributes gradiometric
operation at the long-baseline scale and site-noise characterisation in an underground
environment.

ZAIGA (China)~\cite{Zhan:2019quq} contributes long-baseline underground infrastructure
strategy and multi-site differential acceleration methodology at the km scale.

The longer-term roadmap is framed by ELGAR~\cite{Canuel:2019abg}, the proposed
European large-scale network, and by the space-based proposals AEDGE~\cite{ElNeaj:2019aedge}
and STE-QUEST~\cite{Aguilera:2014stequest}, which share scientific objectives with AICE
and define the natural extension of this programme beyond the ground-based 100\,m scale.

They are not independent parallel efforts: the community behind AICE is in large part the
same community that drives these national programmes, and the scientific objectives,
simulation tools, and hardware approaches have been developed in close coordination.
The TVLBAI Proto-Collaboration, the coordination framework for this community, has been
signed by over 55 institutions in more than 20 countries~\cite{TVLBAIMOU,TVLBAISIG}.

\paragraph{TVLBAI roadmap and European Strategy.}

The TVLBAI programme defines the strategic framework within which AICE is {a} priority
100\,m demonstrator.
Four major international workshops~\cite{TVLBAISummary,abdalla_terrestrial_2025,TVLBAI3},
the most recent hosted by the Laboratorio Subterr\'aneo de Canfranc,
drawing scientists from particle physics, atomic physics, astrophysics, and cosmology,
have developed the scientific case and the roadmap from 10\,m prototypes to km-scale
detectors.
The TVLBAI community has submitted a coordinated input to the 2026 update of the European
Strategy for Particle Physics~\cite{TVLBAIESPP}, identifying AICE as the key next step
towards km-scale TVLBAI detectors in the mid-2030s.

\subsection{Why CERN, and why now?}

\label{sec:whycern}

The transition from the present generation of instruments, which include
tabletop experiments and the 10\,m-scale prototypes now in operation, to an
instrument at the 100\,m scale represents the necessary next step in the
current roadmap for long-baseline atom interferometry. It provides the longer
interrogation times and spatial separations required for a substantial
increase in scientific reach, while remaining an achievable next step
before the construction of future km-scale detectors. The 100\,m stage is
therefore a central near-term objective of the international TVLBAI
roadmap~\cite{TVLBAISummary,abdalla_terrestrial_2025,TVLBAI3,TVLBAIESPP}.
The same transition is being pursued in the United States through MAGIS-100
at Fermilab~\cite{MAGIS-100:2021etm}. AICE would provide a complementary
European facility, building on the established cooperation between the AION
and MAGIS programmes and enabling future coordinated observation strategies
between geographically separated instruments in an international network.

Several possible European locations for long-baseline atom interferometers
have been investigated within this programme. These include the STFC Boulby
Underground Laboratory in the UK and the Laboratorio Subterr\'{a}neo de
Canfranc in
Spain~\cite{TVLBAISummary,abdalla_terrestrial_2025,TVLBAIESPP}.
Such sites may play important and complementary roles in the longer-term
distributed programme. For the immediate 100\,m step, however, the PX46
access shaft at LHC Point~4 is uniquely mature. It combines an existing
vertical baseline of the required scale and sufficient free horizontal
cross-section with completed site-specific feasibility and implementation
studies~\cite{Arduini:2851946, Arduini:2025jhe}, well-characterised environmental conditions, established host-laboratory
infrastructure and an identified near-term construction window.

The maturity of PX46 has been established through two dedicated CERN studies.
The conceptual feasibility study completed in 2023 confirmed that the shaft
could accommodate a 100\,m atom interferometer without fundamental technical
obstacles~\cite{Arduini:2851946}. The subsequent implementation study
completed in 2025 developed the corresponding technical concept, including
the required civil engineering, integration with the existing shaft
functions, radiation protection, access and safety, services, environmental
conditions and compatibility with HL-LHC operations~\cite{Arduini:2025jhe}.
These studies distinguish PX46 from other prospective sites. The central
question is no longer whether an instrument of this type could in principle
be accommodated, but how the developed implementation concept should be
realised. The site and its integration are described in detail in
Section~\ref{sec:site}.

CERN also provides a host environment whose capabilities complement the specialist expertise of the AICE collaboration.
Its extensive expertise in large-scale ultra-high-vacuum systems and its
experience with laser infrastructure, optical systems and beam delivery
address important technical requirements of the facility. These strengths are
supported by established capabilities in magnetic systems and field control,
mechanical engineering, radiation protection, controls, data acquisition and
large-scale technical integration, and by engineering and technical teams
with long experience in the installation of large experimental systems
underground. CERN further provides the organisational framework and
international scientific environment needed for a programme bringing together
particle physics, atomic physics, gravitational physics and precision measurements.

The case for proceeding now follows from the convergence of technological,
infrastructural and strategic timescales. The core experimental principles
are established. Single-photon atom interferometry on the strontium clock
transition and differential operation have been demonstrated by the AION
prototype~\cite{baynham_prototype_2026}, and the national programmes
described in Section~\ref{sec:wider} provide the technological basis for
integration at the 100\,m scale. Further development will be required to
reach the later Baseline and Stretch performance levels, but no new
experimental principle is required to proceed with the engineering,
construction and initial commissioning of the facility.

At the same time, LS3 provides the critical opportunity to carry out the
enabling works at the interface with the LHC infrastructure. Completing these
works during the shutdown would prepare the experimental area and establish
the required separation from the LHC infrastructure, allowing the detector to
be installed subsequently and commissioned without interfering with HL-LHC
operation, as described in Section~\ref{sec:civil}. Deferring these works
would leave the programme dependent on a later suitable accelerator shutdown
and would therefore risk a substantial delay. This opportunity
coincides with the strategic priority given to AICE by the TVLBAI community
as the key European step towards future km-scale
instruments~\cite{TVLBAIESPP}. Proceeding on this timescale would also
preserve opportunities for overlapping or sequential observation campaigns
with other long-baseline facilities, including MAGIS-100, and would thereby
maximise the scientific value of coordinated measurements.

The benefits would flow in both directions. CERN's infrastructure,
engineering expertise and operational experience would facilitate the AICE
physics programme and reduce its implementation risks. In return, AICE would
bring to CERN a substantial international community in cold-atom physics,
quantum sensing, optical frequency metrology and precision measurement. It
would establish a flagship quantum-sensor facility on the CERN site, create
new opportunities for training and technology development and provide a
concrete experimental realisation within the CERN Quantum Technology
Initiative~\cite{DiMeglio:2021QTI} and the Physics Beyond Colliders
programme~\cite{bib:PBC_mandate}.
The endorsement of AICE by the TVLBAI Proto-Collaboration, summarised in
Appendix~\ref{app:tvlbai-endorsement}, reflects the broad community consensus
that PX46 offers the most mature and timely opportunity for the European
100\,m programme. CERN is not simply a suitable host for AICE. The
combination of the PX46 site, the capabilities of the laboratory and the LS3
implementation window therefore makes CERN the unique European location where this next stage
of long-baseline atom interferometry can be realised on the appropriate
timescale.

\subsection{Structure of this document}
\label{sec:docmap}

Section~\ref{sec:concept} describes the experimental concept, including the measurement
principle, facility architecture, operating modes, and upgrade path.

Section~\ref{sec:physics} presents the full physics case for AICE: ultralight dark matter in
the scalar, vector $B{-}L$, and axion/ALP channels (Section~\ref{sec:uldm});
gravitational waves in the mid-frequency band (Section~\ref{sec:gw});
equivalence principle, fine-structure constant, and gravitational physics
(Sections~\ref{sec:ep} and~\ref{sec:alpha}); and a sensitivity summary
(Section~\ref{sec:sensitivity}).

Section~\ref{sec:detector} covers detector and technical systems.
Section~\ref{sec:site} describes the PX46 site and CERN integration.
Section~\ref{sec:civil} covers infrastructure preparation and civil engineering during LS3.
Section~\ref{sec:noise} presents the noise and systematics assessment.
Section~\ref{sec:ops} describes commissioning and operations.
Section~\ref{sec:org} covers organisation, costs, schedule, and risks. Section~\ref{sec:conclusions} summarises our conclusions.

Appendix~\ref{app:tvlbai-endorsement} provides a statement of TVLBAI support and endorsement.

Appendix~\ref{app:Acronyms} provides a list of the acronyms used throughout this document.
\section{Experimental Concept}
\label{sec:concept}

AICE is conceived as a configurable long-baseline atom-interferometry
facility rather than a single-purpose instrument. A single
$140\,\mathrm{m}$ vertical infrastructure realises the scalar, vector and
axion-like ultralight dark matter searches, the embedded
precision-physics campaigns and the gravitational-wave pathfinder
programme outlined in Section~\ref{sec:physics}, by changing the atomic species,
the source configuration and the interferometer pulse sequence rather
than the facility itself. This Section presents the concept underlying
that versatility at the conceptual level appropriate to this proposal.
The measurement principle is described in
Section~\ref{sec:concept:principle}, the facility architecture and the
sense in which AICE is a configurable platform in
Section~\ref{sec:concept:facility}, and the individual
operating modes in Section~\ref{sec:concept:modes}. The technical
realisation of each subsystem is deferred to Section~\ref{sec:detector}.
%This section defines the measurement concepts and the
%operating modes, while the physics reach of each mode, including the
%projected sensitivities and the treatment of backgrounds, is given in the
%physics case of Section~\ref{sec:physics}.

% ---------------------------------------------------------------------
\subsection{Measurement principle}
\label{sec:concept:principle}

\subsubsection{Atom interferometry}
\label{sec:concept:ai}

It is helpful to introduce the measurement principle as a 
progression from a single interferometer to a differential gradiometer,
then to a multi-source facility and finally to a set of configurable
operating modes. The sensing elements of AICE are freely-falling ensembles
of ultracold strontium (or ytterbium) atoms, which combine two properties that make them
powerful quantum sensors. Their free-fall motion provides an inertial
reference that is largely decoupled from the environment, while the
coherence of their internal clock transition provides a precise phase
reference analogous to a clock. A single atomic ensemble therefore
provides both an inertial reference and a clock-like phase reference, and
the measurement consists of reading out the phase difference accumulated between two
atomic paths.

\begin{figure}[t]
  \centering
  \includegraphics[width=0.45\textwidth]{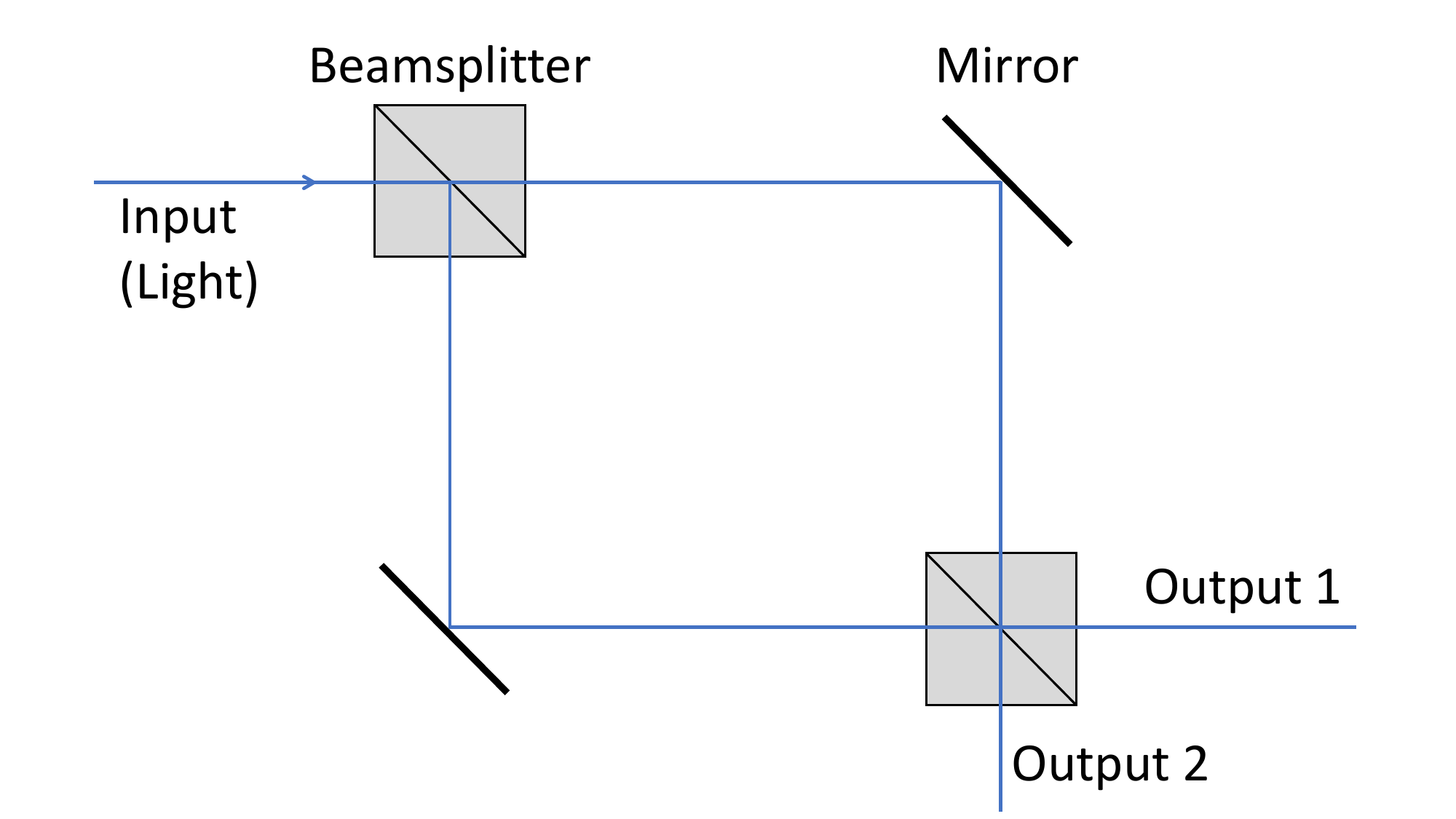}\hfill
  \includegraphics[width=0.45\textwidth]{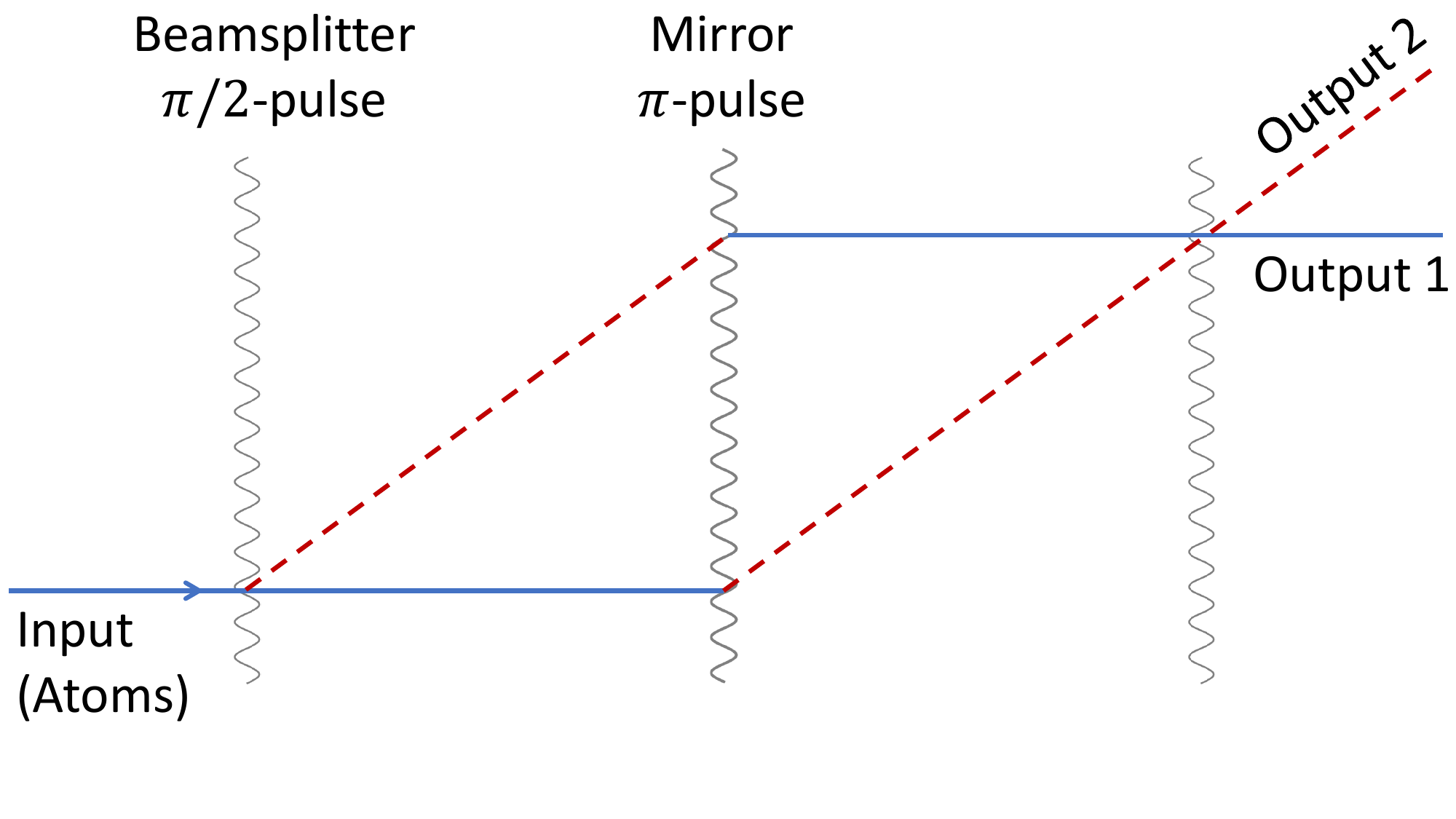}
  \caption{\textit{Left:} a Mach-Zehnder optical interferometer.
  \textit{Right:} the analogous atom interferometer, with the ground
  state $\ket{g}$ represented by solid blue lines, the excited state $\ket{e}$
  by dashed red lines, and laser pulses by wavy lines. The
  $\pi/2$, $\pi$, $\pi/2$ pulse sequence acts as beamsplitter, mirror and
  beamsplitter respectively; the unlabelled rightmost pulse in the
  atom-interferometer schematic is the final $\pi/2$ beamsplitter pulse.
  Adapted from~\cite{Buchmueller2023}.}
  \label{fig:mz}
\end{figure}

Atom interferometers are most easily understood by
analogy with optical interferometers. An atom interferometer replaces the
light waves with the matter waves of cold atoms, and the optical
beamsplitters and mirrors with resonant atom-light
interactions~\cite{Cronin2009}. The basic ingredients are visualised in
the Mach-Zehnder configuration of Fig.~\ref{fig:mz}. A first beamsplitter
divides the matter waves into two components that travel along different paths and
accrue different phases, a mirror brings them back together, and a final
beamsplitter recombines them so that they interfere. The populations at
the two output ports depend on the differential phase
$\Delta\varphi = \phi_1 - \phi_2$ between the two arms.

In the atom-optics implementation the splitters and mirror are provided by pulses that are resonant with a
transition between an atomic ground state and a long-lived excited state,
$\ket{g}\leftrightarrow\ket{e}$. Each pulse both changes the internal
state and imparts a photon momentum kick $\pm\hbar k_0$ with
$k_0 = 2\pi/\lambda$, where $\lambda$ is the laser wavelength, so that the
two states separate in space. A $\pi/2$
pulse acts as a beamsplitter, a $\pi$ pulse after a time $T$ acts as a
mirror, and a final $\pi/2$ pulse at time $2T$ interferes the paths before the
populations are read out. The two components separate spatially and
sample different, position-dependent phases. AICE realises this concept
using single-photon optical clock transitions.

\subsubsection{Single-photon clock-transition interferometry}
\label{sec:concept:singlephoton}

\begin{figure}[t]
  \centering
  \includegraphics[width=0.5\textwidth]{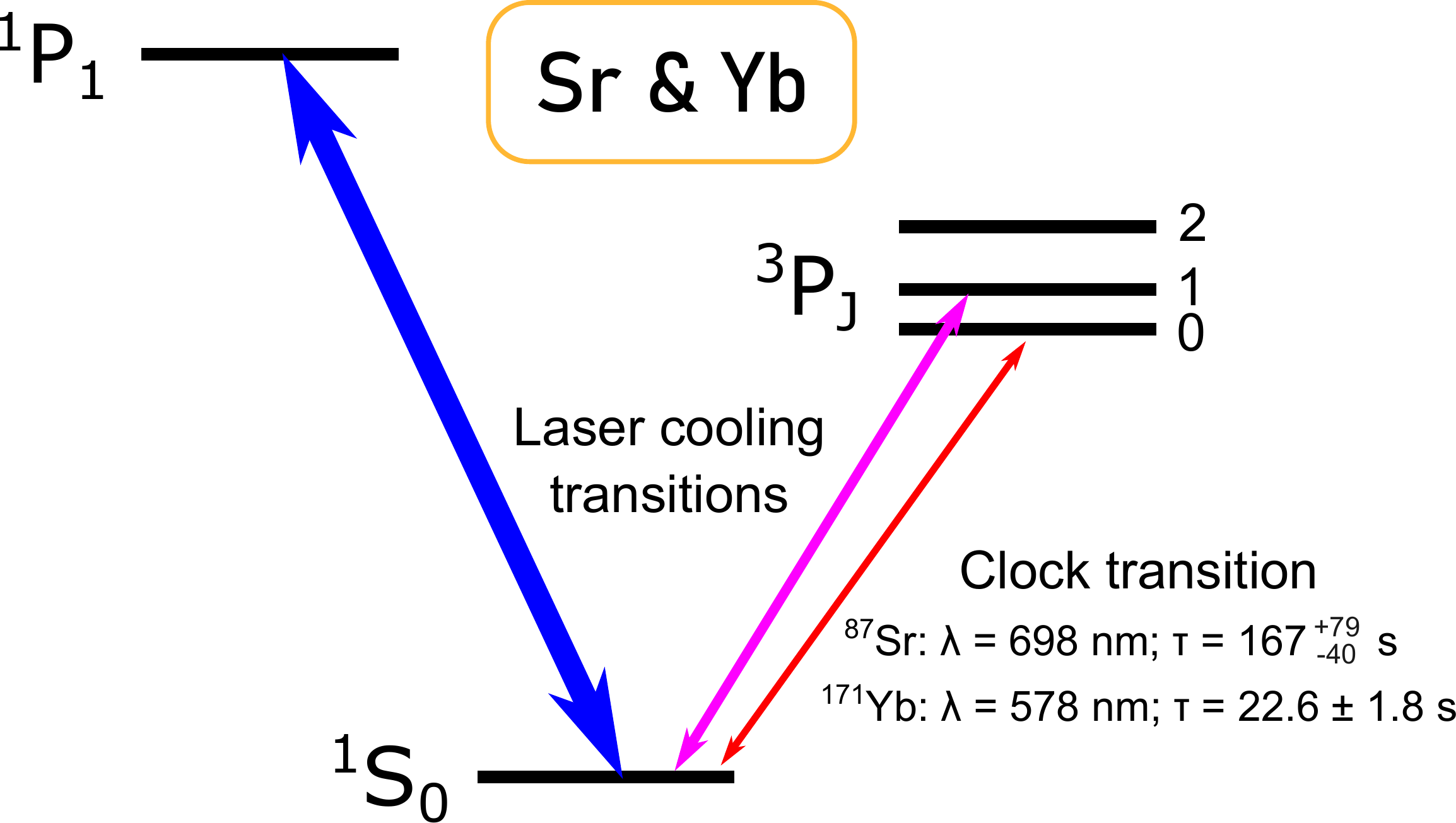}
  \caption{The transitions
  used for laser cooling and single-photon clock-transition interferometry in AICE. The long lifetimes of the
  $^3P_0$ states in Sr~\cite{dolde_direct_2025} and Yb~\cite{xu_measurement_2014} suppress spontaneous-emission decoherence during the
  interferometer sequence. Indirect measurements of the Sr lifetime are in agreement with the direct measurement in Ref.~\cite{dolde_direct_2025} quoted here, but are in conflict with each other~\cite{muniz_cavity-qed_2021,lu_determining_2024}.
  }
  \label{fig:clock}
\end{figure}

For single-photon clock-transition large-momentum-transfer (LMT) atom
optics, the coherent momentum transfer, and hence the sensitivity, is
constrained in part by spontaneous emission from the excited state. AICE
therefore uses the ultra-narrow
$^1S_0\!\leftrightarrow\,^3P_0$ optical clock transitions of strontium at
$698\,\mathrm{nm}$ and ytterbium at $578\,\mathrm{nm}$. As shown in Fig.~\ref{fig:clock}, the metastable $^3P_0$ states in $^{87}$Sr and $^{171}$Yb have lifetimes in excess of $100\,\mathrm{s}$ and $20\,\,\mathrm{s}$ respectively~\cite{dolde_direct_2025,xu_measurement_2014}, so that spontaneous emission is
strongly suppressed during the interrogation. These are the same
transitions that underpin state-of-the-art optical
clocks~\cite{Ludlow2015}, and their use is what makes long interrogation
times and large-momentum-transfer (LMT) sequences compatible at the $140\,\mathrm{m}$
scale.

% \begin{figure}[t]
%   \centering
%   \includegraphics[width=0.5\textwidth]{figures/Level_diagram.pdf}
%   \caption{The $^{87}$Sr $^1S_0\!\leftrightarrow\,^3P_0$ clock transition
%   used for single-photon interferometry in AICE. The long lifetime of the
%   $^3P_0$ state suppresses spontaneous-emission decoherence during the
%   interferometer sequence. Adapted from~\cite{Buchmueller2023}.}
%   \label{fig:clock}
% \end{figure}

A single-photon clock transition imprints the laser phase from a unidirectional laser beam directly onto
the atomic superposition at each pulse, in contrast to the two-photon
Raman or Bragg schemes which inherit the combined phase of two counterpropagating laser beams~\cite{Bott2023,Graham:2012sy,MAGIS-100:2021etm,Bongs:2025rqe}. For
LMT atom optics of order $n$, the effective
momentum transfer is
\begin{equation}
  k_{\mathrm{eff}} = n\,k_0\,, \; {\rm where} \; \quad k_0 = 2\pi/\lambda\,,
  \label{eq:keff}
\end{equation}
where $n$ counts the number of photon recoils; a two-photon interaction
transfers two recoils. We use the definition in Eq.~\eqref{eq:keff}
throughout this proposal.

\subsubsection{Differential operation, large momentum transfer and signal
scaling}
\label{sec:concept:differential}

A single interferometer is intrinsically sensitive to the phase of the
interrogation laser, which competes with the time-dependent signals
targeted by AICE. The standard mitigation is a \emph{gradiometer}, in
which two or more vertically separated interferometers are interrogated by
a common laser pulse train. Laser phase noise then becomes a common-mode effect
that cancels in the differential signal, while a genuine differential
signal survives. In this sense the atomic ensembles provide the phase
reference internally, which allows common laser noise to be rejected along
a single vertical baseline rather than by comparison with a second optical
arm~\cite{Graham:2012sy,MAGIS-100:2021etm}. The differential sensitivity
grows linearly with the separation $\Delta r$ between the interferometers,
which is the main reason why a long vertical baseline is advantageous.

LMT atom optics provide a complementary gain. Driving the atoms through
$n$ successive momentum kicks exposes the interferometer to the laser
phase $n$ times, increasing the accumulated phase, and hence the
sensitivity, by a factor $n$, in analogy with the multiple round trips of
an optical cavity. The same picture of repeatedly sampling the laser phase
underlies the resonant interrogation mode introduced in
Section~\ref{sec:concept:modes}, where the sampling is tuned to a chosen
signal frequency. These two effects set how the AICE signals scale. The
scalar, vector and equivalence-principle channels scale with $n$, whereas the axion channel acts on the transition frequency and its reach is independent of $n$,
being set instead by the interrogation time and the nuclear coupling. In
all cases the accumulated phase is proportional to $\Delta r$, so the same
long-baseline, common-interrogation architecture serves every channel.

Importantly, the laser phase noise cancellation is only complete for \textit{single-photon} interferometry, in which momentum is imparted to the atoms using a sequence of $n$ consecutive pulses from alternating directions, producing $n$ photon recoils. By contrast, the laser phase noise is not fully cancelled for two-photon schemes such as Bragg diffraction or Bloch oscillation (see Table~\ref{tab:LMT_history}), in which momentum is imparted two (or more) recoils at a time by simultaneous upward- and downward-propagating pulses. The residual laser phase noise affecting two-photon interferometry increases with the amount of LMT and with the interferometer separation $\Delta r$, presenting a significant challenge for the detection of ultralight dark matter or gravitational waves. For this reason, the AICE proposal focuses primarily on single-photon interferometry.

\subsubsection{Signal transduction and observable extraction}
\label{sec:concept:transduction}

The final step of the measurement chain is the conversion of a physical
signal into an interferometer phase. In this sense AICE acts as a
transducer, turning the time-dependent fields of interest into the measured observable of phase. An oscillating ultralight scalar
field interacting with atomic constituents acts as a coherent classical background oscillating at its Compton frequency that
modulates fundamental constants, and therefore the atomic clock-transition
frequency $\delta\nu$~\cite{Derr2023}. Similarly, an axion field induces an oscillatory
shift of the clock-transition frequency $\delta\nu$, which accumulates as a phase over the
interrogation time. A
vector field with $B\!-\!L$ couplings exerts a composition-dependent force
that differs between atomic species, producing a differential acceleration $\delta a$
between two co-located species. The same differential-acceleration
observable tests the universality of free fall, and hence the weak
equivalence principle, by comparing the gravitational accelerations of
the two species. A gravitational wave modifies the propagation time of
the laser pulses between two separated interferometers such that a different atom propagation phase $\delta\phi_p$ accumulates between the interferometers~\cite{Schaffrath2025}. These effects ultimately reduce to a
differential phase $\delta\phi$, through the
measurement chain
\begin{equation}
  \text{field} \;\longrightarrow\;
  \{\,\delta\phi_p,\ \delta a,\ \delta\nu\,\} \;\longrightarrow\;
  \text{interferometer phase}\,.
  \label{eq:transduction}
\end{equation}
The correspondences between the physical signals, the detector configurations, and the measurable effects are summarised in Table~\ref{tab:transduction}.

\begin{table}[h]
  \centering
  \caption{Signal transduction in AICE. The detector configurations are described in Section~\ref{sec:concept:modes}.}
  \label{tab:transduction}
  \begin{tabular}{lll}
    \hline
    Physical signal & Configuration & Induced effect \\
    \hline
    Scalar ULDM & Differential gradiometer & Clock-frequency modulation \\
    Vector $B\!-\!L$ ULDM & Dual-species full-baseline & Species-dependent force \\
    WEP / UFF & Dual-species full-baseline & Species-dependent acceleration \\
    Axion & Single-species full-baseline & Clock-frequency modulation \\
    Gravitational wave & Differential gradiometer & Laser propagation delay \\
    Fine structure constant & Single-species Ramsey-Bord\'{e} & Recoil velocity \\
    \hline
  \end{tabular}
\end{table}

The physics effects identified above are realised through
different operating configurations of the same facility. These
configurations are introduced below in Section~\ref{sec:concept:modes} and form the basis of the measurement
modes used throughout the proposal.

% ---------------------------------------------------------------------
\subsection{Facility architecture}
\label{sec:concept:facility}

The distinguishing feature of AICE is that a single infrastructure
supports a broad and largely complementary measurement programme. It is
useful to think of the facility in three layers. The \emph{fixed
infrastructure} comprises the PX46 shaft, the vertical vacuum tower, the
support structure and the site services, established once and unchanged
throughout the programme. The \emph{shared core systems} comprise the
$698\,\mathrm{nm}$ strontium clock laser, the timing and control
architecture, the data acquisition and the atom-source platform. These
systems are common to all science phases and define the core technical
capabilities of the facility, while their technical implementation may
evolve over time as operational experience is gained and technology
advances. The \emph{configurable systems} comprise the
choice of species, the number and arrangement of atom sources, the
addition of the $578\,\mathrm{nm}$ ytterbium laser, and the pulse
sequence, which evolve as the programme advances. This layered design
allows one facility to address several physics goals through incremental evolution. The $140\,\mathrm{m}$ baseline represents a
compromise between sensitivity, interrogation time, infrastructure
complexity and compatibility with the available CERN shaft geometry.

\begin{figure}[t]
  \centering
  \includegraphics[width=0.92\textwidth]{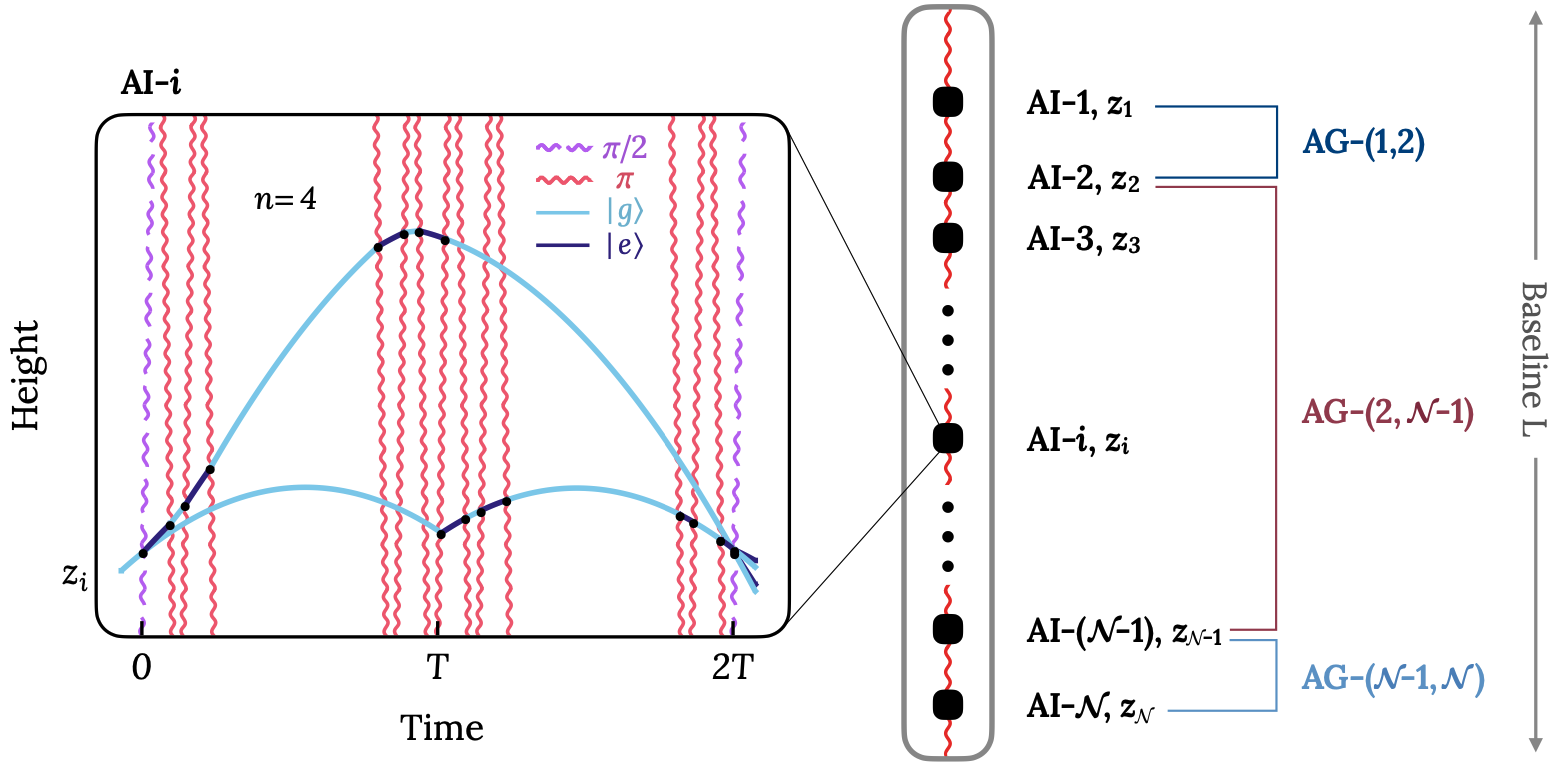}
  \caption{Conceptual illustration of the multi-source facility
  architecture to be adopted by AICE. \textit{Left:} the space-time
  diagram of one interferometer with $n=4$ LMT momentum kicks.
  \textit{Right:} several interferometers spaced within a single vertical
  vacuum tower of length $L$, forming multiple gradiometer combinations
  over different baselines. Multiple source stations permit simultaneous
  measurements over different baselines and provide a scalable
  architecture for future large-scale facilities. In AICE the number of
  sources grows from $N_{\mathrm{source}}\sim3$ at commissioning towards
  $\sim10$ to $20$ at later stages (cf.\ Table~\ref{fig:timeline}). Adapted
  from~\cite{Badurina:2022ngn}.}
  \label{fig:facility}
\end{figure}

A central element of this architecture is that AICE can operate with a
growing number of atom sources, illustrated conceptually in
Fig.~\ref{fig:facility}. Increasing the number of sources is not merely a
way to accumulate statistics. It allows several interferometers to be
operated simultaneously over different baselines within the same tower,
which permits cross-checks between independent measurements, and offers a route to the
characterisation and potential suppression of environmental backgrounds
such as gravity gradient noise, as discussed in
Section~\ref{sec:noise:GGN:mitigation}, and allows a flexible allocation of sources to
different science programmes as the priorities of the facility evolve. The
same architecture extends naturally towards the future distributed and
kilometre-scale atom-interferometer facilities of the TVLBAI roadmap.

The same facility measures different physical effects depending
on how it is configured. A single species in a gradiometer measures a
differential phase sensitive to modulations of fundamental constants, two
co-located species over the full baseline measure a differential
acceleration, and a single species in a full-baseline Mach-Zehnder
measures a modulation of the clock-transition frequency, corresponding to
the three effects listed in Table~\ref{tab:transduction}. Moving between these
measurements requires only a change of species, source arrangement or pulse
sequence, with no modification of the shaft, the vacuum system or the
primary laser. The operating modes that realise them are described in
Section~\ref{sec:concept:modes}.

% ---------------------------------------------------------------------
\subsection{Measurement configurations and operating modes}
\label{sec:concept:modes}

As discussed above, AICE operates in three geometrical configurations, each sensitive to different fields introduced above, and in each case the pulse sequence
may be chosen to be broadband or resonant. This Section defines the
configurations and the interrogation modes at the conceptual level and
serves as the reference for the operating-mode terminology used throughout
the proposal. For each configuration the schematic is shown on the left in the following figures,
and the corresponding space-time diagram on the right. The defining
parameters follow the performance ladder of Table~\ref{tab:roadmap}.

\subsubsection{Differential gradiometer mode}
\label{sec:concept:mode:grad}

\begin{figure}[t]
  \centering
  \begin{subfigure}[t]{0.27\textwidth}\centering
    \includegraphics[width=\linewidth,trim=10pt 35pt 10pt 35pt,clip]{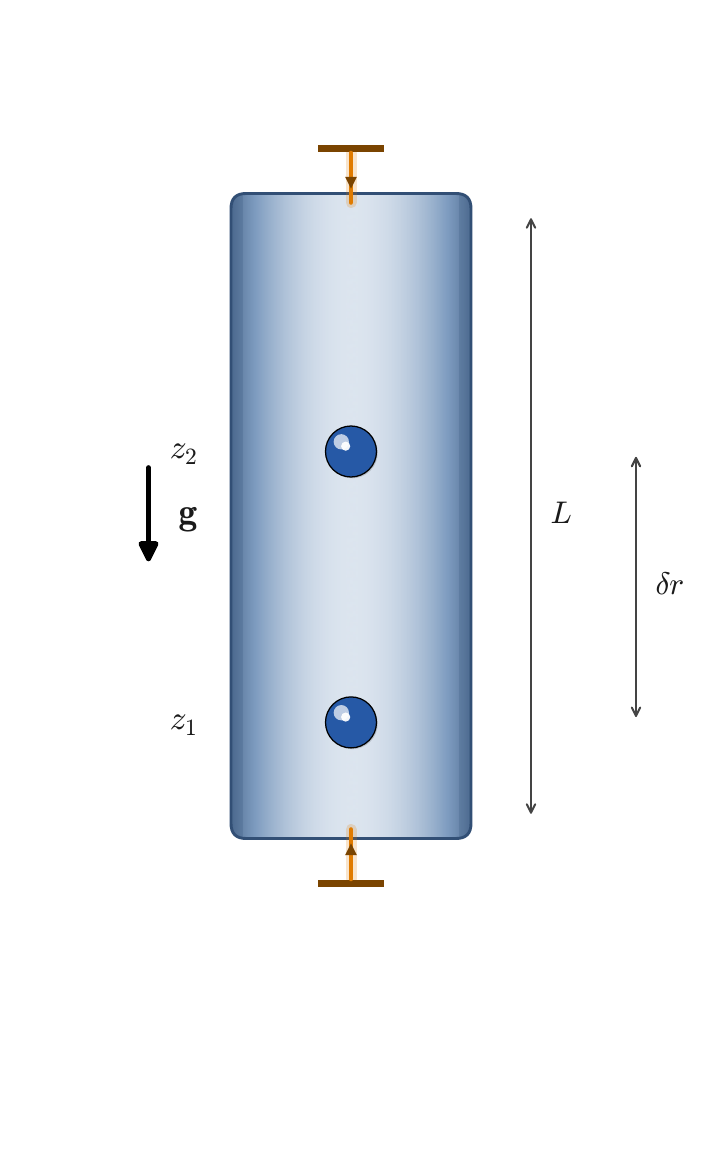}
  \end{subfigure}\hspace{0.015\textwidth}
  \begin{subfigure}[t]{0.69\textwidth}\centering
    \includegraphics[width=\linewidth,trim=5pt 8pt 5pt 8pt,clip]{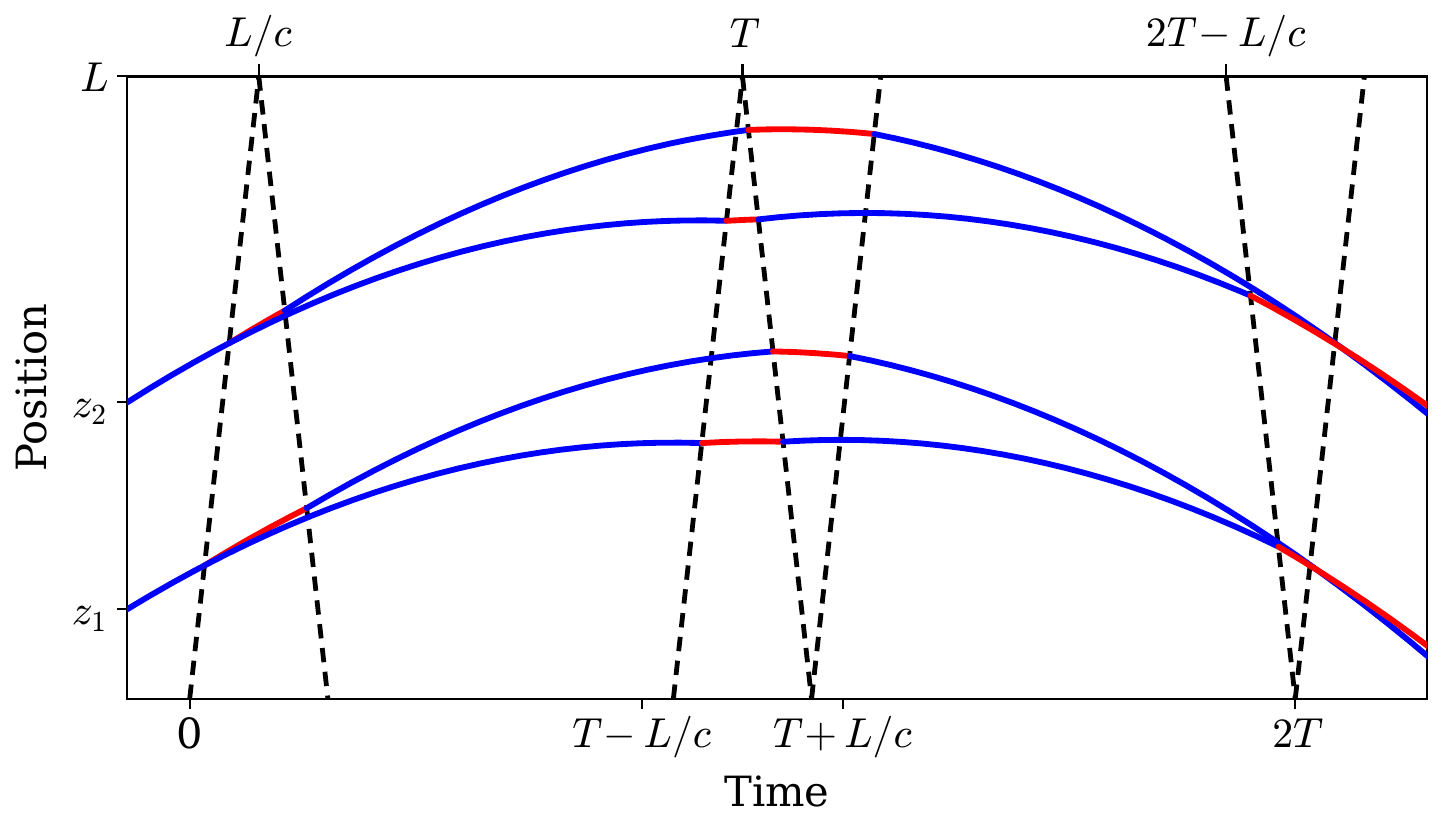}
  \end{subfigure}
  \vspace{-0.5em}
  \caption{Differential gradiometer configuration. \textit{Left:} two
  vertically separated $^{87}$Sr interferometers in the $L=140\,\mathrm{m}$
  AICE shaft, separated by $\delta r=L/2$. \textit{Right:} the
  $\pi/2$, $\pi$, $\pi/2$ Mach-Zehnder pulse sequence interrogates both
  interferometers with the same pulse train; dashed lines denote the laser
  world-lines. The baseline gradiometer uses an interrogation time
  $T=2.5\,\mathrm{s}$ and provides common-mode rejection of laser phase
  noise.}
  \label{fig:mode_grad}
\end{figure}

Two vertically separated $^{87}$Sr interferometers are interrogated by a
common clock-laser pulse train, as shown in Fig.~\ref{fig:mode_grad}.
Common laser phase noise cancels in the differential signal, as demonstrated experimentally in~\cite{AION:2025igp}, while a
modulation of the clock frequency by an ultralight dark matter field,
sampled at the two heights at times separated by the light propagation
delay~\cite{DiPumpo2022}, produces a differential phase. The measured observable is the
differential phase
\begin{equation}
  \Delta\phi = \phi(z_2) - \phi(z_1)\,,
  \label{eq:grad_obs}
\end{equation}
whose response scales as
\begin{equation}
  \Delta\phi \propto k_{\mathrm{eff}}\,\Delta r\,,
  \qquad k_{\mathrm{eff}} = n\,k_0\,,
  \label{eq:grad_resp}
\end{equation}
where $\Delta r$ is the separation of the two interferometers and the
proportionality factor depends on the signal model and the pulse sequence.
The response therefore increases linearly with both the interferometer
separation and the LMT order, which is the basic reason that AICE combines
a long baseline with large momentum transfer. This is the baseline
differential gradiometer configuration.

\subsubsection{Dual-species full-baseline mode}
\label{sec:concept:mode:dual}

\begin{figure}[t]
  \centering
  \begin{subfigure}[t]{0.27\textwidth}\centering
    \includegraphics[width=\linewidth,trim=10pt 35pt 10pt 35pt,clip]{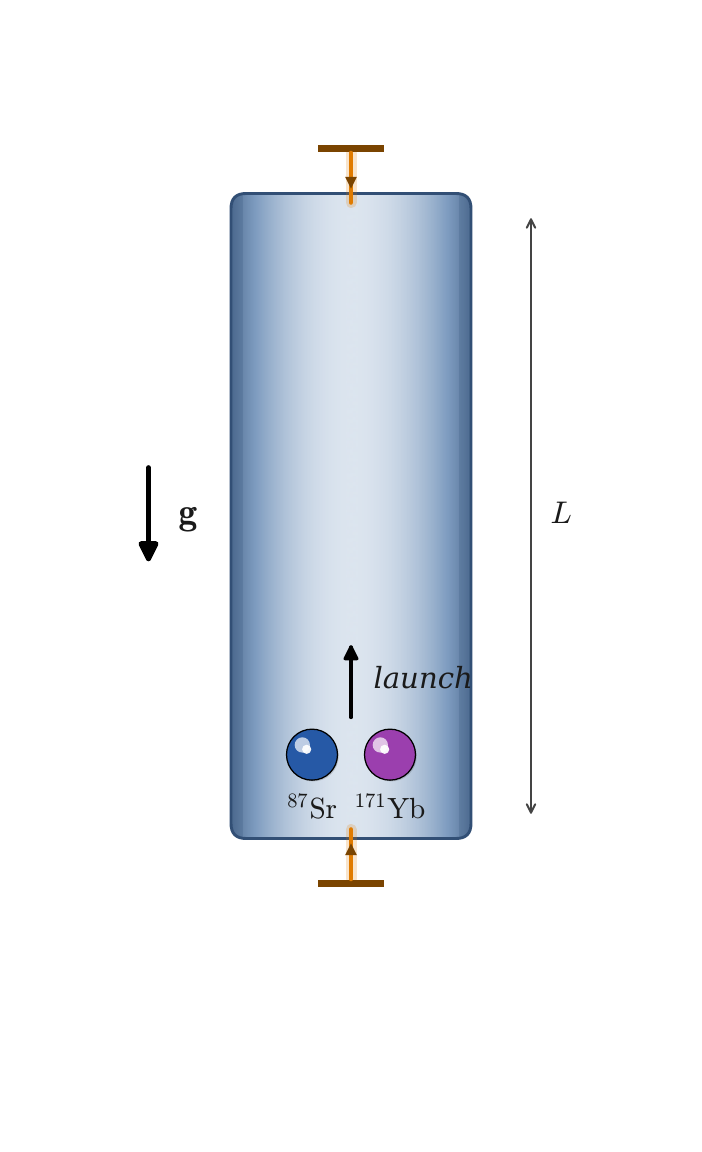}
  \end{subfigure}\hspace{0.015\textwidth}
  \begin{subfigure}[t]{0.69\textwidth}\centering
    \includegraphics[width=\linewidth,trim=5pt 8pt 5pt 8pt,clip]{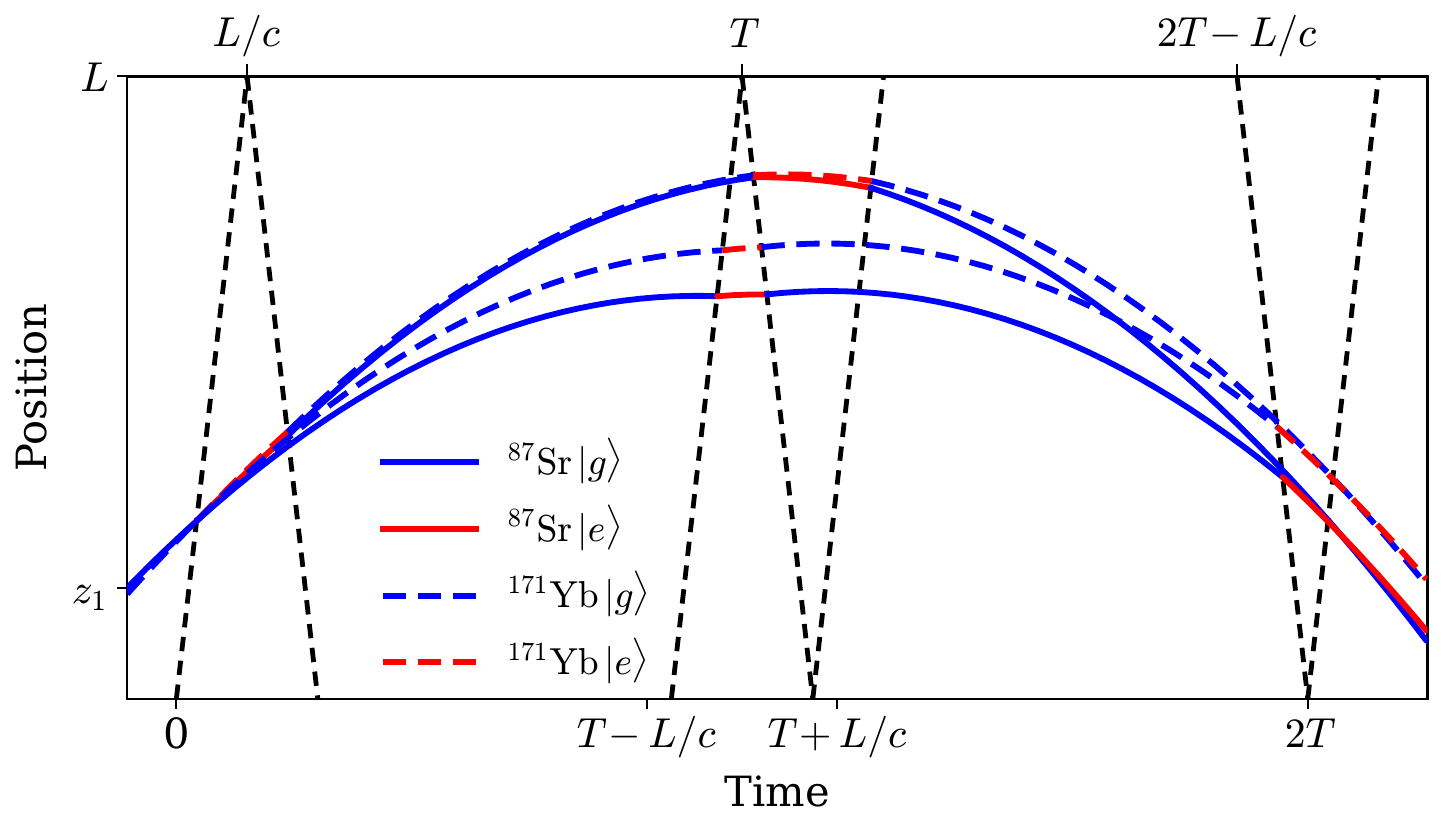}
  \end{subfigure}
  \vspace{-0.5em}
  \caption{Dual-species full-baseline configuration. \textit{Left:}
  co-launched $^{87}$Sr and $^{171}$Yb atom interferometers interrogated
  over the full $L=140\,\mathrm{m}$ AICE shaft. \textit{Right:} the two
  species share the same launch geometry and pulse sequence, while their
  different clock transitions provide sensitivity to species-dependent
  responses. The baseline full-baseline Mach-Zehnder interrogation time is
  $T_{\mathrm{MZ}}=5.0\,\mathrm{s}$.}
  \label{fig:mode_dual}
\end{figure}

Two co-located species are launched from the bottom of the shaft and
interrogated over the full one-way baseline, as shown in
Fig.~\ref{fig:mode_dual}. Because the species carry different
($B-L$)-charge-to-mass ratios, the measured effect is the differential
acceleration between them,
\begin{equation}
  \Delta a = a_A - a_B\,,
  \label{eq:dual_obs}
\end{equation}
with uniform gravitational acceleration and common platform motion cancelling
to leading order. Residual differential coupling can arise from scale-factor
mismatch, imperfect co-location of the two species and spatially varying
gravity gradients, as discussed in Section~\ref{sec:noise:GGN:impact}. The resulting
differential phase scales as
\begin{equation}
  \Delta\phi \propto k_{\mathrm{eff}}\,\Delta a\,T^2\,,
  \label{eq:dual_resp}
\end{equation}
which shows why the full-baseline configuration with the long
interrogation time $T = 5.0\,\mathrm{s}$ is used for this mode. The
$^{87}$Sr and $^{88}$Sr pairing serves as a same-element isotopic
validation configuration. In $^{88}$Sr the clock transition is not
directly accessible with the $698\,\mathrm{nm}$ laser alone, so this
configuration requires a co-propagating multiphoton scheme using
additional laser wavelengths~\cite{carman_collinear_2025}. The
$^{87}$Sr and $^{171}$Yb pairing is the main upgraded configuration, using an additional
$578\,\mathrm{nm}$ laser and yielding a larger compositional contrast. This is
the baseline dual-species differential-acceleration configuration.

\subsubsection{Full-baseline single-species clock mode}
\label{sec:concept:mode:single}

\begin{figure}[t]
  \centering
  \begin{subfigure}[t]{0.27\textwidth}\centering
    \includegraphics[width=\linewidth,trim=10pt 35pt 10pt 35pt,clip]{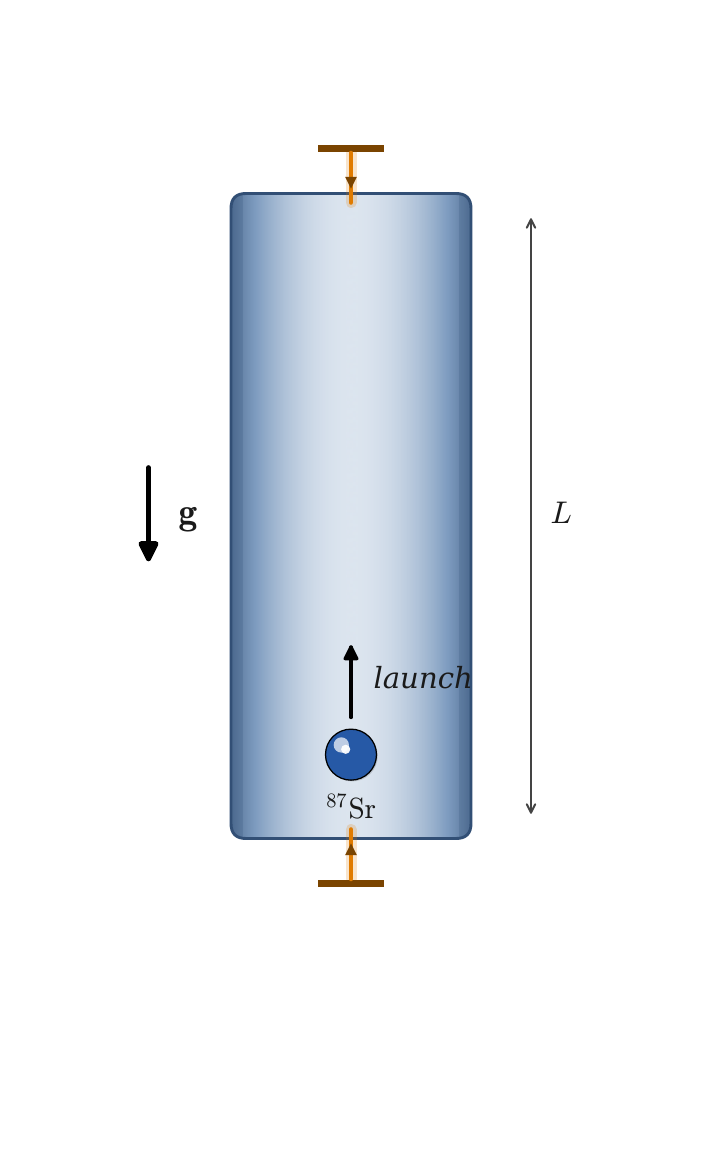}
  \end{subfigure}\hspace{0.015\textwidth}
  \begin{subfigure}[t]{0.69\textwidth}\centering
    \includegraphics[width=\linewidth,trim=5pt 8pt 5pt 8pt,clip]{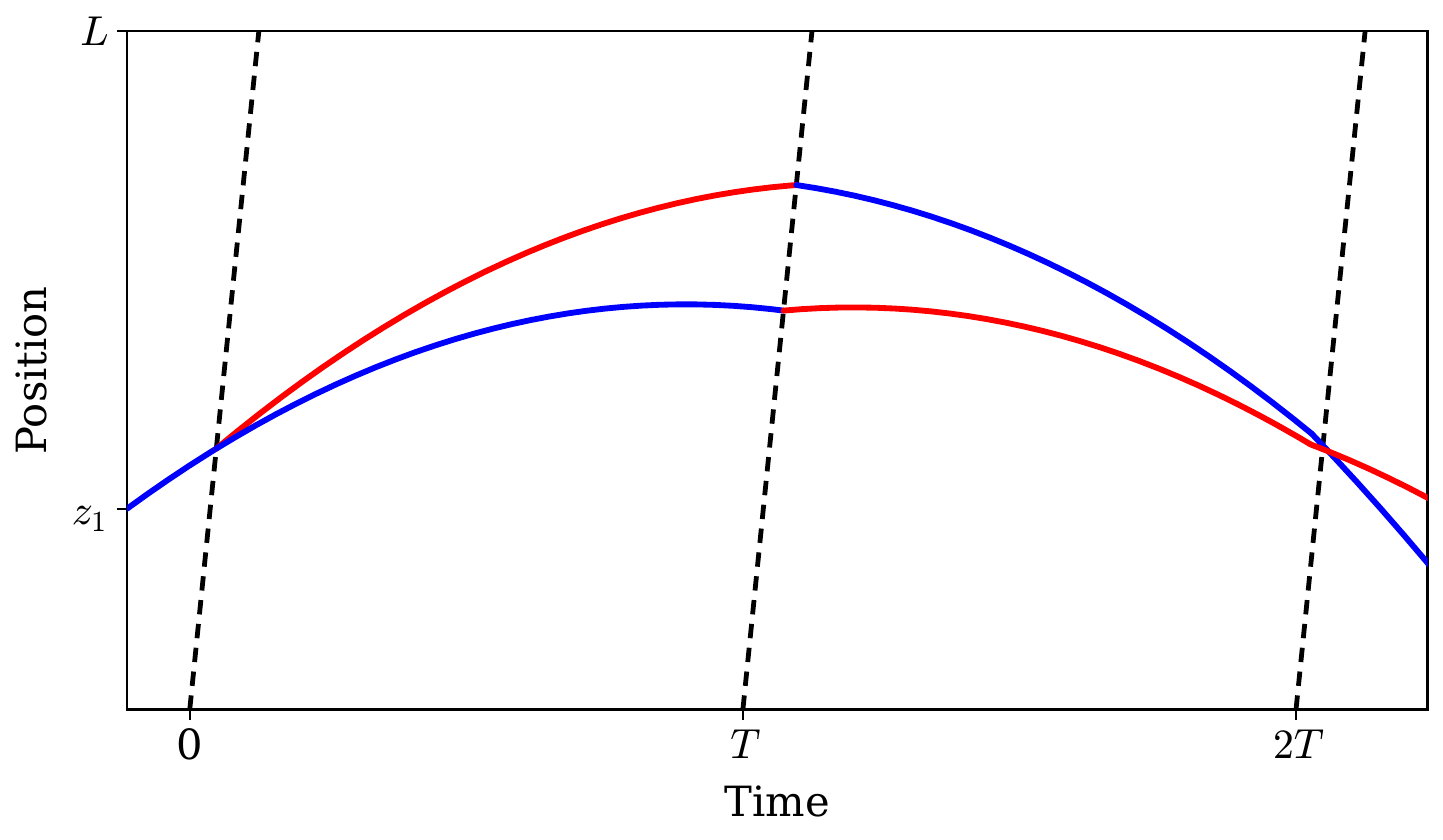}
  \end{subfigure}
  \vspace{-0.5em}
  \caption{Full-baseline single-species clock-interferometer
  configuration. \textit{Left:} a single $^{87}$Sr atom interferometer
  launched from the shaft bottom and interrogated over the full
  $L=140\,\mathrm{m}$ AICE baseline. \textit{Right:} the corresponding
  Mach-Zehnder spacetime diagram. The baseline single-species
  full-baseline interrogation time is $T_{\mathrm{ax}}=5.0\,\mathrm{s}$.}
  \label{fig:mode_single}
\end{figure}

A single $^{87}$Sr cloud is launched from the bottom of the shaft and
interrogated in a standard Mach-Zehnder sequence over the full baseline,
with no gradiometer pair, as shown in Fig.~\ref{fig:mode_single}. The
configuration is sensitive to effects which modulate the clock frequency, causing an accumulated phase
\begin{equation}
  \Delta\phi_{\mathrm{ax}} = \int \delta\omega_{\mathrm{clock}}(t)\,dt\,,
  \label{eq:ax_resp}
\end{equation}
where $\delta\omega_{\mathrm{clock}}(t)$ is the oscillating shift of the
clock transition induced by an axion field. Unlike the gradiometer and
dual-species modes, the response is driven primarily by the
clock-frequency modulation and the interrogation time rather than by the
recoil momentum transfer, so the reach is independent of the LMT
order. To extract an axion signal free from laser phase noise, two simultaneous Mach-Zehnder sequences are implemented using the same source cloud, with equal but opposite stretched nuclear-spin states $M_F$~\cite{Graham:2017pmn}; the difference in phase between these interferometers is sensitive to the axion field, but the laser phase cancels in common mode. This is the baseline
single-species clock-interferometer configuration.

\subsubsection{Pulse-sequence options: broadband, resonant and
Ramsey-Bord\'e modes}
\label{sec:concept:mode:readout}

The three configurations above define the geometry of the measurement.
The choice of pulse sequence sets the response function, and this is an
independent option that applies to each geometry. The same instrument can
therefore be operated in different ways without any hardware change,
simply by altering the sequence of laser pulses.

\paragraph{Broadband interrogation.}
The default mode uses the standard $\pi/2$, $\pi$, $\pi/2$ Mach-Zehnder
sequence outlined above, which maximises the frequency coverage for a given configuration.
Schematically, its phase response to a signal of angular frequency
$\omega$ exhibits the characteristic envelope
\begin{equation}
  R_{\mathrm{bb}}(\omega) \propto \sin^2\!\left(\frac{\omega T}{2}\right)\,,
  \label{eq:broadband}
\end{equation}
which is suppressed at low frequency, with a first maximum near
\begin{equation}
  \omega T \simeq \pi\,, \qquad f \simeq \frac{1}{2T}\,,
  \label{eq:broadband_peak}
\end{equation}
and a declining, oscillatory response at higher frequency, where the
signal completes many cycles during $T$ and its effect largely averages
out. The
interrogation time $T$ therefore sets the characteristic frequency scale
of the response. This broad response suits an initial survey of an unknown
signal. This frequency selectivity can also be used to avoid particularly
noisy portions of the band, although it does not itself subtract gravity
gradient noise, see Section~\ref{sec:noise:GGN:mitigation}.

\begin{figure}[t]
  \centering
  \includegraphics[width=0.475\textwidth]{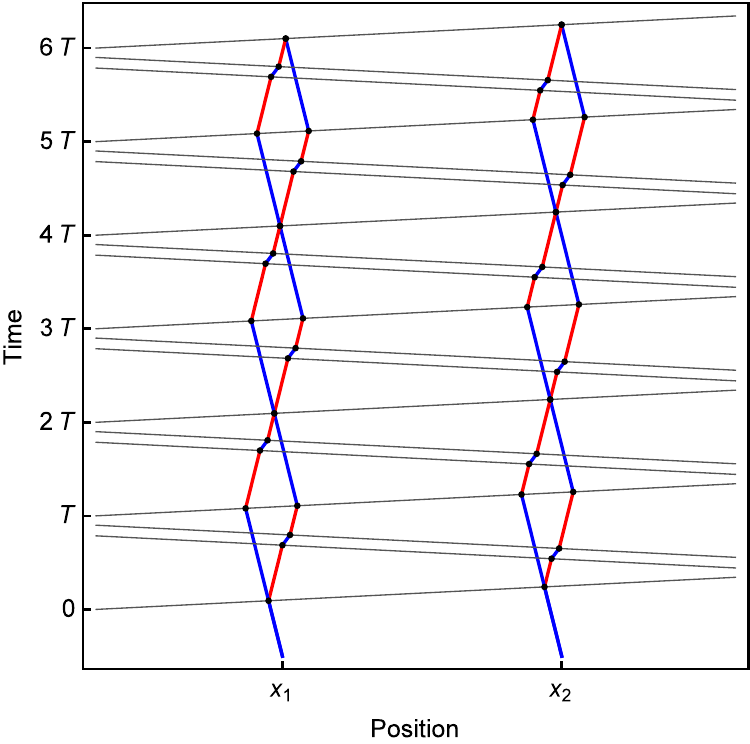}
  \caption{Space-time diagram of a resonant interrogation sequence for a
  gradiometer of two interferometers at positions $x_1$ and $x_2$. A train
  of intermediate $\pi$ pulses replaces the single $\pi$ pulse of the
  Mach-Zehnder sequence and creates $Q$ closed loops, or ``diamonds", here
  shown for $Q=3$. The successive loops add the signal phase coherently,
  enhancing the response in a narrow band around the resonant frequency of
  Eq.~\eqref{eq:resonant}. Ground-state trajectories are shown in blue and
  excited-state trajectories in red. Adapted from~\cite{Graham:2016plp}.}
  \label{fig:resonant}
\end{figure}

\paragraph{Resonant interrogation.}
Resonant interrogation uses the same geometry but chooses the pulse timing
so that the interferometer response is concentrated around a selected
target frequency. The single $\pi$ pulse of the Mach-Zehnder sequence is
replaced by a train of intermediate $\pi$ pulses, $\pi/2$, $\pi$, $\dots$,
$\pi$, $\pi/2$, which creates $Q$ closed ``diamonds", in the
space-time diagram of
Fig.~\ref{fig:resonant}~\cite{Graham:2016plp,DiPumpo:2023hfk}. In broadband
operation the signal contributions acquired during successive portions of
the interferometer sequence eventually cancel once the signal oscillates
rapidly compared with the interrogation time, because the interferometer
arms remain fixed while the signal completes many cycles. Resonant
operation prevents this cancellation by repeatedly exchanging the two
interferometer arms with the intermediate $\pi$ pulses. The pulse timing is
chosen so that the signal changes sign between successive half-loops, so
that the phase contributions add coherently rather than averaging away.
In conceptual terms, broadband interrogation measures many frequencies
simultaneously with moderate sensitivity, whereas resonant interrogation
sacrifices bandwidth in order to concentrate sensitivity into a narrow
frequency interval. The response is then peaked at a resonant frequency
and confined to a narrow band,
\begin{equation}
  \omega_r \simeq \frac{\pi}{\tau}\,, \qquad f_r \simeq \frac{1}{2\tau}\,,
  \qquad \Delta f \sim \frac{f_r}{Q}\,,
  \label{eq:resonant}
\end{equation}
where $\tau$ is the spacing between successive loops, to be distinguished
from the Mach-Zehnder interrogation time $T$ used above. On resonance the
phase response is enhanced relative to the broadband response by a factor
of order $Q$,
\begin{equation}
  \Phi_{\mathrm{res}}(\omega_r) \sim Q\,\Phi_{\mathrm{bb}}(\omega_r)\,,
  \label{eq:resonant_enh}
\end{equation}
so the resonant frequency is set by the loop spacing, the on-resonance
sensitivity grows in proportion to $Q$, and the bandwidth narrows in
inverse proportion as $1/Q$. Switching
between broadband and resonant operation requires only a change of pulse
sequence and no change to the hardware, so the resonant frequency can be
retuned, or the instrument returned to broadband running, in software.
The two modes are alternative response strategies rather than mandatory
frequency regions, and AICE can interleave them within an observing
programme. Broadband running is useful when broad frequency coverage is
desired, whereas resonant running is useful when sensitivity is to be
concentrated at a known or hypothesised frequency.

\paragraph{Frequency-dependent optimisation.}
The resonant enhancement cannot be increased without limit. For a fixed
baseline the total time available for the atom trajectory is finite,
$T_{\mathrm{tot}} \lesssim 2 T_{\mathrm{flight}}$ with
$T_{\mathrm{flight}}$ of order $5\,\mathrm{s}$ for the full-baseline AICE
cold-atom `fountain', and the number of loops that fit is approximately
$Q \sim T_{\mathrm{tot}}/2\tau$. Targeting a higher frequency reduces the
loop spacing $\tau$ and therefore allows more loops, but at the same time
the pulse timing becomes more demanding, the finite pulse duration matters
more, and the LMT efficiency is harder to maintain, so the sequence cannot
be packed arbitrarily densely. The achievable enhancement is therefore a
compromise, and for a given target frequency there exists an optimal
combination of loop spacing, number of loops and LMT order. The optimal
values of $Q$, the LMT order $n$ and the loop spacing are 
frequency-dependent~\cite{DiPumpo:2023hfk,Schach_2025}. Broadband operation is the default survey mode, while
resonant operation is a frequency-targeted mode in which the pulse
sequence is optimised for a chosen signal frequency.

Coherent ultralight-dark-matter signals are natural candidates for
resonant interrogation because of their narrow intrinsic linewidth, of
order $\Delta f / f \sim 10^{-6}$. Depending on the source characteristics
and the search strategy, gravitational-wave searches may employ either
broadband or resonant interrogation.

\paragraph{Ramsey-Bord\'e and auxiliary modes.}
AICE may also be operated with Ramsey-Bord\'e and related
clock-interferometry sequences~\cite{Borde1989,Schelfhout2024}. Unlike the
Mach-Zehnder geometries used for the primary AICE science programme,
Ramsey-Bord\'e interferometers are directly sensitive to photon recoil and
therefore provide access to precision measurements of recoil-related
quantities and fundamental constants. In particular they enable
determinations of the fine-structure constant $\alpha$ through the
atom-recoil measurement, while also offering complementary handles on
systematic uncertainties. They use the same single-photon clock-transition
infrastructure and are introduced through a change of pulse sequence rather
than of hardware. The corresponding precision-physics programme is
described in Section~\ref{sec:physics}.

\clearpage

%% sec_physics.tex

\providecommand{\GeV}{\,\mathrm{GeV}}
\providecommand{\eV}{\,\mathrm{eV}}
\providecommand{\nm}{\,\mathrm{nm}}
\providecommand{\keff}{k_{\mathrm{eff}}}
\providecommand{\phiDM}{\delta\varphi}
\providecommand{\rDM}{\rho_{\mathrm{DM}}}
\providecommand{\mstar}{m^{*}}

\section{Physics Case}
\label{sec:physics}

%\textcolor{red}{Badurina, Blas, Carlton, Di Pumpo, Ellis, Gaaloul, Guellati-Khelifa, Lellouch, Lopez Asamar, Maggiore, McCabe, Mitchell, Vaskonen}

%%------------------------------------------------------------
\subsection{Overview}
%%------------------------------------------------------------

The operating parameters used for all sensitivity projections in this
Section are those of the AICE performance ladder defined in
Table~\ref{tab:roadmap} of the Introduction.
The Baseline and Stretch entries correspond to the two successive AICE
science phases, and the 1\,km entry is shown for context and
corresponds to a future TVLBAI facility~\cite{TVLBAISummary}.
The scalar channel uses the compact top-bottom gradiometer
configuration with cloud separation $\delta r = L/2$ and interrogation
time $T_{\mathrm{grad}}$. The vector $B{-}L$ and equivalence-principle
channels use a full-Baseline dual-species Mach-Zehnder configuration,
in which each species runs a single Mach-Zehnder interferometer over
the full one-way drop, with interrogation time $T_{\mathrm{MZ}}$. The
axion/ALP channel uses the
full-shaft single-species Mach-Zehnder configuration with interrogation
time $T_{\mathrm{ax}}$, as defined in Section~\ref{sec:concept}. Because the
vector/EP and axion modes both interrogate over the full Baseline, they
share the same free-fall time, $T_{\mathrm{MZ}} = T_{\mathrm{ax}}
\simeq \sqrt{2L/g}$; only the scalar gradiometer, whose two clouds
split the shaft, runs at the shorter $T_{\mathrm{grad}}$.
The single-photon character of the 698\,nm $^{1}S_{0}$--$^{3}P_{0}$
clock transition gives the effective momentum transfer
$\keff = n\,k_{0}$ with $k_{0}=2\pi/\lambda$, which differs by a factor
of two from the convention used for two-photon Raman or Bragg
interferometers.
Here $n$ is the large-momentum-transfer (LMT) order of the atom optics,
corresponding to the
$n\hbar k_{0}$ momentum splitting between the two interferometer arms; it is
independent of the specific pulse-sequence topology, which may be a broadband
Mach-Zehnder gradiometer or a multi-loop (``diamond'') geometry with resonant
enhancement at a chosen frequency~\cite{Graham:2016plp,DiPumpo:2023hfk,Schach_2025}.
The pulse sequences used for the sensitivity projections in this Section, and
the resonant-mode option, are described in Section~\ref{sec:concept:mode:readout}. %3.4. \textcolor{red}{Should be a hyperlink. Section number incorrect?}
Unless otherwise stated, the projections assume shot-noise-limited operation,
using the phase-noise amplitudes and one-year integration times listed in
Table~\ref{tab:roadmap}. The estimated impact of unmitigated seismic gravity gradient noise is
quantified in Section~\ref{sec:noise:GGN:impact}. The principal exception to
the shot-noise convention is the GW strain figure of Section~\ref{sec:gw},
which also displays that benchmark.
Here
the statistical treatment of the ULDM field is specified in
Section~\ref{sec:uldm_framework}.

%%------------------------------------------------------------
\subsection{Ultralight dark matter (ULDM)}
\label{sec:uldm}
%%------------------------------------------------------------

The nature of dark matter remains unresolved despite decades of
experimental effort across a wide range of energy scales. Collider
searches at the LHC and elsewhere, as well as direct- and indirect-detection experiments, have placed
strong constraints on weakly-interacting particles at and above the
electroweak scale. At much lower masses, a broad and well-motivated
class of candidates is the focus of many laboratory experiments and astrophysical searches.

Bosonic ultralight dark matter (ULDM) fields with masses in the range $10^{-22}$--$10^{-12} \ \mathrm{eV}$
arise naturally in several extensions of the Standard Model.
For example, ultralight scalar fields appear as moduli or dilaton-like states in supersymmetric and extra-dimensional scenarios including string theory~\cite{Arvanitaki:2014faa}, and axion-like pseudoscalar particles (ALPs) are a generic prediction of string-theoretic
constructions, with the QCD axion being a specific
realisation~\cite{Graham:2015gua}. Ultralight vector bosons can
arise from anomaly-free gauge extensions such as
$B{-}L$~\cite{Graham:2015ifn}, and massive spin-2 dark matter candidates have also been proposed~\cite{deRham:2014zqa}. 

In all of these cases the dark matter
field behaves, on laboratory scales, as a coherently oscillating
classical background with a frequency set by the particle mass.
The corresponding experimental signatures differ qualitatively from
those targeted by conventional particle searches, e.g., at colliders. Instead of rare
production events, one may consider small,
time-dependent perturbations of atomic systems, including effective
forces and shifts of fundamental constants, induced by ultralight dark matter. Detecting these effects
requires precision measurement of phase, frequency, or acceleration
at low frequencies, typically in the range $10^{-7}$--$1\,\mathrm{Hz}$.

Atom interferometers are particularly well suited for exploring this regime.
Their sensitivity to phase accumulation over long free-fall times,
combined with large effective momentum transfers and intrinsic
common-mode rejection, enables access to parameter space that is
difficult to probe with other techniques. In the AICE facility,
several distinct ultralight dark matter channels can be explored using
different operating configurations of the same instrument: scalar
dark matter through modulation of fundamental constants in an atom
gradiometer, vector $B{-}L$ dark matter through differential
acceleration in a dual-species configuration, and axion-like
particles through spin-dependent phase accumulation in a full-Baseline
single-species Mach-Zehnder interferometer. Add-ons to the basic instrument could be used to explore local modifications of the dark matter field that occur in, e.g., chameleon and symmetron models.

A key feature of AICE is that these channels do not require separate
instruments. They are implemented by switching between operating
modes of a common infrastructure, as described in
Section~\ref{sec:concept}: see, in particular, the operating modes illustrated in
Section~\ref{sec:concept:modes}. The 140\,m PX46 shaft, the common-vacuum
interferometry tower, and the 698\,nm strontium clock-transition
laser system are shared across all three campaigns. The addition of
a $^{88}$Sr atom source for the vector campaign requires no change
to the shaft geometry or the vacuum system, but the primary clock laser
source must be extended with high-power lasers at 679 nm, 688 nm, and 707 nm to drive the transition for atom interferometry~\cite{carman_collinear_2025}. Operation with $^{171}$Yb requires a different laser system and atom sources but no other modifications to the instrument. This allows a single facility to
probe a broad and largely complementary region of the ultralight dark
matter parameter space across scalar, vector, and axion scenarios
from $10^{-22}\ \mathrm{eV}$ to above $10^{-14}\ \mathrm{eV}$. The projections presented
in this Section are therefore best understood not as independent
experiments, but as different measurements within a unified
experimental programme.

The scalar and tensor channels reach peak sensitivity near
$m \approx 8 \times 10^{-16}\ \mathrm{eV}$ for the gradiometer interrogation time
$T_{\mathrm{grad}} = 2.5$\,s. The vector $B{-}L$ channel, which runs on the
full-Baseline Mach-Zehnder mode at $T_{\mathrm{MZ}} = 5.0$\,s, together with the
pseudoscalar ALP channel at $T_{\mathrm{ax}} = 5.0$\,s, reaches its optimum near
$m \approx 4 \times 10^{-16}\ \mathrm{eV}$. A candidate signal in one channel
would imply correlated signatures in the others, providing a
built-in cross-check.

The sensitivity projections in this Section use the Baseline, Stretch,
and 1\,km parameter sets of Table~\ref{tab:roadmap} of the
Introduction. The instrument configurations for the Baseline and
Stretch phases are detailed in Sections~\ref{sec:detector}
and~\ref{sec:laser-upgrades}, while the 1\,km scenario is an extrapolation to the
proposed TVLBAI programme~\cite{TVLBAISummary}.

%% ── Common framework ──────────────────────────────────────────────────────

\subsubsection{Common framework for ultralight dark matter}
\label{sec:uldm_framework}

For dark matter masses in the range $10^{-22}$--$1\ \mathrm{eV}$, the occupation
number per de Broglie volume is large and the field behaves as a
coherently oscillating classical background~\cite{Hui:2021tkt}. The
coherence time $\tau_c$ is set by the mass and the local virial velocity $\beta \simeq
10^{-3}$:
\begin{equation}
  \tau_c \;\simeq\; \frac{2\pi}{m\beta^2}\,,
  \label{eq:coherence}
\end{equation}
which exceeds any laboratory interrogation time for
$m \lesssim 10^{-14}\ \mathrm{eV}$. The field amplitude is fixed by the local
dark matter energy density
$\rDM \simeq 0.4\,\mathrm{GeV\,cm}^{-3}$:
\begin{equation}
  \Phi_0 \; \simeq \; \frac{\sqrt{2\rDM}}{mc^2}\,,
  \label{eq:amplitude}
\end{equation}
and the oscillation frequency is $\omega = mc^2/\hbar$, implying a
AICE sensitivity band $f_{\rm DM} \sim 10^{-7}$--$1\,\mathrm{Hz}$.

The sensitivity of any atom interferometer reaches its peak at the
mass for which the field completes approximately half a cycle within
the interrogation time $T$:
\begin{equation}
  \mstar \;\simeq\; \frac{\pi\hbar}{T}\,.
  \label{eq:mstar}
\end{equation}
For $m \gg \mstar$, the phase accumulates over many oscillation cycles and
averages towards zero, so the sensitivity degrades as
$(m/\mstar)^{3/2}$. At $m \ll \mstar$ the signal decreases towards
lower masses, producing the characteristic V-shaped sensitivity
envelopes visible in Figures~\ref{fig:ULDM_scalar}--\ref{fig:ULDM_axion}.
This common framework is applicable to all
the benchmark channels discussed below, though
the detailed shape differs between channels because the response
functions are not identical.

\medskip
\noindent\textit{Sensitivity estimator and stochastic factor.}\quad
The sensitivity projections in this Section are obtained with the
effective-time prescription given in Eq.~(50) of~\cite{Gue:2024onx}, applied identically to the scalar, vector and axion
channels so that the projections are mutually consistent and directly
comparable. The interferometer phase noise is treated as an amplitude
spectral density $\phiDM$ in $\mathrm{rad}/\sqrt{\mathrm{Hz}}$ rather than
as a per-shot phase, and is combined with an effective averaging time
that interpolates between the coherent regime
($T_{\mathrm{int}} \le \tau_c$), in which the reach on a coupling improves
as $T_{\mathrm{int}}^{-1/2}$, and the decoherent regime
($T_{\mathrm{int}} > \tau_c$), in which it improves as
$(T_{\mathrm{int}}\,\tau_c)^{-1/4}$, with the one-year integration time
$T_{\mathrm{int}}$ of Table~\ref{tab:roadmap}. The projections are quoted
at a signal-to-noise ratio of unity (SNR${}=1$) at the atom shot-noise
limit, with no additional statistical factor folded into the curves; the
statistical conventions adopted, and the refinements that are discussed
rather than applied, are set out in Section~\ref{sec:uldm_caveats}. The
local dark-matter density is fixed throughout to
$\rDM = 0.4\,\mathrm{GeV\,cm}^{-3}$, the value adopted in
Ref.~\cite{Gue:2024onx}, and the virial velocity dispersion to
$\beta = 10^{-3}$.

%% ── Scalar DM ────────────────────────────────────────────────────────────

\subsubsection{Statistical conventions and caveats}
\label{sec:uldm_caveats}

All the SNR = 1 sensitivity projections in this Section 
%are quoted at a signal-to-noise ratio of unity (SNR = 1) against an atom-shot-noise-limited phase resolution, \textcolor{red}{repetition} 
are evaluated using the effective-time prescription of Eq.~(50)
of~\cite{Gue:2024onx} applied identically to the scalar, vector and
axion channels. Gravity gradient noise is excluded from these curves
altogether, rather than being included in an unmitigated form, and no
additional statistical factor is folded into them. The same statistical
convention is used for the gravitational-wave reach (Section~\ref{sec:gw}),
so that all AICE projections in this proposal are mutually consistent and
directly comparable, with the single exception that the GW strain figure
additionally overlays the estimated unmitigated seismic-GGN benchmark of
Section~\ref{sec:noise:GGN:impact}. We adopt this single transparent convention because the
external constraints overlaid for comparison are  derived under a
range of differing statistical prescriptions; matching
each of them individually would not provide a uniform basis for comparison. The purpose
of the projected sensitivities is to illustrate the intrinsic experimental
reach under a common convention, while statistical refinements are
discussed separately.

A stochastic amplitude correction of approximately 1.5 has been discussed
in the literature for coherent measurements~\cite{Gue:2024onx}. As the
external constraints shown here are derived under heterogeneous statistical
prescriptions, we do not incorporate this correction into the projected
reach curves.

A further, confidence-level-dependent degradation can be defined following
Appendix~C.4 of~\cite{528c-xs6p}, which can further weaken the projected
reach in the short-coherence regime. For the same reason we do not apply
such corrections to the projected curves.

For the vector $B{-}L$ channel a factor of $1/\sqrt{3}$ may arise for an
isotropically polarised field. As the treatment of this effect is not
necessarily uniform across the existing experimental limits and projected
sensitivities, no such correction is applied here.

The overlaid constraints, taken largely from the AxionLimits
compilation~\cite{OHare:2020wah}, retain the confidence levels and
statistical treatments of their original sources, which are not uniform;
direct comparisons should therefore be interpreted with the statistical
conventions of the original references in mind. Astrophysical energy-loss
bounds, such as the SN\,1987A limit in the axion channel, are subject to
$\mathcal{O}(1)$ modelling uncertainties and carry no experimental
confidence level at all; their treatment, including the illustrative
modelling-uncertainty band shown in Figure~\ref{fig:ULDM_axion}, is
described in Section~\ref{sec:uldm_axion}. Gravity-gradient
(Newtonian) noise is likewise not included in the ULDM projected
sensitivities. Its impact is quantified in
Section~\ref{sec:noise:GGN:impact} and the mitigation programme is set out
in Section~\ref{sec:noise:GGN:mitigation}.

\subsubsection{Ultralight scalar dark matter}
\label{sec:uldm_scalar}

We begin with the scalar channel, which provides the most natural
first implementation of the gradiometer mode.

\medskip
\noindent\textit{Coupling and signal.}\quad
A cosmological scalar field $\phi$ oscillating at frequency
$\omega_\phi = m_\phi c^2/\hbar$ can couple to Standard Model fields
through operators that modulate fundamental
constants~\cite{Graham:2015ifn, Arvanitaki:2014faa}. The two
couplings of primary interest are
\begin{equation}
  \mathcal{L} \;\supset\;
  \frac{\phi}{\Lambda}
  \Bigl[
    d_{m_e}\,m_e\,\bar{e}e
    \;+\;
    \frac{d_e}{4}\,F_{\mu\nu}F^{\mu\nu}
  \Bigr]\,,
  \label{eq:scalar_lag}
\end{equation}
where $\Lambda = M_{\rm Pl}\sqrt{8\pi\alpha}$ and $d_{m_e}$, $d_e$
are dimensionless coupling constants for the electron-mass and photon
channels, respectively. The modulations of fundamental constants generated by
these couplings induce a differential phase in the gradiometer through an oscillating atomic transition frequency~\cite{Derr2023}.
In the signal model adopted here, this response is written in terms
of a coupling-dependent effective acceleration $g_{\rm sc}(\omega)$
following the treatment of Refs.~\cite{Graham:2015ifn,
Badurina:2021lna}. A non-zero differential signal arises already for
a single atomic species in the gradiometer configuration, so
the scalar search does not rely on dual-species operation.

\medskip
\noindent\textit{Interferometer response.}\quad
The scalar measurement uses a gradiometer with cloud
separation $\delta r = L/2 = 70\,\mathrm{m}$ at the 140\,m Baseline and
LMT atom optics of order $n$ on the $^{87}$Sr clock
transition at $\lambda = 698\ \rm nm$. As noted in the Overview, the
single-photon character of the clock transition gives the effective
momentum transfer
\begin{equation}
  \keff \;=\; n\,k_0\,,
  \qquad
  k_0 = 2\pi/\lambda\,,
  \label{eq:keff2}
\end{equation}
which evaluates to $\keff = 8.996 \times 10^{9}\,\mathrm{rad\,m}^{-1}$
at the Baseline value $n = 10^{3}$ and $\keff = 3.598 \times
10^{11}\,\mathrm{rad\,m}^{-1}$ at the Stretch and 1\,km value
$n = 4\times10^{4}$. The differential
phase for a scalar field at frequency $\omega_\phi$ follows from
Ref.~\cite{Graham:2015ifn} with the sign and normalisation corrections
of Ref.~\cite{Badurina:2021lna}:
\begin{equation}
  \delta\phi_{\rm scalar}
  \; \simeq \;
  \frac{\keff\,g_{\rm sc}\,T^2}
       {\sqrt{2}\,\omega_\phi^2\,\delta r}
  \Bigl[
    3\sin\!\left(\tfrac{\omega_\phi T}{2}\right)
    -
    \sin\!\left(\tfrac{3\omega_\phi T}{2}\right)
  \Bigr]\,,
  \label{eq:scalar_phase}
\end{equation}
where $g_{\rm sc}$ encodes the coupling-dependent perturbation to
the free-fall acceleration. The minimum detectable coupling is
obtained by setting $\delta\phi_{\rm scalar} = \phiDM$ and inverting
for $d_{m_e}$ or $d_e$.

\begin{figure}[ht]
  \centering
  \includegraphics[width=0.48\linewidth]{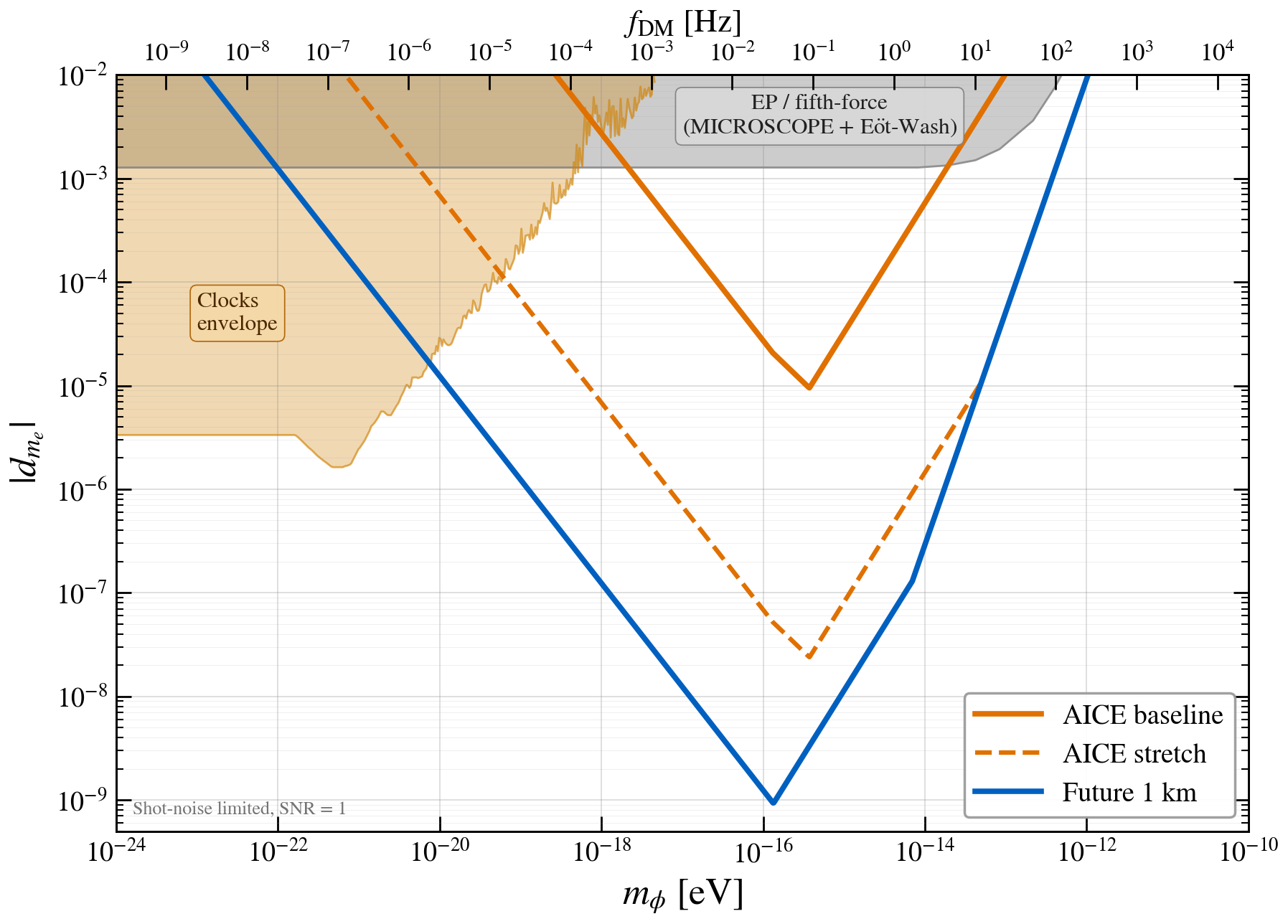}
  \hfill
  \includegraphics[width=0.48\linewidth]{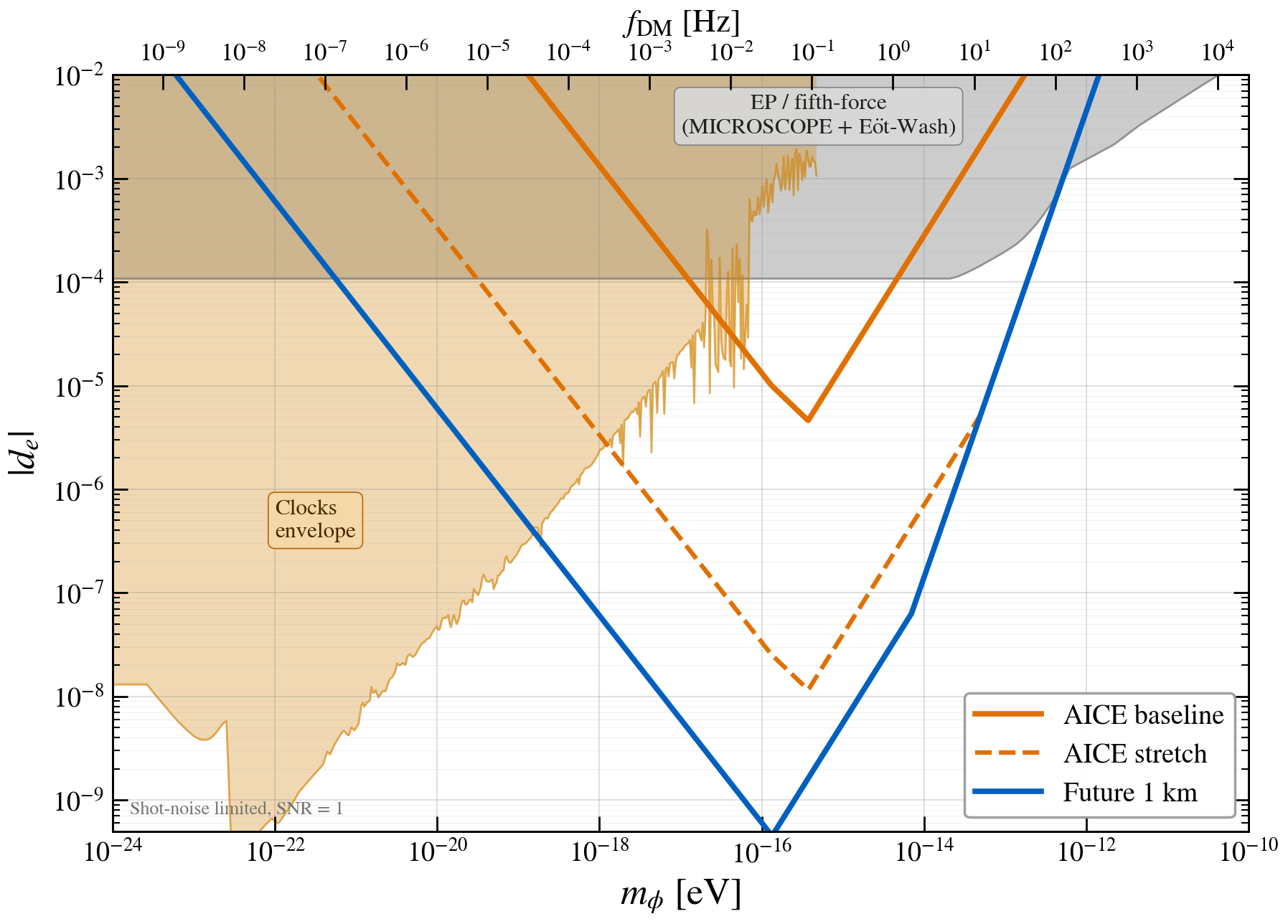}
  \caption{Projected AICE sensitivity to the scalar ULDM electron-mass
    coupling $|d_{m_e}|$ (left) and photon coupling $|d_e|$ (right)
    versus scalar mass $m_\phi$, with DM oscillation frequency
    $f_{\rm DM}$ on the upper axis.  Three scenarios are shown (see
    Table~\ref{tab:roadmap}): AICE Baseline, AICE Stretch, and a
    future 1\,km extension.  Shaded regions mark existing constraints,
    taken from the AxionLimits compilation~\cite{OHare:2020wah}.
    The grey band, labelled \textit{EP/fifth-force}, is the
    pointwise envelope of the final MICROSCOPE result with Ti/Pt test
    masses~\cite{Touboul:2022yrw} and the E\"ot-Wash equivalence-principle
    results~\cite{Schlamminger:2007ht,Wagner:2012ui}; MICROSCOPE
    dominates over most of the displayed mass range.  The tan band,
    labelled \textit{Clocks envelope}, is the pointwise minimum across
    atomic-clock and cavity-stabilised laser searches in each coupling
    channel; the dominant individual contributions in the 
    mass range relevant to AICE are the H-maser/Si-cavity comparison in~\cite{Kennedy:2020bac} for $|d_{m_e}|$ and the
    $^{171}$Yb$^+$/Sr optical-clock comparison in~\cite{Filzinger:2023zrs} for $|d_e|$.
    {The emergence of clock comparison studies based on the $^{229}$Th optical clock expand the limits in the tan bands to heavier 
masses~\cite{DeCol:2026hnl}. Synchronous clock comparison measurements in the valley of overlap with AICE open the prospect of combined limits of still greater sensitivity.}
   {The corresponding impact of the estimated unmitigated seismic gravity gradient noise is shown in Fig.~\ref{fig:ULDM_GGN} and discussed in Section~\ref{sec:noise:GGN:impact}.}  }
  \label{fig:ULDM_scalar}
\end{figure}

\medskip
\noindent\textit{Sensitivity projections.}\quad
The dominant existing bounds on scalar ULDM couplings in the low-mass
regime come from the E\"ot-Wash torsion-balance
experiment~\cite{Schlamminger:2007ht} and the MICROSCOPE
satellite~\cite{Touboul:2022yrw}; the final MICROSCOPE result is
$\eta(\mathrm{Ti,Pt}) = [-1.5 \pm 2.3\,(\mathrm{stat}) \pm 1.5\,(\mathrm{syst})]\times10^{-15}$,
consistent with zero at the $10^{-15}$ level~\cite{Touboul:2022yrw}.
At higher mediator masses the equivalence-principle bounds weaken because of
the finite interaction range and the spatial extent of the Earth as the source,
and atomic-clock comparisons become the leading laboratory constraints. For $d_{m_e}$, the tightest current bound in the mass range relevant to AICE
comes from the hydrogen-maser to silicon-cavity frequency comparison in~\cite{Kennedy:2020bac}. For $d_e$ the tightest
bound is the Yb$^+$/Sr optical-clock comparison in~(2023)~\cite{Filzinger:2023zrs}, which improves on
earlier spectroscopic results by several orders of magnitude.
AICE surpasses the MICROSCOPE bound by several orders of magnitude
in both coupling channels over the range $10^{-19}$--$10^{-14}\ \mathrm{eV}$.
Near the optimal mass $m_\phi \sim 10^{-16}\ \mathrm{eV}$ the AICE Baseline
reaches $|d_{m_e}| \approx 9.5 \times 10^{-6}$ (approximately one order
of magnitude below the clock bound in~\cite{Kennedy:2020bac})
and $|d_e| \approx 4.6 \times 10^{-6}$. The Stretch plateau reaches
$|d_{m_e}| \approx 2.4 \times 10^{-8}$ and $|d_e| \approx 1.2 \times
10^{-8}$, and the 1\,km scenario reaches $|d_{m_e}| \approx 9.3 \times
10^{-10}$ and $|d_e| \approx 4.5 \times 10^{-10}$. The projected reach
and existing constraints are shown in Figure~\ref{fig:ULDM_scalar}. 
%\textcolor{red}{Rounded to 2 sig. figs}

The loss of sensitivity below $10^{-15}$~eV is due to the limited atom flight time, whereas the loss of sensitivity above $m_\phi \sim 10^{-15}$~eV is due to shot noise. We recall that a prototype laboratory study~\cite{AION:2025igp} has found that laser noise can be suppressed below the level of atom shot noise in a gradiometer configuration. The impact of gravity gradient noise on this channel is shown in Fig.~\ref{fig:ULDM_GGN} and discussed in Section~\ref{sec:noise:GGN:impact}, where the scalar channel is used as the representative quantitative example.

\medskip
\noindent\textit{Realisation in AICE.}\quad
The scalar search is implemented in AICE using the single-species
 gradiometer mode of Section~\ref{sec:concept:mode:grad}, with $^{87}$Sr
and the Baseline LMT order $n = 10^{3}$ on the 698\,nm laser system,
rising to $n = 4\times10^{4}$ at the Stretch level.
$^{87}$Sr is the default atomic species for the AICE gradiometer
and full-shaft configurations: its fermionic $I = 9/2$ nuclear spin
makes the ${}^{1}S_{0}$--${}^{3}P_{0}$ clock transition naturally
allowed through hyperfine mixing, and the same atom source and
laser system are re-used in the axion campaign.
$^{88}$Sr enters the programme only as the bosonic partner in the
vector dual-species configuration.
This mode provides the natural entry point for the scalar ULDM
programme and can be commissioned using the single-species
infrastructure of Section~\ref{sec:detector}, without modification
to the atom source assembly.

{Wavelike dark 
matter could also be detected gravitationally, 
as discussed in Sec.~\ref{sec:otherfundphys}. 
The possibility of 
time-dependent effects, particularly as a 
consequence of localized, overdense regions 
formed through dark-matter self-interactions in the
vicinity of heavy, celestial objects in the solar system, such as
the Sun or planets, 
can make these effects appreciable.}

%% ── Axion-like particles ─────────────────────────────────────────────────

\subsubsection{Axion-like particles (ALPs)}
\label{sec:uldm_axion}

The ALP channel is addressed using a distinct operating mode, using a
single-species Mach-Zehnder interferometer over the full 140\ m Baseline.

\begin{figure}[ht]
  \centering
  \includegraphics[width=0.78\linewidth]{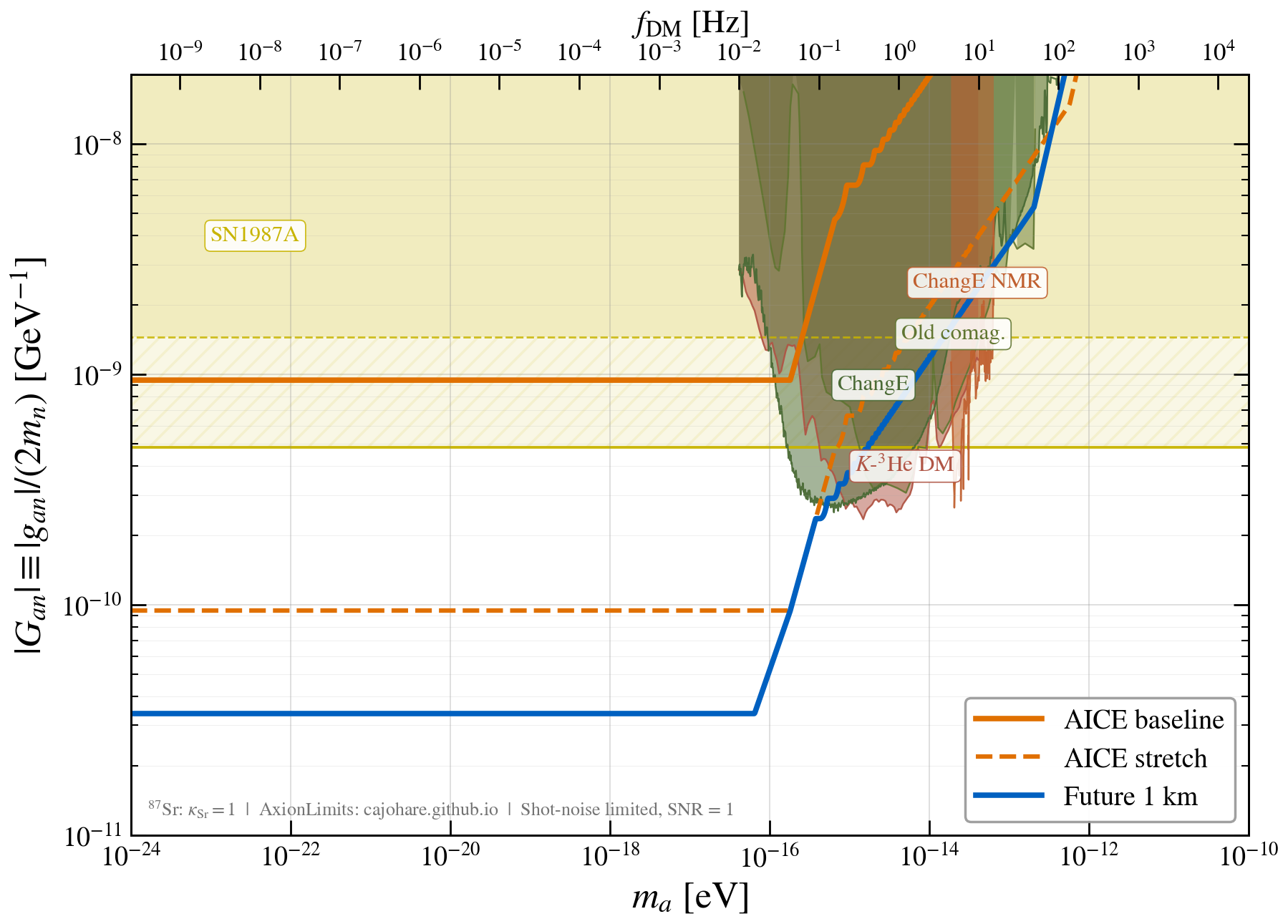}
  \caption{Projected AICE sensitivity (SNR${}=1$, one year of
    integration, atom shot-noise limited) to the dimensionful
    axion-neutron coupling $|G_{an}| \equiv |g_{an}|/(2m_n)$ as a
    function of axion mass $m_a$.  The yellow region is the SN\,1987A
    energy-loss exclusion in the pure-neutron-coupling limit,
    $g_{ap}=0$, of the two-coupling constraint of
    Ref.~\cite{Carenza:2019pxu}: $|g_{an}| < 9.09\times10^{-10}$ in the
    dimensionless convention, corresponding to
    $|G_{an}| < 4.84\times10^{-10}\,\mathrm{GeV}^{-1}$ (solid line),
    treated as a mass-independent plateau across the displayed mass
    range.  The hatched band above the nominal edge extends to three
    times the nominal coupling and indicates the spread associated with
    supernova-modelling systematics~\cite{Carenza:2019pxu,
    Chang:2018rso, Bar:2019ifz}; it is a systematic envelope shown for
    illustration and carries no statistical confidence interpretation.
    For this proposal we treat the region above the dashed edge as
    conservatively excluded under the adopted envelope.
    The coloured laboratory bands show
    the K--$^3$He atomic comagnetometer dark-matter
    search~\cite{Lee:2022vvb}, ChangE (both resonant hybrid-spin and NMR
    realisations)~\cite{Wei:2023rzs,Xu:2023vfn}, and old comagnetometer
    data recast as axion constraints~\cite{Bloch:2019lcy}, at their
    published confidence levels.  All exclusion curves are taken from
    the AxionLimits compilation~\cite{OHare:2020wah} and converted
    consistently to the $G_{an}$ convention; the statistical conventions
    of the projections and of the overlaid constraints are discussed in
    Section~\ref{sec:uldm_caveats}.  All dark-matter limits and
    projections assume axions constitute the entire local dark-matter
    density, $\rDM = 0.4\,\mathrm{GeV\,cm}^{-3}$.
    The annotation $^{87}\mathrm{Sr}:\kappa_{\rm Sr}=1$ denotes the adopted
    Stretched-state shell-model neutron-spin projection factor.}
  \label{fig:ULDM_axion}
\end{figure}

\medskip
\noindent\textit{Coupling and signal.}\quad
ALPs and the QCD axion couple to nucleons through the derivative
interaction
\begin{equation}
  \mathcal{L} \;\supset\;
  \frac{g_{aN}}{2m_N}\,\partial_\mu a\,\bar{N}\gamma^\mu\gamma^5 N\,,
  \label{eq:axion_L}
\end{equation}
where $g_{aN}$ is the dimensionless axion-nucleon coupling ($N = p, n$) and
$m_N$ the nucleon mass. Throughout we express the sensitivity in terms of
the dimensionful axion-neutron coupling $G_{an}\equiv g_{an}/(2m_n)$ (mass
dimension $-1$), the convention used throughout this section and in
Fig.~\ref{fig:ULDM_axion}.\footnote{Note a symbol clash in the literature:
Refs.~\cite{Graham:2017pmn, Lee:2022vvb} denote this dimensionful
$\mathrm{GeV}^{-1}$ coefficient $g_{aNN}$, whereas the dimensionless
coupling $g_{an}$ follows the convention of
Refs.~\cite{Carenza:2019pxu, OHare:2020wah}.} The oscillating axion field couples to the
nucleon spin and induces a spin-dependent interferometric phase between two
arms prepared in different nuclear-spin states of the
$^{1}S_{0}$--$^{3}P_{0}$ clock manifold. Over an interrogation time $T$,
following the atom-interferometer treatment of Ref.~\cite{Graham:2017pmn}, the
differential phase can be expressed as
\begin{equation}
  \delta\phi_{\rm ax}
  \;=\;
  \kappa_{\rm Sr}\,G_{an}\,\frac{\beta\sqrt{2\rDM}}{m_a}\,
  \sin\!\bigl(m_a T/\hbar\bigr)\,,
  \label{eq:axion_freq}
\end{equation}
where $\beta \simeq 10^{-3}$ is the local virial velocity and $\kappa_{\rm Sr}$
is the difference in neutron-spin projection between the two arms; we adopt the
Stretched-state shell-model value $\kappa_{\rm Sr}=1$ for $^{87}$Sr
($I=9/2$, $m_I=\pm9/2$). The signal is a spin-dependent energy difference
rather than a recoil effect, so the reach is independent of the LMT order $n$
and is set by $\phiDM$, $T$ and $\kappa_{\rm Sr}$ alone.

\medskip
\noindent\textit{Sensitivity formula.}\quad
The minimum detectable coupling is obtained by inverting
Eq.~(\ref{eq:axion_freq}) against the phase-noise floor, evaluated
numerically using the AICE operating parameters of Table~\ref{tab:roadmap}. The interferometer response follows
Graham et al.~\cite{Graham:2017pmn}, while the treatment of finite coherence
and the transition between coherent and incoherent averaging follows the
common ULDM statistical framework introduced in
Section~\ref{sec:uldm_framework}. On the low-mass plateau
($m_a \lesssim \mstar = \pi\hbar/T$) the reach scales as
\begin{equation}
  G_{an}^{\rm sens}
  \;\simeq\;
  \frac{\phiDM\,\hbar}{\kappa_{\rm Sr}\,\beta\sqrt{2\rDM}\,T\,\sqrt{T_{\mathrm{int}}}}\,,
  \label{eq:axion_sens}
\end{equation}
rising above $\mstar$ through the dimensionless response-and-averaging factor
$\mathcal{G}(m_a,T)$, which combines the single-shot Mach-Zehnder response
with the finite-coherence averaging of that framework. The LMT order $n$ does
not appear, confirming that the axion reach is
independent of momentum transfer.

\medskip
\noindent\textit{Sensitivity projections.}\quad
The dominant existing constraint on $|G_{an}|$ across most of the
AICE-relevant mass range is the astrophysical bound from the
SN\,1987A supernova neutrino signal. In the pure-neutron-coupling limit,
$g_{ap}=0$, the two-coupling constraint of Ref.~\cite{Carenza:2019pxu}
gives $|g_{an}| < 9.09 \times 10^{-10}$ in the dimensionless convention,
corresponding to
$|G_{an}| < 4.84 \times 10^{-10}\,\mathrm{GeV}^{-1}$.
This bound is treated as a mass-independent plateau across the
displayed mass range, in line with the standard supernova energy-loss
argument.  The strongest laboratory exclusions, reaching below the
SN\,1987A bound in narrow windows within the range
$10^{-16}$--$10^{-14}\ \mathrm{eV}$, come from a K--$^3$He atomic
comagnetometer dark-matter search~\cite{Lee:2022vvb}, the ChangE searches
(resonant hybrid-spin and NMR realisations)~\cite{Wei:2023rzs, Xu:2023vfn},
and comagnetometer data recast as constraints on axion-like
relics~\cite{Bloch:2019lcy}. Further searches, including the nEDM
experiment~\cite{Abel:2017rtm}, CASPEr-ZULF~\cite{Garcon:2019inh} and a
liquid-state nuclear-spin comagnetometer~\cite{Wu:2019exd}, constrain this
mass range at substantially weaker couplings, above the range displayed in
Figure~\ref{fig:ULDM_axion}.  The exclusion data
shown in
Figure~\ref{fig:ULDM_axion} are taken from the AxionLimits
compilation~\cite{OHare:2020wah}, where additional astrophysical constraints are also displayed, and are converted consistently to the
$G_{an}$ convention.

These estimates indicate that AICE would provide a powerful terrestrial probe
of axion parameter space currently bounded primarily by SN\,1987A.
The AICE Baseline, with a plateau sensitivity of
$|G_{an}| \approx 9.5 \times 10^{-10}\,\mathrm{GeV}^{-1}$, approaches the
nominal SN\,1987A bound to within a factor of two, while the
Stretch scenario surpasses the nominal bound by a factor of about five and
the 1\,km extension by a factor of about fourteen.

The SN\,1987A bound is an astrophysical energy-loss estimate subject to
$\mathcal{O}(1)$ modelling uncertainties and does not correspond to a
well-defined statistical confidence level. Published analyses of this
bound have shifted it by factors of approximately two to
five~\cite{Carenza:2019pxu, Chang:2018rso}, and its overall robustness has
also been questioned~\cite{Bar:2019ifz}. For illustration,
Figure~\ref{fig:ULDM_axion} therefore supplements the nominal exclusion
edge with a hatched band extending to three times the nominal coupling,
indicating the spread associated with supernova-modelling systematics;
this band is a systematic envelope for illustrative purposes and carries
no statistical confidence interpretation. The AICE Baseline probes
couplings within this modelling-uncertainty band, and a terrestrial
measurement at this level would provide an independent laboratory
cross-check of a limit otherwise derived entirely from supernova physics.

\medskip
\noindent\textit{Realisation in AICE.}\quad
The axion search is performed in the full-Baseline single-species
Mach-Zehnder configuration outlined in \cref{fig:mode_single}, using
$^{87}$Sr with the full free-fall time $T_{\mathrm{ax}} = 5.0$\,s. No
dual-species capability is required. This configuration is distinct
from the gradiometer mode used for commissioning and for the
Baseline scalar programme, and is expected to be deployed at a later
stage of the AICE science programme once the required level of
interferometric control has been achieved.
Alternative implementations based on $^{171}$Yb are also possible and remain under investigation~\cite{Zhou:2024ygw}.

%% ── Vector B-L DM ────────────────────────────────────────────────────────

\subsubsection{Vector \texorpdfstring{$B{-}L$}{B-L} dark matter}
\label{sec:uldm_vector}

The vector $B{-}L$ channel introduces a qualitatively different
requirement: simultaneous interrogation of two atomic species.

\medskip
\noindent\textit{Coupling and signal.}\quad
A massive vector boson $A^\mu$ associated with a gauged $B{-}L$
symmetry exerts an oscillating force on any particle carrying
$B{-}L$ charge~\cite{Graham:2015ifn}. Because different
atomic species carry different $B{-}L$ charge-to-mass ratios,
the simultaneous measurement of two different co-located species produces a
differential acceleration directly proportional to the coupling. We define $g_{B-L}$ through the interaction $\mathcal{L}\supset -g_{B-L}\,A_\mu J^\mu_{B-L}$ for the unit $B{-}L$ current, giving
\begin{equation}
  \Delta a(t)
  \;=\;
  g_{B-L}\,\eta_{\rm eff}\,\omega_V\,\Phi_0\,
  \cos\!\bigl(\omega_V t\bigr)\,,
  \qquad
  \omega_V = m_V c^2/\hbar\,,
  \label{eq:vec_accel}
\end{equation}
where $\Phi_0 = \sqrt{2\rDM}/m_V$ and
\begin{equation}
  \eta_{\rm eff}
  \;=\;
  \frac{(B{-}L)_1}{A_1}
  -
  \frac{(B{-}L)_2}{A_2}
  \label{eq:eta_eff}
\end{equation}
is the differential $B{-}L$ charge per unit mass, with $A_1$ and $A_2$ the atomic weights of the two species.%TODO physics editors: Change Record wording is ``atomic weights''; confirm vs ``mass numbers'' if preferred. The differential acceleration and the accumulated phase are linear in the coupling $g_{B-L}$, since the two co-located species are driven by the same dark-matter field. Uniform gravitational
acceleration and common platform motion cancel to leading order in the
inter-species phase difference, which is the key advantage of the
dual-species configuration. Residual differential coupling can arise from
scale-factor mismatch, imperfect co-location of the two species and spatially
varying gravity gradients, as discussed in
Section~\ref{sec:noise:GGN:impact}. The accumulated interferometer phase over time $T$
is~\cite{Graham:2015ifn}
\begin{equation}
  \delta\phi_{B-L}
  \;=\;
  \keff\,\eta_{\rm eff}\,\omega_V\,\Phi_0\,g_{B-L}\,T^2\,
  \mathcal{F}(\omega_V T)\,,
  \label{eq:vec_phase}
\end{equation}
where $\keff = nk_0$ from Eq.~(\ref{eq:keff2}) and $\mathcal{F}$ is
the dimensionless frequency response that peaks at
$\omega_V T \approx \pi/2$. For the benchmark sensitivity estimate
adopted here, the minimum detectable coupling scales approximately as
$g_{B-L}^{\rm min} \propto \keff^{-1}$, highlighting the
importance of the single-photon transition for the projected reach.

\begin{figure}[ht]
  \centering
  \includegraphics[width=0.78\linewidth]{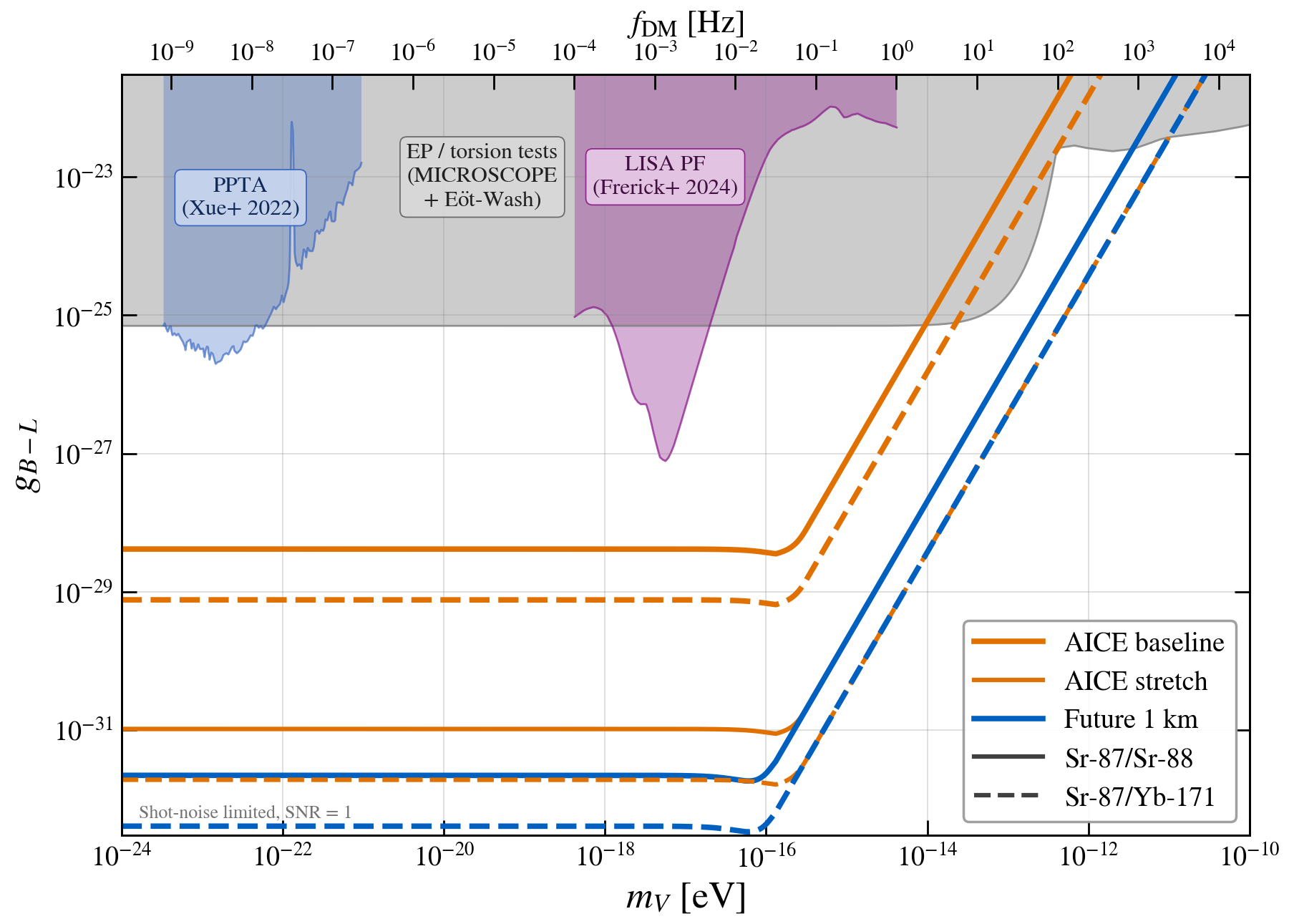}
  \caption{Projected AICE sensitivity to the vector $B{-}L$ coupling
    $g_{B-L}$ versus vector boson mass $m_V$, with the Compton
    frequency $f_{\rm DM}$ on the upper axis.
    Solid (dashed) lines of each scenario colour show the sensitivities of the
    $^{87}$Sr/$^{88}$Sr Baseline ($^{87}$Sr/$^{171}$Yb upgrade)
    species pairs.
    Shaded regions are existing exclusions.
    The grey ``EP/torsion tests'' envelope is the pointwise
    minimum of MICROSCOPE~\cite{Touboul:2022yrw}, evaluated from the final
    MICROSCOPE result as a static fifth-force constraint following~\cite{Amaral:2024tjg},
    together with the E\"ot-Wash~\cite{Wagner:2012ui} bound and subdominant
    contributions from~\cite{Shaw:2021laz,Ross:2025uly} that are
    included in the envelope but do not set its shape at any mass.
    The magenta LISA Pathfinder band~\cite{Frerick:2023hjh} is
    world-leading from $4\times10^{-19}$ to $10^{-17}\,\mathrm{eV}$,
    with a best limit $g_{B-L}\sim 1.5\times10^{-27}$.
    The blue PPTA band~\cite{Xue:2021gyq} shows results from a Parkes pulsar
    timing array dark-photon dark-matter search covering
    $3\times10^{-24}$ to $10^{-21}\,\mathrm{eV}$.
    Natural units are used throughout
    ($g_{B-L}/\sqrt{\hbar c}=g_{B-L}$).
    Exclusion curves are sourced from the AxionLimits
    compilation~\cite{OHare:2020wah}.}
  \label{fig:ULDM_vector}
\end{figure}

\medskip
\noindent\textit{Species configurations.}\quad
Two species pairings are projected. The first pairs $^{87}$Sr with
$^{88}$Sr. Both isotopes share the 698\,nm clock transition. In $^{87}$Sr the
698\, laser system directly interrogates the transition, while in $^{88}$Sr additional lasers at 679~nm, 688~nm and 707~nm drive the transition indirectly but from a single direction~\cite{carman_collinear_2025}. The one additional neutron of $^{88}$Sr gives
$|\eta_{\rm eff}| \simeq 7.2 \times 10^{-20}$, and the solid curves in
Figure~\ref{fig:ULDM_vector} correspond to this pairing. The second
configuration pairs $^{87}$Sr with $^{171}$Yb; the substantially
different $B{-}L$ charge-to-mass ratios produce a larger
$|\eta_{\rm eff}|$ and hence greater sensitivities, shown as dashed curves. This option requires an
additional 578\,nm ytterbium clock laser and is an upgrade-path
configuration.

\medskip
\noindent\textit{Sensitivity projections.}\quad
The existing constraints on $g_{B-L}$ in the ULDM mass range are
shown in Figure~\ref{fig:ULDM_vector}. At low masses
($m_V \lesssim 2\times10^{-18}\ \mathrm{eV}$), the tightest direct-detection
bounds come from rotating torsion-balance measurements~\cite{Ross:2025uly}, which achieve
$g_{B-L} \leq 9\times10^{-26}$ at peak sensitivity, improving on
the earlier torsion-balance result of~\cite{Shaw:2021laz}
by a factor of three. In the intermediate range
$4\times10^{-19}$--$3\times10^{-17}\ \mathrm{eV}$, the world-leading
constraint is from the LISA Pathfinder free-fall data analysed in~\cite{Frerick:2023hjh}, which reach
$g_{B-L} \sim 1.5\times10^{-27}$ at the optimal mass
$m_V \approx 5\times10^{-18}\ \mathrm{eV}$. At higher masses above
$\sim3\times10^{-17}\ \mathrm{eV}$, the leading bounds are from the
E\"ot-Wash equivalence-principle test~\cite{Schlamminger:2007ht}
and the final MICROSCOPE result~\cite{Touboul:2022yrw} with
the final MICROSCOPE result~\cite{Touboul:2022yrw}. Treating the measured
$\eta(\mathrm{Ti,Pt})$ asymmetrically, this yields
$|g_{B-L}| \lesssim 1.1 \times 10^{-25}$ at low mediator masses
($95\%$~CL), evaluated as a static fifth-force constraint from the final
MICROSCOPE result following Appendix~E of~\cite{Amaral:2024tjg}~\cite{Fayet:2025pf}.
The AICE Baseline achieves
$g_{B-L} \approx 1.31 \times 10^{-29}$ with $^{87}$Sr/$^{88}$Sr,
an improvement of approximately $10^3$ over the best current
torsion-balance bound, driven by the effective momentum
transfer at the Baseline LMT order $n = 10^{3}$ over the full 140\,m Baseline.
The plateau sensitivities for the $^{87}$Sr/$^{88}$Sr pair are:
AICE Baseline $g_{B-L} \approx 1.31 \times 10^{-29}$;
AICE Stretch $g_{B-L} \approx 3.27 \times 10^{-32}$;
future 1\,km TVLBAI $g_{B-L} \approx 4.18 \times 10^{-33}$.
For the $^{87}$Sr/$^{171}$Yb upgrade pair the corresponding values are
$g_{B-L} \approx 2.41 \times 10^{-30}$ (Baseline),
$6.03 \times 10^{-33}$ (Stretch), and $7.67 \times 10^{-34}$ (1\,km).

\medskip
\noindent\textit{Realisation in AICE.}\quad
The vector search is implemented using the dual-species
Mach-Zehnder mode of Section~\ref{sec:concept:mode:dual}. For the
$^{87}$Sr/$^{88}$Sr configuration the primary hardware addition is a
second atom source for $^{88}$Sr and addition of high-power lasers at 679~nm, 688~nm and 707~nm. The dual-species common-mode rejection
methodology has been validated using the AION-10
prototype~\cite{baynham_prototype_2026}. This configuration provides the natural route to the vector ULDM
programme once dual-species operation has been established.

%% ── Axion - gluon coupling ─────────────────────────────────────────────────
\subsubsection{Axion-gluon coupling}

Additionally to the axion-nucleon coupling explored in Section~\ref{sec:uldm_axion} the AICE experiment can also be used to constrain the axion-gluon coupling by interrogating two isotopes or species simultaneously.

%The axion-gluon coupling can also be tested by interrogating two species simultaneously.
\medskip
\noindent\textit{Coupling and Signal.}\quad
We consider axion-like particles coupling to the gluon field strength via
\begin{equation}
    \mathcal{L} = \frac{ \sqrt{\hbar c} g_s^2}{32 \pi^2}\frac{a}{f_a} G_{\mu \nu}^a \tilde{G}^{a \mu \nu} \, ,
\end{equation}
with $f_a$ being a decay constant, $g_s$ the strong coupling and $\tilde{G}^{a \mu \nu}$ the dual gluon field strength.
The oscillating axion field modulates the masses of nucleons and atomic binding energies through a quadratic coupling to the pion mass~\cite{Kim:2024oscillationsatomicenergylevels}, inducing a time dependence in the atomic mass $m_A$ and the transition frequency $w_A$: 
\begin{equation}
    \begin{split}
        m_A(t) = m_A^0 (1 + Q_M^A \cos(2\omega_a t + \phi_0)), \\
        w_A(t) = w_A^0(1 + Q_M^A \cos(2\omega_a t + \phi_0)),
    \end{split}
\end{equation}
%$\text{d}m_A (t) = m_A^0 [Q_M^A]_a \text{d} \theta^2(t)$ 
with the dimensionless axionic mass charge $[Q_M^A]_a = \frac{\partial \ln m_A}{\partial (\theta^2)}$, where $\theta = \sqrt{\hbar c} a/f_a$ with the axion field $a$ \cite{Gue:2024onx, Kim:2024oscillationsatomicenergylevels}. 
This modulation induces an acceleration dependent on the atomic mass that leads to a differential phase in the interferometric sequence between different atomic isotopes or species. 

In a dual-isotope (DI) configuration, the wave number $k_0^A \approx k_0^B \approx k_0$ can be assumed as equal for both isotopes, so the interferometer phase is
\begin{equation}
    \delta \phi^{\rm DI}_\text{axion-gluon} = \frac{2 k_\text{eff}  \rho_\text{DM} v_\text{DM}}{\omega_a^3 f_a^2} |[Q_M^A]_a - [Q_M^B]_a| \sin (\omega_a T)^2 \, , 
\end{equation}
with $\omega_a$ the frequency of the axion field and $k_\text{eff} = n k_0$ \cite{Gue:2024onx}.
For the dual-species (DS) case $k_0^A \neq k_0^B$ and the phase is
\begin{equation}
    \delta \phi^{\rm DS}_\text{axion-gluon} = \frac{2 n  \rho_\text{DM} v_\text{DM}}{\omega_a^3 f_a^2} |[Q_M^A]_a k_0^A - [Q_M^B]_a k_0^B| \sin (\omega_a T)^2 \, ,
\end{equation}
where the larger difference in coupling and $k_0$ factors leads to enhanced sensitivities. However, a more careful consideration of the configuration is required to mitigate laser phase and other noise sources. A concrete study of a dual-species configuration for axion searches can be found in Ref.~\cite{VLBAI:2026} along with realistic noise projections.
%Therefore, the differential phase is calculated with the ratio of the two wave numbers $\delta \phi_\text{axion-gluon} = \phi_{A} - k_0^A / k_0^B \phi_{B}$, so that the interferometer phase for the dual species case is the same with $k_\text{eff} = n k_0^A$ \cite{VLBAI:2026}. 

\begin{figure}[ht]
  \centering
  \includegraphics[width=0.78\linewidth]{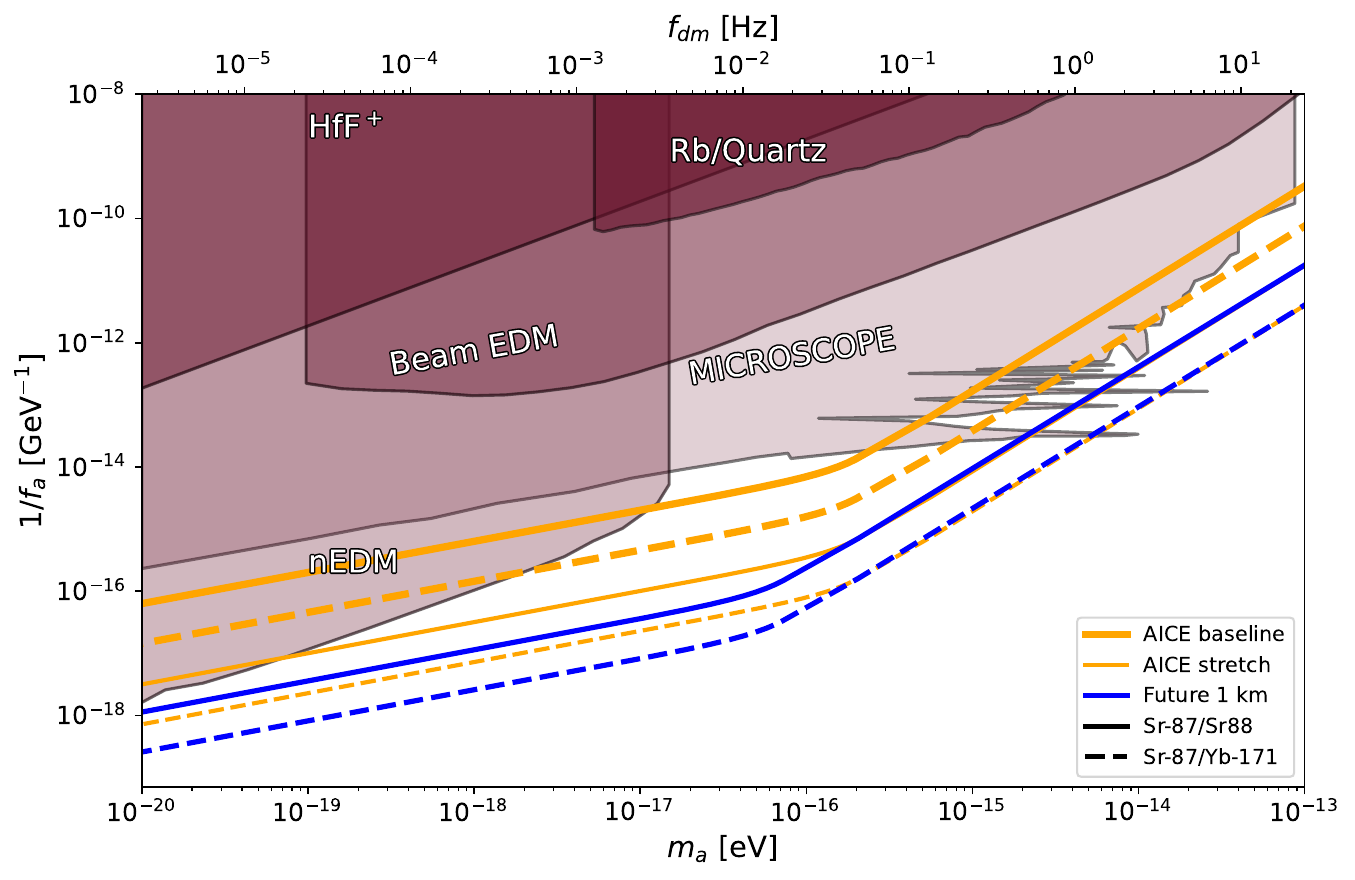}
  \caption{Projected AICE sensitivity to the axion-gluon coupling $1/f_a$ as a function of axion mass $m_a$; the Compton frequency $f_\text{DM}$ is displayed on the upper x-axis.
  Solid lines show the $^{87}$Sr/$^{88}$Sr configuration of each scenario, dashed lines the $^{87}$Sr/$^{171}$Yb configuration. 
  The AICE experiment sensitivities are shown in orange for the Baseline (solid line) and the Stretch (dashed line) scenarios, the future $1 \, \text{km}$ experiment is shown in blue.
  The shaded regions are the existing constraints for the axion-gluon coupling from the MICROSCOPE \cite{Touboul:2022yrw}, nEDM \cite{Abel:2017rtm}, Beam EDM \cite{Schulthess:2022}, HfF$^+$ \cite{Roussy:2021} and Rb/Quartz \cite{Zhang:2023} measurements.
  The exclusion curves are sourced from the AxionLimits compilation~\cite{OHare:2020wah}.}
  \label{fig:ULDM_axion_gluon}
\end{figure}

\medskip
\noindent\textit{Sensitivity projections.}\quad
The leading laboratory constraint on the axion-gluon coupling at low masses $m_a \leq 10^{-17} \, \text{eV}$ is the bound from nuclear spin precession in electric and magnetic fields (nEDM) \cite{Abel:2017rtm}.
For higher masses $10^{-17}\, \text{eV} \leq m_a \leq 10^{-13} \, \text{eV}$, the leading laboratory constraint is the final MICROSCOPE result \cite{Touboul:2022yrw}.
The existing constraints in Fig.~\ref{fig:ULDM_axion_gluon} are from the AxionLimits compilation \cite{OHare:2020wah}, where further astrophysical bounds are also displayed. However, in this plot we focus on the laboratory constraints as complementary direct probes of axion dark matter.

The AICE Baseline sensitivity improves the MICROSCOPE bounds in the mass range from  $10^{-17} \, \text{eV} \leq m_a \leq 10^{-16}\, \text{eV}$ in the dual-isotope $^{87}$Sr/$^{88}$Sr configuration. 
For the dual-species $^{87}$Sr/$^{171}$Yb configuration the AICE Baseline improves the bounds of nEDM and MICROSCOPE in approximately the mass range $ 10^{-18} \, \text{eV} \leq m_a \leq 10^{-15}\, \text{eV}$. 
The AICE Stretch and future $1\, \text{km}$ experiments improve further the nEDM and MICROSCOPE bounds in both configurations in the observed mass range.
The dual-species configuration leads to better sensitivities, as the difference $k_0^A Q_M^A - k_0^B Q_M^B$ is bigger for the two different species. 
Furthermore, the improvement from the AICE Baseline to AICE Stretch configuration is greater than that from AICE Stretch to a future $1 \, \text{km}$ experiment, as the phase largely depends on the LMT number $n$ and the time $T$, so increasing both of these leads to a higher increase in the sensitivity than going to larger Baseline $L$.
%In the dual-species $^{87}$Sr/$^{171}$Yb configuration the AICE Baseline sensitivity improves the MICROSCOPE results in the mass range from $10^{-17} \, \text{eV} \leq m_a \leq 10^{-16}\, \text{eV}$. 
%In this configuration, the AICE Stretch and the future $1 \, \text{km}$ interferometers improve the nEDM and the MICROSCOPE results in the entire mass range. 
%In the dual-isotope $^{87}$Sr/$^{88}$Sr configuration, the AICE Stretch and future $1 \, \text{km}$ experiments improve upon the existing constraints in the mass range of $10^{-17} \, \text{eV} \lesssim m_a \lesssim 10^{-16}\, \text{eV}$.

\medskip
\noindent\textit{Realisation in AICE.}\quad
The search for the axion-gluon coupling is carried out in the dual-species Mach-Zehnder mode described in Section~\ref{sec:concept:mode:dual}. 
Similarly to the vector search explained in Section~\ref{sec:uldm_vector}, this search can be implemented for the $^{87}$Sr/$^{88}$Sr configuration or the $^{87}$Sr/$^{171}$Yb configuration.
For the $^{87}$Sr/$^{171}$Yb configuration the larger difference in the masses and the wave numbers lead to larger $|[Q_M^A]_a k_0^A - [Q_M^B]_a k_0^B|$ and, therefore, to higher sensitivities. 
A concrete realisation of this configuration can be seen in Ref.~\cite{VLBAI:2026}, for a 10 m scale dual-species experiment which may act as a pathfinder to be further developed by AICE.
\subsubsection{Other ULDM possibilities}

%% ── Remaining ULDM stubs ─────────────────────────────────────────────────

\paragraph{Tensor dark matter.}

\begin{figure}[ht]
  \centering
  \includegraphics[width=0.8\linewidth]{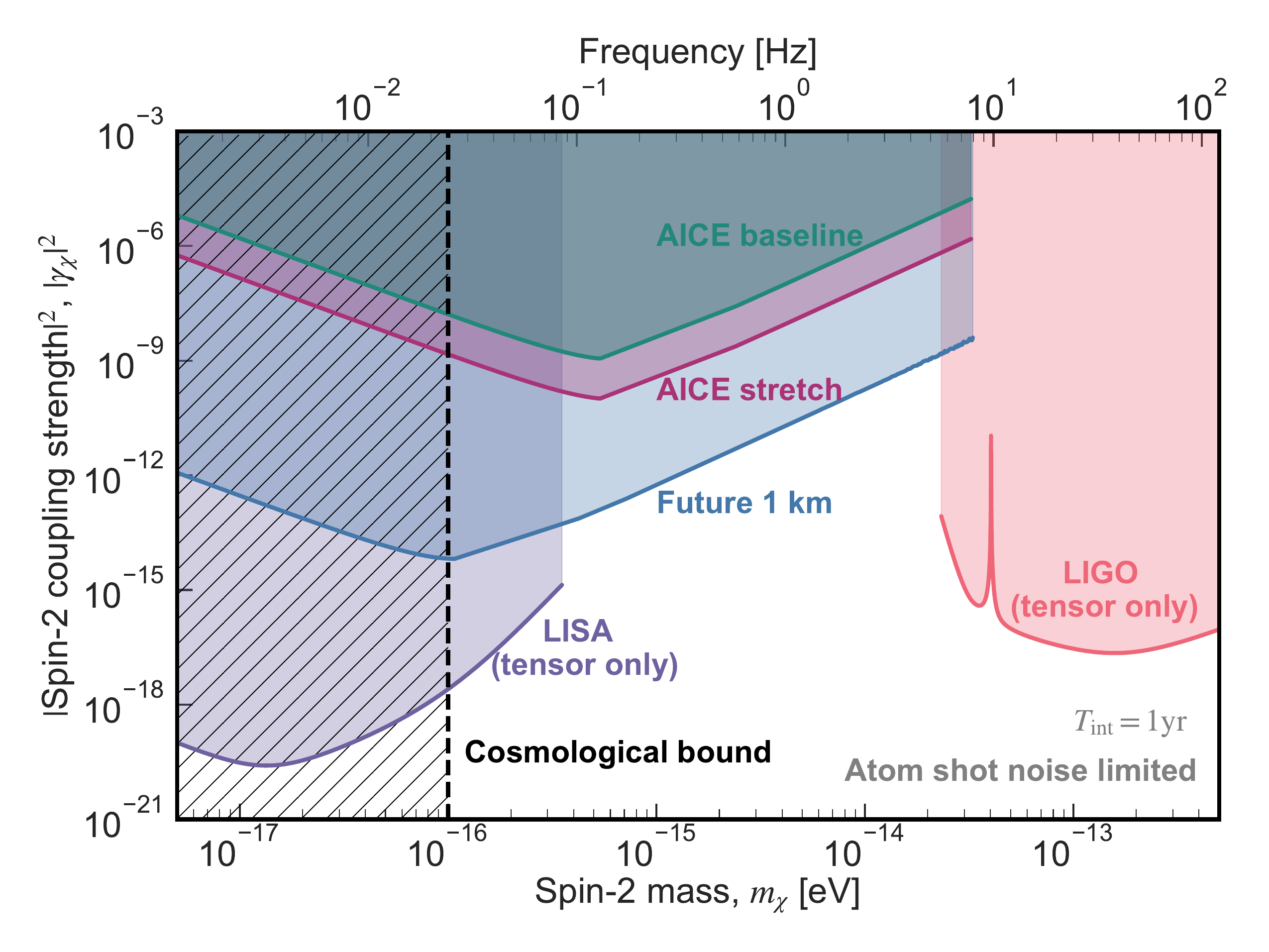}
  \caption{Projections for the SNR = 1 sensitivity to the spin-2 ULDM coupling strength $|\gamma_\chi|^2$ of AICE and a future 1~km atom interferometer experiment~\cite{Blas:2024kps} are compared with the sensitivities of LIGO~\cite{LIGOScientific:2014pky} and LISA~\cite{LISA:2017pwj}. One-year measurement campaigns and operation at the atom shot noise limit are assumed for the atom interferometer experiments. The model-dependent cosmological stability bound~\cite{HIGUCHI1987397} is also displayed.}
  \label{fig:Tensor}
\end{figure}

AICE may also be used to search for tensor (spin-2) dark matter~\cite{deRham:2014zqa} in the form of waves of massive graviton-like bosonic fields~\cite{Blas:2024kps}. This search can be conducted in parallel with the search for ultralight scalar dark matter discussed above. Fig.~\ref{fig:Tensor} displays projections for the SNR = 1 sensitivities of AICE and a future 1~km atom interferometer experiment to the spin-2 ULDM coupling strength $|\gamma_\chi|^2$. The sensitivity curves have been calculated assuming one-year measurement campaigns using $^{87}$Sr, the Baseline and Stretch AICE experimental parameters given in Table~\ref{tab:roadmap}, and operation at the atom shot noise limit. The atom interferometer sensitivities are compared to those of LIGO~\cite{LIGOScientific:2014pky} and LISA~\cite{LISA:2017pwj}, and cover the range of mass where they lose their sensitivities. The model-dependent cosmological stability bound on a lower-mass spin-2 field~\cite{HIGUCHI1987397} is also displayed. 

%% --------------------------------------------------------------------
%% BLANKED OUT pending completion of Section 2.2.5 (Tensor dark matter).
%% Figure kept in source for future reinstatement with updated numbers
%% consistent with the AICE performance ladder in Table~\ref{tab:TVLBAI-performance}.
%% To reinstate: delete the \iffalse line below and the matching \fi line.
%% --------------------------------------------------------------------
%% --------------------------------------------------------------------

\paragraph{Quadratic couplings.} 
Quadratic couplings of spin-0 fields to Standard Model particles are expected to be generic features of ULDM models, appearing at the loop level~\cite{Beadle:2023flm} if not already at the tree level. Prospective atom interferometer sensitivities to quadratic ULDM couplings have been considered in~\cite{Beadle:2023flm} and~\cite{Gue:2024onx}. As shown in Fig.~5 of the latter reference, searches for quadratic couplings with a ${\cal O}(100)$~m atom interferometer may be more sensitive to such models than current experimental constraints for ULDM masses $\gtrsim 10^{-17}\ \mathrm{eV}$. \\
%\textcolor{red}{Recalculate~\cite{Gue:2024onx} for AICE or drop?}

%% --------------------------------------------------------------------
%% BLANKED OUT pending completion of Section 2.2.6 (Quadratic couplings).
%% Figure kept in source for future reinstatement with updated numbers
%% consistent with the AICE performance ladder in Table~\ref{tab:TVLBAI-performance}.
%% To reinstate: delete the \iffalse line below and the matching \fi line.
%% --------------------------------------------------------------------
\iffalse
\begin{figure}[ht]
  \centering
  \includegraphics[width=0.8\linewidth]{figures/Gue_Hees_Wolf.png}
  \caption{AICE sensitivity to quadratic axion couplings~\cite{Gue:2024onx}. \textcolor{red}{Any chance of generating a similar plot for AICE, or shall we drop this topic?}}
  \label{fig:Quadratic}
\end{figure}
\fi
%% --------------------------------------------------------------------

\begin{figure}[ht]
  \centering
  \includegraphics[width=0.45\linewidth]{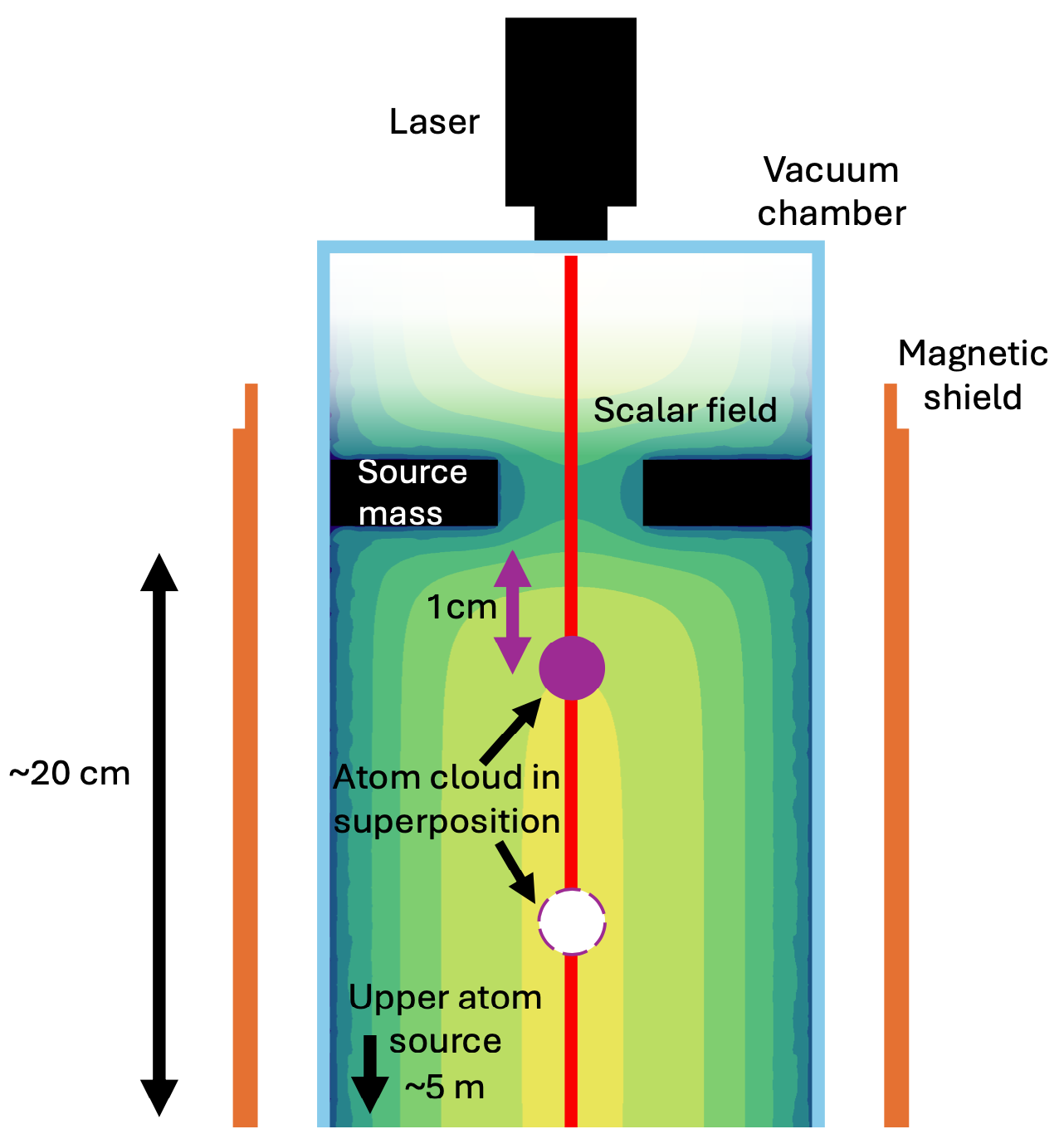}
\includegraphics[width=0.45\linewidth]{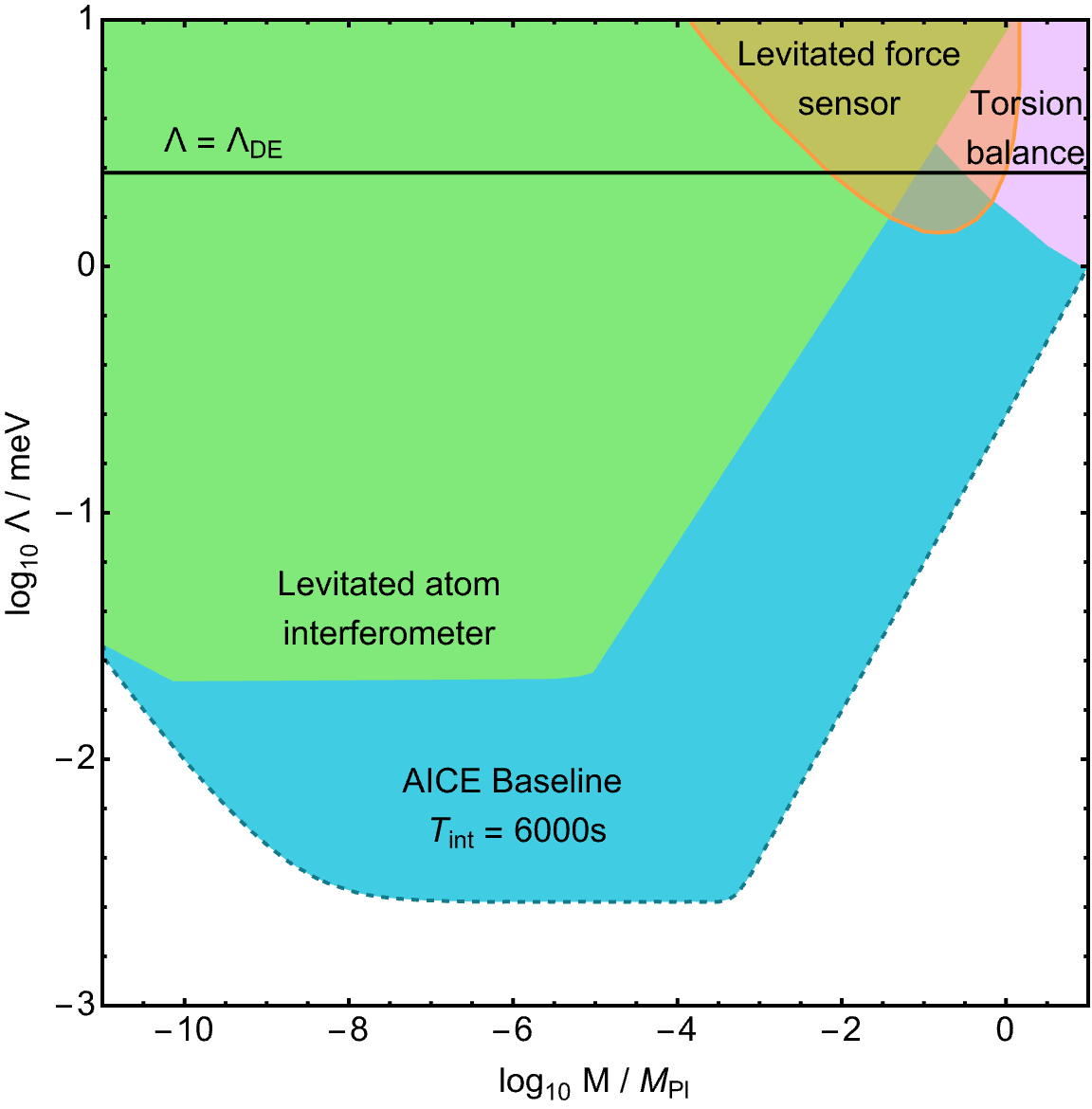}
\caption{Left panel: Schematic of the top of the proposed experimental setup~\cite{Banks:2025vvz} to search for screened scalar fields such as chameleons and symmetrons, showing the upper interferometer of the gradiometer (not to scale). A 1 cm thick annular source mass is positioned at the top of the vacuum chamber,
with a central hole of radius 40\% of that of the vacuum chamber to allow for laser transmission. A superposition of atom clouds is shown, with the upper cloud approaching the plate to within 1 cm and the lower cloud remaining distant.  Right panel: The sensitivity of AICE to chameleon parameters is compared to current experimental constraints on chameleon models: Baseline operation for 6000~s is assumed.} 
   %The scalar force drops off exponentially with distance, barely affecting the lower cloud. A second, lower identical interferometer (not shown) is manipulated simultaneously using common laser pulses (depicted by the red line).}
  \label{fig:Chameleon}
\end{figure}

\paragraph{Chameleons and symmetrons.}
Several dark-sector models predict scalars that give rise to fifth-force interactions subject to density-dependent screening mechanisms, enabling them to evade many experimental constraints. Examples include chameleons~\cite{Khoury:2003aq} and symmetrons~\cite{Hinterbichler:2010es}. AICE may be used to search for such particles by equipping it with an annular planar source mass inside the vacuum chamber~\cite{Banks:2025vvz}. As illustrated in Fig.~\ref{fig:Chameleon}, one of the clouds in the upper interferometer approaches to within ${\cal O}(1)$~cm of the source mass, while the other cloud is sufficiently distant for its influence to be negligible. This induces a phase shift that is compared with that in a second interferometer that is not affected by the source mass. The planned operational procedure to distinguish the static fifth force from backgrounds and from the plate’s Newtonian gravity, calibrate the measurement, characterise the plate’s Newtonian gravity and reach shot-noise-limited sensitivity is described in~\cite{Banks:2025vvz}. As seen in the right panel of Fig.~\ref{fig:Chameleon}, the proposed measurement could probe chameleon parameter space well beyond the existing bounds, and analogous projections for symmetron models are given in~\cite{Banks:2025vvz}.
%% --------------------------------------------------------------------
%% BLANKED OUT pending completion of Section 2.2.7 (Chameleons and symmetrons).
%% Figure kept in source for future reinstatement with updated numbers
%% consistent with the AICE performance ladder in Table~\ref{tab:TVLBAI-performance}.
%% To reinstate: delete the \iffalse line below and the matching \fi line.
%% --------------------------------------------------------------------

%% --------------------------------------------------------------------

%% ── Programme summary ────────────────────────────────────────────────────

\subsubsection{Programme summary and {phased} implementation}
\label{sec:uldm_programme}

The channels described above correspond to distinct operating
modes of the same facility. Their implementation within AICE is
summarised in Tables~\ref{tab:uldm_campaigns} and~\ref{tab:uldm_sensitivity}.

\begin{table}[ht]
  \centering
  \caption{AICE operating modes for the three primary ULDM searches.
    The interrogation time $T$ corresponds to $T_{\mathrm{grad}} = 2.5$\,s for the
    scalar gradiometer, to $T_{\mathrm{MZ}} = 5.0$\,s for the vector full-Baseline
    Mach-Zehnder, and to $T_{\mathrm{ax}} = 5.0$\,s for the axion full-shaft
    Mach-Zehnder; the vector and axion modes share the same full one-way drop.
    The LMT order $n$ is quoted at the Baseline value;
    at Stretch and 1\,km, $n = 4\times10^{4}$ is used for the scalar and
    vector channels (see Table~\ref{tab:roadmap}).
    The effective momentum transfer $\keff$ uses the single-photon
    convention $\keff = nk_0$ for the 698\,nm $^1S_0$--$^3P_0$
    transition.}
    \vspace{3mm}
  \label{tab:uldm_campaigns}
  \smallskip
  \begin{tabular}{p{2.4cm}p{2.7cm}p{3.1cm}ccp{1.8cm}}
    \toprule
    Channel & Species & Configuration & $T$ [s] & $n$ & $\keff$ \\
    \midrule
    Scalar ($d_{m_e}$, $d_e$)
    & $^{87}$Sr (single)
    &  gradiometer
    & 2.5 & $10^{3}$ & $nk_0$ \\
    Vector ($g_{B-L}$)
    & $^{87}$Sr/$^{88}$Sr or $^{87}$Sr/$^{171}$Yb
    & Dual-species Mach-Zehnder
    & 5.0 & $10^{3}$ & $nk_0$ \\
    Axion ($G_{an}$)
    & $^{87}$Sr (single)
    & Full-shaft Mach-Zehnder
    & 5.0 & ${\geq}1$ & $k_0$ \\
    \bottomrule
  \end{tabular}
\end{table}

\begin{table}[ht]
  \centering
  \caption{Plateau sensitivities for each channel and scenario,
    corresponding to the parameter ladder of Table~\ref{tab:roadmap}.
    The roll-off mass $\mstar = \pi\hbar/T$ is evaluated at
    $T_{\mathrm{ax}}$ for the axion channel, at $T_{\mathrm{MZ}}$ for the
    vector channel, and at $T_{\mathrm{grad}}$ for the scalar channel.
    Vector sensitivities are quoted separately for the two species
    pairs, with $^{87}$Sr/$^{88}$Sr as the Baseline pair and
    $^{87}$Sr/$^{171}$Yb as the upgrade pair.
    Scalar $|d_{m_e}|$ and $|d_e|$ values are quoted at the optimal
    mass near $\mstar$; the full mass dependence is shown in
    Figure~\ref{fig:ULDM_scalar}.}
  \label{tab:uldm_sensitivity}
    \vspace{3mm}
  \begin{tabular}{p{2.4cm}p{3.2cm}p{3.2cm}p{2.6cm}}
    \toprule
    Scenario & Channel & Plateau sensitivity & Roll-off mass \\
    \midrule
    AICE Baseline & $|G_{an}|$
      & $9.45\times10^{-10}\,\mathrm{GeV}^{-1}$
      & $4.1\times10^{-16}\ \mathrm{eV}$ \\
    AICE Stretch  & $|G_{an}|$
      & $9.45\times10^{-11}\,\mathrm{GeV}^{-1}$
      & $4.1\times10^{-16}\ \mathrm{eV}$ \\
    1\,km         & $|G_{an}|$
      & $3.38\times10^{-11}\,\mathrm{GeV}^{-1}$
      & $1.5\times10^{-16}\ \mathrm{eV}$ \\[4pt]
    AICE Baseline & $g_{B-L}$ ($^{87}$Sr/$^{88}$Sr)
      & $1.31\times10^{-29}$
      & ${\sim}4\times10^{-16}\ \mathrm{eV}$ \\
    AICE Stretch  & $g_{B-L}$ ($^{87}$Sr/$^{88}$Sr)
      & $3.28\times10^{-32}$
      & ${\sim}4\times10^{-16}\ \mathrm{eV}$ \\
    1\,km         & $g_{B-L}$ ($^{87}$Sr/$^{88}$Sr)
      & $4.18\times10^{-33}$
      & ${\sim}1.5\times10^{-16}\ \mathrm{eV}$ \\[4pt]
    AICE Baseline & $g_{B-L}$ ($^{87}$Sr/$^{171}$Yb)
      & $2.41\times10^{-30}$
      & ${\sim}4\times10^{-16}\ \mathrm{eV}$ \\
    AICE Stretch  & $g_{B-L}$ ($^{87}$Sr/$^{171}$Yb)
      & $6.03\times10^{-33}$
      & ${\sim}4\times10^{-16}\ \mathrm{eV}$ \\
    1\,km         & $g_{B-L}$ ($^{87}$Sr/$^{171}$Yb)
      & $7.70\times10^{-34}$
      & ${\sim}1.5\times10^{-16}\ \mathrm{eV}$ \\[4pt]
    AICE Baseline & $|d_{m_e}|$
      & $9.53\times10^{-6}$
      & ${\sim}8\times10^{-16}\ \mathrm{eV}$ \\
    AICE Stretch  & $|d_{m_e}|$
      & $2.39\times10^{-8}$
      & ${\sim}8\times10^{-16}\ \mathrm{eV}$ \\
    1\,km         & $|d_{m_e}|$
      & $9.35\times10^{-10}$
      & ${\sim}3\times10^{-16}\ \mathrm{eV}$ \\[4pt]
    AICE Baseline & $|d_e|$
      & $4.62\times10^{-6}$
      & ${\sim}8\times10^{-16}\ \mathrm{eV}$ \\
    AICE Stretch  & $|d_e|$
      & $1.16\times10^{-8}$
      & ${\sim}8\times10^{-16}\ \mathrm{eV}$ \\
    1\,km         & $|d_e|$
      & $4.54\times10^{-10}$
      & ${\sim}3\times10^{-16}\ \mathrm{eV}$ \\
    \bottomrule
  \end{tabular}
\end{table}

The AICE ultralight dark-matter programme is realised through a
sequence of operating configurations of the same instrument, each
optimised for a different class of signals.
The representative operating parameters for each configuration and
the corresponding plateau sensitivities are summarised in
Tables~\ref{tab:uldm_campaigns} and~\ref{tab:uldm_sensitivity},
which define the benchmark scenarios used in the sensitivity
projections throughout this section.
While the detailed scheduling will depend on the technical evolution
of the facility and the outcome of commissioning, an indicative
progression can be outlined.

The initial period of operation is expected to focus on commissioning
the instrument in a strontium-based  gradiometer configuration.
The primary objectives are to establish stable operation, characterise
noise sources, and validate interferometric performance over the full
baseline. Physics sensitivity during this period is not expected to be
competitive, although parasitic measurements, for example related to
precision determinations of fundamental constants or equivalence-principle
tests, may already be possible.

Following commissioning, the programme naturally transitions into
sustained operation in the gradiometer configuration, which provides
the Baseline sensitivity for scalar ultralight dark matter in both the
$d_{m_e}$ and $d_e$ channels. This configuration is also closely related to the
gravitational-wave sensitivity described in Section~\ref{sec:gw},
as the gradiometer geometry provides the common-mode rejection
relevant for both signal classes.

A natural second stage of the programme would be to introduce a second atomic species
with dual-species operation, giving access to vector $B{-}L$ dark-matter and other
signatures. The choice of species---for example $^{87}$Sr/$^{88}$Sr or
$^{87}$Sr/$^{171}$Yb---and the detailed implementation will depend on
ongoing developments in source preparation and interferometric control.

The search for pseudoscalar (axion-like) dark matter would be realised in a distinct operating
mode, employing a full-baseline single-species Mach-Zehnder interferometer
using $^{87}$Sr with $T_{\mathrm{ax}} = 5.0$\,s. This configuration differs from the
gradiometer and is expected to be deployed at a later stage, once
the required level of interferometric control has been achieved.

In the longer term, improvements to the phase noise---e.g., through a higher atom
flux and/or quantum-projection noise reduction via squeezing---would benefit all three ULDM
channels proportionally without requiring changes to the Baseline geometry
or species programme. A future km-scale detector within the TVLBAI
framework~\cite{TVLBAISummary} would extend the reach of all three
channels by approximately one order of magnitude in coupling, accessing
axion parameter space currently bounded primarily by SN\,1987A and reaching
the $10^{-32}$ level in $g_{B-L}$.

This phased approach is intended as a flexible framework rather than a
fixed timeline. The relative duration and ordering of individual stages
will ultimately be determined by technical progress, commissioning
outcomes, and the evolving scientific priorities of the AICE programme {including the opportunities for networking with other long-baseline atom interferometers {\it \`a la} LIGO-Virgo-KAGRA}.

%%------------------------------------------------------------

%%------------------------------------------------------------

%% --------------------------------------------------------------------
%% BLANKED OUT pending completion of Section 2.3 (Gravitational waves in the mid-frequency band).
%% Figure kept in source for future reinstatement with updated numbers
%% consistent with the AICE performance ladder in Table~\ref{tab:TVLBAI-performance}.
%% To reinstate: delete the \iffalse line below and the matching \fi line.
%% --------------------------------------------------------------------
\iffalse
\begin{figure}[ht]
  \centering
  \includegraphics[width=0.8\linewidth]{CharacteristicStrain_CERN_Sandstone-1.pdf}
  \caption{AICE gravitational-wave sensitivity to characteristic
    strain, covering the mid-frequency band
    (from~\cite{Arduini:2851946}).}
  \label{fig:GWs}
\end{figure}
\fi
%% --------------------------------------------------------------------

%%------------------------------------------------------------
\subsection{Precision physics measurements}
%%------------------------------------------------------------

\subsubsection{Equivalence principle}
\label{sec:ep}

The universality of free fall (UFF), a cornerstone of the Einstein
equivalence principle (EP), states that the gravitational acceleration of two
bodies is independent of their composition. {Violations are predicted in many extensions of the Standard Model, including
scalar-tensor theories, and light ultralight scalar
dark matter such as dilatons} with non-universal couplings to Standard Model
fields~\cite{Gue:2024onx}. The phenomenological figure of merit is the
E\"otv\"os ratio
\begin{equation}
  \eta_{AB} \;=\; 2\,\frac{a_A - a_B}{a_A + a_B}\,,
  \label{eq:eta}
\end{equation}
where $a_A$ and $a_B$ are the accelerations of two test bodies $A$ and
$B$ in the same gravitational field. Macroscopic torsion balances have
established $\eta \lesssim 2\times 10^{-13}$ for
Be/Ti~\cite{Schlamminger:2007ht}, and the MICROSCOPE space mission has
since improved this bound to
$\eta(\mathrm{Pt},\mathrm{Ti}) \lesssim 10^{-15}$~\cite{Touboul:2022yrw},
following its first results in~\cite{Touboul:2017grn}.
Atom-based tests of UFF have progressed from $\eta \sim 10^{-7}$
\cite{Schlippert2014,Tarallo2014} and $10^{-8}$ \cite{Zhou2015} to a
current best of
$\eta(\,^{85}\mathrm{Rb},\,^{87}\mathrm{Rb}) = (1.6\pm 1.8_{\mathrm{stat}}
\pm 3.4_{\mathrm{syst}})\times 10^{-12}$ in a 10\,m
tower~\cite{Asenbaum2020}. For comparison, a first in-orbit atom interferometer test of the WEP has obtained $\eta(\,^{85}\mathrm{Rb},\,^{87}\mathrm{Rb}) = (-3.1 \pm 4.6) \times 10^{-7}$~\cite{Zhang:2026lzu}, while proposed space missions such as STE-QUEST target
$\eta \lesssim 10^{-17}$ with $^{85}$Rb\,/\,$^{41}$K quantum test
masses~\cite{Aguilera:2014stequest,Ahlers:2022stequest}. A ground-based atom-interferometric
experiment at the 140\,m scale taking place between the 10\,m towers and
the proposed space missions, is a natural next step. {The AICE facility} realises this
opportunity by using its dual-source shaft geometry, long interrogation
time, and single-photon clock-transition interferometry to deliver a
staged EP programme embedded within the same timeline as the
ULDM and $\alpha$ campaigns, and even allows for tests of other aspects of the EP beyond UFF~\cite{DiPumpo2021,DiPumpo_2023-T3}.

\paragraph{Measurement principle at AICE.}
A dual-species Mach-Zehnder configuration co-locates two atomic
species $A$ and $B$ and interrogates them with matched
$\pi/2$--$\pi$--$\pi/2$ pulse sequences over the full Baseline. The
differential acceleration
$\Delta a = a_A - a_B$ is extracted from the differential interferometer
phase
\begin{equation}
  \Delta\phi_{AB} \;=\; k_{\mathrm{eff}}\,\Delta a\,T^{2}_{\mathrm{MZ}}\,,
  \label{eq:dphi-ep}
\end{equation}
where $k_{\mathrm{eff}} = n \cdot 2\pi/\lambda$ is the effective
momentum transfer due to LMT of order $n$ on the clock transition of
wavelength $\lambda$, and $T_{\mathrm{MZ}}$ is the full-baseline
interrogation time.
For the AICE dual-species parameters ($T_{\mathrm{MZ}} = 5.0$\,s on the
140\,m shaft, with $n = 10^{3}$ at Baseline and $n = 4\times 10^{4}$ at
Stretch), the scale factor $k_{\mathrm{eff}}\,g\,T_{\mathrm{MZ}}^{2}$
reaches $\sim 2\times 10^{12}$\,rad at Baseline and increases to
$\sim 9\times 10^{13}$\,rad at Stretch, so that a differential phase
resolution $\delta\phi \sim 10^{-4}$\,rad at Baseline translates into
\begin{equation}
  \eta \;\approx\; \frac{\delta\phi}{k_{\mathrm{eff}}\,g\,T_{\mathrm{MZ}}^{2}}
  \;\approx\; 5\times 10^{-17}\,
  \left(\frac{\delta\phi}{10^{-4}\,\mathrm{rad}}\right)\,.
  \label{eq:eta-phase}
\end{equation}
The shot-noise-limited sensitivity for AICE Baseline (Stretch)
parameters is formally far below this figure, of order $10^{-21}$
($10^{-23}$), so the EP measurement is systematics-limited rather than
statistics-limited. The defensible targets quoted below are therefore
set by the achievable control of species-differential effects rather
than by counting statistics, and are unchanged by the move to the
full-baseline Mach-Zehnder time. The EP observable is intrinsically differential
and does not depend on the absolute calibration of the gravitational
field, which enters as a common factor in both interferometers and
cancels to the level set by species-differential systematics.

\paragraph{Species configurations.}
AICE implements two dual-species modes, used in sequence across the
facility timeline.

\emph{Isotopic validation configuration $^{87}$Sr\,/\,$^{88}$Sr.} Both
isotopes are addressed on the same 698\,nm $^{1}S_{0}\!\to\!{}^{3}P_{0}$
clock transition, sharing laser beams at similar wavelengths in the 679-707~nm range. The fermion-boson pair
$^{87}\mathrm{Sr}\,(I=9/2)$ and $^{88}\mathrm{Sr}\,(I=0)$ rejects some systematics in common-mode.
This configuration serves as a systematics-validation
campaign. The small expected signal makes it an ideal null test of the
dual-species readout, species co-location alignment, and
gravity-gradient cancellation protocols, at an AICE target sensitivity
$\eta \lesssim 10^{-13}$.

\emph{Flagship configuration $^{87}$Sr\,/\,$^{171}$Yb.} The two species
are addressed on their respective clock transitions, 698\,nm for
$^{87}$Sr and 578\,nm for $^{171}$Yb, using the dual-wavelength laser
{beams} deployed for the vector ULDM campaign. The nuclear-spin-carrying
fermionic isotopes $^{87}$Sr\,($I=9/2$) and $^{171}$Yb\,($I=1/2$)
provide large compositional contrast through their neutron-to-proton
ratios ($N/Z = 49/38$ for $^{87}$Sr vs.\ $100/70$ for $^{171}$Yb), enhancing sensitivity to
composition-dependent violations. This is the main AICE EP physics
result. The dual-species Yb/Sr pairing has an analogue in the Yb/Rb
proposal at VLBAI~\cite{Hartwig2015} and is complementary to the Rb/K
approach of STE-QUEST~\cite{Aguilera:2014stequest,Ahlers:2022stequest},
extending the search to a single-photon clock-transition {pairing}.

\paragraph{Staged AICE EP programme - positioning in the landscape.}
The EP programme is distributed across the main facility phases and
reuses the dual-source, long-Baseline infrastructure without dedicated
facility time. The sequence of AICE campaigns and their positioning
within the equivalence-principle landscape is summarised in
Table~\ref{tab:ep_landscape}. An isotopic Sr/Sr campaign during
commissioning and the scalar ULDM phase validates the dual-species
systematics chain. The vector ULDM phase introduces the
$^{87}$Sr/$^{171}$Yb configuration, which simultaneously serves as the
vector $B{-}L$ ULDM search and the flagship EP measurement at
$\eta \lesssim 10^{-14}$. A dedicated Ultimate EP campaign at Stretch
parameters targets $\eta \lesssim$ few$\,\times 10^{-16}$, reaching
beyond the MICROSCOPE level~\cite{Touboul:2022yrw} with a systematically
distinct atom-interferometric technique and test-mass composition.
Beyond the composition-dependent test, the $^{87}$Sr/$^{88}$Sr isotopic
pair also probes spin-gravity couplings, since $^{87}$Sr carries nuclear
spin while $^{88}$Sr does not~\cite{Tarallo2014}.

\begin{table}[htbp]
\centering
\caption{Equivalence-principle landscape and the AICE programme. The
  upper block lists three established benchmarks, namely macroscopic
  torsion balances, the MICROSCOPE space mission, and the best current
  ground-based atom-interferometric result. The middle block gives the
  AICE campaigns, distributed across the facility phases defined in
  Table~\ref{tab:roadmap}. The lowest row summarises the sensitivity of a representative
  proposed space mission. Intermediate AICE targets are indicative
  placeholders derived from the AICE performance ladder. The Ultimate
  AICE target of $\eta \lesssim$ few$\,\times 10^{-16}$ is the firm
  programme objective.}
  \hspace{3mm}
\label{tab:ep_landscape}
\small
\begin{tabular}{@{}p{3.1cm}p{3.9cm}p{2.3cm}p{4.6cm}@{}}
\toprule
Experiment / phase & Configuration & Target $\eta$ & Notes \\
\midrule
Torsion balance            & Be/Ti                                & $\lesssim 2\times 10^{-13}$ & Achieved~\cite{Schlamminger:2007ht} \\
MICROSCOPE (space)         & Pt/Ti                                & $\lesssim 10^{-15}$         & Achieved~\cite{Touboul:2022yrw} \\
Atom interferometer, 10\,m & $^{85}$Rb/$^{87}$Rb                  & ${\sim}\,10^{-12}$          & Achieved~\cite{Asenbaum2020} \\
\midrule
\multicolumn{4}{@{}l}{\textit{AICE programme, 140\,m dual-source shaft}} \\
\midrule
Commissioning              & $^{87}$Sr/$^{88}$Sr, Initial         & $\lesssim 10^{-12}$         & Isotopic validation, same-wavelength common-mode rejection \\
Scalar ULDM                & $^{87}$Sr/$^{88}$Sr, Baseline        & $\lesssim 10^{-13}$         & Refined isotopic validation at full Baseline control \\
Vector $B{-}L$             & $^{87}$Sr/$^{171}$Yb, Baseline       & $\lesssim 10^{-14}$         & Flagship dual-wavelength EP result, co-located with vector ULDM \\
Ultimate EP                & $^{87}$Sr/$^{171}$Yb, Stretch        & few$\,\times 10^{-16}$      & Terrestrial bridge to space missions \\
\midrule
STE-QUEST (space)          & $^{85}$Rb/$^{41}$K                   & $\lesssim 10^{-17}$         & Proposed~\cite{Aguilera:2014stequest,Ahlers:2022stequest} \\
\bottomrule
\end{tabular}
\end{table}

\paragraph{Systematics and traceability.}
The systematic budget for the AICE EP programme is inherited from the ULDM
programme (Section~\ref{sec:uldm}) and is extended by the
species-differential sensitivity specific to a dual-species test.
Common-mode systematics (gravity gradients, Coriolis, wavefront
distortion, laser phase noise, vibration) are mitigated to leading
order by the dual-source dual-species common-mode
readout, to be first characterised in the $^{87}$Sr/$^{88}$Sr isotopic campaign.
Species-differential systematics are dominated by initial co-location
offsets and differences in launch velocities (mitigated by a common-trap
preparation stage followed by species-resolved release), differential
gravity-gradient coupling (mitigated by the gravity-gradient-nulling
frequency offset technique used in the ULDM Baseline configuration),
differential scale factors for Sr and Yb clock transitions (calibrated
in situ via repeated single-species gravimetry), and magnetic and
blackbody-radiation shifts, calibrated on the clock transitions. The frequency
references for both clock transitions are traceable to the SI second
through optical-frequency combs, and the consistency of the AICE EP
result with MICROSCOPE~\cite{Touboul:2022yrw} and with independent
atom-interferometric tests is itself a cross-check of the
dual-species systematics control.

\subsubsection{Fine-structure constant}
\label{sec:alpha}

\begin{figure}
    \centering
    \includegraphics[width=0.99\textwidth]{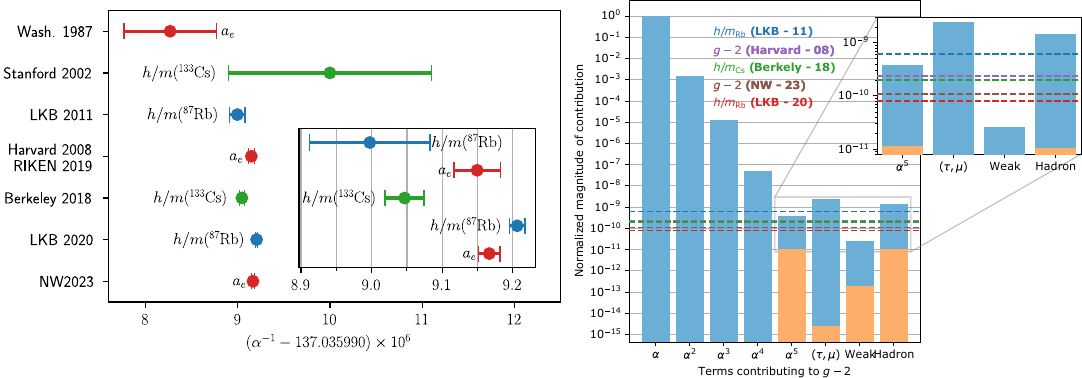}
    \vspace{-0.2cm}
    \caption{{Left panel: Comparison of the most precise determinations of the
    fine-structure constant $\alpha$. The red points $a_e$ are from $g_e-2$ measurements
    and QED calculations, the green and blue points are obtained from measurements of
    caesium and rubidium atomic recoils, respectively. The data are from
    \cite{VanDyck1987} (Washington 1987), \cite{Wicht_2002} (Stanford 2002),
    \cite{Bouchendira2011} (Paris 2011), \cite{Hanneke2008} (Harvard 2008),
    \cite{aoyama2020,atomsAoyama} (RIKEN 2019), \cite{parker2018} (Berkeley 2018),
    \cite{morel2020} (Paris 2020) and \cite{fan2023} (Northwestern 2023). The inset
    compares the most accurate values. Right panel: The relative magnitudes of different
    contributions to the electron anomalous magnetic moment $a_e$ (blue) with their
    uncertainties (orange) showing, from left to right: electronic QED contribution
    calculated to ${\cal O}(\alpha^5)$, known exactly at the lowest four orders, muon and tau
    contributions, weak interaction and hadronic contributions. The dashed lines
    represent the experimental precision achieved by the caesium and rubidium
    experiments as well as the Harvard and Northwestern $g_e-2$ measurements.}}
    \label{fig:alpha_values}
\end{figure}

The AICE precision programme addresses the more-than-$5\sigma$ tension between the two
most precise atom-interferometry determinations of the fine-structure constant, obtained
with caesium in 2018~\cite{parker2018} and rubidium in 2020~\cite{morel2020}. AICE
provides an independent determination of $\alpha$ using single-photon clock-transition
interferometry on strontium and ytterbium, a systematically distinct measurement family
whose consistency with the alkali values is a probe of the origin of the tension.
A determination of $\alpha$ at the sub-$10^{-11}$ level has direct consequences for
precision tests of the Standard Model (SM). The three principal contributions to the SM
prediction for $a_e \equiv (g_e-2)/2$ are
\begin{equation*}
a_e(\mathrm{theo}) = a_e(\mathrm{QED}) + a_e(\mathrm{Hadron}) + a_e(\mathrm{Weak})
\ \ \text{with} \ \
a_e(\mathrm{QED}) = \sum_{n=1}^{\infty} C_{2n}\left(\frac{\alpha}{\pi}\right)^{n} + a_{\mu,\tau}\,,
\end{equation*}
where the QED contribution is an asymptotic series in powers of $\alpha$ together with
the muon and tau contributions $a_{\mu,\tau}$. The QED contribution is dominant for the
electron and well understood in principle, but its comparison with experiment requires an
independent and precise value of $\alpha$. The most recent measurement of the electron
magnetic moment~\cite{fan2023} has reached a precision at which the QED prediction for
$a_e$ is limited by the uncertainty in $\alpha$ itself, as illustrated in
the right panel of Figure~\ref{fig:alpha_values}. Driving $\sigma_\alpha/\alpha$ below $10^{-11}$
lowers {the uncertainty in} the QED contribution to $a_e$ below the hadronic contribution, so that $a_e$
becomes limited by hadronic physics rather than by $\alpha$, opening sensitivity to
physics beyond the Standard Model at the level of the hadronic uncertainty.

\paragraph{Measurement principle at AICE.}
Atom interferometry determines $\alpha$ via the recoil method~\cite{Clade_metrologia_2016},
which combines the photon-recoil measurement $h/m(X)$ with the Rydberg constant and the
relative atomic masses of the species $X$ and the electron,
\begin{equation}
  \alpha^{2} \;=\; \frac{2 R_\infty}{c}\,\frac{A_r(X)}{A_r(e)}\,\frac{h}{m(X)}\,.
  \label{eq:alpha-recoil}
\end{equation}
With $h$ fixed by the SI definition and $R_\infty$ known at $1.1\times 10^{-12}$, the
uncertainty at the sub-$10^{-11}$ level is shared between $h/m(X)$ and the mass ratio
$A_r(X)/A_r(e)$. The electron relative mass $A_r(e)$ is currently known at
$1.8\times 10^{-11}$ and is being improved independently~\cite{CODATA2022}, while
$A_r(X)$ for Sr and Yb is delivered by the PENTATRAP Penning-trap
spectrometer~\cite{Repp2012} at the few-$\times 10^{-12}$ level, as recently demonstrated
for bosonic Yb isotopes~\cite{Door2025}. The atom-interferometry measurement of $h/m(X)$
is realised at AICE on the $^{1}S_{0}\rightarrow{}^{3}P_{0}$ clock transition, using the
same 698\,nm ($^{87}$Sr) and 578\,nm ($^{171}$Yb) laser systems deployed for the ULDM programme (Section~\ref{sec:uldm}).

\paragraph{Advantages of single-photon clock transitions.}
Some of the dominant systematics in the existing caesium and rubidium determinations are
two-photon effects, such as residual
Raman phase shifts or Bloch-oscillation light shifts~\cite{parker2018,morel2020}.
Single-photon interferometry on the clock transition eliminates the two-photon
differential light shifts by construction, removes lattice-related phases associated
with Bloch-oscillation enhancement, and replaces the multi-port Raman and Bragg dynamics
with simpler two-level dynamics. Other important shifts are still present with single-photon interferometry, such as Gouy phase, wavefront distortion, and differential AC-Stark shifts, but will be somewhat different---and potentially controllable to smaller uncertainty---in AICE compared with the existing two-photon, smaller-scale experiments. The LMT Ramsey-Bord\'e scheme developed
in~\cite{Schelfhout2024} applies directly to the AICE geometry, with the dual-source
shaft configuration enabling the trajectories of that scheme at the 140\,m baseline.
Strontium and ytterbium provide two independent single-photon realisations, constraining
species- and transition-specific systematics. LMT orders $n > 100$ have been demonstrated
on the Sr $689$\,nm intercombination line~\cite{rudolf2020_689LMT}, and $n > 60$ on the Sr $698\,$nm clock line in recent unpublished work extending results from the UK AION project~\cite{baynham_prototype_2026}.

\paragraph{Staged AICE $\alpha$ programme.}
The AICE $\alpha$ programme is distributed across the campaigns shown in
Figure~\ref{fig:timeline}, each embedded within a ULDM phase of the main programme
without requiring dedicated facility time. The sequence of measurements and their
indicative precision targets is summarised in Table~\ref{tab:alpha_campaigns}. A first
$\alpha(\mathrm{Sr})$ result appears during commissioning as a tool-commissioning
campaign. Refined Sr determinations follow during the scalar ULDM phase at Baseline
parameters. The vector phase introduces $^{171}$Yb as a second species, yielding the
first $\alpha(\mathrm{Yb})$ single-photon determination and a Sr--Yb comparison that is
itself a test of $\alpha$ universality across two clock-transition species, complementary
to the alkali Rb--Cs comparison; the co-located equivalence-principle test is described
in Section~\ref{sec:ep}. An $\alpha$ follow-up campaign is embedded within the axion
phase, benefiting from the upgraded optical configuration and improved long-Baseline
systematics control. The final Ultimate $\alpha$ campaign runs at Stretch parameters,
targeting $\sigma_\alpha/\alpha \lesssim 10^{-11}$.

\begin{table}[htbp]
\centering
\caption{AICE $\alpha$ campaigns, indicative precision targets, and configurations.
  The intermediate targets are indicative placeholders derived from the AICE performance
  ladder of Table~\ref{tab:roadmap}; the Ultimate target of
  $\sigma_\alpha/\alpha \lesssim 10^{-11}$ is the firm programme objective.}
\label{tab:alpha_campaigns}
\vspace{3mm}
\scriptsize
\begin{tabular}{@{}llll@{}}
\toprule
Phase & Configuration & Target $\sigma_\alpha/\alpha$ & Notes \\
\midrule
Commissioning       & $^{87}$Sr gradiometer, Initial                & $\lesssim 10^{-10}$      & First $\alpha(\mathrm{Sr})$ from single-photon clock transition \\
Scalar ULDM         & $^{87}$Sr gradiometer, Baseline               & few $\times 10^{-11}$    & Refined $\alpha(\mathrm{Sr})$ at full Baseline systematics control \\
Vector $B{-}L$      & $^{87}$Sr/$^{171}$Yb dual-species Mach-Zehnder & few $\times 10^{-11}$    & First $\alpha(\mathrm{Yb})$; Sr--Yb universality; co-located EP test \\ %(Section~\ref{sec:ep}) \\
Axion / ALP         & Sr or Yb, single-species                      & ${\sim}\,2\times 10^{-11}$ & $\alpha$ follow-up with upgraded optical configuration \\
Ultimate $\alpha$   & Sr or Yb, Stretch                             & $\lesssim 10^{-11}$      & Final AICE $\alpha$ determination \\
\bottomrule
\end{tabular}
\end{table}

% ---------------------------------------------------------------------------
% Figure 11 of the AICE Technical Proposal -- AICE alpha timeline
% Caption clarifies that AICE points are projections drawn centred on the
% CODATA 2022 reference value, with error bars indicating projected
% precision rather than represented uncertainties on a measured value.
% ---------------------------------------------------------------------------
\begin{figure}[htbp]
  \centering
  \includegraphics[width=0.85\textwidth]{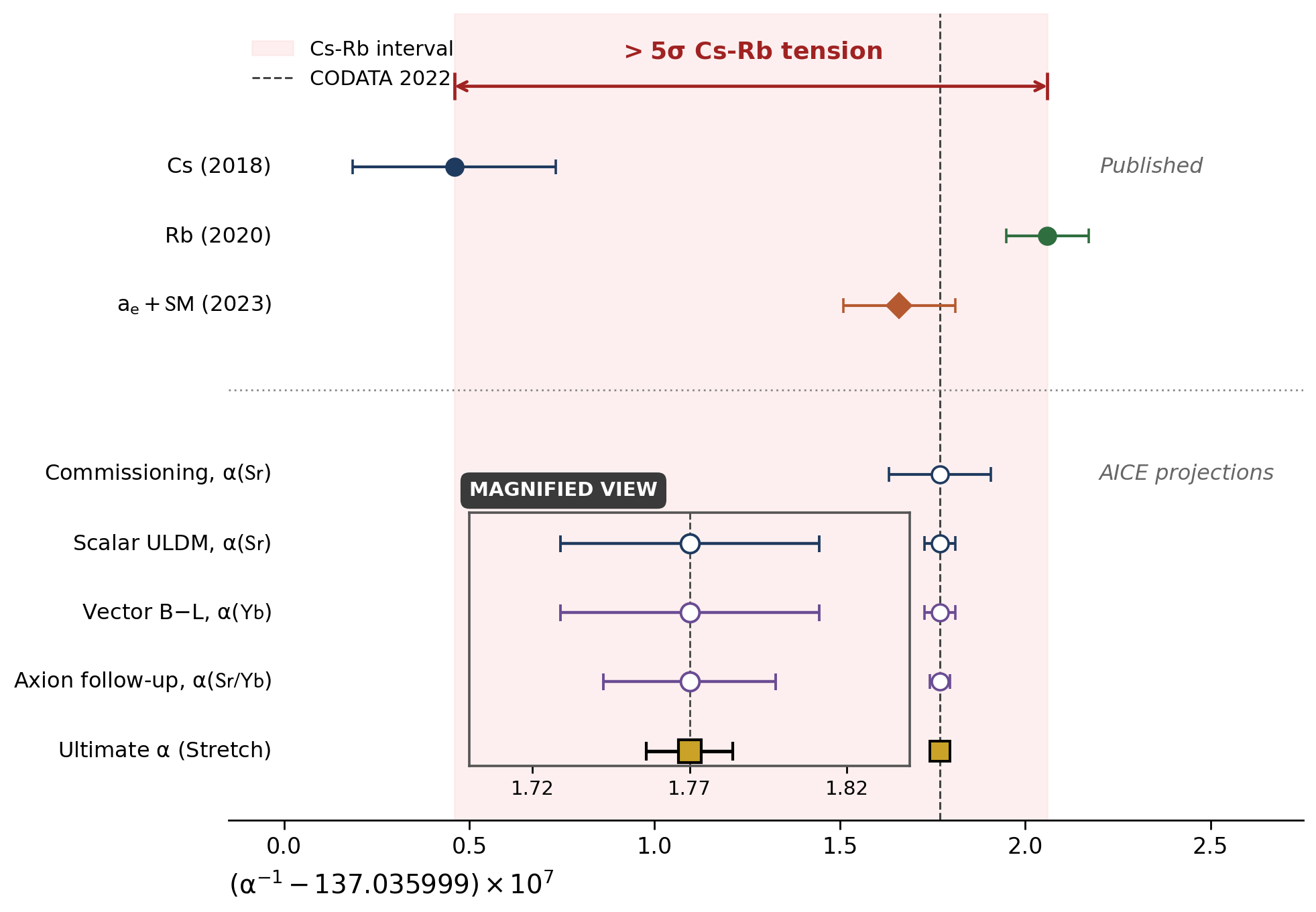}
  \caption{Indicative evolution of the AICE fine-structure constant
    programme. Published recoil determinations from caesium in
    2018~\cite{parker2018} and rubidium in 2020~\cite{morel2020} exhibit
    a discrepancy at more than $5\sigma$. The shaded band indicates the
    interval spanned by these alkali measurements. The projected AICE
    strontium and ytterbium measurements are shown at the operating
    stages listed in Table~\ref{tab:alpha_campaigns}, with precisions
    improving from commissioning to Stretch operation. As these are
    projected precisions rather than actual measurements, the AICE
    points are drawn centred on the CODATA 2022 reference
    value~\cite{CODATA2022}, with error bars indicating the projected
    $\sigma_\alpha/\alpha$ at each stage. Their comparison with the
    alkali values will provide a systematic test of the
    origin of the present tension. The vertical dashed line marks the CODATA 2022
    reference, and the $a_e + \mathrm{SM}$ point derived
    from~\cite{fan2023} is shown for context. The framed magnified view
    inside the AICE projection block zooms on
    $(\alpha^{-1}-137.035999)\times 10^{7}\in[1.70,1.84]$, with each
    advanced-stage row aligned to its main-panel counterpart so that
    the Ultimate $\sigma_\alpha/\alpha \lesssim 10^{-11}$ error bar is
    visible.}
  \label{fig:alpha_timeline}
\end{figure}

\paragraph{Systematics, traceability, and interpretation.}
The systematic programme for $\alpha$ is inherited directly from the AICE ULDM programme
described in Section~\ref{sec:uldm}. The dual-source geometry required for
gravity-gradient nulling in the Baseline ULDM configuration provides the same
mitigation for the $\alpha$ campaigns. In-vacuum beam-cleaning optics over the 140\,m
shaft suppress high-spatial-frequency wavefront distortion, active Coriolis compensation
via piezo-steered mirrors addresses Earth-rotation effects at the long Baseline, and
atom-based in-situ wavefront tomography characterises the residual Gaussian beam
parameters. Frequency references are traceable to the SI second through
optical-frequency combs referenced to the clock transitions. The PENTATRAP mass-ratio
cadence is matched to the AICE campaign schedule so that $A_r(X)$ remains subdominant
to $h/m$ in the $\alpha$ budget at every stage. The resulting $\alpha$ values feed the
interpretation of $a_e$ and correlated precision observables through global-fit
frameworks such as MasterCode~\cite{MasterCode}, consistent with CODATA
inputs~\cite{CODATA2022}.

\paragraph{Relation between the $\alpha$ and EP programmes.}
The $\alpha$ and EP programmes at AICE are not separate add-ons but two
complementary uses of the same 140\,m dual-source facility. The
$\alpha$ campaigns validate the facility-level shared systematics.
These include wavefront flatness over the long baseline, Coriolis and
rotation compensation, gravity-gradient control, timing and laser-phase
stability, species-resolved readout, and LMT phase control on the clock
transition. Achieving $\sigma_\alpha/\alpha \lesssim 10^{-11}$ on a
single species demonstrates that these shared systematics are under
metrology-grade control. The EP measurement uses the same
infrastructure in a dual-species differential configuration, where the
shared systematics largely cancel in the differential signal, and the
residual budget is dominated by species-differential effects such as
co-location offsets, differential launch velocities, differential
gravity-gradient coupling, differential light shifts, and scale-factor
mismatches between the Sr and Yb clock transitions. The Sr/Sr isotopic
campaign will characterise these differential effects in a
configuration with minimum compositional contrast before the flagship
Sr/Yb measurement is deployed. The quantitative bridge is
Eq.~(\ref{eq:eta-phase}). For AICE Baseline parameters, a differential
phase control of $\delta\phi \sim 10^{-4}$\,rad corresponds to
$\eta \sim 5\times 10^{-17}$, below the Stretch-phase
Ultimate EP target in Table~\ref{tab:ep_landscape} and reinforcing that
the measurement is limited by species-differential systematics rather
than by phase statistics. AICE is designed
to be the first ground-based platform capable of simultaneously
delivering a sub-$10^{-11}$ determination of $\alpha$ using
single-photon interferometry and a dual-species EP test at the
few$\,\times 10^{-16}$ level, bridging the domains of precision
metrology and fundamental tests of gravity.

%%------------------------------------------------------------
\subsubsection{Sensitivity summary}
\label{sec:sensitivity}
%%------------------------------------------------------------

\paragraph{Precision physics: fine-structure constant and equivalence principle.}
The AICE precision-physics programme delivers two staged campaigns
embedded within the ULDM facility timeline. The fine-structure constant
sequence summarised in Table~\ref{tab:alpha_campaigns} yields a first
$\alpha(\mathrm{Sr})$ result at the ${\sim}10^{-10}$ level during
commissioning, refined Sr and first Yb determinations at the
few-$\times 10^{-11}$ level across the Baseline scalar and vector
phases, and an Ultimate $\alpha$ run at Stretch parameters targeting
$\sigma_\alpha/\alpha \lesssim 10^{-11}$ (Section~\ref{sec:alpha},
Figure~\ref{fig:alpha_timeline}). In parallel, the equivalence-principle
sequence summarised in Table~\ref{tab:ep_landscape} delivers a Sr/Sr
isotopic validation at $\eta \lesssim 10^{-13}$, a flagship Sr/Yb
dual-wavelength result at $\eta \lesssim 10^{-14}$ during the vector
phase, and an Ultimate EP campaign at Stretch parameters targeting
$\eta \lesssim$ few$\,\times 10^{-16}$ (Section~\ref{sec:ep}). The two
programmes share the dual-source, long-baseline infrastructure
developed for the ULDM programme and are connected through the relation described at
the end of Section~\ref{sec:alpha}. The $\alpha$ result certifies the
facility-level shared systematics. The EP result exploits differential
cancellation to reach the few$\,\times 10^{-16}$ regime competitive
with MICROSCOPE.

\subsubsection{Other possible fundamental-physics opportunities}\label{sec:otherfundphys}

%\subsubsection{Gravitational physics}

\paragraph{Gravitational Aharonov-Bohm effect.}

{The gravitational Aharonov-Bohm effect is a phase shift acquired by a coherent superposition traversing regions of different gravitational potential, even where the local gravitational force on each path is negligible, in direct analogy with the electromagnetic Aharonov-Bohm effect. A first measurement of this effect was reported in~\cite{overstreet2022observation}. A similar measurement could be made using AICE if it was fitted with an annular planar source mass inside the vacuum chamber, such as that proposed to search for chameleons and symmetrons~\cite{Banks:2025vvz}.}

\paragraph{Measurement of G.}

{There is continuing controversy about the value of the gravitational constant G, see~\cite{schlamminger2026redetermination} and ~\cite{gibney2026big}. Several atom interferometer measurements have been reported in the last 20 years \cite{Bertoldi2006,fixler2007atom,Lamporesi2008,Rosi2014}, but their uncertainty is still uncompetitive when compared to the CODATA \cite{CODATA2022}. Due to its unparalleled LMT and free-fall times, AICE offers in-principle improvements to existing measurements of $G$, but this possibility has not been studied in detail.}

\paragraph{Gravitational redshift.}
%{Likewise, AICE might enable a competitive measurement of the gravitational redshift.}
The best ground-based measurements of the gravitational redshift, with an accuracy of $9 \times 10^{-5}$, involve the comparison of static atomic clocks located at the top and bottom of the Tokyo Skytree tower \cite{Takamoto2020}. The two clocks rely on the optical transition between the two clock states of $^{87}\mathrm{Sr}$ atoms trapped in an optical lattice and are separated by a height difference of 450~m.
In contrast, the sensitivity of microwave clocks using freely-falling Cs or Rb atoms in atomic fountains is not sufficient (by more than one order of magnitude) to detect the gravitational redshift of freely falling atoms compared to a static frequency reference at the bottom of the fountain \cite{Roura2020,Roura_2021,Ashby2021}.

By employing an interferometric scheme similar to a gradiometric configuration but where the atoms are launched with different velocities at the top and bottom sources, one can perform an unprecedented local measurement of the gravitational redshift with freely-falling atoms \cite{Roura2025}. The improvement by many orders of magnitude compared to atomic fountain clocks using Cs or Rb atoms is a consequence of the longer baseline (a factor of $10^2$) and the much higher energy difference between the two clock states (close to a factor of $10^5$), corresponding to an optical rather than a microwave transition frequency. The attainable sensitivity will be competitive with the comparison of static atomic clocks if the systematic effects associated with temperature gradients along the baseline can be sufficiently mitigated~\cite{Roura2025} through multiple temperature sensors distributed along the baseline and combined with suitable modelling and post-correction.

Interestingly, the maximum sensitivity can be achieved with Mach-Zehnder interferometers involving just three laser pulses and with no need for LMT pulse sequences. Thus, this kind of non-null measurement can be performed at the early stages of the facility timeline, and it can also make a valuable contribution to the characterization of overall systematic effects.

Extending the scheme to double-loop configurations~\cite{DiPumpo_2023-T3} or using symmetric transitions~\cite{Ufrecht2020} allows for competitive tests for additional aspects of the equivalence principle~\cite{DiPumpo2021,Roura2020,Roura_2021}.

%% --------------------------------------------------------------------
%% BLANKED OUT pending completion of Section 2.4.3 (Gravitational physics)
%% subparagraph on the gravitational Aharonov-Bohm effect.
%% Figure kept in source for future reinstatement with updated numbers
%% consistent with the AICE performance ladder in Table~\ref{tab:TVLBAI-performance}.
%% To reinstate: delete the \iffalse line below and the matching \fi line.
%% --------------------------------------------------------------------
\iffalse
\begin{figure}[ht]
  \centering
  \includegraphics[width=0.7\linewidth]{Gravitational_Aharonov-Bohm.png}
  \caption{Gravitational Aharonov-Bohm effect measurement with atom
    interferometry~\cite{overstreet2022observation}.}
  \label{fig:Aharonov-Bohm}
\end{figure}
\fi
%% --------------------------------------------------------------------

\paragraph{Gravitational curvature.} 
{A protocol for measuring the local gravitational curvature has been proposed in~\cite{Werner:2024yxw}, and could be considered for implementation in AICE.}

%% --------------------------------------------------------------------
%% BLANKED OUT pending completion of Section 2.4.3 (Gravitational physics)
%% subparagraph on gravitational curvature.
%% Figure kept in source for future reinstatement with updated numbers
%% consistent with the AICE performance ladder in Table~\ref{tab:TVLBAI-performance}.
%% To reinstate: delete the \iffalse line below and the matching \fi line.
%% --------------------------------------------------------------------
\iffalse
\begin{figure}[ht]
  \centering
  \includegraphics[width=0.7\linewidth]{Gravitational_curvature.png}
  \caption{Configuration of cloud trajectories for measuring gravitational curvature with a long-Baseline
    atom interferometer~\cite{Werner:2024yxw}.}
  \label{fig:Gravitational_curvature}
\end{figure}
\fi
%% --------------------------------------------------------------------

\paragraph{Tests of quantum mechanics.}
{Long-Baseline interferometers such as AICE offer possibilities for powerful probes of quantum mechanics, including testing scenarios for wave-function collapse such as continuous spontaneous localisation models~\cite{bassi2003dynamical}, measuring gravitational tidal effects on wave-functions~\cite{Asenbaum:2017}, and probing the quantum relativity principle~\cite{overstreet2023inference} that the laws of physics should take the same form in all reference frames, including quantum reference frames~\cite{giacomini2019quantum}.}\\

\paragraph{Gravitational effects of Solar System dark matter.} 
{Ultra-light dark matter in the AICE target mass range is gravitationally focused by the Sun and planets in the Solar System~\cite{Kim:2021yyo}. This leads to distinctive time-dependent signatures and enhancements in both single atom interferometer and gradiometer phases from these purely 
gravitational effects that AICE can potentially detect or limit as well.  Sufficiently strongly self-interacting ultra-light dark matter may also form bound states around the Sun and planets~\cite{Budker:2023sex}. This will create over-dense sub-halos of dark matter that overlap with the Earth and may allow AICE and future atom interferometry experiments to detect dark matter purely from its gravitational effects~\cite{Baum:2022duc,Badurina:2025xwl,Carlton:2026SSDM}.}\\

\subsection{Gravitational waves in the mid-frequency band}
\label{sec:gw}

A prominent long-term objective of long-baseline atom interferometer experiments is the search for gravitational waves (GWs) in the range $\sim 0.03–3$~Hz that is intermediate between the peak sensitivities of present and proposed terrestrial laser interferometer experiments such as LIGO~\cite{LIGOScientific:2014pky}, Virgo~\cite{VIRGO:2014yos} and KAGRA~\cite{Aso:2013eba} (LVK), the Einstein Telescope (ET)~\cite{ET:2025xjr} and Cosmic Explorer (CE)~\cite{Reitze:2019iox}, and planned space-borne laser interferometers such as LISA~\cite{LISA:2017pwj}, TaiJi~\cite{Hu:2017mde} and TianQin~\cite{TianQin:2015yph}. AICE will demonstrate the potential of atom interferometers for GW detection and pioneer the exploration of the spectrum of GWs in this intermediate frequency range, serving as a pathfinder for future longer-baseline detectors~\cite{TVLBAISummary,abdalla_terrestrial_2025}. 

\paragraph{Exploration scheme. } The AION prototype experiment has demonstrated the principle of differential interferometry using the 698~nm strontium transition~\cite{AION:2025igp}, directly validating the core interferometric technique used by AICE and its applicability to GW detection. The long baseline of AICE and the corresponding free-fall time of several seconds opens up to exploration the intermediate-frequency interrogation window around ${\cal O}(1)$~Hz. AICE can carry out this pioneering programme of exploration in parallel with the ULDM search programme described above, benefiting from the same projected staged increases in sensitivity.

\paragraph{Potential GW sources. } Among the targets of GW experiments in this intermediate frequency range are mergers of black holes (BHs) with masses intermediate between those whose mergers have been detected by LIGO and Virgo, which have masses $\lesssim {\cal O}(100)$ solar masses, and the supermassive black holes (SMBHs) observed in the centres of galaxies~\cite{EventHorizonTelescope:2019dse,EventHorizonTelescope:2022wkp}, which have masses $\gtrsim {\cal O}(10^6)$ solar masses. Searches for such intermediate-mass BH mergers could cast light on the mechanisms that form SMBHs, an outstanding problem in astrophysics. 

There would also be synergies with GW detectors operating at higher and lower frequencies. Intermediate-frequency GW measurements would be sensitive to the early infall stages of binary BH systems whose mergers would be measurable subsequently by terrestrial laser interferometers, enabling their directions, timing and redshifts to be predicted, and thereby facilitating precursor and afterglow multimessenger astrophysical observations. Conversely, prior lower-frequency observations by space-borne laser interferometers such as LISA could be used to predict future intermediate-mass BH mergers to be observed by long-baseline atom interferometers. Thus, these devices would be integrated into a global network of GW detectors. 

Atom interferometers operating in the intermediate frequency range may also be sensitive to a stochastic background of GWs produced by a fundamental physics process such as a first-order phase transition in the early Universe or the evolution of a network of cosmic strings~\cite{Badurina:2019hst}.

\begin{figure}
    \centering
    \includegraphics[width=0.8\textwidth]{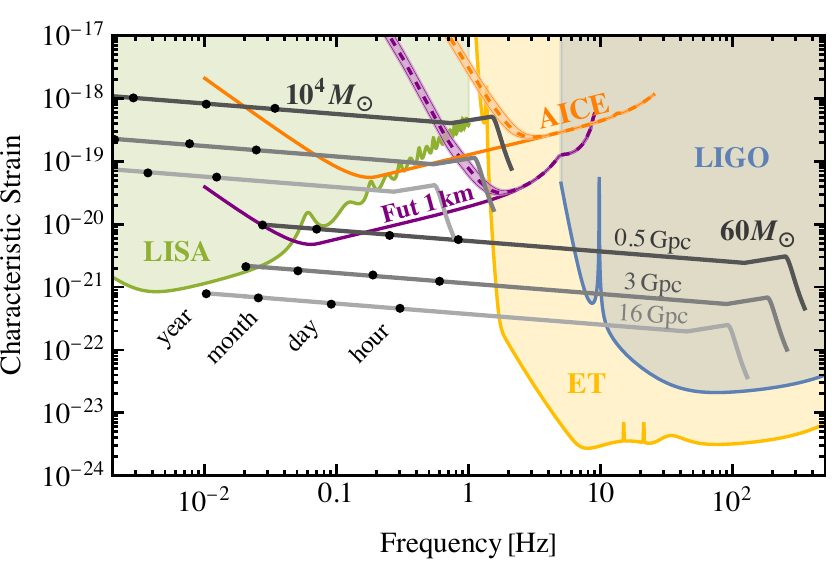}
    \vspace{-0.2cm}
    \caption{Comparison of the GW strain sensitivities of the AICE Stretch scenario and that of a prospective 1~km atom interferometer with those of LIGO and the prospective sensitivities of the Einstein Telescope (ET) and LISA. The orange and purple solid lines show the shot-noise-limited sensitivities of the AICE Stretch scenario and a possible future 1\ km atom interferometer. The dashed lines and shaded bands show the estimated unmitigated seismic-GGN benchmark and its modelling uncertainty, as defined in Section~\ref{sec:noise:GGN:impact}. Atmospheric GGN, mitigation and subtraction are not included. These sensitivities are compared with the GW signals expected for mergers of equal-mass binary BHs with total masses of $10^4$ and 60 solar masses at the indicated distances. Dots mark the calculated signals at the indicated times before the final merger phase.}
    \label{fig:GWs}
\end{figure}

\begin{figure}
    \centering
    \includegraphics[width=0.8\textwidth]{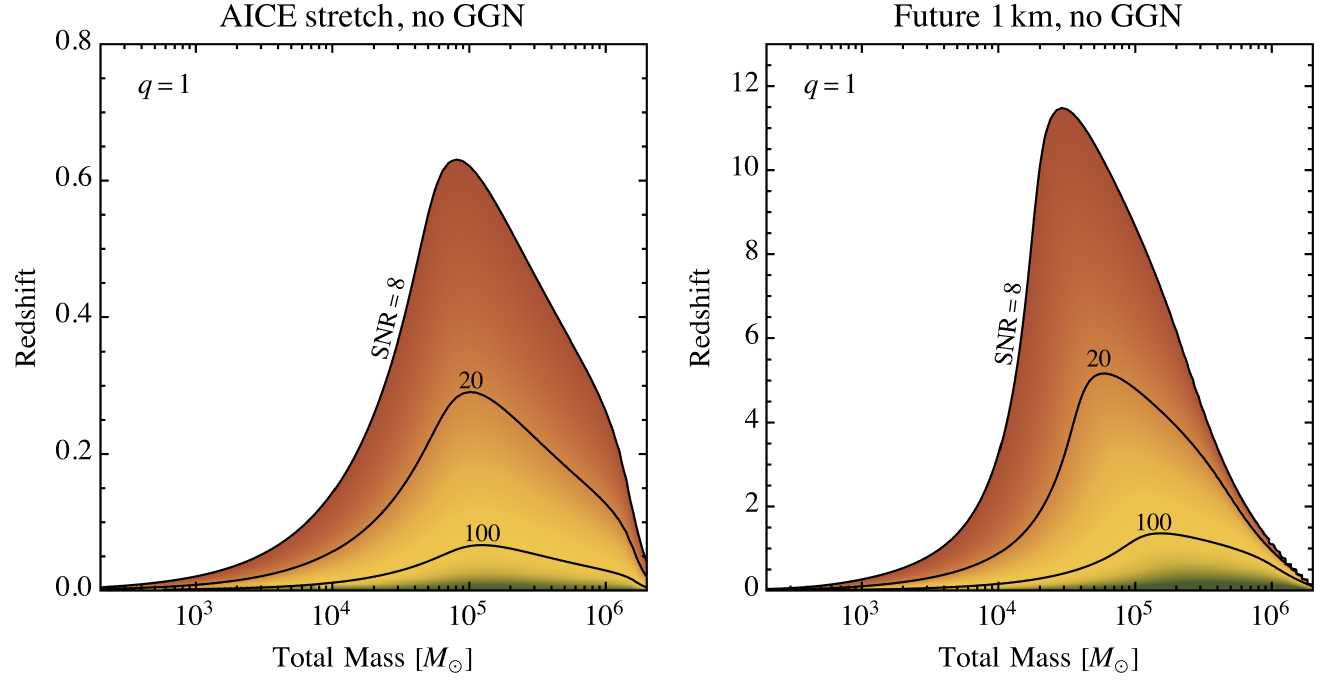}
    \vspace{-0.2cm}
    \caption{Contours of the shot-noise-limited (i.e., neglecting GGN) signal-to-noise ratios (SNRs) for Stretch AICE (left panel) and a 1~km atom interferometer (right panel) for detections of the GWs from mergers of equal-mass BH binaries as functions of mass and redshift.}
    \label{fig:GWs2}
\end{figure}

\paragraph{Sensitivity projections. } Figure~\ref{fig:GWs} illustrates the shot-noise-limited GW strain sensitivities of the AICE Stretch scenario and a possible future 1\ km atom interferometer as solid orange and purple lines, respectively. The dashed lines and shaded bands show the estimated unmitigated seismic-GGN benchmark and its modelling uncertainty, as defined in Section~\ref{sec:noise:GGN:impact}. Atmospheric GGN, mitigation and subtraction are not included. Also shown are the sensitivity of LIGO (which falls off for frequencies $\lesssim 10$~Hz) and the prospective sensitivities of the Einstein Telescope (ET) (which falls off sharply for frequencies $\lesssim 2$~Hz), and of LISA (which falls off gradually at frequencies $\gtrsim 10^{-2}$~Hz). We see that Stretch AICE and a prospective 1~km atom interferometer could contribute interesting sensitivity in the frequency range intermediate between the most sensitive ranges of LIGO, ET and LISA.  The GW sensitivities are compared to the prospective signals of  mergers of pairs of  5000 solar mass BHs at distances of 0.5, 3 and 16 billion parsecs. Fig.~\ref{fig:GWs2} displays contours of the shot-noise-limited (i.e., neglecting GGN) signal-to-noise ratios (SNRs) for detections of the GWs from mergers of equal-mass BH binaries as functions of mass and redshift for Stretch AICE (left panel) and a 1~km atom interferometer (right panel).

%\newpage

%\newpage

We also see in Fig.~\ref{fig:GWs} that a 1~km atom interferometer could potentially detect a typical LVK binary pair of 30 solar-mass BHs weeks before the final merger phase at a distance of 0.5 Gpc (for comparison, the closest LIGO event occurred at a distance of 0.2~Gpc). Conversely, LISA could be able to detect the inspiral of a pair of intermediate-mass BHs in the earlier stages before the final merger, it would miss the infall, merger and ringdown stages, which could only be measured by AICE or a 1-km atom interferometer. More broadly, precise measurements of the accumulated GW phase in this band can reveal departures from vacuum binary evolution induced e.g.\ by dark matter overdensities, accretion discs, and superradiant clouds, turning compact binary signals into probes of the matter surrounding black holes~\cite{Bertone:2019irm}.
\section{Detector and Technical Systems}
\label{sec:detector}

\subsection{Overview and design philosophy}
\label{sec:detector:overview}
The detector design of AICE follows directly from the configurable facility
concept introduced in Section~\ref{sec:concept}. Rather than being conceived as
a detector with a fixed technical implementation, AICE is designed as a
long-lived experimental facility whose capabilities are expected to evolve
throughout its operational lifetime. The planned science
programme spans more than a decade and provides the natural stepping stone
towards future kilometre-scale detectors within the TVLBAI roadmap. The
detector architecture therefore reflects not only the requirements of the
initial experimental programme, but also the expectation that atom
interferometry, laser technology and quantum sensing will continue to advance
substantially during the lifetime of the facility.

The underlying design philosophy is to separate the long-lived infrastructure
from the technical systems that realise the experiment. The PX46 shaft, the
vacuum envelope, the primary support structure and the associated civil
engineering form the permanent backbone of the facility. These elements are
installed once and are dimensioned from the outset for the ultimate Stretch
configuration. Around this backbone, the detector systems are designed to
remain accessible, modular and upgradeable. Standardised mechanical, optical,
electrical and ultra high vacuum interfaces allow individual subsystems to
evolve without modification of the underlying infrastructure.

This philosophy applies throughout the detector. The atom sources are modular
assemblies that can be exchanged or expanded as improved source technology
becomes available. This allows the facility to grow from the Initial
configuration towards the higher source multiplicities envisaged for Baseline
and Stretch operation, while accommodating additional atomic species,
beginning with $^{87}$Sr and subsequently adding $^{88}$Sr and $^{171}$Yb for the dual-species programme. The modular ultra-high-vacuum interfaces also allow future
source technologies to be incorporated without modification of the main
interferometer tube.

The laser systems follow the same philosophy. They are housed in the surface
laser laboratory and coupled to the interferometer through optical fibres and
the beam delivery telescope, separating laser development from the detector
installed in the shaft. The $698\,\mathrm{nm}$ strontium clock laser system is
a core capability of the facility and is present throughout all science
phases, but its technical implementation is expected to evolve over time.
Future improvements in optical power, frequency stability, beam delivery and
control can therefore be incorporated without intervention on the vacuum
system or the civil infrastructure. The addition of the
$578\,\mathrm{nm}$ ytterbium laser system for the dual-species programme
represents the first planned extension of these capabilities, and the same
architecture naturally accommodates further developments as performance
requirements evolve.

The same philosophy extends to the controls, electronics, diagnostics,
magnetic systems and other technical subsystems, all of which are expected to
develop as operational experience is gained and new technologies become
available. Future upgrades may require additional power, cooling,
instrumentation or control hardware, and the facility architecture provides
the flexibility needed to accommodate such developments without modification
of the primary infrastructure.

This architecture maximises the long-term scientific return of the facility.
The principal long-term investment is the permanent infrastructure, while the
technical implementation of the detector is expected to improve continuously
through advances in atom sources, laser systems and quantum technologies.
Scientific capability can therefore evolve throughout the operational lifetime
of AICE without major modifications to the facility itself.

The following Sections describe the individual detector subsystems, namely the
vacuum system, atom sources, laser systems, magnetic systems, support
structure and vibration diagnostics. Throughout the detector, emphasis is
placed on clearly-defined subsystem interfaces that allow future technical
developments to be incorporated through incremental upgrades rather than major
redesigns of the facility.

\subsection{Vacuum system}
\label{sec:vacuum}

%\textcolor{red}{Carlo Scarcia, Jack Sander - input from Jim Kowalkowski/MAGIS}

%\textcolor{red}{Carlo: The idea is to present the general detector layout and do a breakdown structure decomposition on requirements, assumptions and design description (including alternatives).}

%\textcolor{red}{Carlo: Most of the things we sorted out with Jack. Waiting for feedback on experience, technical details and improvements by the MAGIS colleagues.}

%\textcolor{orange}{Jack: I will work on an updated schematic of the layout. It will be fairly different to AION with the laser lab at the top, and some nomenclature of the sub-systems may need to change.}

%\noindent \textcolor{red}{Working on the definition of the noises, but confirmation for the magnitude of effects is needed - Carlo.}

%\textcolor{orange}{Jack: possible sections related to vac system: \begin{itemize}
%    \item laser lab
%    \item Telescope optics
%    \item Interconnect chambers
%    \item 5m beam pipe concept with frame and mounting
%    \item pumping stations between beam pipe sections
%    \item retro-reflective mirror (design shared between AION and MAGIS)
%\end{itemize}}

%\begin{wrapfigure}{R}{0.55\textwidth}
%\centering
%    \includegraphics[width=0.53\textwidth]{figures/instrumentschematic.png}
%    \caption{Schematic drawing of the atom interferometer instrument, showing conceptual designs for each key element. The number and location of atom sources is to be defined: they can be added all along the beam pipe. BRM denotes the Bottom Retro-reflecting Mirror.}
    %\vspace{-12cm}
%    \label{fig:instumentschematic}
%\end{wrapfigure}

\noindent The AICE vacuum system can be decomposed into four main subsystems:
\begin{itemize}
    \item Transfer beam pipe for the interferometry laser;
    \item Interferometry beam telescope;
    \item Main interferometer tube;
    \item Retroreflective mirror.
\end{itemize}

\noindent Each subsystem’s requirements and design features are presented below, together with possible alternative approaches to be verified at a later stage of the design. Where possible, the interferometer design is conceived as modular, allowing future upgrades as the experiment evolves without disrupting the overall layout and with reduced downtime. Unless otherwise stated, all the components foreseen to be used for the vacuum system assembly are off-the-shelf. To meet the stringent magnetic permeability requirements, the pipes, chambers, and bellows will be made of AISI 316L, while all connecting flanges will be made of AISI 316LN. All components will be vacuum fired at 950$^\circ$C for 2 h. This treatment will also help to reduce the H$_2$ outgassing rate.

The atom sources described in Section~\ref{sec:atom-sources} are also part of the vacuum system, and will be connected to the interconnect sections of the main interferometer tube. However, the atom sources will be installed for specific AICE experiments, rather than as a fixed part of the AICE facility.

\subsubsection{Transfer beam pipe for the interferometry laser}

\noindent The interferometry laser input to the vacuum system, together with its propagation path through the transfer beam pipes from the laser source to the telescope, will be maintained in a dust-minimised environment to ensure efficient injection. At this first stage, the required vacuum level is 1$\times$10$^{-8}$ mbar or lower. Handling, assembly, commissioning, and operation of the vacuum assembly will be carried out to minimise hydrocarbon contamination. To reduce the risk of dust transport and hydrocarbon contamination, the roughing phase will be performed using dry pumps equipped with a soft-start inlet valve.

%\noindent 
The laser beam is injected from the laser laboratory into the vacuum line through an optical fibre feedthrough\footnote{If the beam path from the laser source is sufficiently short and stable, the optical fibre may not be required, allowing compatibility with higher laser power.}. The vacuum line consists of DN100 ConFlat (CF) flanged tubes and fittings. The total length of the transfer beam pipes depends on the chosen laser transport method and on the distance between the laser laboratory and the telescope stage. If most of the distance can be covered using low-attenuation fibre-optic cables, the length of the transfer beam pipes may be reduced. Consequently, the final pipe length and, where required, the sectorisation of the line will be defined during a later design review phase.

%\noindent 
Commissioning of the vacuum line will be carried out from atmospheric pressure down to 1$\times10^{-6}$ mbar using a mobile pumping group equipped with an 80 l/s turbomolecular pump backed by dry multi-Roots pumps. The pumping group will be connected to the system through a DN63 FKM-sealed angle valve. Once the target pressure has been reached, the mobile pumping group will be removed, and the operational vacuum will be maintained by a 75 l/s sputter ion pump.

\subsubsection{Interferometry beam telescope}

%\noindent 
The telescope, positioned above the main interferometer tube, houses the optical components required to control the laser profile and to compensate for Coriolis effects through dedicated beam-steering optics. As for the laser laboratory and transfer beam lines, the required pressure level is 1$\times$10$^{-8}$ mbar or lower. The dust cleanliness requirements during handling and assembly will be assessed at a later stage; however, an ISO 7 cleanliness level or better may conservatively be assumed to preserve the system's optical performance. 
All pipes and fittings forming the telescope assembly will be DN100 CF-flanged.

%\noindent 
As the central element of the interferometer vacuum layout, the telescope assembly will be installed so that it can be isolated from both the transfer beam line and the main interferometer tube by means of gate valves. To allow differential pumping between the telescope assembly and the main interferometer tube, a calibrated orifice will be installed downstream of the gate valve connecting the two systems. The conductance of this orifice will be defined once the final pressure requirements, gas-load budget and pumping configuration of the telescope assembly have been established.

%\noindent 
As the pointing direction of the lenses in the telescope is critical to instrument performance, it is anticipated that they will be mounted directly to flanges and adjusted or actuated ex-vacuo using a mechanism attached to the external frame.

%\noindent 
The commissioning of the telescope assembly will follow the same approach foreseen for the transfer beam line. Pumpdown from atmospheric pressure to 1$\times10^{-6}$ mbar will be performed using a mobile pumping group equipped with an 80 l/s turbomolecular pump backed by dry multi-Roots pumps and connected through a DN63 FKM-sealed angle valve. Once the target pressure has been reached, the mobile pumping group will be removed.

\begin{wrapfigure}{R}{0.42\textwidth}
\centering
    \includegraphics[width=0.41\textwidth]{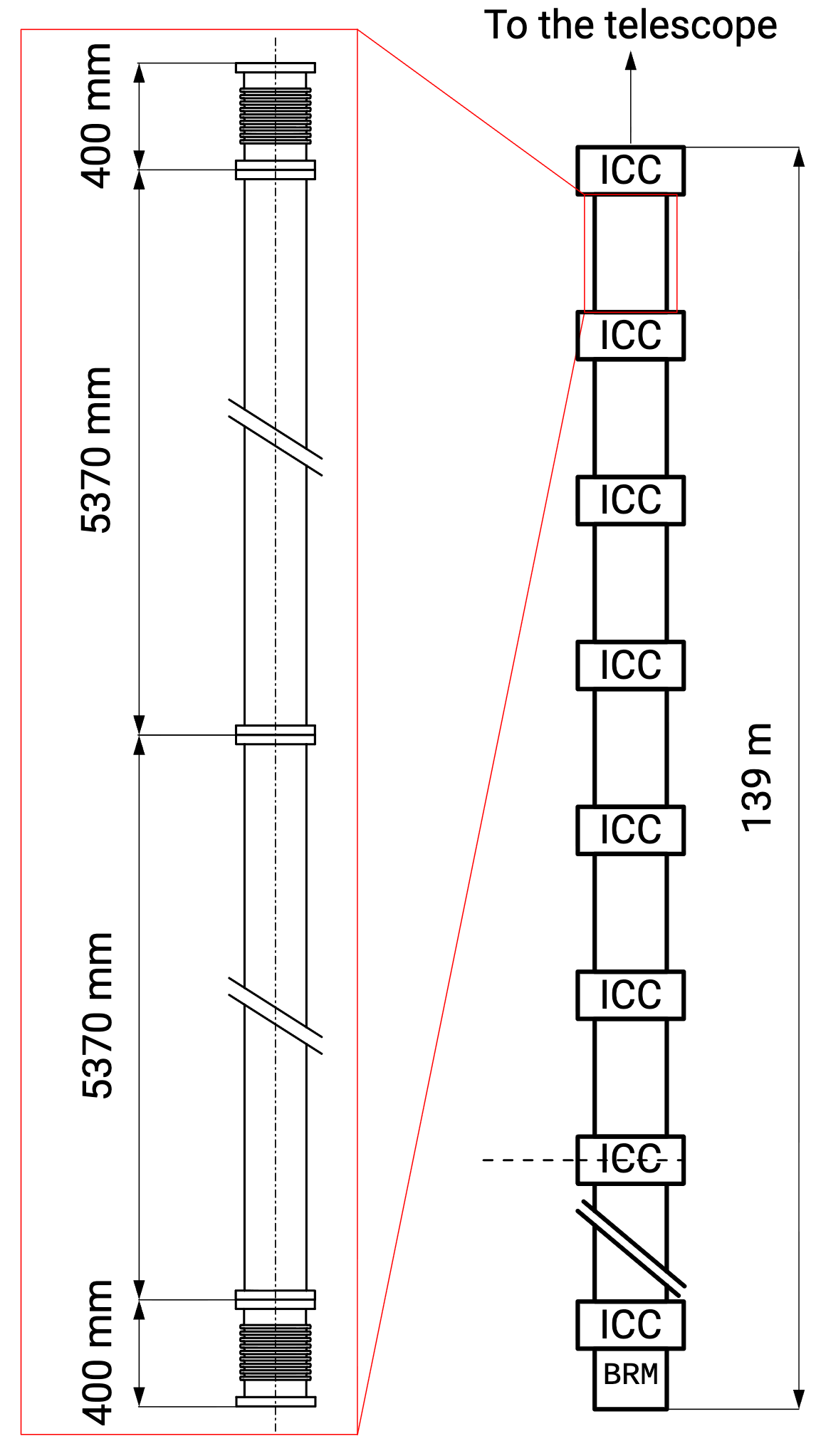}
    \caption{Schematic drawing of the main interferometer tube. The number and location of atom sources are to be defined: they can be added all along the beam pipe. BRM denotes the Bottom Retro-reflecting Mirror.}
    \vspace{-1.4cm}
    \label{fig:instrumentschematic}
\end{wrapfigure}

%\begin{wrapfigure}{R}{0.42\textwidth}
%\centering
%    \includegraphics[width=0.41\textwidth]{figures/beampipestack.jpg}
%    \caption{Illustration of the stack of modular sections and interconnect chambers. The middle interconnect node includes an atom source (no structure included). \textcolor{red}{???? - Are these ?s asking for more clarity in the figure?}}
%    \vspace{-1.5cm}
 %   \label{fig:beampipestack}
%\end{wrapfigure}
%I've done my best to format this, improvements welcome! JS
%Take a look at my version. JE Thanks John much better!

%\noindent 
Although the outgassing budget of the telescope assembly has not yet been defined, a 75 l/s sputter ion pump is expected to be sufficient to maintain the operational vacuum. Once the optical materials, internal components and final assembly configuration are known, the pumping capacity will be re-evaluated and, if necessary, increased. \\

\subsubsection{Main interferometer tube}

%\noindent 
The baseline design of the main interferometer tube vacuum system consists of an approximately 140 m-long vertical vacuum pipe hosting the interconnecting chambers connected to the atom sources. The retroreflective mirror is located at the bottom of the system.

%\noindent 
To minimise atom losses and preserve coherence by reducing collisions with residual gas molecules, the total pressure in the main interferometer tube will be of the order of 1 $\times$ 10$^{-12}$ mbar.

%\noindent 
To preserve the optical properties of the internal surfaces and to minimise unwanted particulate-light interactions, which could lead to phase front distortions or scattering, the main interferometer tube components will be prepared, installed and commissioned in an ISO 6 environment.

As depicted in Figure~\ref{fig:instrumentschematic}, the main interferometer tube is located between the telescope (top) and retroreflective mirror (bottom), and is composed of:

\begin{itemize}
\item 24 modular sections;
\item 13 interconnecting chambers.
\end{itemize}

%(\textcolor{red}{total length? - 140m given at the start of the section, do we want the length of each component?})

\noindent \textbf{Modular sections}

\noindent Each modular section consists of one DN200 CF $\times$ 5370 mm pipe connected to one DN200 $\times$ 400 mm expansion bellows.

%\noindent 
To meet the vacuum requirements without interrupting the geometry of the magnetic shielding, the use of a non-evaporable getter (NEG) coating is preferred over lumped pumping solutions along the modular sections. Dust release from lumped NEG pumps fabricated with sintered material has been assessed in the extremely sensitive context of superconducting radio-frequency (SRF) accelerating cavities, and found to have no measurable impact in terms of operational performance \cite{CIOVATI201792}. Moreover, niobium thin films are commonly used for SRF cavities achieving a very high level of cleanliness \cite{CALATRONI200695}.  The potential release of dust from the NEG coating will be assessed, but it is expected to have no operational impact for AICE. The NEG coating will be activated at 210$^\circ$C for 48 h.

%\noindent 
At this preliminary design stage, no specific low-reflectivity surface treatment is foreseen to mitigate possible stray light from the retroreflective mirror. This aspect will be studied and quantified at a later stage. Possible solutions include local surface treatment, additional low-reflectivity coatings, or installing short baffle sections with vacuum-compatible low-reflectivity coatings, such as Vantablack or Acktar. The vacuum compatibility of these solutions will be verified before implementation.
%\textcolor{red}{Can we be more definite?}

%\noindent 
The vacuum pipes of each modular section will be surrounded by layers that satisfy the other performance requirements of the instrument. The vacuum pipe will be wrapped in bake-out tape with an insulated layer to allow in-situ bakeout. Outside this are magnetic field coils and shielding, detailed below in Section~\ref{sec:shielding}. Finally, an external frame will be assembled around each pipe section. This will provide structure, and allow the modules to be transported and attached to the shaft walls. \\

\begin{figure}[h]
\centering
\includegraphics[width=\linewidth]{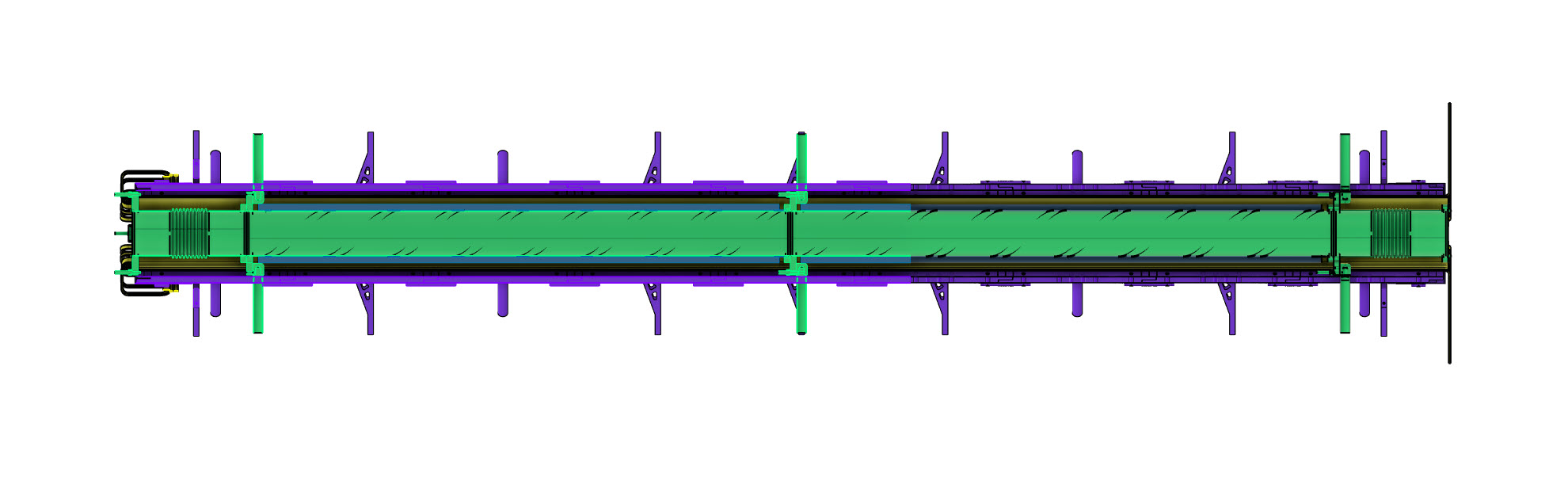}
\caption{Cross-section of the modular sections, showing their layered structure. The vacuum components and their mounting structures are shown in green, the bakeout tape and insulation are shown in blue, the magnetic field coils are shown in yellow and the magnetic shield in purple.}
\label{fig:beampipelayers}
\end{figure}

\noindent \textbf{Interconnecting chambers (ICCs)}

\noindent The interconnecting chambers are installed every two modular sections, as shown in Fig.~\ref{fig:instrumentschematic}. Each interconnecting chamber is a custom-made assembly composed of:

\begin{itemize}
\item Two custom-made DN200 CF to DN300 CF reducer fittings, 60 mm long, hosting the launch lattice;
\item One custom-made DN300 CF $\times$ 300 mm multi-port chamber hosting cameras, viewports and ancillary components.
\end{itemize}

%\noindent 
To ease operation and future modifications, each interconnecting chamber will be connected to the upstream and downstream modular sections by means of DN200 all-metal gate valves. In the next design stage, the number of valves might be reduced to only two units at the top and bottom of the main interferometer arm to reduce the cost. The reduction of the sector valves will be studied in the future as a cost-benefit exercise, with particular attention paid to a risk mitigation strategy focused on the bakeout design and venting procedure.

%\noindent 
Each interconnecting chamber will also be equipped with a NEG cartridge/SIP pumping set. This system will cope with the gas load from the atom source assembly, which is not yet known, and with species not efficiently pumped by the NEG coating in the modular sections, such as noble gases and CH$_4$. The bakeout temperature will be verified once the materials used for the ancillary components have been defined, while the bakeout duration will remain the same as that foreseen for the modular sections.
\begin{figure}[h!]
\centering
\includegraphics[width=1\linewidth]{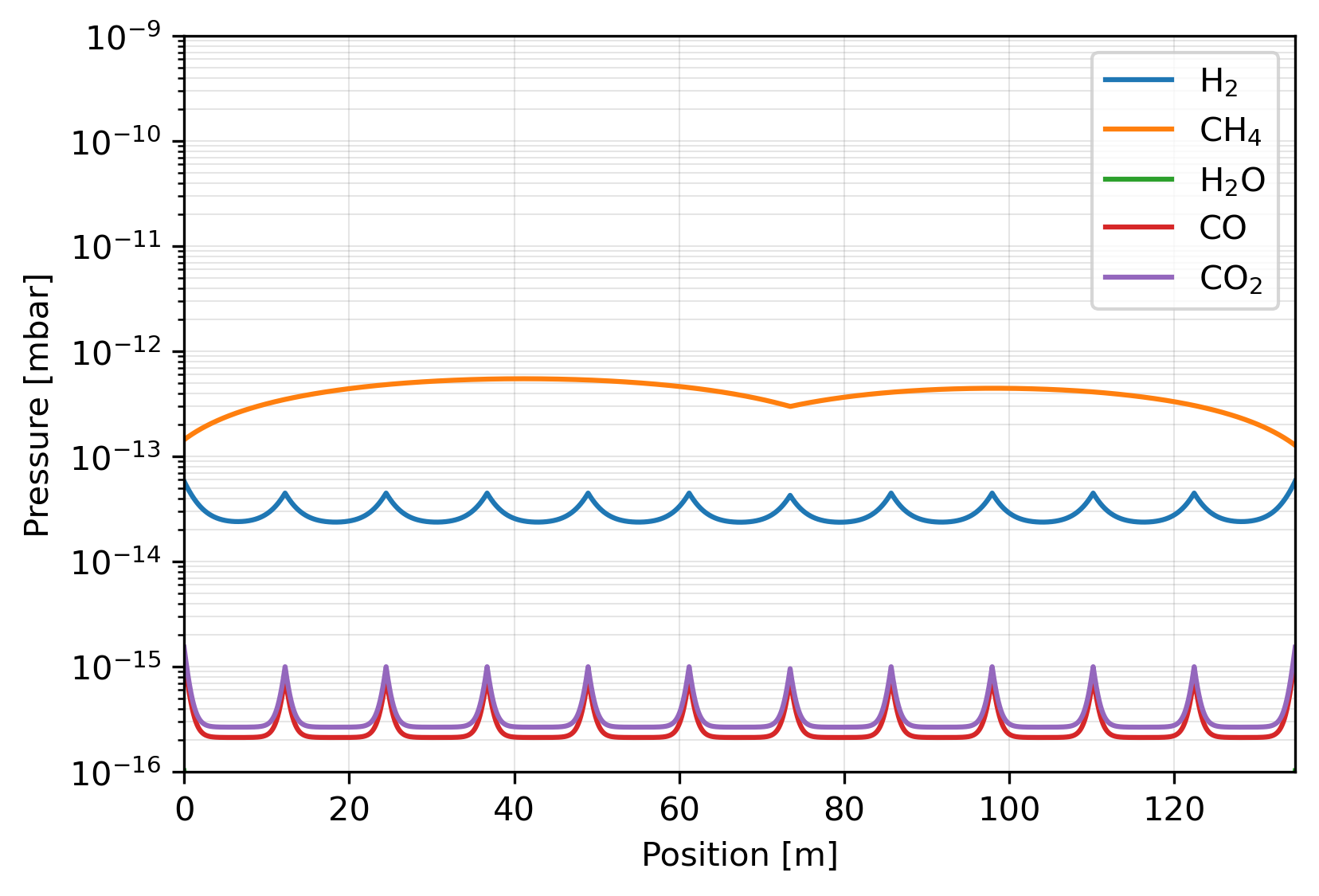}
\caption{Partial pressures distribution along the main interferometer tube. The scenario depicts the distribution of gases released by the non-NEG coated components, namely the ICC, bellows and valves, without including the contribution of ancillaries and atom sources. Three 300 l/s SPIs are installed at the extremities and centre of the vacuum system length. The gas load is assumed from vacuum-fired stainless steel and the NEG pumping speeds are conservatively assumed degraded to the lowest sticking factors.}
\label{fig:pressuredistr}
\end{figure}

%\noindent 
As with the modular sections, an external frame will provide structure to the interconnect chamber assemblies. Within this frame, an optics platform can be installed, with the option of active isolation to meet the vibrational optics requirements. While the baseline configuration of the interferometer contains 3 atom sources, the modular interconnect chambers allow for maximum flexibility and installation of atom sources anywhere on the tower at a future date without breaking vacuum on the main interferometer beam pipe. \\

\begin{figure}[h!]
\centering
\includegraphics[width=0.6\linewidth]{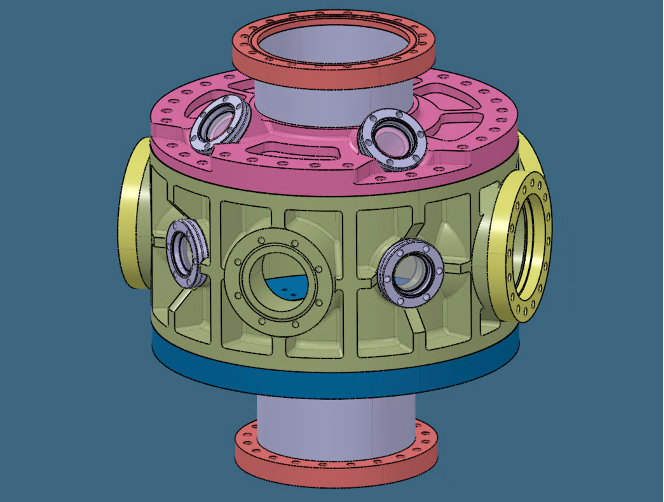}
\caption{Conceptual design for an interconnect chamber~\cite{Bongs:2025rqe}.}
\label{fig:interconnectchamber}
\end{figure}

\subsubsection{Retroreflective mirror section}

\begin{figure}%[h!]
\centering
\includegraphics[width=\linewidth]{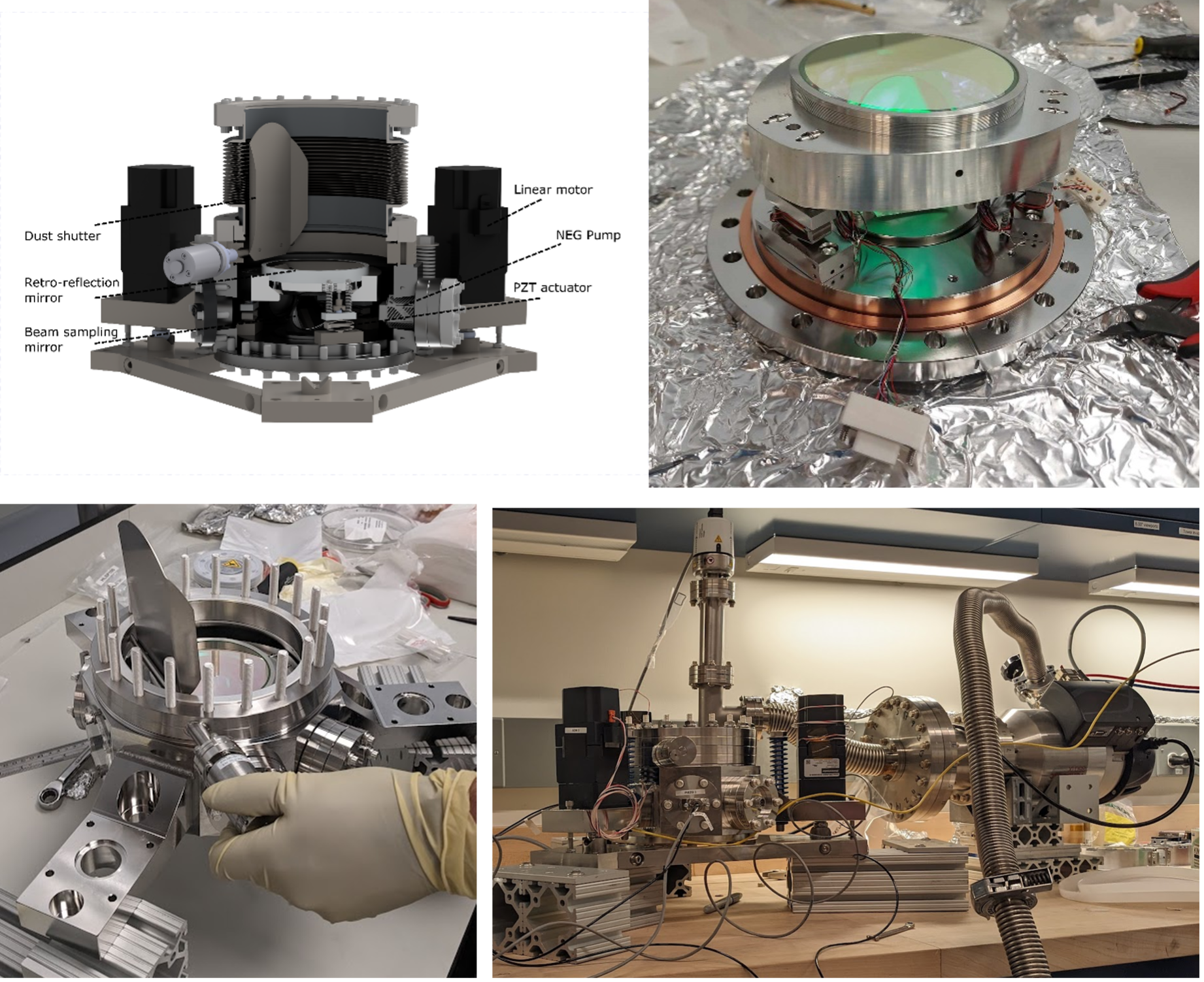}
\caption{The AION/MAGIS retroreflecting mirror: design and images during commissioning of the first unit. Figures taken from~\cite{Bongs:2025rqe}.}
\label{fig:retromirror}
\end{figure}

%\noindent 
The retroreflective mirror, whose initial design is shared with AION-10 and MAGIS (see Figure~\ref{fig:retromirror}), is connected to a DN200 CF $\times$ 1 m pipe equipped with a DN63 CF port. This section will have the same vacuum characteristics as the modular sections. 

%\noindent 
The DN63 CF port will be equipped with an all-metal angle valve to allow connection of the mobile pumping groups required for commissioning the complete main interferometer tube assembly.

%\noindent 
Commissioning of the vacuum line will be carried out using a mobile pumping group equipped with an 80 l/s turbomolecular pump backed by dry multi-Roots pumps. As for the upper vacuum sections, the roughing phase will be performed using a soft-start valve to minimise dust transport. The mobile pumping group will also be equipped with a mass spectrometer to perform leak detection and to support the vacuum qualification of the main interferometer tube.

\subsubsection{Installation}
\begin{figure}[h!]
\centering
\includegraphics[width=0.6\linewidth]{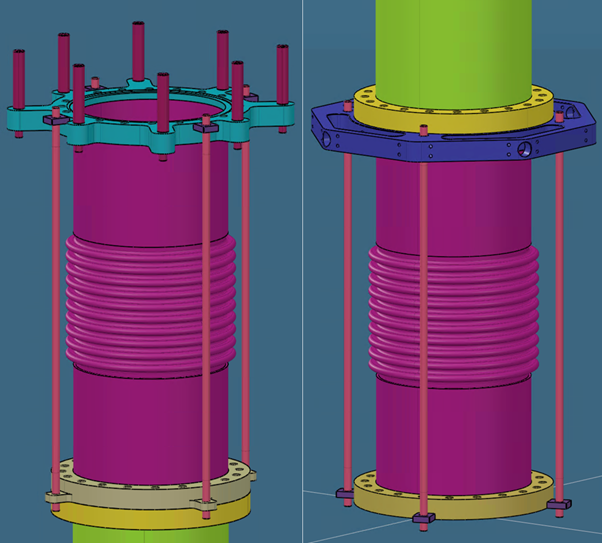}
\caption{Conceptual design for the retractable bellows~\cite{Bongs:2025rqe}.}
\label{fig:BPbellows}
\end{figure}

A detailed plan for installation of modular sections was developed for the AION-10 project~\cite{Bongs:2025rqe}, and can be adapted to the 140 m baseline. The plan involves building the discrete modular sections off-site, first by assembling and capping the vacuum pipes in an ISO6 clean environment. The rest of the module assemblies can be built in a general assembly area, before transport to the shaft. Each modular interferometry beam pipe section includes retractable bellows, shown in Fig.~\ref{fig:BPbellows}, which allows installation of each module's frame to the shaft walls without the delicate process of connecting the vacuum system. It will also allow a staged commissioning of the vacuum system, as detailed above. Assembly of the vacuum system is foreseen from bottom to top. Once the modules are in place on the shaft walls, the vacuum connections can be made.

\subsection{Atom sources} \label{sec:atom-sources}

As described in Sections \ref{sec:staged} and \ref{sec:concept:principle}, the AICE facility will initially use Sr and Yb atoms---species with histories of use in optical atomic clocks~\cite{Ludlow2015} that have atomic structures amenable to laser cooling and clock-transition interferometry~\cite{baynham_prototype_2026}. However, if a strong physics case were to emerge requiring a different atomic species, the AICE facility is compatible with a range of future experiments, e.g., gate valves at each connection node allow the isolation and replacement of any atom source along the baseline (see Section \ref{sec:vacuum}).

Long-baseline atom interferometry requires advanced atom sources: not only is the rate of production of cold atoms a key parameter in determining the phase noise limit $\delta\phi_\mathrm{noise}\sim 1/N_\mathrm{1s}$ (see Section~\ref{sec:staged}), but the atoms must also be extremely cold to limit cloud expansion during interferometry flight~\cite{MAGIS-100:2021etm,Hartwig2015}. To prepare cold Sr and Yb samples for AICE, the laser cooling techniques originally pioneered in degenerate gas experiments listed in Table \ref{tab:atom_production_rates}~\cite{campbell_fermi-degenerate_2017,stellmer_production_2013,roy_rapid_2016,vaidya_degenerate_2015} will be combined with matter-wave lensing to manipulate the atoms into the pK regime~\cite{kovachy_matter_2015,gaaloul_space-based_2022,herbst_matter-wave_2024}.

\begin{figure}
    \centering
    \includegraphics[width=0.9\linewidth]{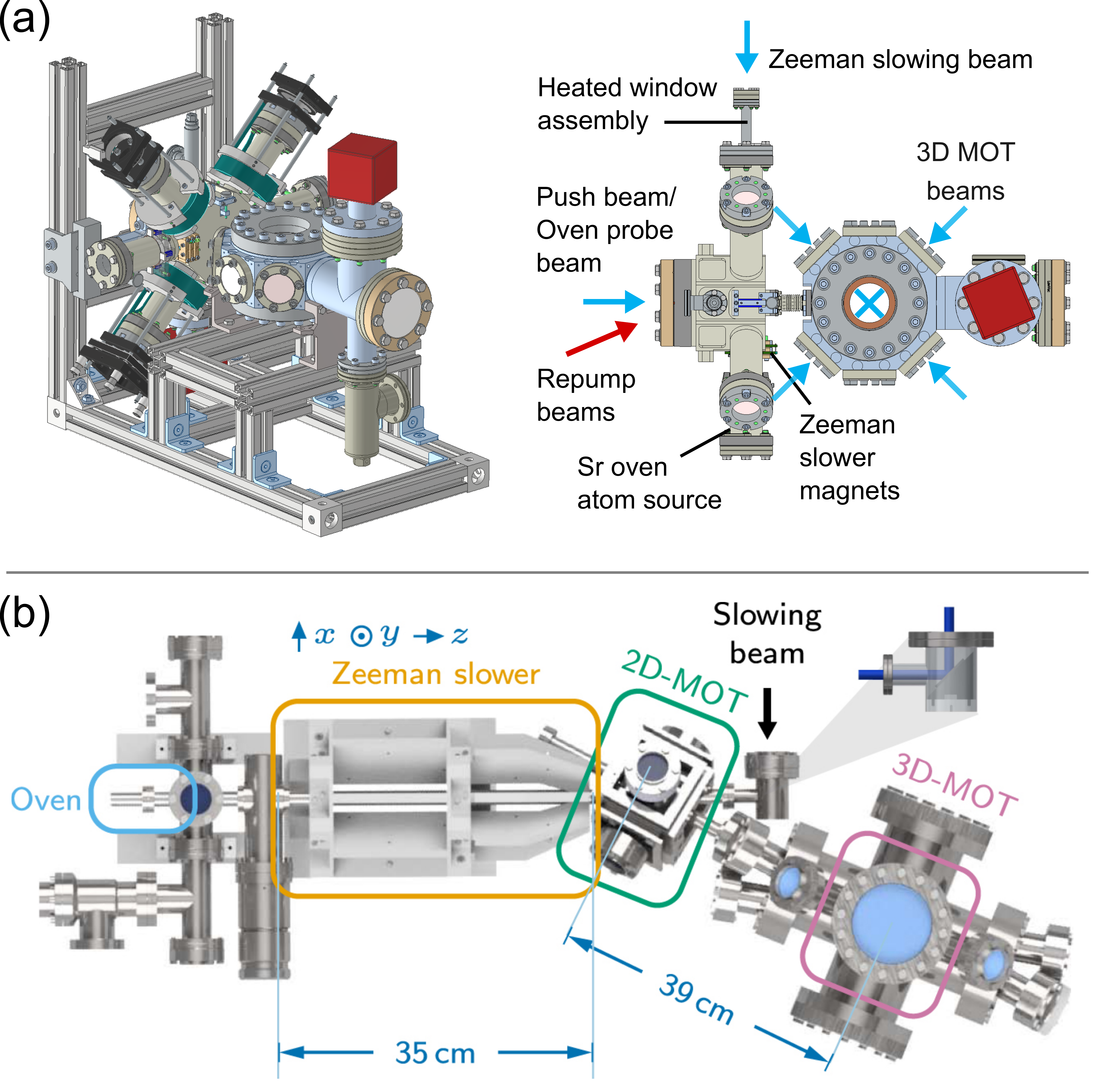}\\
    \caption{Compact cold-atom sources developed by AICE collaborators for (a) Sr and (b) Yb. Adapted from ~\cite{walker2026high,wodey_robust_2021}.}
    \label{fig:atom_sources}
\end{figure}

The Initial parameters of AICE have been chosen to be within the capabilities of existing Sr and Yb cold-atom technologies, targeting the production of $10^5$--$10^6\,$atoms/s. The first atom sources will be based on the compact, high-flux Sr~\cite{walker2026high} and Yb sources~\cite{wodey_robust_2021} shown in Fig.~\ref{fig:atom_sources}, which were developed by members of the AION project in the UK~\cite{Badurina:2019hst,baynham_prototype_2026} (inspired by the Sr source~\cite{feng2024} from the MIGA project in France~\cite{Canuel:2017rrp}) and the VLBAI project in Germany~\cite{Hartwig2015,schilling_gravity_2020,schlippert2020matter}. The engineering design and assembly of atom-source vacuum chambers can take advantage of the technical resources available at CERN, allowing centralized production of multiple units following a model similar to that used for the AION project~\cite{Stray2024}.

R\&D to upgrade atom sources will be crucial to meet the Baseline and Stretch phase noise targets, $\delta\phi_\mathrm{noise}\sim10^{-4}$--$10^{-5}/\sqrt{\mathrm{Hz}}$. Related R\&D in cold-atom sources is already underway for a range of quantum experiments, with recent highlights including the creation of the first continuous BEC~\cite{chen_continuous_2022}, the first continuous reservoir of \textsuperscript{87}Sr atoms at the few-\qty{}{\micro\kelvin} scale~\cite{escudero_steady-state_2021}, and the first continuous outcoupling of cold Sr atoms into an environment suitable for atomic clock interrogation~\cite{chen_narrow-line-mediated_2025}. These continuous sources could unlock very high atom flux and atom-interferometer repetition rates. It is not yet clear which laser cooling configuration will be preferred, so the AICE facility will be kept compatible with any option.

\subsection{Laser systems}
\label{sec:laser}

\begin{figure}
    \centering
    \includegraphics[width=0.9\linewidth]{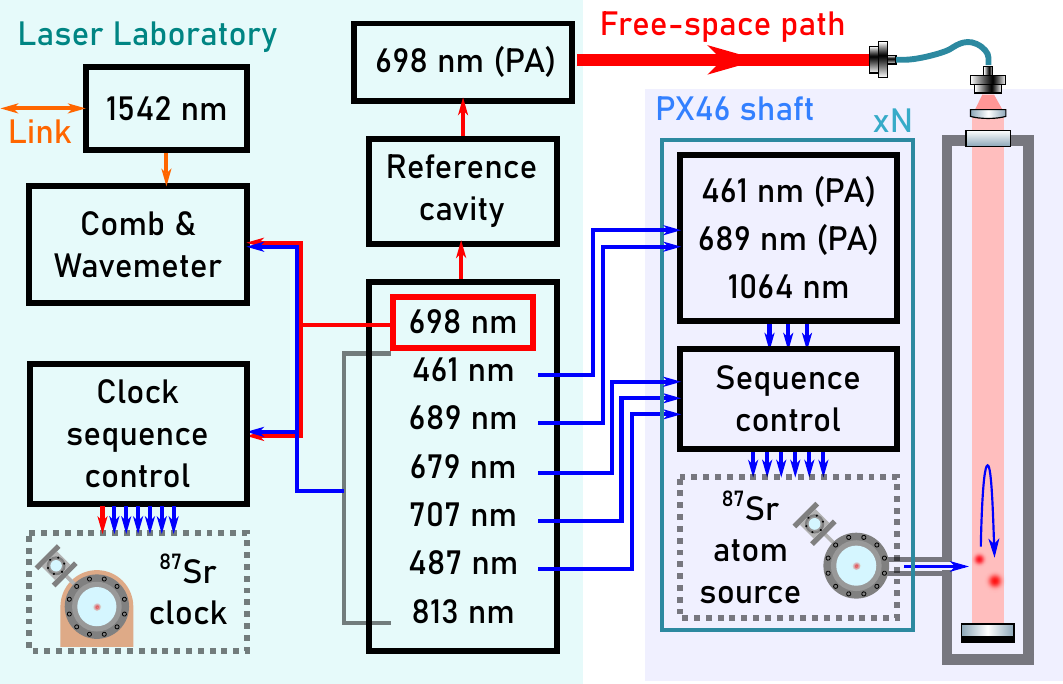}\\
    \caption{The laser systems for the strontium AICE baseline configuration. The master lasers are housed in a surface-level laser laboratory within SX4, along with frequency stabilisation systems and a 1542~nm link to CERN and the European fibre network~\cite{pizzocaro_international_2026}. The \qty{698}{\nano\meter} optical fibres (red) are phase-stabilised~\cite{ma_delivering_1994}. Power amplifiers (PAs)---or slave lasers---follow the frequency of the stabilised master lasers. The PAs have short fibre connections to the instrument in PX46 to minimise stimulated Brillouin scattering~\cite{kobyakov_stimulated_2010}. The top-to-bottom configuration for the atom interferometry beam is shown; an alternative bottom-to-top configuration is also under consideration. The height of the atom source is indicative. Analogous laser systems are foreseen for the ytterbium configuration.}
    \label{fig:Lasers}
\end{figure}

The AICE facility relies on a suite of high-performance laser systems for atom cooling, trapping, state preparation, interferometric manipulation and detection. The baseline configuration employs single-photon atom interferometry on the ultra-narrow $^{1}$S$_{0}\rightarrow{}^{3}$P$_{0}$ clock transition of $^{87}$Sr at 698 nm. %This approach benefits from the intrinsic suppression of the wavefront-aberration systematics that would be associated with two-photon Raman transitions, while providing a direct route to large-momentum-transfer atom optics. Similar advantages are shared by the $^{171}$Yb clock transition that will be used in searches for vector-like dark matter and violations of the Equivalence Principle.
The laser architecture is shown in Fig.~\ref{fig:Lasers}, and is divided into three subsystems:

\begin{enumerate}
\item Master lasers and frequency stabilisation systems in a dedicated surface laboratory within SX4;
\item Local power amplification, beam conditioning, diagnostics and delivery (with control systems) next to the AICE vacuum tube within the PX46 shaft;
\item Optical frequency distribution and beam transport between the SX4 laser laboratory and the PX46 shaft.
\end{enumerate}

The laser systems required for the baseline instrument (Sr) and the planned upgrade (Yb) are listed in Table~\ref{tab:laser_overview}. Lasers and power amplifiers that meet the baseline requirements for Sr and Yb are commercially available from multiple suppliers, having been developed since the early 2000s for various quantum experiments including optical lattice clocks~\cite{Ludlow2015}. The supplier choice will be made at the procurement stage, but the laser technologies will include extended-cavity diode lasers (ECDLs), injected laser diodes, fibre lasers and fibre amplifiers. %The baseline strontium system comprises lasers for first-stage cooling at 461 nm, second-stage cooling at 689 nm, transparency at 487~nm, repumping at 679 nm and 707 nm, and the ultra-stable 698 nm clock laser used for interferometry. Operation with ytterbium will require the introduction of laser systems with wavelengths 399 nm, 556 nm and 578 nm for first-stage cooling, second-stage cooling, and clock interferometry.

Frequency stabilisation systems are a large part of the laser infrastructure, with the strontium clock interferometry laser having the most stringent specifications. The master 698~nm laser is referenced to a high-finesse ultra-low-expansion optical cavity providing sub-Hz linewidth and sub-$10^{-15}$ fractional frequency stability over averaging times up to a few seconds. To provide long-term stability below $10^{-15}$, at timescales of hours to days for which the cavity stability is insufficient, the interferometry laser is further stabilised to a strontium atomic clock within the laser lab. An optical frequency comb is used to transfer the fractional frequency stability from the interferometry laser to all other lasers, except for the far-off-resonant transparency and dipole-trapping lasers, for which wavemeter-based steering is sufficient. A 1542~nm laser stabilised to the optical frequency comb is also used to link the AICE factility to other CERN sites and to the European fibre network~\cite{pizzocaro_international_2026}, allowing frequency distribution to nearby experiments and clock-clock comparison measurements for fundamental physics~\cite{roberts_search_2020}.

Laser light is transferred to the experimental shaft through polarisation-maintaining, single-mode optical fibres. The optical fibres must be up to 200~m to reach atom sources towards the bottom of the PX46 shaft, leading to fibre-induced phase noise~\cite{falke_delivering_2012}, significant for the atom interferometry laser, and stimulated Brillouin scattering~\cite{kobyakov_stimulated_2010}, significant for any laser beams $\gg\qty{100}{\milli\watt}$. For the interferometry laser, the fibre-induced phase noise is suppressed using standard optical-path-length noise cancellation techniques~\cite{ma_delivering_1994}. Stimulated Brillouin scattering is avoided by sending $<\qty{100}{\milli\watt}$ through the long optical fibres, and then amplifying to the W-scale near the atom sources where needed.

The interferometry beam requires several Watts (initial) or more than a hundred Watts (upgrade) at the atoms, delivered either to the top or the bottom of the baseline tube. Both configurations are under consideration for AICE; the top-down configuration is depicted in Fig.~\ref{fig:Lasers}, and follows the MAGIS approach of free-space vacuum beam delivery from the laser laboratory to the entrance of the shaft, followed by a short optical fibre to clean up the spatial mode and beam pointing stability~\cite{MAGIS-100:2021etm}. The alternative bottom-up configuration would avoid the long free-space link to the shaft entrance and the short fibre, but would require space for a high-power slave laser (or power amplifier) at the bottom of the shaft.

The interferometry beam is expanded and mode-cleaned before entering the vacuum system. The optical design emphasises excellent wavefront quality, beam-pointing stability and long-term alignment reproducibility. These quantities directly affect interferometric contrast and phase stability and therefore enter the overall noise budget described in Section \ref{sec:noise}.

\subsubsection{Upgrades} \label{sec:laser-upgrades}

The baseline Sr instrument and Yb implementation will need upgrades to reach the full physics potential of the AICE facility. We list some foreseeable upgrades here.

The available power in the interferometry beam determines the rate at which $\pi$-pulses can be driven on the clock transition, imposing limits on the possible number of LMTs. To progress towards the stretch parameters for AICE outlined in Table~\ref{tab:roadmap} ($n \sim 10^4$ LMTs), we outline an upgrade path for the clock laser in Table~\ref{tab:clock_laser_requirements}. The upgrade is not accessible with commercially available laser technology, but will be a key R\&D activity for AICE and other TVLBAI experiments.

Three-photon clock interferometry in Sr~\cite{carman_collinear_2025} would open up the use of bosonic isotopes of Sr, enabling experiments such as the search for vector B-L dark matter discussed in Section~\ref{sec:uldm_vector}. This would require an additional master laser at 688~nm, sub-Hz stabilisation of 688~nm and 679~nm, and power amplifiers for 689~nm, 688~nm and 679~nm.

The 578~nm clock transition in $^{171}$Yb has long been used in optical lattice clocks~\cite{Ludlow2015}, and the transition has been measured to high precision in the other stable fermionic isotope $^{173}$Yb \cite{Clivati2016}, but clock interferometry with bosonic isotopes $^{168}$Yb, $^{170}$Yb, $^{172}$Yb and $^{174}$Yb would be possible using the narrow-linewidth 507~nm $^{1}$S$_{0}\rightarrow{}^{3}$P$_{2}$ transition~\cite{yamaguchi_high-resolution_2010} or f-shell transitions~\cite{safronova_two_2018,dzuba_testing_2018}. This would open up more vector $B-L$ dark matter searches, but introduces additional anti-reflective coating and chromatic aberration specifications for the interferometry beam delivery path.

In order to reach the Stretch parameters for atom-interferometer phase noise in Table~\ref{tab:roadmap} ($\delta\phi_\mathrm{noise} \sim \qty{1e-5}{\radian}$), upgraded techniques of laser cooling, atom transport and readout will eventually be required - see Section~\ref{sec:atom-sources}. These techniques will require additional laser beams, possibly at new laser wavelengths to address various transitions in Sr and Yb. In combination with efforts to increase in atom number, the phase noise could be further reduced by using controlled atom-atom interactions~\cite{cassens_entanglement-enhanced_2025} or cavity-enhanced non-destructive measurements~\cite{hosten_measurement_2016} to prepare entanglement-enhanced atomic states~\cite{pezze_quantum_2018} such as squeezed states optimised for differential interferometry~\cite{corgier_optimized_2025}.

Sufficient laser lab space, laser links and services will be installed (see Section~\ref{sec:laser-lab-infrastructure}) to allow these upgrades to be implemented without modification of the infrastructure and beam-distribution network.

\begin{table}[htbp]
\centering
\begin{minipage}{\textwidth}
\caption{Principal laser systems foreseen for AICE. The baseline configuration uses
$^{87}$Sr only. The ytterbium system is required for the vector ULDM and
equivalence-principle programme.}
\vspace{3mm}
{\small
\begin{tabular}{llllll}
\hline
Function & Wavelength & Master & Power amp(s) & Linewidth & Status \\
\hline
Sr 1\textsuperscript{st}-stage cooling
    & 461 nm
    & \qty{500}{\milli\watt}
    & \qty{500}{\milli\watt}($\times N$)
    & $<\qty{1}{\mega\hertz}$
    & Baseline \\
Sr 2\textsuperscript{nd}-stage cooling \& launch
    & 689 nm \footnote{To generate sufficient power, a 1379~nm master laser with Raman fibre amplifiers and second harmonic generation may be required.}
    & \qty{200}{\milli\watt}
    & \qty{1}{\watt}($\times N$)
    & $<\qty{1}{\kilo\hertz}$
    & Baseline \\
Sr clock interferometry
    & 698 nm
    & \qty{25}{\milli\watt}
    & \qty{8}{\watt}
    & $<\qty{1}{\hertz}$
    & Baseline \\
Sr detection / repumping
    & 679, 707 nm
    & \qty{100}{\milli\watt}
    & -
    & $<\qty{1}{\mega\hertz}$
    & Baseline \\
Sr magic-wavelength lattice
    & 813 nm
    & \qty{1}{\watt}
    & -
    & $<\qty{10}{\kilo\hertz}$
    & Baseline \\
Sr transparency
    & 487 nm
    & \qty{200}{\milli\watt}
    & -
    & -
    & Baseline \\
Sr/Yb dipole trapping
    & 1064 nm
    & None
    & \qty{5}{\watt}($\times N$)
    & -
    & Baseline \\
Yb 1\textsuperscript{st}-stage cooling
    & 399 nm
    & \qty{500}{\milli\watt}
    & \qty{500}{\milli\watt}($\times N$)
    & $<\qty{1}{\mega\hertz}$
    & Upgrade \\
Yb 2\textsuperscript{nd}-stage cooling \& launch
    & 556 nm
    & \qty{1}{\watt}
    & \qty{1}{\watt}($\times N$)
    & $<\qty{1}{\kilo\hertz}$
    & Upgrade \\
Yb clock interferometry
    & 578 nm
    & \qty{25}{\milli\watt}
    & \qty{8}{\watt}
    & $<\qty{1}{\hertz}$
    & Upgrade \\
Yb detection / repumping
    & 649, 770 nm
    & \qty{100}{\milli\watt}
    & -
    & $<\qty{1}{\mega\hertz}$
    & Upgrade \\
\hline
\end{tabular}
}
\label{tab:laser_overview}
\end{minipage}
\end{table}

\begin{table}[htbp]
\centering
\caption{Preliminary performance requirements for the AICE strontium clock laser systems.}
\vspace{3mm}
\begin{tabular}{lll}
\hline
Parameter & Baseline requirement & Stretch target \\
\hline
Clock wavelength & 698 nm & 698 nm \\
Fractional frequency stability & $<10^{-15}$ & $<10^{-16}$ \\
Laser linewidth & $<1$ Hz & $<0.1$ Hz \\
Residual fibre phase noise & $<1$ mrad & $<0.3$ mrad \\
Beam-pointing stability & $<1\,\mu$rad RMS & $<0.3\,\mu$rad RMS \\
Wavefront error & $\lambda/100$ & $\lambda/300$ \\
Average Power & 8 W &  {100 W}\\
Availability & $>90\%$ & $>95\%$ \\
\hline
\end{tabular}
\label{tab:clock_laser_requirements}
\end{table}

The requirements in Table~\ref{tab:clock_laser_requirements} are provisional
engineering targets, set to keep the associated technical noise contributions
below the atom shot-noise level for the corresponding phase-noise budget in
Table~\ref{tab:roadmap}, rather than final specifications. The detailed
mapping of each requirement onto a specific sensitivity target and operating
mode requires dedicated modelling and simulation of the interferometer
response, which will be carried out as part of the design work towards the Technical Design Report (TDR); the values quoted here are intended to demonstrate feasibility against
the AICE performance ladder and to guide the laser R\&D, not to fix the final
budget.

\subsection{Magnetic system}
\label{sec:shielding}

%\textcolor{red}{Jack Sander - with input from RAL, MAGIS}\\

The fermionic \textsuperscript{87}Sr atom has a complex Zeeman structure, due to its nuclear angular momentum $I = 9/2$. This creates multiple possible clock transitions, which are all sensitive to magnetic fields. In particular, the $m_F = 9/2$ to $9/2$ clock transition exhibits a frequency shift of 489~Hz/G~\cite{boyd_nuclear_2007}.

In order to avoid parasitic excitations via unwanted transitions, such as the $m_F = 9/2$ to $7/2$ transition, two requirements must be fulfilled:\\
 (1) A well-defined constant quantisation axis along the relevant parts of the interferometer beam pipe must be provided by a horizontal magnetic bias field of up to 10~G in order to resolve neighbouring Zeeman sublevels and optimise state preparation and interrogation.\\
(2) The interferometer laser must be linearly polarised with its polarisation angle less than $10\,\textrm{mrad}$  from the applied magnetic field, so that the $m_F = 9/2$ to $9/2$ transition is driven while the parasitic $m_F = 9/2$ to $7/2$ ($\sigma^-$) transition is strongly suppressed.
 A $10\,\textrm{mrad}$ misalignment between magnetic field and polarisation direction would already correspond to a $10^{-4}$ ratio of Rabi frequencies on the undesired vs the desired Zeeman transitions, corresponding to a $10^{-4}$  pulse infidelity.
 
The light polarisation will be set by the in-vacuo optics, which will be difficult to align within the required tolerance to a single set of bias coils.
Hence a tunable horizontal bias field  with homogeneity better than $5\,\textrm{mG}$ and $5\,\textrm{mrad}$ (angular) is required. It must be applicable in both horizontal axes so as to be able to align its direction  with the linear light polarisation.

In addition, time-dependent magnetic field noise must remain below 1~$\mu$G/$\sqrt{\text{Hz}}$ in order to suppress phase noise in the interferometric signal. 

These requirements are met by the magnetic shielding and field design described in~\cite{Bongs:2025rqe}, as summarised below. \\

\noindent
{\bf Magnetic Shielding:} Various geometries of mu-metal magnetic shielding were modelled in~\cite{Bongs:2025rqe} using the COMSOL electromagnetic modelling software package~\cite{COMSOL}, based on a FEA algorithm to model the magnetic fields. The modelling has been done for a single 5 m beam pipe section, using continuous shields (neglecting any holes for mounting and fixings). The modelling was done assuming a 160\ mm beam pipe diameter, whereas AICE has taken the decision to use 200\ mm pipes, in light of experience on AION and MAGIS. It is not anticipated that this change will affect materially the modelling below, as the interferometry beam waist is unchanged.

The magnetic field noise density amplitude in PX46 has been measured (see below), and can be reduced to the required level for frequencies $> 10^{-4}\,\textrm{Hz}$ when shielded by a factor of 1000, as could be provided by a double layer of shielding with staggered octagonal geometry. A simple race-track coil configuration system will provide the required field strength in the interferometry region, but the homogeneity and divergence requirements will not be met. For this, saddle-type coil configurations are required, as discussed below.

\begin{figure}
    \centering
    \includegraphics[width=0.65\linewidth]{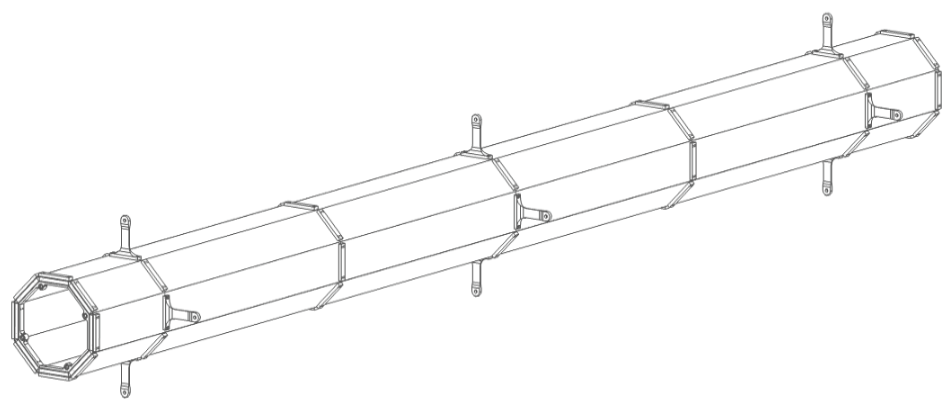}
    \includegraphics[width=0.65\linewidth]{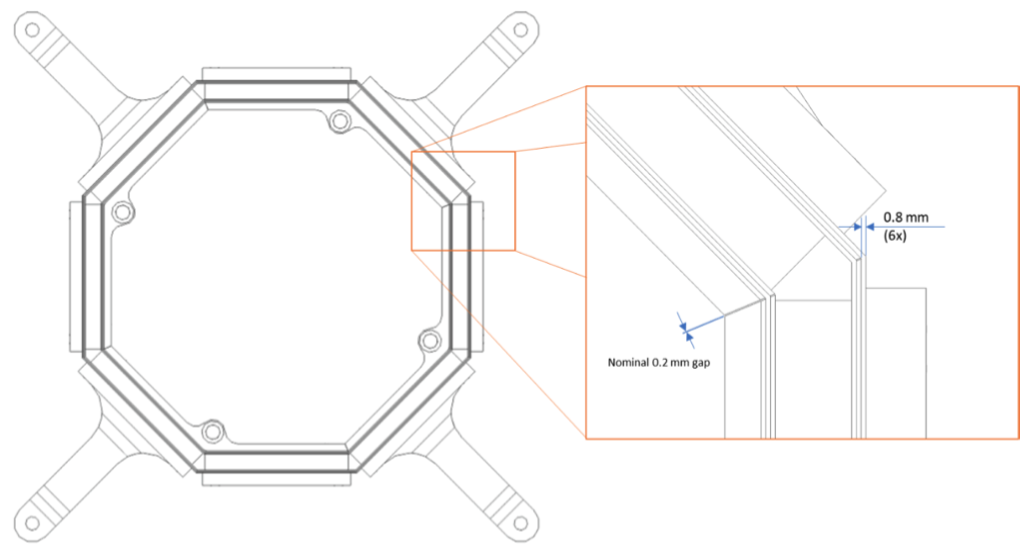}
     \includegraphics[width=0.65\linewidth]{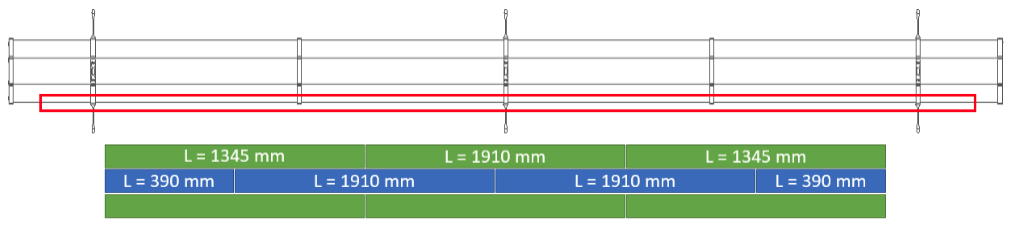}
   \caption{Top panel: CAD image of the magnetic shielding.
   Middle panel: Zoomed-in view of the transverse cross-section of the shielding. Note the staggered V approach to avoid gaps through the full shielding thickness.
   Bottom panel: Representative sketch showing the longitudinal staggering between shielding layers 1-3 (inner shield). An identical scheme is used in shielding layers 4-6 (outer shield). Figures taken from~\cite{Bongs:2025rqe}.}
    \label{fig:Shielding}
\end{figure}  

The design of the magnetic shielding proposed in~\cite{Bongs:2025rqe} is shown in Fig.~\ref{fig:Shielding}. It has been driven not only by the shielding requirements, but also by practical requirements such as ease of installation and assembly, annealing needs and de-Gaussing capabilities. With the interferometer region being located in long cylindrical tubes, the magnetic field modelling has been carried out for cylindrical and octagonal shields of mu-metal. These have a typical thickness of $2-3\,\textrm{mm}$, in both a single-layer configuration and a double-layer configuration.

The design converged to octagonal shields, which is also the design choice of the MAGIS collaboration~\cite{MAGIS-100:2021etm} and of the VLBAI interferometer in Hannover~\cite{schlippert2020matter, wodey2020scalable}. This allows mounting the mu-metal in a staggered geometry (removing seams at the joints) and allows the use of flat sheets of mu-metal.

A single magnetic shield consists of three layers of $0.8\,\textrm{mm}$ thick mu-metal, arranged in an octagonal prism, for a total thickness of $2.4\,\textrm{mm}$. To compress the layers, ensure minimal gaps and provide regular interfaces to both the tower and inner vacuum chamber, the layers are fixed regularly using a modular mild-steel collar and clamping system. The second magnetic shield has the same general design, with increased guiding dimensions to allow a nominal $15\,\textrm{mm}$ gap between the inner and outer shields. The assembly method is discussed below.

As mu-metal cannot typically be sourced in lengths longer than $3\,\textrm{m}$, and to promote good assembly practice, the assembly must contain both longitudinal and transverse splits. Longitudinally, the octagon is formed by 4 vee-shaped cross-sections for each layer. The middle layer is then offset by 45 degrees to both the outer and inner layers to minimise the impact of gaps between mu-metal plates. In the transverse direction, the middle layer is similarly staggered, with the intent of minimising all shielding leaks.

It was envisaged in~\cite{Bongs:2025rqe} that the shield would be fixed to the UHV chamber on the upper flange of the tied bellow assembly on the lower end of each beam pipe assembly. At the middle and upper interfaces, it will be fixed in 5 degrees of freedom using low-friction linear bearings to bespoke UHV flanges. This provides mechanical stability, and provides a pathway to stabilise the UHV chamber at the centre of the unsupported $4.6\,\textrm{m}$ span.\\
%It may be possible to implement a further radial mount in the centre of each $2\,\textrm{m}$ chamber, though initial stability simulation indicates that this will make a negligible overall contribution to the stiffness of the structure or component subassemblies.
  
\begin{figure}
    \centering
    \includegraphics[width=0.49\linewidth]{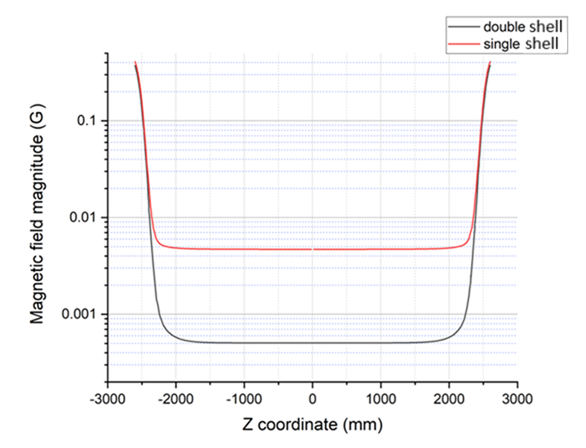}
    \includegraphics[width=0.49\linewidth]{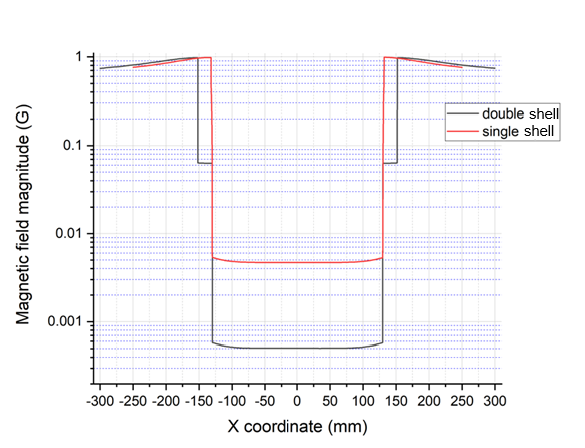}
   \caption{{\it Left panel}: The magnetic field magnitude in the centre of single layer (red line) and double layer (black line) octagonal $2\,\textrm{mm}$ shielding as function of vertical position ($z$). The Earth magnetic field ($0.6\,\textrm{G}$ outside the shields) is seen to penetrate a distance of $\sim 500\,\textrm{mm}$ from the ends. The magnetic field is $4.4\,\textrm{mG}$ at the centre of the shield for the single-layer shielding and $0.5\,\textrm{mG}$ for the double-layer shielding configuration.
   {\it Right panel}: The magnetic field magnitude as function of distance to the central axis of the tube for a single-layer (red line) and double-layer (black line) shield configuration. An Earth magnetic field of $0.6\,\textrm{G}$ is applied outside the shields. Figures taken from~\cite{Bongs:2025rqe}.}
    \label{fig:SingleDouble}
\end{figure}

%\subsubsection*{Shielding Analysis}
\noindent
{\bf Analysis of single- and double-layer octagonal shields:}
Results of modelling single- and double-layer octagonal geometries are shown in Fig.~\ref{fig:SingleDouble} ($B$ in the centre of the tube as function of height $z$ in the left panel and $B$ as function of the distance to the central axis of the tube in the right panel). In this representation, the $x$-axis is purely horizontal, with the model assuming a field of $0.6\,\textrm{G}$ in this direction. The field inside the single layer shields is shown to be at the level of $4.4\,\textrm{mG}$, so we have a static shielding factor of 130 in this configuration. Considering the requirement of setting magnetic fields to mG level homogeneities and anticipating compromises to the shielding when allowing for access ports, etc., we conclude that this single-layer shielding configuration will not be sufficient. The results for double-layer octagonal shield geometry modelling show an increased shielding down to $0.5\,\textrm{mG}$ inside the shields and field gradients $< 0.01\,\textrm{G/m}$. We can conclude that a double-layer octagonal shield would be suitable for the AICE set-up.\\
 
\begin{figure}
    \centering
    \includegraphics[width=0.6\linewidth]{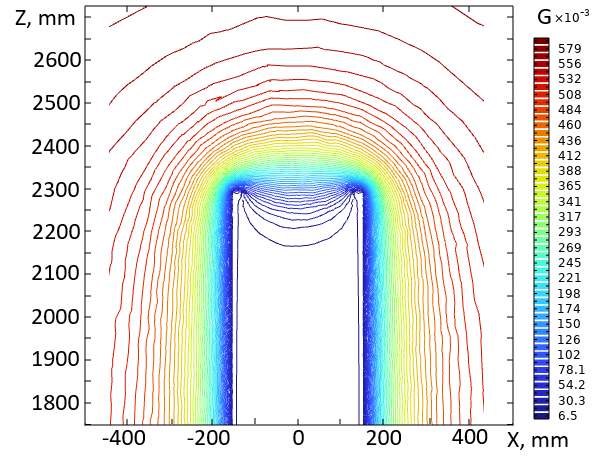}
   \caption{The magnetic field lines near the top section of the shields illustrating how the magnetic field will penetrate into the top section of the shields. Figure taken from~\cite{Bongs:2025rqe}.}
    \label{fig:Bleak}
\end{figure}

\noindent
{\bf Magnetic field near the end-sections of the shielding:}
The magnetic field near the tops of the shields will leak inside the shields, as can be seen in Fig.~\ref{fig:Bleak}, showing that the field penetrates $\sim150\,\textrm{mm}$ into the shields to a level of $6.5\,\textrm{mG}$ for the double-layer shielding configuration. The effects of this can be mitigated by end-cap screening and compensation coils, but we anticipate that there will be lengths at the top and bottom ends of the shields that cannot be used for interferometry purposes, so the shields will have to be slightly longer than the designed interferometry length.\\

\noindent
{\bf Magnetic field measurements and noise:}
Magnetic field measurements have been made at the top and bottom of PX46 over a period of 6 days \cite{Arduini:2851946}, see Sections~\ref{sec:electromagnetic} and~\ref{sec:machine_cycle}. %{\it allowing us to construct the ambient magnetic noise spectral density in a frequency range of $1\,\textrm{$\mu$Hz}$ to $50\,\textrm{Hz}$. Fig.~\ref{fig:BBeecroft} shows a set of magnetic field measurements at various heights in the Beecroft building.} %There is a noticeable fluctuation in the ambient magnetic field as a function of height, which can be explained by the difference in surroundings over the full length of the experimental location, ranging from a basement environment, through the vicinity of the (ferromagnetic) stairwell structures and closeness to large windows at the top. The measurements have been repeated and were found to be stable in time.} 
The designed static shielding factor of 1000 is sufficient to ensure that both the overall frequency-dependent ambient magnetic noise and the LHC-produced magnetic field, also considering its variations during a machine cycle, as illustrated in Figs.~\ref{fig:MagField_cycle_freq} and~\ref{fig:MagField_cycle_time}, would not represent issues for interferometry purposes in the frequency range $> 1\mu$G/$\sqrt{\rm Hz}$ where AICE would take data.\\

\begin{figure}
    \centering
    \includegraphics[width=0.75\linewidth]{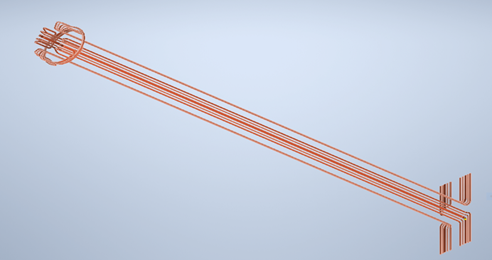}
   \caption{CAD image of a guide field coil assembly. Figure taken from~\cite{Bongs:2025rqe}.}
    \label{fig:FieldCoil}
\end{figure}

\noindent
{\bf Magnetic guide field:}
In order to provide a stable and tuneable horizontal magnetic field the design in~\cite{Bongs:2025rqe} contains field coil assemblies for each interferometry region. Fig.~\ref{fig:FieldCoil} shows a CAD image of a field coil assembly. The wires are assembled onto a semi-cylindrical shell measuring $4.6\,\textrm{m}$ in length, and a maximum envelope diameter of $272\,\textrm{mm}$, including the wires. Due to the size of the shell, it is expected that this will be fabricated in parts before being mechanically clamped together. At such breaks, the connection between the wires will be specified carefully to maintain magnetic field homogeneity. \\

\begin{figure}
    \centering
    \includegraphics[width=0.6\linewidth]{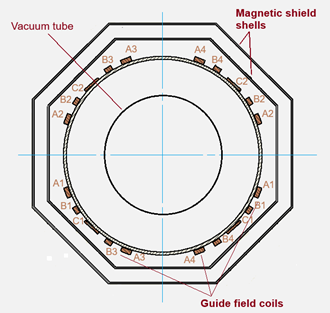}
   \caption{Field coil geometry showing the two sets of coils. Figure taken from~\cite{Bongs:2025rqe}.}
    \label{fig:CoilGeom}
\end{figure}

\begin{figure}
    \centering
    \includegraphics[width=0.9\linewidth]{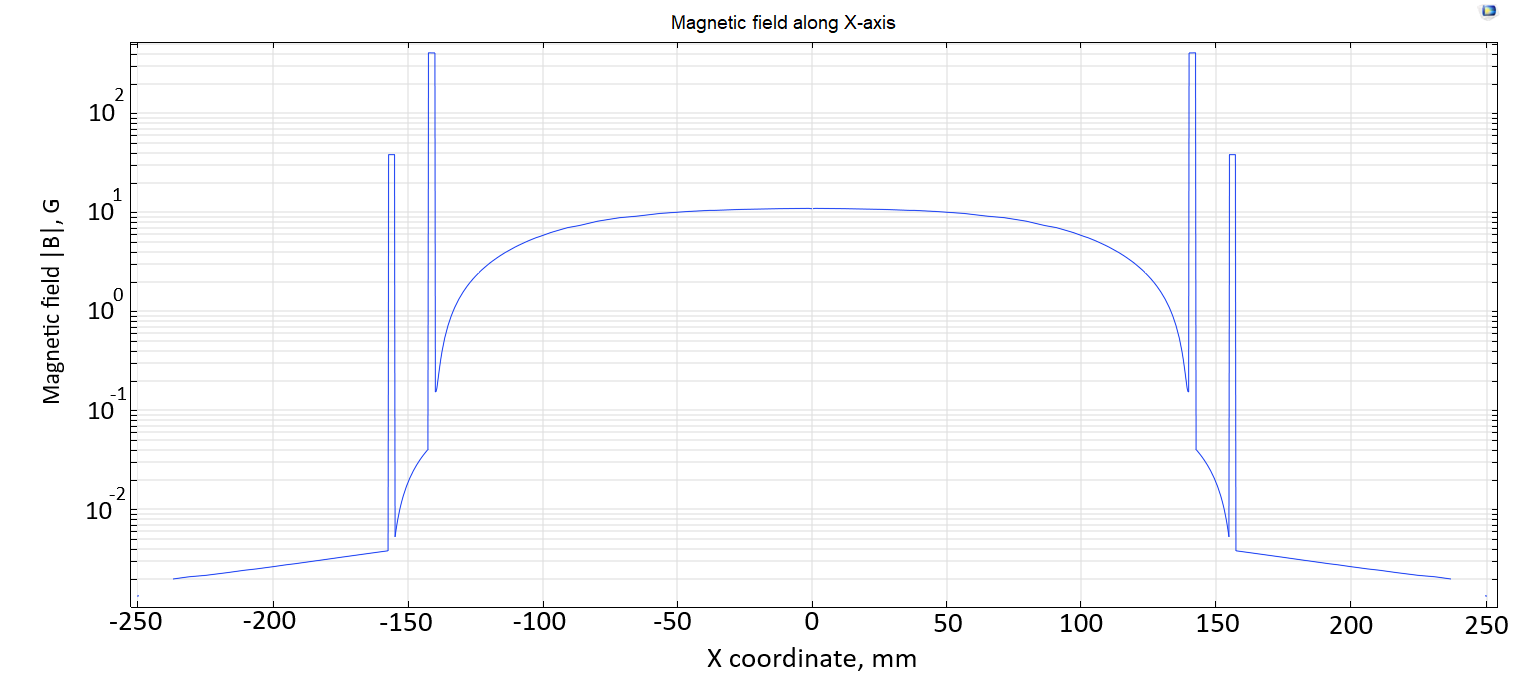}
    \caption{Modelling of the field generated by the magnetic guide field coils inside the double layer magnetic shield. Figure taken from~\cite{Bongs:2025rqe}.}
    \label{fig:fieldmodelled}
\end{figure}

\noindent
{\bf Field homogeneity and field divergence modelling:}
The magnetic guide field has been modelled based on a set of five pairs of coils $4.6\,\textrm{m}$ long, as sketched in Fig.~\ref{fig:CoilGeom}. By selecting specific coil pairs the horizontal orientation of the resulting magnetic field can be set. Fig.~\ref{fig:fieldmodelled} shows the magnetic field generated in the tube, inside the double magnetic shields. The shielding for a cylindrical shield geometry is known analytically (see, e.g.,~\cite{wodey2020scalable}), we have used COMSOL multiphysics v6.3 in all the shielding and magnetic guide field calculations~\cite{COMSOL}. The current in coils A1 and A2 are set to $4.6\,\textrm{A}$, in coils B1 and B2 to $3.5\,\textrm{A}$ and in coils C1 and C2 to $3.3\,\textrm{A}$. A field strength of $10.8\,\textrm{G}$ is generated. The field homogeneity in the interferometry region is below $5\,\textrm{mG}$. 
 
\begin{figure}
    \centering
    \includegraphics[width=0.65\linewidth]{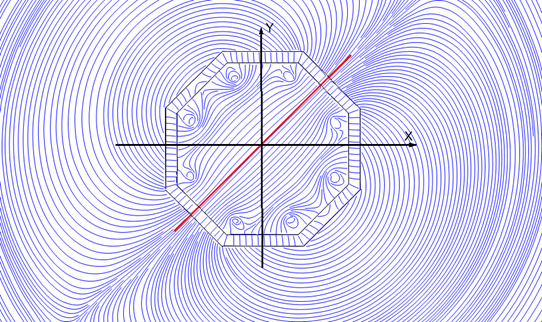}
    \hspace{5mm}
    \includegraphics[width=0.95\linewidth]{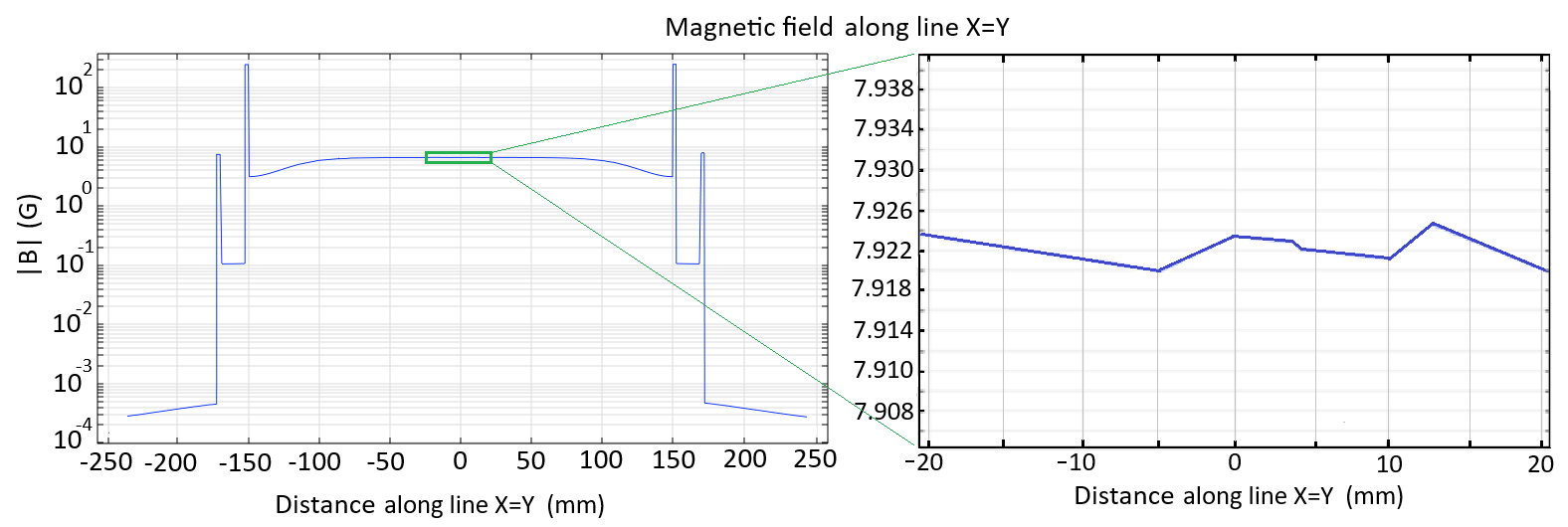}
    \caption{Upper panel: Magnetic field lines generated with equal currents through the coil sets. The magnetic field orientation is along the $x = y$ line. Lower panel: The magnetic field magnitude generated along the $x = y$ line. The inset shows the field in the central region with a homogeneity below $5\,\textrm{mG}$. Figures taken from~\cite{Bongs:2025rqe}.}
    \label{fig:fieldlines}
\end{figure}

{\bf Magnetic Field Orientation:}
The orientation of the field generated by this set of coils has also been modelled, and is well below $5\,\textrm{mrad}$ in the $19\,\textrm{mm}$ range from the centre of the tube.
The field direction can be adjusted in the $(x, y)$ plane by activating both sets of coils. 
The upper panel of Fig.~\ref{fig:fieldlines} shows the field lines as modelled with the coils arranged for a $45^\circ$ angle in the $(x, y)$ plane. This field orientation is obtained by applying the same current settings to each coil set. The corresponding field magnitude along the $x = y$ axis is shown in the lower panel of Fig.~\ref{fig:fieldlines}. As one would expect in this symmetrical layout, the field homogeneity stays below $5\,\textrm{mG}$.

\begin{figure}
    \centering
    \includegraphics[width=0.75\linewidth]{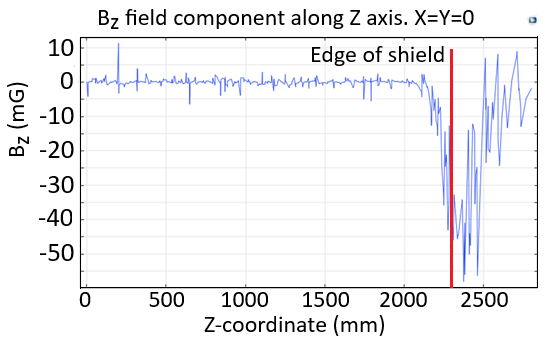}
    \caption{The magnetic field $B_z$ inside the shields: near the edge of the shield the external field will penetrate into the shields. The spike at $z\approx 200\,\textrm{mm}$ is due to mesh refinement. Figure taken from~\cite{Bongs:2025rqe}.}
    \label{fig:fieldpen}
\end{figure}

{\bf Correction Coils:}
The magnetic field near the shield edges will naturally be affected by the external magnetic field (the Earth's field and the additional magnetic environment surrounding the experiment). As this effectively reduces the available length over which interferometry can be performed, it is of importance to minimise the region where the field is affected. Fig.~\ref{fig:fieldpen} shows the magnetic field as modelled near the shield edges: the field is disturbed by the external environment up to a distance of $100\,\textrm{mm}$ from the edge, which would reduce the available interferometry length by $200\,\textrm{mm}$.

\begin{figure}
    \centering
    \includegraphics[width=0.4\linewidth]{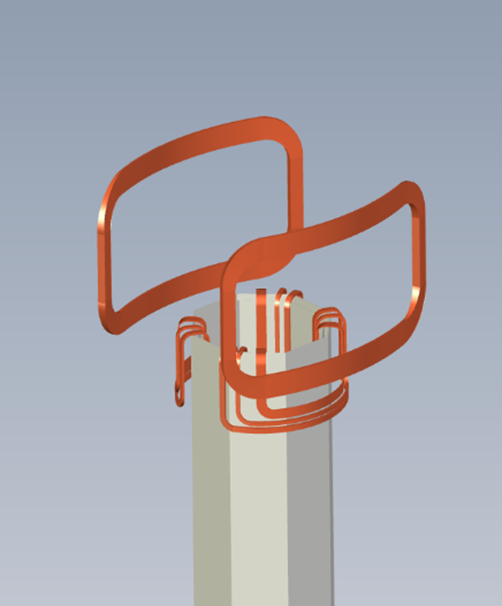}
    \includegraphics[width=0.55\linewidth]{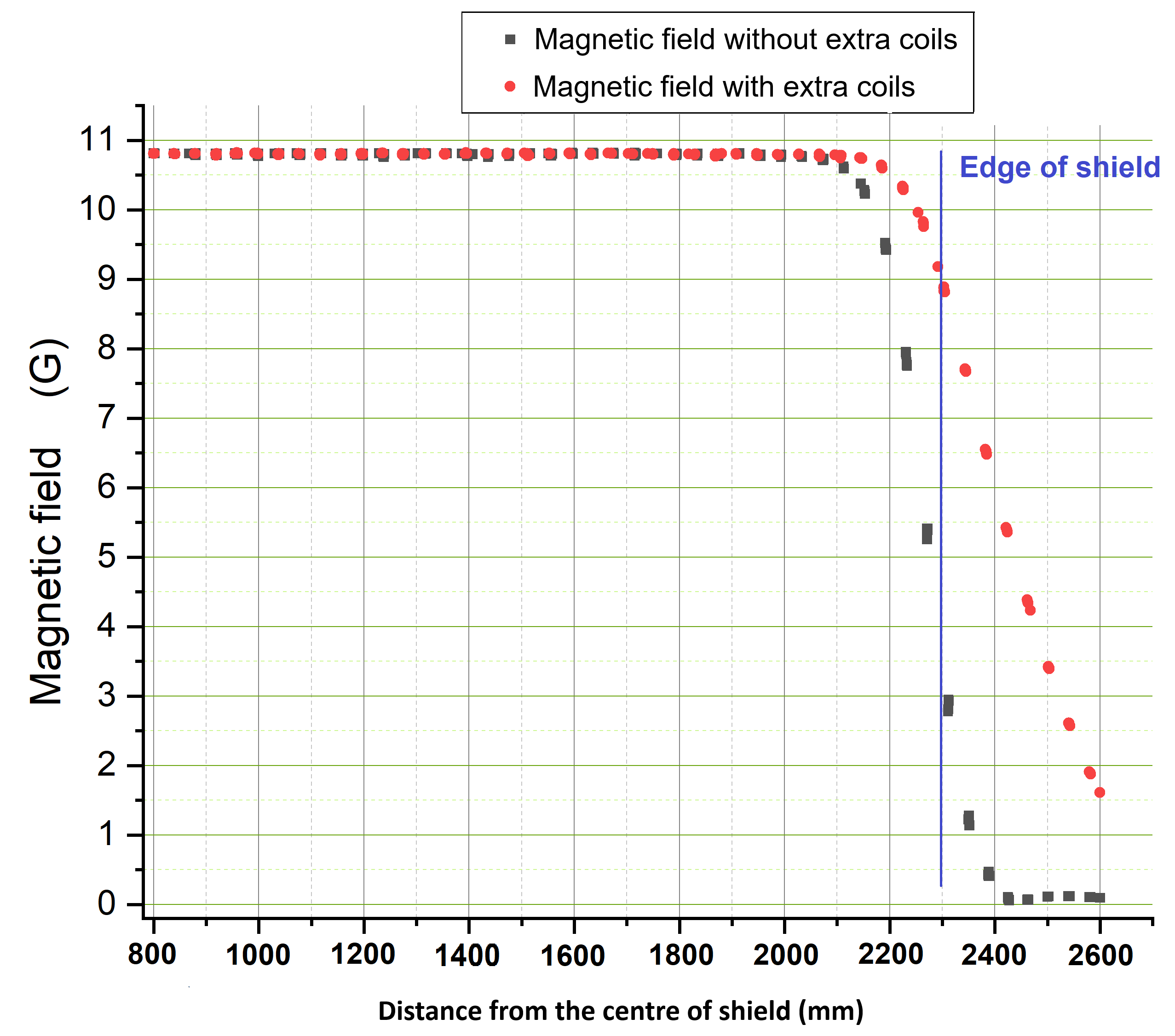}
    \caption{Left panel: A pair of correction coils placed near the field edge. Right panel: The magnetic field in the shields modelled with (red) and without (black) the compensation coil energised. Figures taken from~\cite{Bongs:2025rqe}.}
    \label{fig:compcoils}
\end{figure}

To maximise the available interferometry region in the tube, correction coils will need to be added near the edges of the shields. A pair of correction coils has been modelled to provide the correction to the external magnetic field near the shield edges (see Fig.~\ref{fig:compcoils}). Modelling the guide field with the correction coils activated shows that the guide field is extended closer to the shield edge, providing a longer interferometry region.

\begin{figure}
    \centering
    \includegraphics[width=1\linewidth]{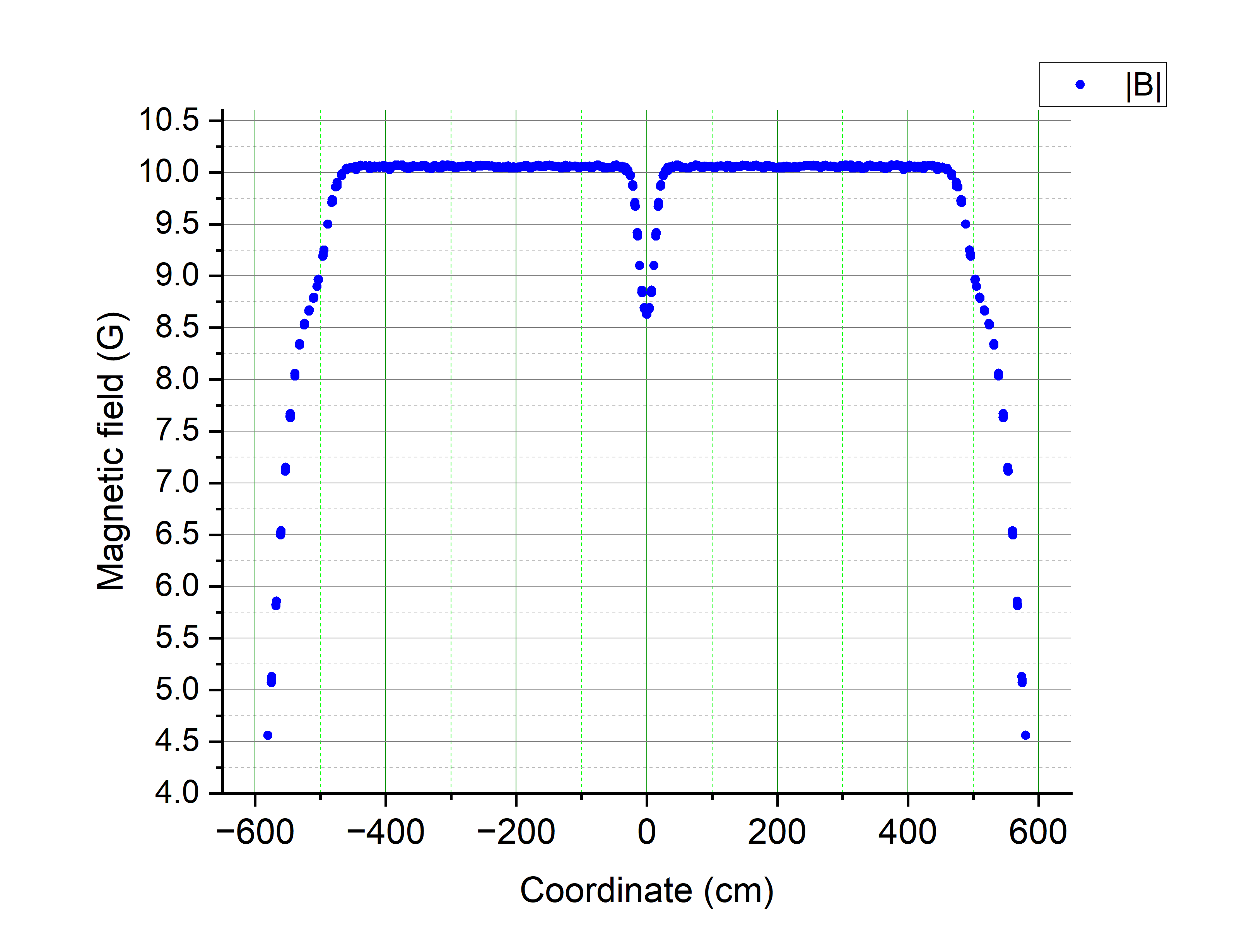}
    \caption{Magnetic field over a representative 10\ m length of the interferometer beam pipe, applying guide fields and compensation coils. Figure taken from~\cite{Bongs:2025rqe}. 
    %\textcolor{red}{The figure covers only a short distance. Requires modification.}
    }
    \label{fig:fullfield}
\end{figure}

The magnetic field was modelled in~\cite{Bongs:2025rqe} over the length of two separate interferometer and shield sections. The magnetic field, with the guide field coils and the compensation coils applied, is shown in Fig.~\ref{fig:fullfield}. In this model there is an interconnect chamber between the two 5 m pipe sections. The drops in field on either end of the figure are outside the shield and coil effects.

\subsection{Support structure and vibration diagnostics}
%\textcolor{red}{Damien L., Marc Timmins, Carlo S., Michael G.}\\
%\textcolor{red}{summarise analysis in AION-10 TDR}
\noindent The experience gained during AION studies~\cite{Bongs:2025rqe} will provide essential input for the development of the AICE support structure and vibration diagnostics strategy. At this stage, the detailed support design is still to be defined, and the frequency ranges that may limit the system performance will need to be confirmed through dedicated studies. \noindent Once the support concept is established, further vibration measurements will be performed, with seismometers and accelerometers installed along the shaft to characterise the relevant seismic and mechanical noise. These data will guide the definition of suitable passive and, if required, active damping solutions.

\begin{figure}[!tbp]
  \centering
    \begin{minipage}[b]{0.3\textwidth}
    \includegraphics[width=\textwidth]{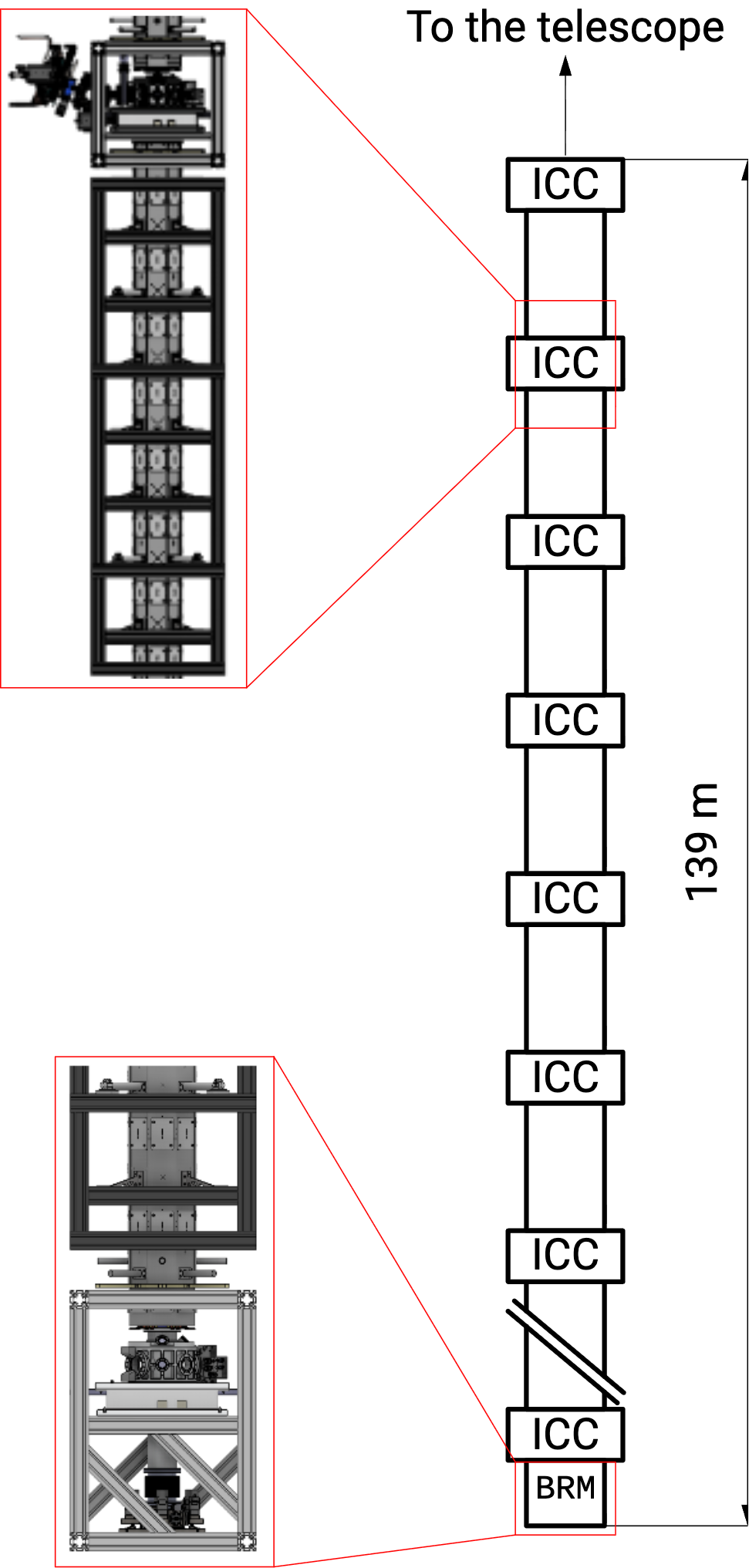}
    \caption{Schematic drawing of the main arm supporting system. The AICE main interferometer will implement a similar reticular support structure as AION-10~\cite{Bongs:2025rqe} (reported in the picture) or MAGIS.}
    \label{fig:Schematicsupport}
  \end{minipage}
  \hfill
    \begin{minipage}[b]{0.6\textwidth}
    \includegraphics[width=\textwidth]{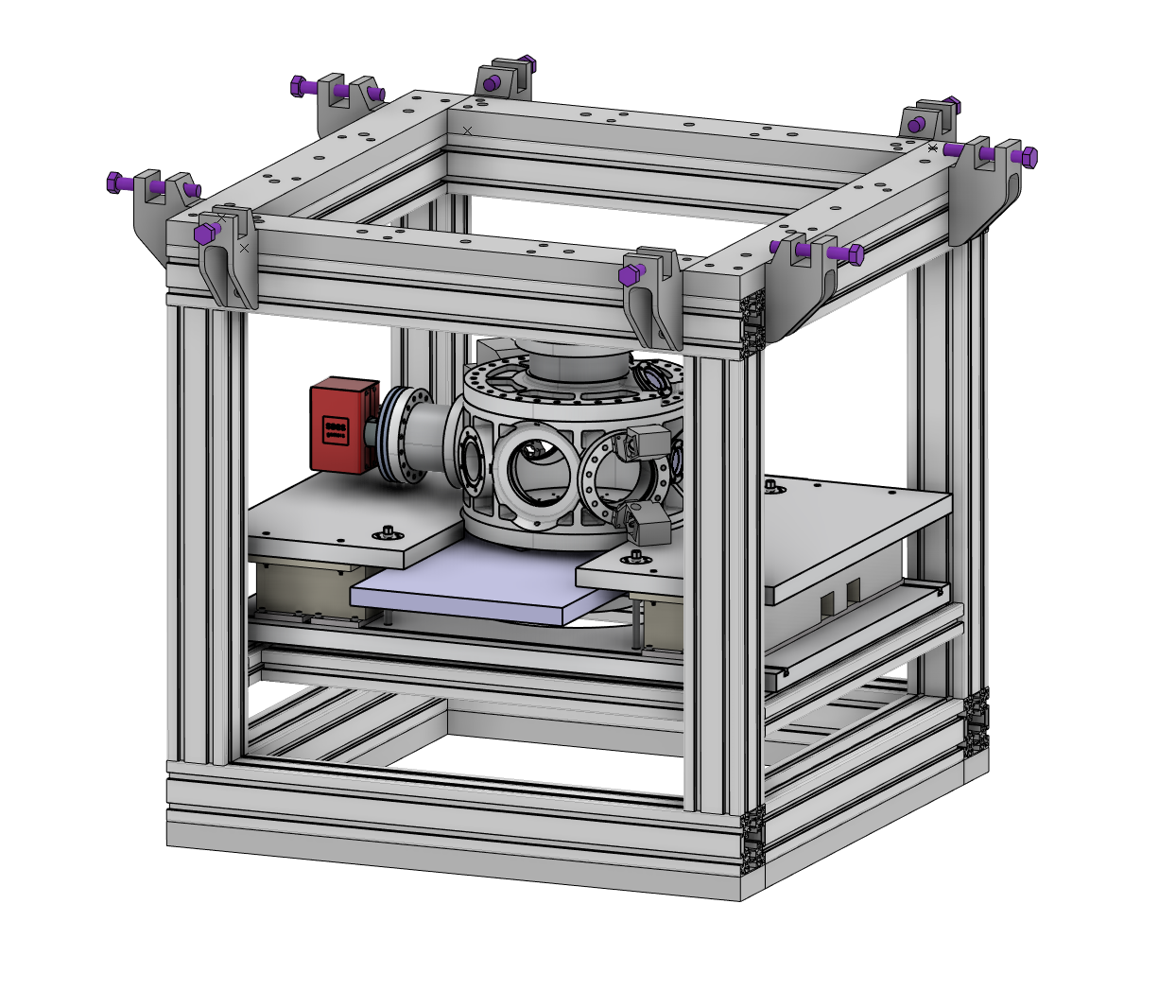}
    \caption{Concept of interconnect chamber in a support frame, including active isolation platforms (grey). The configuration can be modified to isolate either the optics and cameras only, or the whole interconnect chamber. Also, the isolation units can be removed and replaced with solid blocks if not required.}
    \label{fig:AIONsupport}
  \end{minipage}
\end{figure}

In a first approximation, the modular sections and interconnecting chamber supports will be mounted separately as done in MAGIS (see Figs.~\ref{fig:Schematicsupport} and \ref{fig:AIONsupport}). Particular attention to the transfer function studies will target the in-vacuum optics (i.e., the retroreflective mirror) and ancillaries. The fixed in-vacuum optics must be simultaneously compatible with the Sr and Yb interferometry wavelengths, which is non-trivial in combination with the wavefront-flatness and pivot-point requirements. The clear aperture of the modular sections must also be maximised, both to limit diffraction of the interferometry beam and to maximise the achievable LMT and pulse efficiency.

AION studies~\cite{Bongs:2025rqe} show that vibration isolation, if required, can be implemented within the module frames directly. Isolation is likely only required at the key optics: telescope lenses, atom source laser optics, launch lattice assembly, and the retroreflective mirror. These optics are at a few discrete locations along the beam pipe; top, bottom, and at each atom source. Therefore the isolation strategy can be simplified greatly compared with a full isolation of the entire instrument.

% \textcolor{red}{To what extent are the figures taken from the AION-10 study?}

% \textcolor{red}{Comment on compensating the varying ambient magnetic field during the LHC magnet cycle?}

\newpage

\section{PX46 Site and CERN Integration}
\label{sec:site}

The integration of AICE into the LHC infrastructure has been the subject of extensive evaluation by the CERN Physics Beyond Colliders (PBC) Study Group. These assessments, comprising both a Conceptual Feasibility Study \cite{Arduini:2851946} and a subsequent Implementation Study \cite{Arduini:2025jhe}, have concluded that there are no fundamental technical obstacles or ``showstoppers" to hosting a vertical long-baseline atom interferometer at the LHC site. The analysis has identified the PX46 access shaft at Point 4 as the most suitable location due to its total vertical height of approximately 143 meters and an internal diameter of 10.1 meters. 

A central finding of these studies is the feasibility of concurrent operation with the High-Luminosity LHC (HL-LHC), provided that essential infrastructure modifications are completed during the Long Shutdown 3 (LS3) window. By decoupling these preparatory works from the experiment's later construction, AICE can be installed and commissioned during Run 4 without impacting CERN’s core scientific programme.

The following subsections describe the multi-faceted technical analysis performed for this integration. The site overview characterizes the PX46 access shaft at Point 4, confirming its suitability based on its 143\ m vertical height and the results of environmental surveys that found ambient seismic, magnetic, and electromagnetic noise levels to be within acceptable limits for high-precision quantum sensing.

The next Section describes the experiment layout and its installation in the PX46 shaft. In particular, the staging of the infrastructure and experiment installation is described. It is worth underlining here that transport studies confirmed that the installation of the experiment is compatible with the shaft's primary function of moving LHC equipment, as a sufficient horizontal cross-section remains available for concurrent use.

The Section on safety describes the main sources of hazard in PX46 and the risk mitigation strategies pursued. Subsection~\ref{sec:site:safety:fire} details fire safety measures designed to protect personnel from potential smoke propagation from the UX45 cavern through the PX46 shaft. It outlines the implementation of two independent escape routes to ensure redundant egress paths. Central to this strategy is the specialized elevator platform, engineered to facilitate emergency evacuation to either the surface or the shaft base within a mandatory two-minute window. Operational protocols include the mandatory use of self-rescue masks as PPE and the provision of localized fire-extinguishing equipment directly on the mobile platform. These measures are supported by site-specific training and a robust alarm matrix that integrates with existing LHC safety infrastructure to ensure rapid emergency response. The same safety measures and PPE equipment address also the risk of helium release from the LHC machine. It should be noted that helium release from RF cryomodules is considered low hazard due to the cavern's large volume, while more significant incidents in dipole strings are diverted away from the site via pressure-resistant relief doors. This integrated confinement and routing strategy ensures that even a Maximum Credible Incident in adjacent sectors does not create a significant safety hazard within the PX46 shaft.
Finally, studies on radiation protection outline the necessary civil engineering interventions and mitigation measures, most notably the construction of a 0.8-meter thick concrete shielding wall in the TX46 gallery, and the construction of a mobile elevator platform suitable also for evacuation in case of emergency. The shielding is designed to protect personnel during standard operation and in case of accidental beam loss minimizing exposure risk to ionizing radiation, as detailed in Subsection~\ref{sec:site:safety:RP}.

The analysis of the access system addresses the stringent safety requirements for deep-shaft operations. This system guarantees routine access to the experiment's side-arms and is adapted to the management of a controlled evacuation within a mandatory two-minute window in the event of fire or cryogenic helium hazards. Modiﬁcations include extending the LHC Access Control System (LACS) and LHC Access Safety System (LASS) to the shaft, and installing ﬁre detectors, alarms, and emergency communications.

The electrical infrastructure subsection evaluates the project's requirements, confirming that existing electrical power at Point 4 is sufficient to meet the estimated ultimate electrical load foreseen for the experiment of $\approx$ 450~kW. This includes also a detailed analysis of the different power loads, and a preliminary evaluation of the staging of works required during LS3 and for the whole experiment construction.

The following section confirms that no major modifications are required to the Heating, Ventilation, and Air Conditioning (HVAC) systems, and in particular that there is no operational impact to the LHC accelerator. It also describes the creation of a specialized cooling distribution network for the atom sources and a dedicated HVAC system for the surface laser laboratory to ensure the required thermal stability. 

%The laser laboratory and its requirements are described in the following subsection. \textcolor{red}{Which?}

Section~\ref{sec:laser-lab-infrastructure} describes the surface laser laboratory in building SX4, which provides a stable, low-vibration environment for the experiment’s master lasers and frequency references. It houses the strontium clock laser, frequency comb infrastructure and control electronics, with sufficient floor space reserved for future ytterbium system upgrades. The laboratory features a dedicated HVAC system and an access airlock to preserve the thermal stability and pressure cascade required for underground ventilation. High-precision beams are delivered to the 143m shaft via phase-stabilised optical fibres and specialized free-space links for the clock interferometry beam. This centralized modular architecture simplifies maintenance and allows for advanced atom-optics development without interfering with the installed detector infrastructure

Finally, we review the compatibility of AICE with the HL-LHC, confirming the possibility of concurrent operation. Some final considerations on the sustainability assessment of AICE conclude this discussion.

\subsection{Site overview}

The PX46 shaft is the deepest at CERN, featuring a total vertical height of approximately 143 meters from the floor of the SX4 surface building down to the level of the LHC tunnel, with an internal diameter of 10.1 meters. Architecturally, PX46 is integrated into a complex underground network: it is connected directly to the UX45 cavern via the horizontal TX46 gallery and the TU46 ventilation gallery, while its surface access is managed through the SX4 building, which is already equipped with essential heavy-lifting infrastructure including a transport crane.

\begin{figure}[h!]
    \centering
    \includegraphics[width=0.9\textwidth]{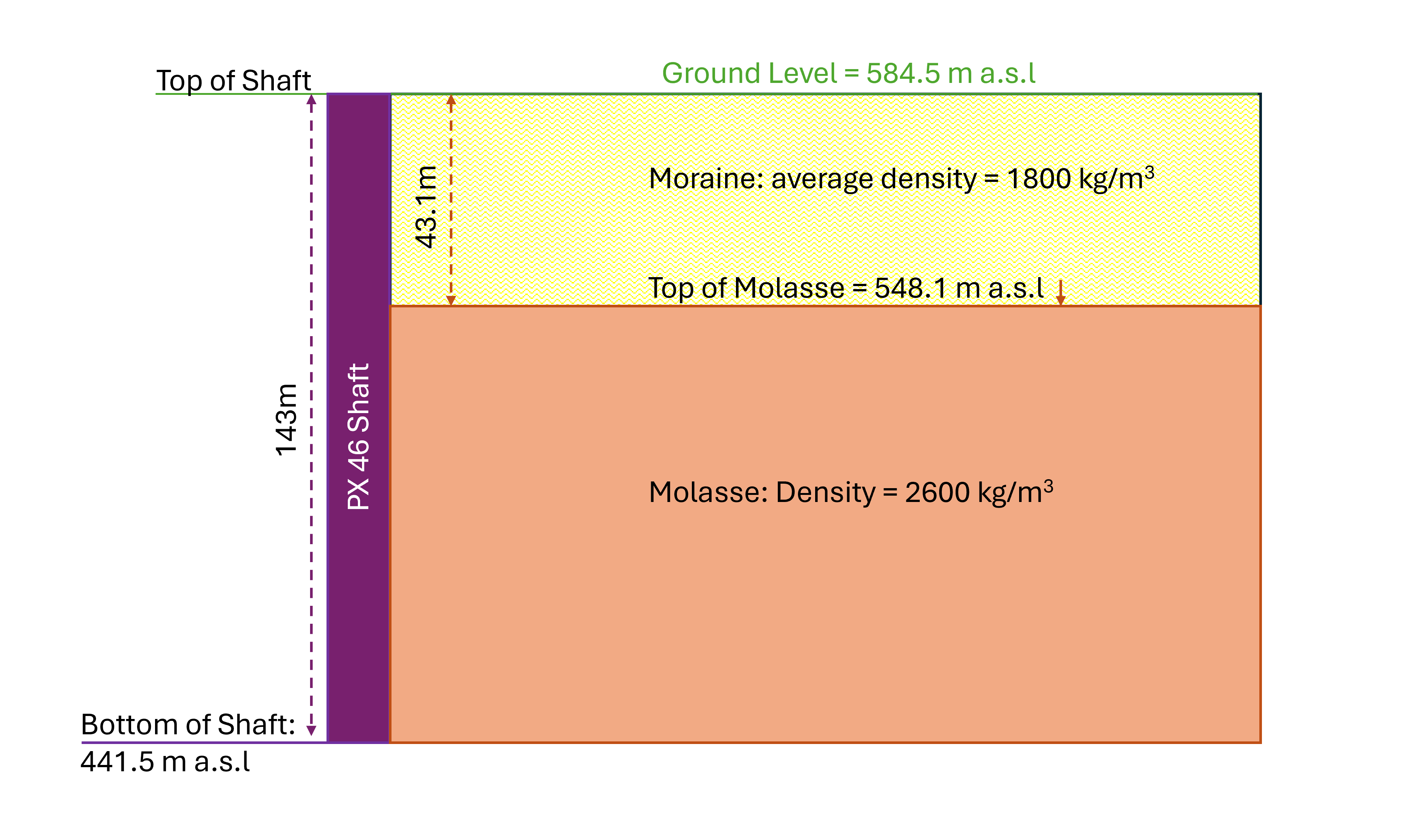}
    \caption{\label{fig:CE:Stratigraphy} Sketch of geological layers in the vicinity of the PX46 shaft, as taken from boreholes made in 1984, whose record can be viewed in full in~\cite{EDMS-943114:1984}.}
\end{figure}

Geologically, the site is well-characterized by core samples that revealed a thin layer of topsoil over a 43\ m thick layer of glacial till (moraine), resting upon a stable substrate of sandstone (molasse), as can be seen in Fig.~\ref{fig:CE:Stratigraphy}. This geological stability is reflected in exploratory seismic measurements conducted at both the top and bottom of the shaft. These studies concluded that the ambient seismic activity and resultant Gravity Gradient Noise (GGN) are within the required limits for the operation of an atom interferometer. Furthermore, despite the proximity of high-power LHC systems, electromagnetic (EM) noise levels—specifically those associated with the 400.8 MHz LHC radiofrequency (RF) power plant in the adjacent UX45 cavern—have been measured and found not to be a concern for the experiment’s sensitive quantum measurements. Slow variations in the ambient magnetic field occurring during LHC magnet ramps are also manageable through the experiment's dedicated magnetic shielding. These studies are recalled in Section~\ref{sec:noise}.

A critical aspect of the site overview is the functional compatibility of PX46 with ongoing LHC operations. The shaft’s primary purposes are the transport of major technical equipment, such as superconducting RF cryomodules and HL-LHC magnets, and as a core component of the UX45 cavern ventilation infrastructure. Detailed transport studies have confirmed that part of the horizontal cross-section of the shaft is not required for these operations. Specifically, an estimated 17 square meters of free area remains available on the side of the shaft, as is illustrated in Fig.~\ref{fig:PX46}, which is more than sufficient to accommodate the vacuum tube, support structures and side-arms of the AICE experiment, as well as the elevator access platform, without compromising the transport of LHC components. Similarly, the presence of the AICE infrastructure is not anticipated to have any significant impact on the airflow speed, pressure and temperature drop along the shaft, and does not obstruct the primary smoke extraction path for the UX45 cavern. To maintain these conditions, the top of the shaft must be airtight. A proposed 20~m$^2$ ventilation room (which could later be connected to the laser laboratory) will be built in the SX4 building to enclose the elevator platform landing. This room will feature cascading doors to minimize airflow disruptions during access.

\subsection{Experiment layout and installation}
\label{sec:site:layout}

% \textcolor{red}{Damien L., Carlo S.}

The AICE experiment will be installed against the wall of the PX46 access shaft, taking advantage of the existing shaft geometry while preserving the clearance required for the transport of LHC equipment. The experimental infrastructure is based on a vertical vacuum system extending over approximately 140~m and supported by a dedicated mechanical structure attached to the shaft wall. The detailed description of these detector subsystems is given in Section~\ref{sec:detector}, while the associated CERN infrastructure and services are presented in Sections~\ref{sec:site:safety} to~\ref{sec:site:CV}.

The installation strategy has been defined to minimise interference with the HL-LHC programme. All civil engineering works and CERN infrastructure modifications are to be completed during LS3, as will be discussed in Section~\ref{sec:civil}, allowing the detector installation and commissioning to proceed subsequently without requiring further modifications to the LHC infrastructure.

\subsubsection{Phase~1 -- Infrastructure preparation during LS3}

The first phase consists of preparing the PX46 infrastructure to accommodate the experiment. These activities are performed during LS3 and provide all the prerequisites required for the subsequent installation of the detector.

The main activities include:

\begin{itemize}
    \item Construction of the radiation shielding infrastructure described in Section~\ref{sec:civil:LS3};
    \item Installation of the elevator platform and handling equipment described in Section~\ref{sec:civil:platform};
    \item Modification of the lid on top of the shaft, construction of the ventilation room to enclose the landing of the platform described in~\ref{sec:civil:platform};
    \item Implementation of the access control, safety and monitoring systems described in Sections~\ref{sec:site:safety} and~\ref{sec:site:access};
    \item Installation of the electrical distribution, ventilation and other technical services described in Sections~\ref{sec:site:electrical} and~\ref{sec:site:CV}.
%    \item Preparation of the surface infrastructure, including the laser laboratory and beam-delivery interfaces described in Section~5.6. \textcolor{red}{How much of this needs to be done during LS3?}
\end{itemize}

At the completion of LS3, the PX46 facility will be fully prepared for the installation of the AICE detector.

\subsubsection{Phase~2 -- Detector installation after LS3}
\begin{figure}
\centering
\includegraphics[width=.6\textwidth]{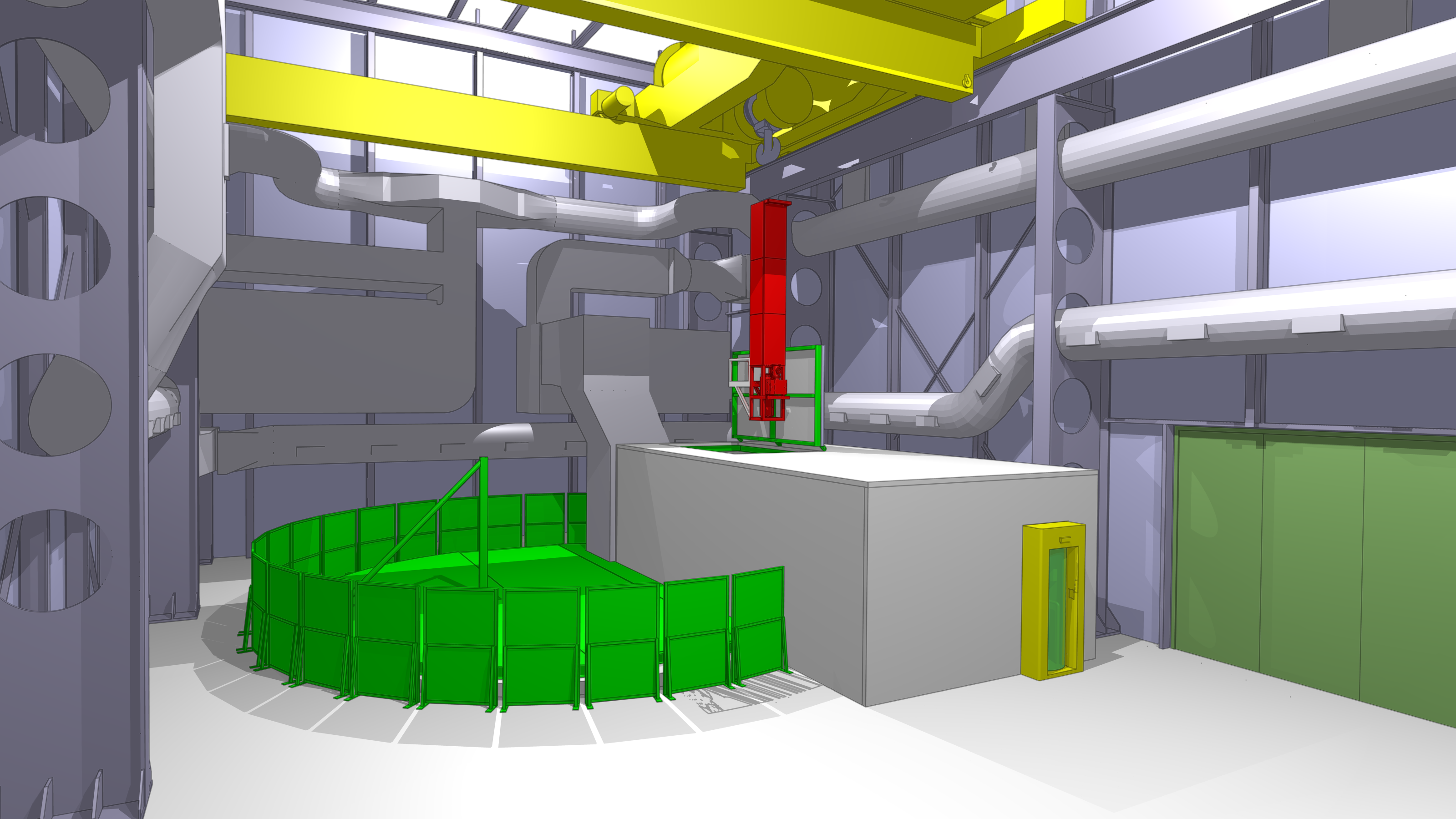}
\caption{Handling concept for the installation of the 5~m-long detector sub-assemblies inside the PX46 shaft (example highlighted in red).}
% \textcolor{red}{Highlight detector component?}
\label{fig:installation_sequence} 
\end{figure}

The second phase starts after LS3 and consists of the installation, integration and commissioning of the AICE detector, of the laser laboratory and beam-delivery interfaces, and of the related services such as power supply and cooling infrastructures. Detector components are delivered to the SX4 surface building, where each approximately 5~m-long standard module is transferred from the horizontal transport configuration to the vertical position using the SX4 building overhead crane.
Prior to the installation of the sub-assemblies, supports will be installed on the shaft wall using the platform after a  geometric survey to identify their positions.
Once the supports are installed, the vertical sub-assemblies are lowered through the hatch of the ventilation room into the PX46 shaft. Assembly of the detector proceeds progressively from the bottom of the shaft upwards. Each sub-assembly is first lowered by the overhead crane towards its final position against the shaft wall and then fixed to the permanent support structure using the platform to access the workplace. This sequence is repeated until the complete detector structure has been assembled. Fig.~\ref{fig:installation_sequence} illustrates the handling concept adopted for the installation of the individual 5~m-long detector sub-assemblies.

The installation of each detector subsystem follows the technical procedures described in Section~4. Following mechanical installation, the experiment undergoes alignment, vacuum commissioning, controls integration and functional verification before entering the commissioning programme described in Section~~\ref{sec:ops}.

\subsection{Safety and hazard mitigation}
\label{sec:site:safety}

% \textcolor{red}{Arnaud D., Simon M., Rui S. for the Fire Brigade, Fabio C.}

The main requirements for the operation of AICE were set out in Section~\ref{sec:detector}: in particular access to the side-arms should in principle be possible 24 hrs/day, 365 days/year. This leads to a comprehensive assessment of the safety hazards that might be encountered in the PX46 shaft, in particular in the event of a fire or of a release of helium from the LHC machine, or relative to the production of ionising radiation. This Section addresses in some detail these cases.
From a general point of view, escape of the operators of AICE must be possible in the shortest time and safest way possible. Given the height of the PX46 shaft, it would be reasonable to assume the use of an elevator for normal access to the experiment and escape in case of emergency, with a parallel staircase for emergency evacuation in case of hazards and in case of failure of the elevator. However, the limited space available in the cross section of the PX46 shaft prevents the installation of both. This led to the choice discussed in the following sections of having a single special lifting platform that can be used for normal access and for escape in all cases of emergencies, and which complies with all relevant requirements, regulations and rules.

\subsubsection{Fire safety measures}
\label{sec:site:safety:fire}

%\textcolor{red}{Missing links, etc.}

\begin{figure}
    \centering
    \includegraphics[width=0.95\linewidth]{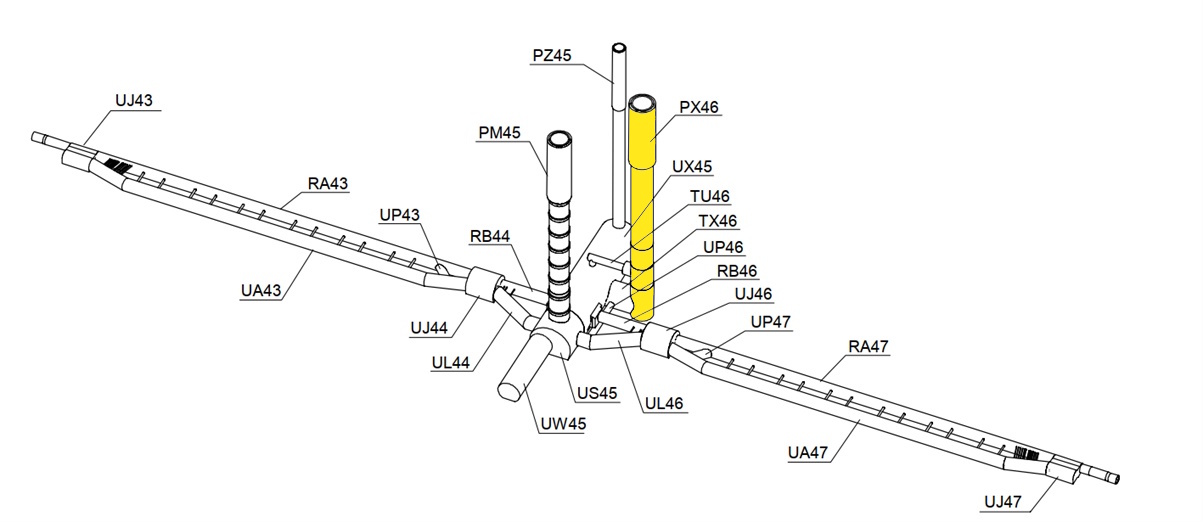}
    \caption{Isometric view of the underground caverns, shafts and tunnels, showing the identifiers of the different structures.}
    \label{fig:iso-view-PX46}
\end{figure}

The UX45 cavern contains high-power RF equipment such as the main klystrons and power converters for the LHC RF system. In the event of fire, the appropriate smoke detection and alarm systems are already in place, and evacuation routes are already prescribed, either through the PZ45 access shaft or through the PM45 access shaft via the UP46 connecting gallery, see Fig.~\ref{fig:iso-view-PX46}. To ensure two independent escape routes from PX46 (through UX45 towards PZ45 or UP46 towards PM45), it could be beneficial to consider a fire door at the intersection of UP46 and TX46, to be assessed in the context of the entire underground facilities.  
All relevant available information regarding fire safety prescriptions is documented in~\cite{fire-safety}.

The AICE experiment in the PX46 shaft might also be impacted by any fire developing in the UX45 cavern. After the construction of a radiation shielding wall in the TX46 gallery as discussed in Section~\ref{sec:civil:shielding}, smoke produced in the UX45 cavern will be evacuated primarily via the TU46 ventilation gallery and the PX46 shaft.  To comply with the risk for the operators associated with the presence of smoke, it is anticipated that the access platform as described in Section~\ref{sec:civil:platform} is used also as a means for escape, following the triggering of a fire alarm. Alarms from any emergency occurring in the UX45 cavern must be made visible and audible for the occupants of the platform and at the boarding stations on the surface and at the base of the shaft, so that they can initiate the emergency evacuation or refrain from using the access platform if they are still on the surface or at the base.  A visible indicator (red flashing light) of an evacuation alarm on the platform, or a set of flashing lights along its run shall be considered (sirens and flashlights located at bottom  or top level can often be neglected). Operators shall put on appropriate Personal Protection Equipment (PPE — self-rescue masks) to protect them from any smoke in PX46, which they shall carry with them at all times when accessing the AICE experiment, or which shall be made available on the mobile platform in a suitable cabinet, and proceed with the evacuation.

The mobile platform must be designed in such a way that in the event of an emergency it can reach, in full autonomy, both the surface and the lower level of the UX46 shaft within approximately 2 minutes.  
A battery backup is mandatory, to allow this operation also in case of loss of the mains electrical supply. In the case of a failure of both the mains electrical supply and of the battery backup, it is also mandatory that the platform can still descend, in a controlled, autonomous and safe way to the lower level of the PX46 shaft with a mechanical failsafe system, within the same timeline of about 2 minutes mentioned before. This will leave enough time for the operators to safely escape through the access doors in the shielding wall in the TX46 gallery and then via the pre-established evacuation paths. The platform itself shall be equipped with at least basic fire extinguishing equipment. Bidirectional voice communication systems between the platform and the Safety Control Room shall also be foreseen, as is the case for all lifts at CERN. These means shall be hardened against power outage, which is an extremely likely scenario during a fire emergency response (they shall for example be able to operate after triggering  general emergency stops -- AUGs -- in nearby zones on the surface and underground).
 
Specific auxiliary means of access for the fire brigade rescue teams will also be implemented, in order to allow the occupants to be reached and evacuated in the event of mechanical blockage of the platform and failure of the manoeuvre for reaching the ground level. 
%Different options, to be investigated, include access via a rescue cage, made either using the existing crane bridge, or a dedicated crane system to be installed. Moreover, i
It is recommended that anchor points for rope rescue  be installed and that annual exercises or operational reconnaissance  be programmed so as to increase firefighters’ readiness in this facility.
 
Smoke from a fire in the LHC machine tunnel in sectors 34 and 45 or from the cryogenic plant in the US45 cavern is expected to remain confined in the LHC tunnel region thanks to the physical separation from the UX45 cavern (due to the fact that the ventilation doors in the RB44, UL44, RB46 and UL46 galleries are normally closed), and either be evacuated from the PM45 access shaft or be blown away from LHC point 4 in the direction of the odd LHC points 3 and 5, following the standard air flow in the LHC tunnel and depending on the fire action matrix for such events.
In any case, alarms will be triggered, and the atom interferometer operators will escape following the procedure described above.
 
The AICE experiment itself and its ancillary infrastructure are foreseen to consist of equipment with a relatively low combustible load; a complete assessment will have to be carried out at the stage of the detailed technical design. As for any CERN equipment, the materials used for fabrication will have to comply with the relevant Safety Rules~\cite{CERNsafety} so as to minimize fire-induced hazards. As an example, the vacuum pumping stations should preferably rely upon dry pumps instead of oil-sealed pumps. 
Specific measures such as thermal protection of live electronics, positioning of smoke sampling pipes near the equipment along the run of the lifting platform initiating depowering of the electronics have to be considered, given the potential impact of a fire on the safety of occupants of the platform.  No major differences from the escape procedures mentioned above are anticipated, but the installation of other smoke detectors and fire alarms will have to be assessed, and a detailed matrix of events and conditions that should trigger alarms, evacuation and depowering should be defined. 

Adequate operator training will be mandatory (and relevant courses must be defined) before granting access to the experiment, so that the final decision on the preferred directions and routes to take for evacuation can be left to the operators (thus avoiding automatic servo triggering of the movement to the mobile platform, for example).

Concerning the surface laser laboratory, standard fire and smoke detectors will have to be installed depending on the assessed risks, as for other laboratories in CERN technical buildings.
Fire extinguishers, evacuation plans and other similar fire mitigation devices will be provided.

\subsubsection{Helium release safety hazard mitigation}
\label{sec:site:safety:helium}

Safety management studies for the PX46 site have addressed specifically the potential for uncontrolled helium releases from the Large Hadron Collider (LHC) cryogenic systems, which include the superconducting radiofrequency (RF) cryomodules at Point 4 and the main superconducting dipole strings in the adjacent arcs. Implementation studies have categorized these into two primary scenarios based on their origin and routing. A release from the RF cryomodules in the RUX45 section would flow through the TU46 ventilation gallery and eventually into the PX46 shaft. However, because the total gaseous volume of helium in these cryomodules (approximately 972 m$^3$) is small compared to the 18,000 m$^3$ volume of the UX45 cavern, such an incident is not considered a significant hazard. This assessment was validated during a real helium venting event in August 2022, where oxygen levels in the TU46 gallery remained within acceptable limits~\cite{Hakulinen:2022}. 
In the more significant scenario of a Maximum Credible Incident (MCI) within the LHC dipole strings, where up to 40 kg/s of liquid helium could be released, safety is managed through a pre-existing sophisticated system of confinement and routing. Pressure-resistant doors in the UL44, UL46, RB44, and RB46 galleries act as relief valves designed to sustain overpressures and ensure that helium is routed away from the machine towards the PM45 shaft. This containment strategy ensures that an MCI in the adjacent sectors does not create a hazard within the PX46 shaft. 

\subsubsection{Radiation protection}
\label{sec:site:safety:RP}

Radiation protection studies have been performed based on the AICE shielding proposal, described in more detail in Section~\ref{sec:civil:shielding} and designed with the requirement to have PX46 classified as Supervised Radiation Area~\cite{EDMS-810149:2007}. 

The following sources of radiation streaming toward PX46 during standard operation of the LHC/HL-LHC have been considered in the assessment~\cite{Devienne:2026}: 
\begin{itemize}
    \item X-ray emission from the RF cavities (including during commissioning periods);
    \item Beam-gas interactions from the LHC tunnel (only during beam operation).
\end{itemize}

\begin{figure}
    \centering
    \makebox[\textwidth][c]{
        \includegraphics[width=0.6\textwidth]{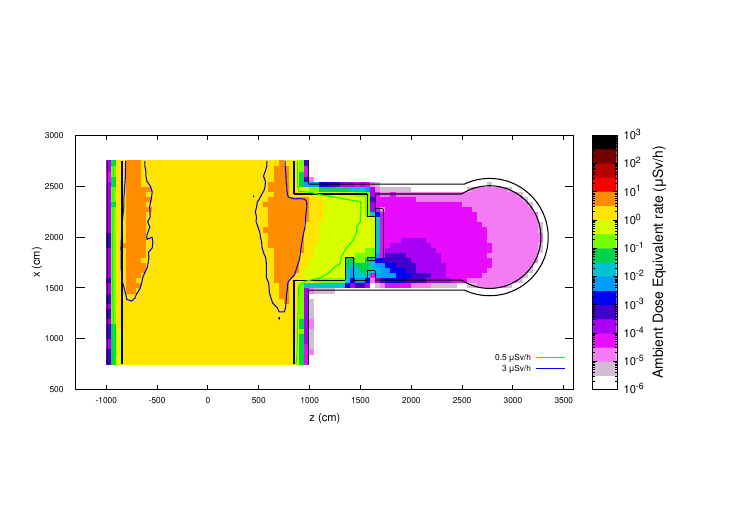}
    }
    \makebox[\textwidth][c]{
        \includegraphics[width=1.0\textwidth]{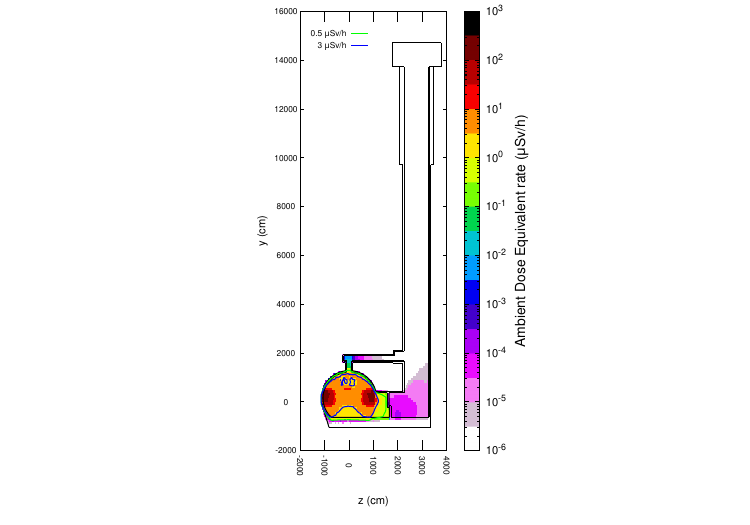}
    }
    \caption{Ambient dose equivalent rate in TX46/PX46 during normal operation of the LHC/HL-LHC at Point~4 with the shielding design proposed for AICE in TX46. {\it Upper panel}: a transversal cut at 1 meter height from TX46/PX46 floor. {\it Lower panel}: a longitudinal cut centred in the middle of TX46/PX46. The ambient dose equivalent rate limit for Supervised Radiation Areas is 3~$\mu\text{Sv/h}$.}
    \label{fig:lhc_ir4_AION-100_RP_sumop}
\end{figure}

\begin{figure}
    \centering
    \makebox[\textwidth][c]{
        \includegraphics[width=0.6\textwidth]{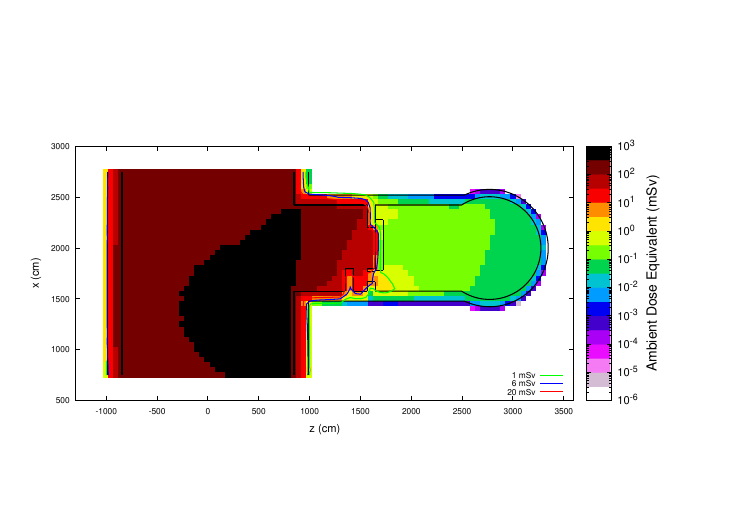}
    }
    \makebox[\textwidth][c]{
        \includegraphics[width=1.0\textwidth]{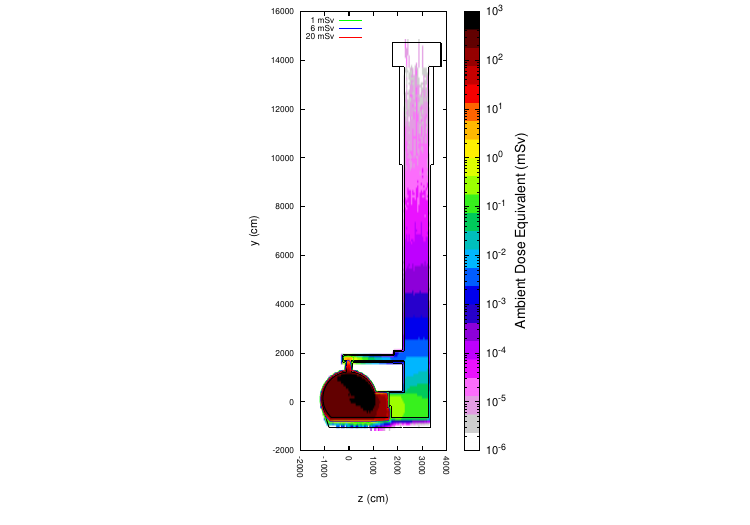}
    }
    \caption{Ambient dose equivalent in TX46/PX46 from an accidental HL-LHC beam loss at Point~4 with the shielding design proposed for AICE in TX46. {\it Upper panel}: a transversal cut at 1 meter height from TX46/PX46 floor. {\it Lower panel}: a longitudinal cut centred in the middle of TX46/PX46. The yearly effective dose limit for Supervised Radiation Areas is 6~mSv.}
    \label{fig:lhc_ir4_AION-100_RP_accidental}
\end{figure}

In addition, an accidental HL-LHC beam-loss close to point~4 has also been evaluated using FLUKA~\cite{Battistoni:2015,Ahdida:2022}. The summed-up contribution of the radiation produced by the X-ray emission from the RF cavities and by the beam-gas interaction during operation results in ambient dose equivalent rates at PX46 well below the limits for a Supervised Radiation Area (see Fig.~\ref{fig:lhc_ir4_AION-100_RP_sumop}). Furthermore, the ambient dose equivalent at the base of PX46 in the case of an accidental HL-LHC beam loss at point~4 remains below the yearly limit of a Supervised Radiation Area (see Fig.~\ref{fig:lhc_ir4_AION-100_RP_accidental}).

Regarding air activation, fresh air in the LHC machine tunnel is injected in even points (2,4,6,8) and extracted in odd points (1,3,5,7), so no air is extracted from the LHC tunnel toward UX45. The air activation in the interconnected volumes UX45, TX46, TU46 and PX46 during standard LHC/HL-LHC operation is negligible, hence the PX46 shaft remains well below the limit of airborne contamination in a Supervised Radiation Area.

Finally, a radiation monitor IG5-H20 including an alarm unit and uninterruptible power supply will be needed to monitor the radiation hazard in the PX46 shaft during LHC operation. A Request for Instrumentation (RFI) has been prepared for the installation of this radiation monitor in PX46. The installation can be completed during LS3 within 1~week, requiring beforehand the availability of sockets for electrical and data networks. Afterwards, the installation should be followed by 2~weeks of commissioning and a source test of the monitor.

\subsection{Access control and alarm systems}
\label{sec:site:access}

%\textcolor{red}{Timo H.}

\subsubsection{Surface rooms}

The required modifications to the access control system fall into two distinct categories: access control to the PX46 shaft, which will normally be classified as a Supervised Radiation Area, and access control to the surface building SX4 and the shaft access room and laser laboratory adjacent to the shaft.

Access control to the SX4 building is managed by the CERN standard site surveillance system (SUSI), which would also control access to the laser room, as is customary in other laser installations of CERN. It is not yet defined if the laser room will be integral with the shaft access enclosure or separate from it, but in any case a specific access model and control to the laser room will be needed.

%\begin{itemize}
%\item Access control to the enclosure on top of PX46 pit:
%
%\begin{itemize}
%\item Identification with regular CERN badge (SUSI) or dosimeter badge (LACS) depending on whether there will be a separate access control to the lifting platform or not.
%\item Lifting platform access will require a dosimeter badge due to the possibility of having to evacuate through the LHC in an emergency.
%\item A small turnstile or PAD for personnel access (requirements vs. cost).
%\item Probably no MAD (overkill?). A separate door for material access (possibly simple key operated or separate reader).
%\end{itemize}

%\item A Red Telephone unit inside the enclosure.

%\item Evacuation system (siren or voice alarm) to inform personnel of danger.

%\item Laser room:
%\begin{itemize}
%\item Separate access control (safety requirements/courses may differ).
%\item Laser safety in the laser room using the usual CERN Laser safety solution (if regular installation) or a bespoke PLC-based safety system (if installation more complicated).
%\item ODH detection (if cryogenic gases).
%\item Fire / smoke detection either by aspiration tubes or point detectors.
%\end{itemize}

%\item Video surveillance of the access point.
%\end{itemize}

\paragraph{Enclosure on top of PX46 shaft:}

\begin{figure}[htbp]
     \centering
     \includegraphics[width=.30\textwidth]{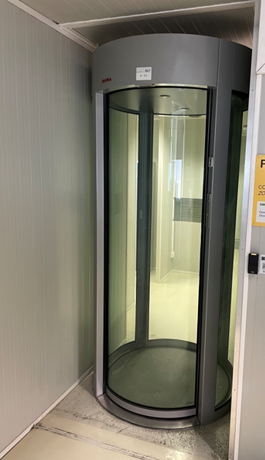}
     \quad
     \includegraphics[width=.50\textwidth]{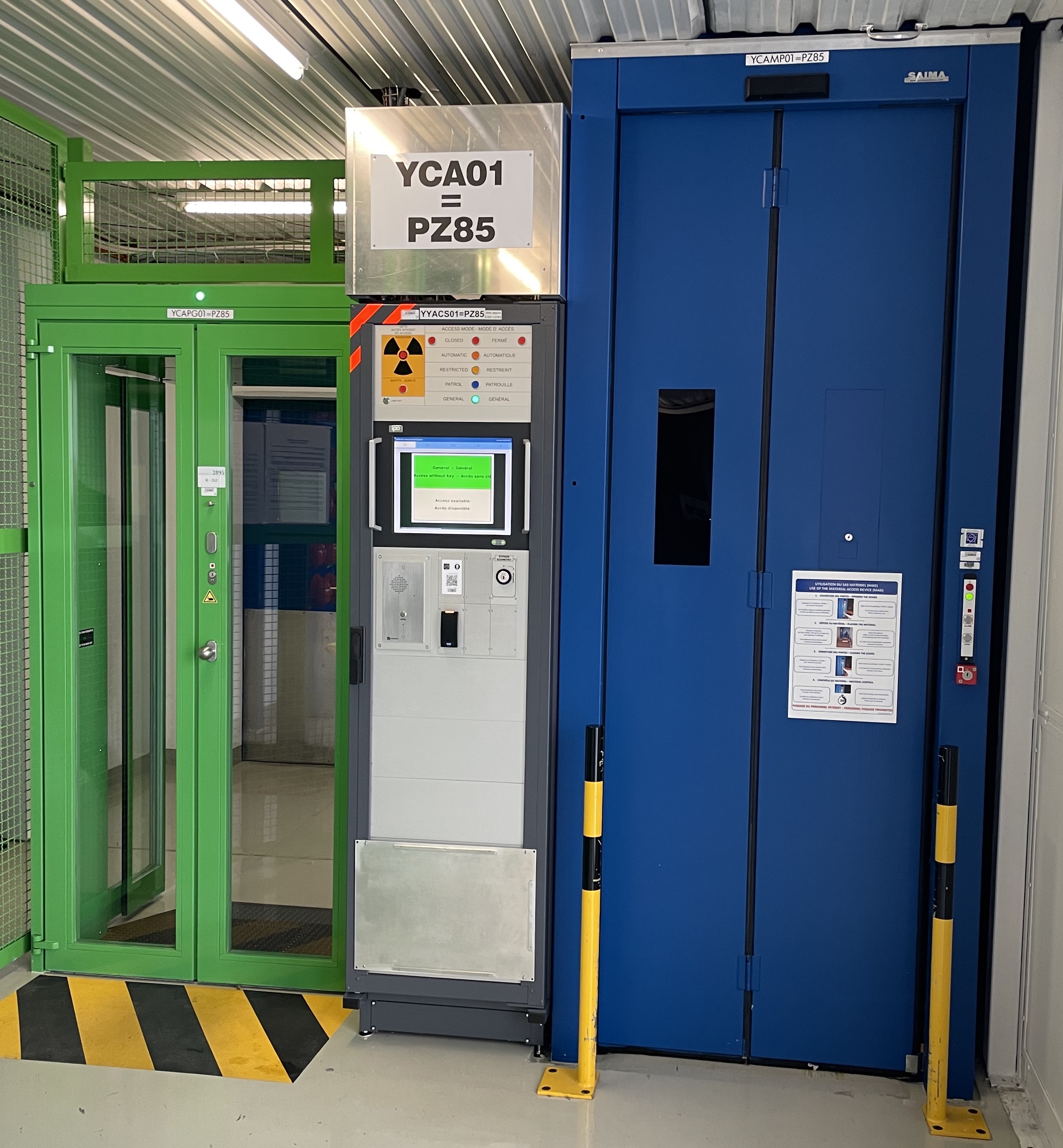}
     \caption{\label{fig:AC:badgepadmad} Examples of a simple airlock with a badge reader ({\it left}), PAD and MAD ({\it right}).}
\end{figure}

From the top of the shaft it will be possible to descend to the bottom of the shaft via the elevator, to which access will need to be appropriately controlled. Depending on the required level of rigour, technical options for access control range from a simple airlock with a badge reader to a full non-interlocked area Personnel Access Device (PAD) and an optional Material Access Device (MAD) (see Figure~\ref{fig:AC:badgepadmad}).

In both solutions access would be controlled by a badge reader attached to the LHC Access Control System (LACS) allowing accurate counting of personnel accessing the shaft. If full biometric identification of the person is required, as is the case with other non-interlocked areas of the LHC, a full PAD solution will be necessary with a somewhat larger footprint and expense.

If normally only relatively small items need to be transported infrequently into the top of the shaft enclosure, a regular door with a simplified access control (either a badge reader or key) can be installed. Frequent transport requirements of that kind usually indicate a need for a full MAD to be installed.

The required access badge would be the standard CERN dosimeter badge used to access radiological areas and a specific access model would be created for the AICE area in LACS with an assigned person responsible for granting this access. Also, specific on-line safety training will need to be prepared for this area to inform users on the specific safety and access particularities in all situations, and a set of required PPE will need to be defined (e.g., helmet, safety shoes, lamp, dosimeter, etc.).
LACS-specific video surveillance of the access device will also be installed, as is the case at all other access points.

Normal fire detection, evacuation, and emergency communications systems will also be installed inside the top of the shaft enclosure:

\begin{itemize}
\item A Red Telephone unit to call CERN emergency services;
\item An evacuation system (siren or voice alarm) to inform personnel of danger;
\item A manual evacuation call point to trigger an evacuation alarm in the enclosure and the shaft.
\end{itemize}

\paragraph{Laser laboratory:}

The laser laboratory may be installed either within the top of the shaft enclosure or in a separate location within the SX4 building.
Therefore, separate access control requirements to the shaft access would normally apply, as the safety requirements and courses differ.

Laser safety in the laser laboratory is foreseen using the standard CERN laser safety solution, if a regular type of laser installation is sufficient, or a bespoke PLC-based safety system, if the  installation is more complicated. As different gases will likely be used in the laser installation, local Gas / ODH detection will be necessary. Also, fire/smoke detection either by aspiration tubes or point detectors will be installed.

\subsubsection{Underground areas}

%\begin{itemize}
%\item Access Safety System (LASS):
%
%\begin{itemize}
%\item Instrumentation of the underground end-of-zone and %2nd barrier door and the movable shielding wall.
%\item Interlocks with the LHC beam. Opening a door %during LHC run will cause beam dump.
%\end{itemize}
  
%\item Fire/smoke detection using aspiration tubes. The %aspiration units located at the top of the pit.
%
%\item A Red Telephone unit at the bottom of the pit.
%
%\item Manual Evacuation call point (break-the-glass) at the bottom of the pit.
%
%\item Siren or voice alarm to inform personnel of danger.
%\end{itemize}

\paragraph{Access safety:}

As it will be possible to descend from the top of the shaft  to the bottom using the elevator, all the sectorization changes in the underground areas UX45, TX46 and PX46 will need to be finalized during LS3.

\begin{figure}[htbp]
     \centering
     \includegraphics[width=0.8\textwidth]{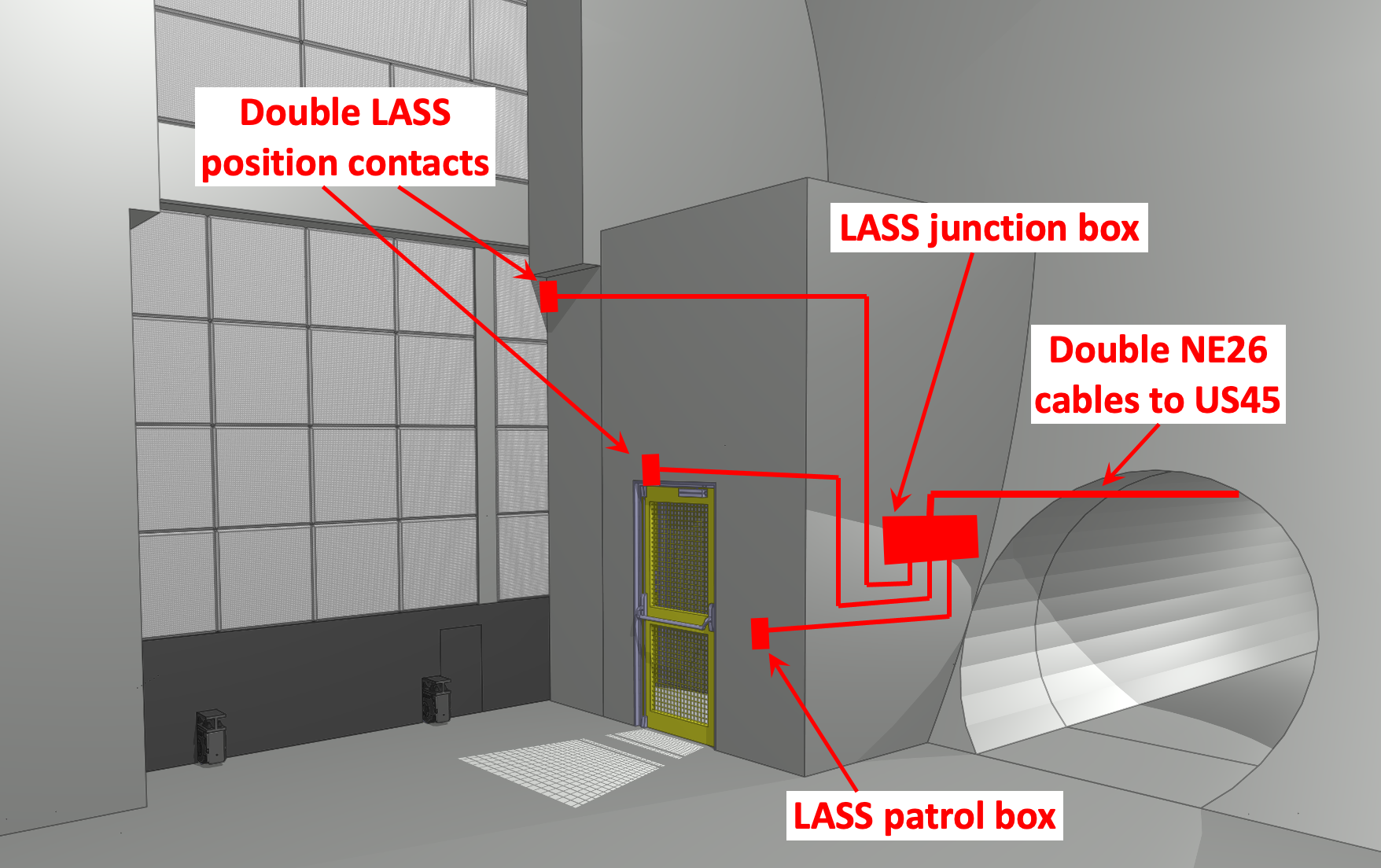}
     \caption{\label{fig:AC:lasscontacts} LASS instrumentation of access doors.}
\end{figure}

\begin{figure}[h!]
     \centering
     \includegraphics[width=.4\textwidth]{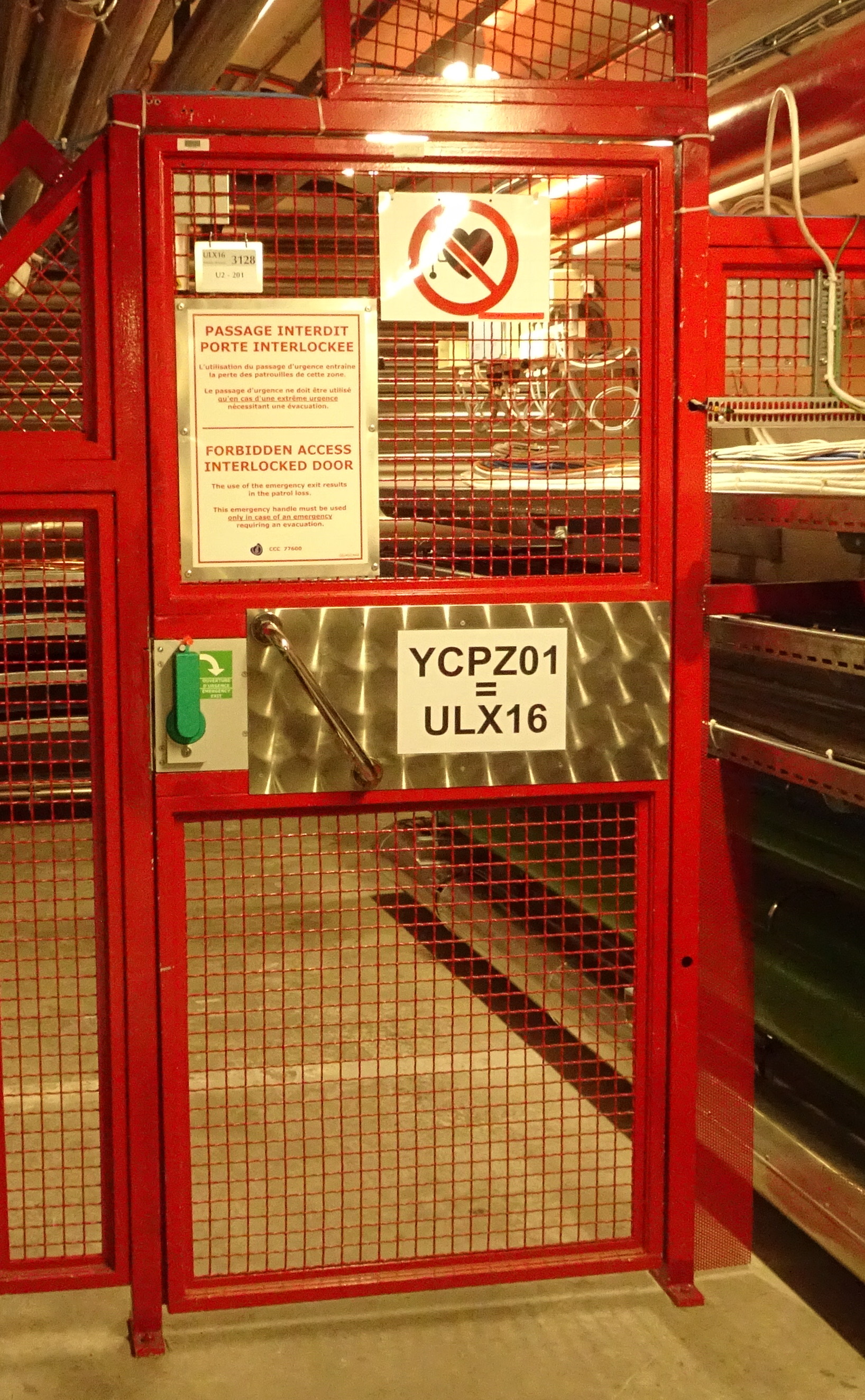}
     \quad
     \includegraphics[width=.4\textwidth]{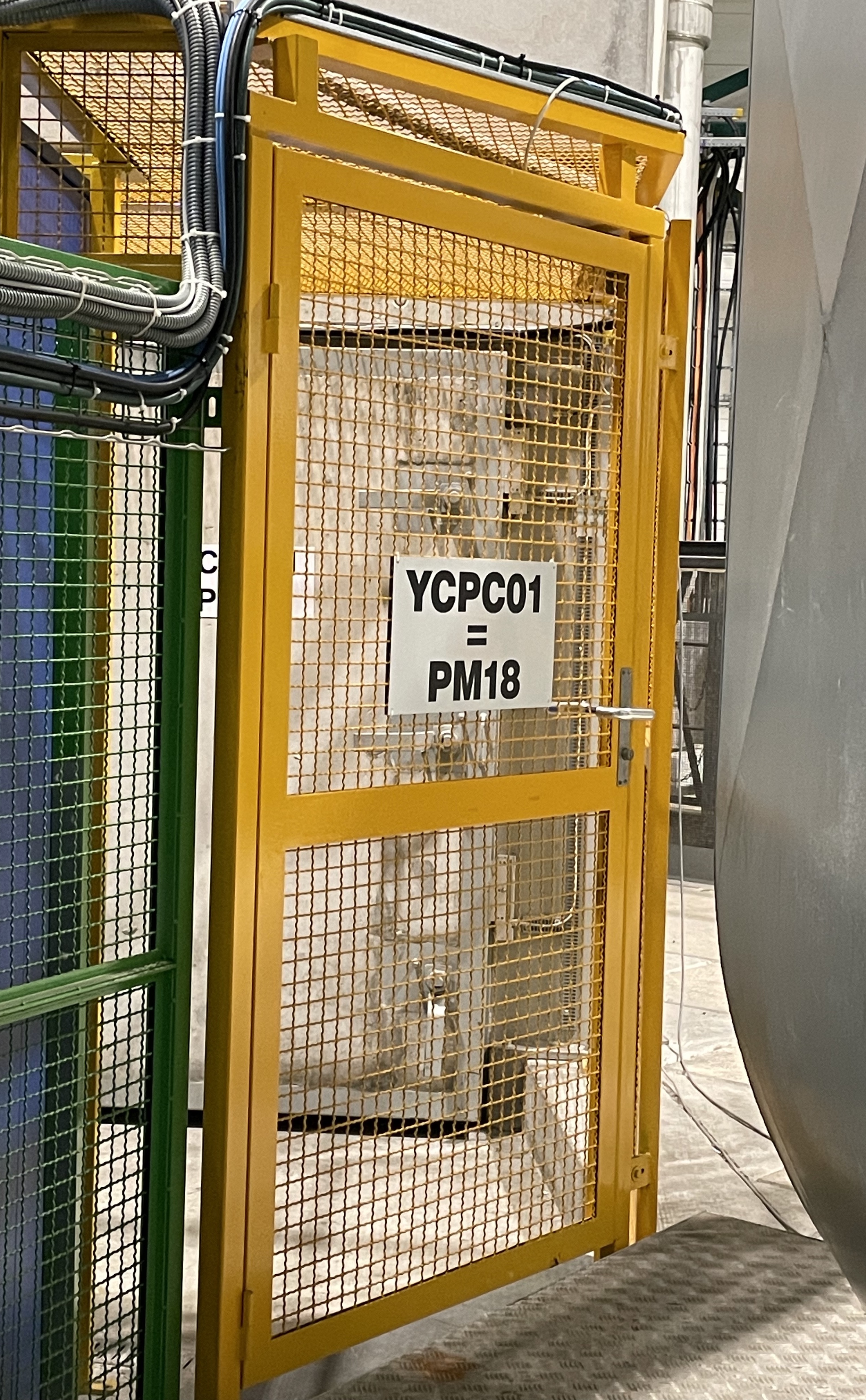}
     \caption{\label{fig:AC:eozventdoors} LASS end-of-zone door (left), and an example of a door for the second barrier (right).}
\end{figure}

The LHC Access Safety System (LASS) will be modified to implement the changes to the sectorization of the UX45 area.
Currently the entire PX46 shaft belongs to the interlocked LASS zone PZ45 together with the UX45 cavern. During an LHC run the top of the shaft is covered to prevent access and to contain any prompt radiation in case of an accidental beam loss close to UX45.
The new end of the PZ45 zone will be at the TX46 connecting gallery, where sufficient radiological shielding will be installed as per the analysis by the Radiation Protection team.
Two new access doors are to be installed successively in the passage through the shielding in TX46, an end-of-zone door and a second barrier door.
This arrangement follows the general principle of the LASS that any access from a non-interlocked area to an area where a radiological risk is present must pass through a minimum of two interlocked doors with separate safety contacts and cabling.
The end-of-zone door is a standard LASS door: red in colour, gridded, with double position contacts (connected to both LASS safety chains, PLC- and relay-based), and including an emergency opening handle on both sides (see Fig.~\ref{fig:AC:lasscontacts}).
The second barrier door is also gridded, like the end-of-zone door, with regular opening handles and LASS double position contacts (see Fig.~\ref{fig:AC:eozventdoors}).
If either of these doors is opened, or even if the end-of-zone door emergency handle is turned, the LHC beam will be dumped and the system put into a safe state.

As a fully movable shielding wall, similar to the large motorized shielding walls of the LHC experiments, will be installed, LASS safety contacts are to be installed also on this device. However, as shielding walls are not considered as emergency exit paths, but rather de-activated (with motors de-energized) during LHC operation, installation of a second barrier is not required.

Installation of an end-of-zone door in TX46 means that PX46 shaft can be classified as a non-interlocked area, no longer supervised by the LASS and allowing access from the top via a simplified procedure by the LHC Access Control System (LACS), to be implemented following the requirements of the experiment. 
This status is similar to other non-interlocked experiment areas at PX15 (ATLAS), PM54 (CMS), PZ85 (LHCb), as well as HL-LHC PM17 and PM57.
TX46 will act as an evacuation route via the UX45 cavern if access to the top of the shaft is unavailable or dangerous.

\paragraph{Alarm systems:}

\begin{figure}[h!]
     \centering
     \includegraphics[width=.45\textwidth]{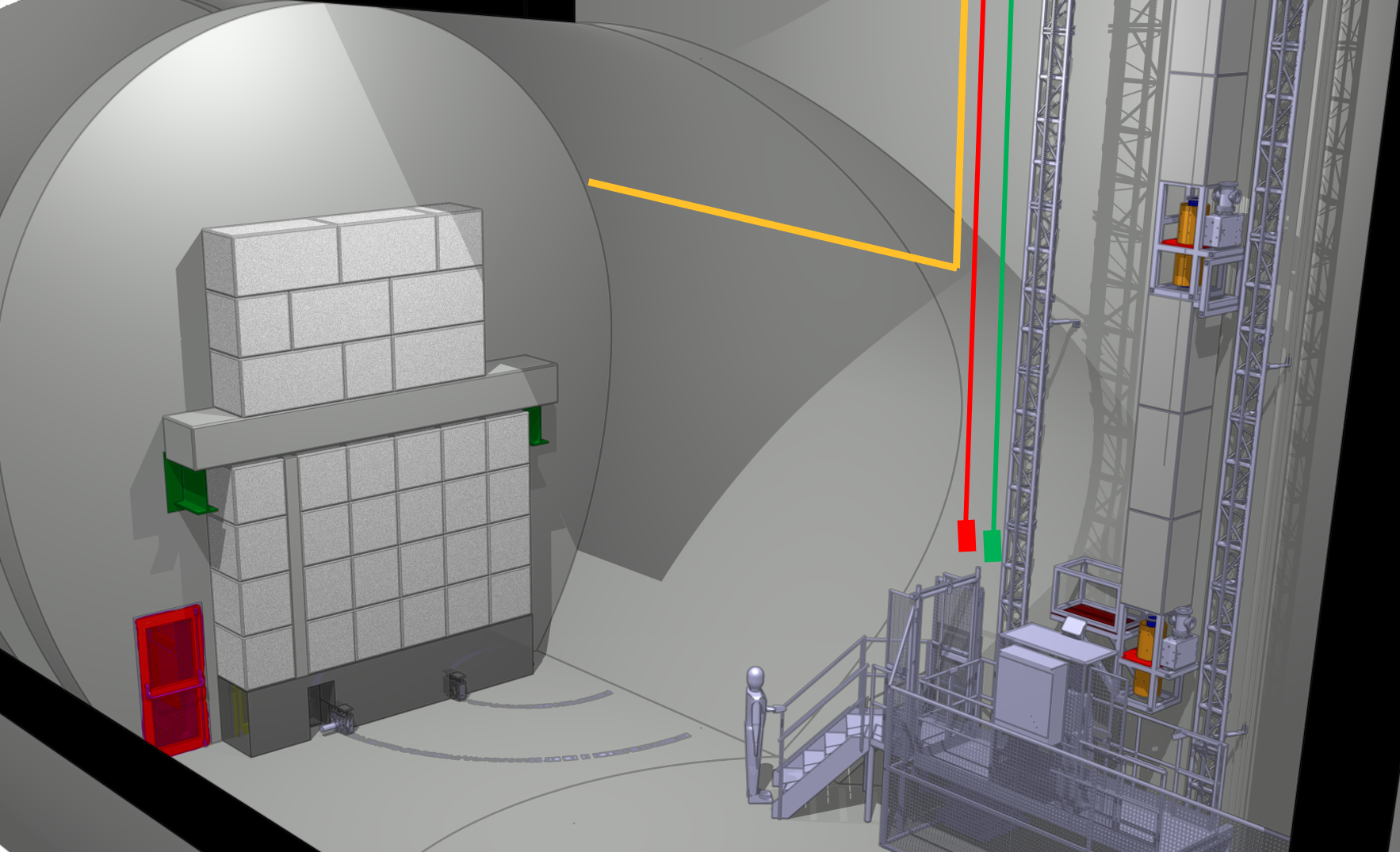}
     \quad
     \includegraphics[width=.45\textwidth]{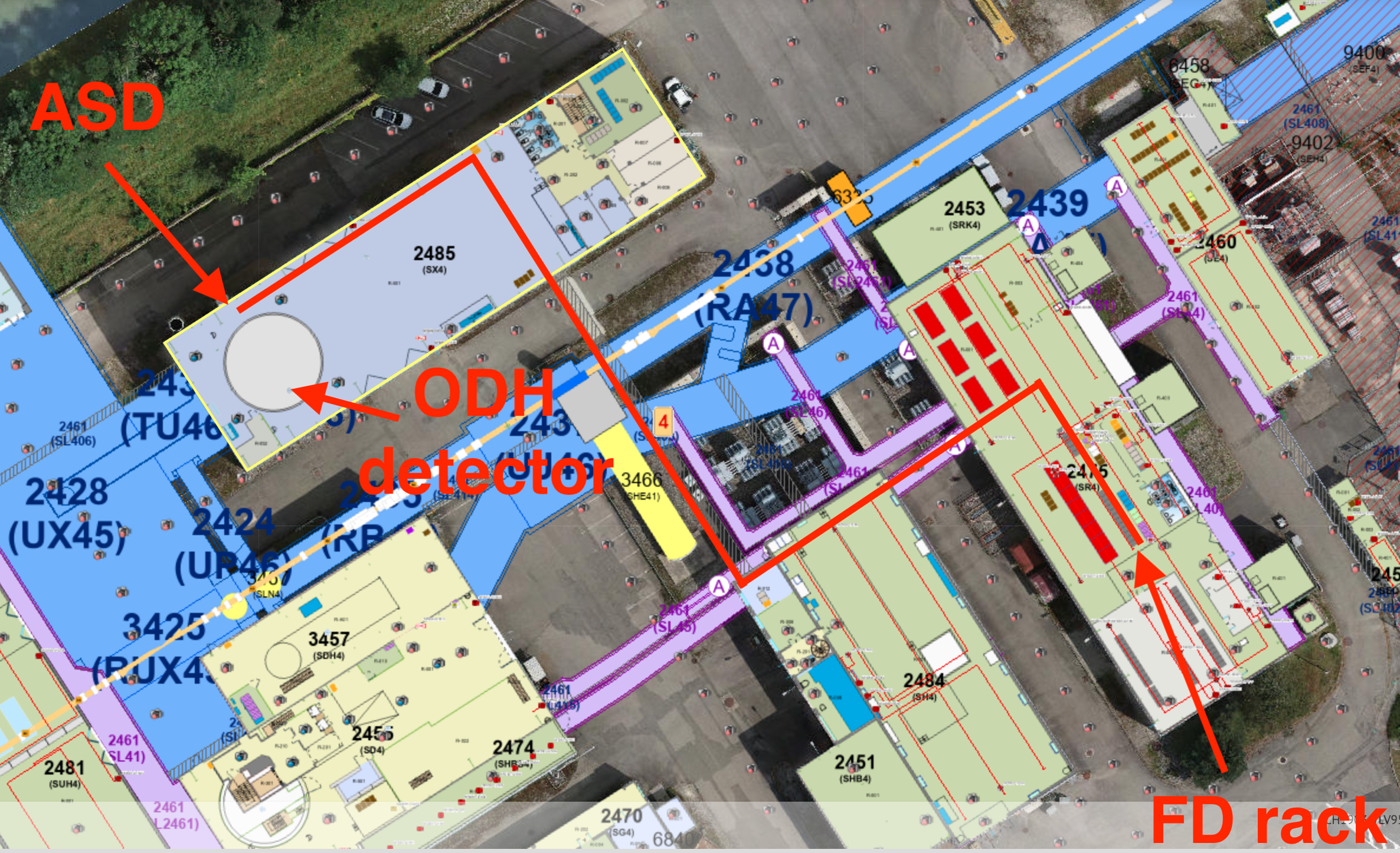}
     \caption{\label{fig:AC:firedet} Layouts of the fire detection aspiration tube (orange), red emergency telephone (red), and manual call point (green) at the bottom of the PX46 shaft (left panel), and cabling from the fire detection rack to the air aspiration device (ASD) (right panel).}
\end{figure}

\begin{figure}[h!]
     \centering
     \includegraphics[width=0.8\textwidth]{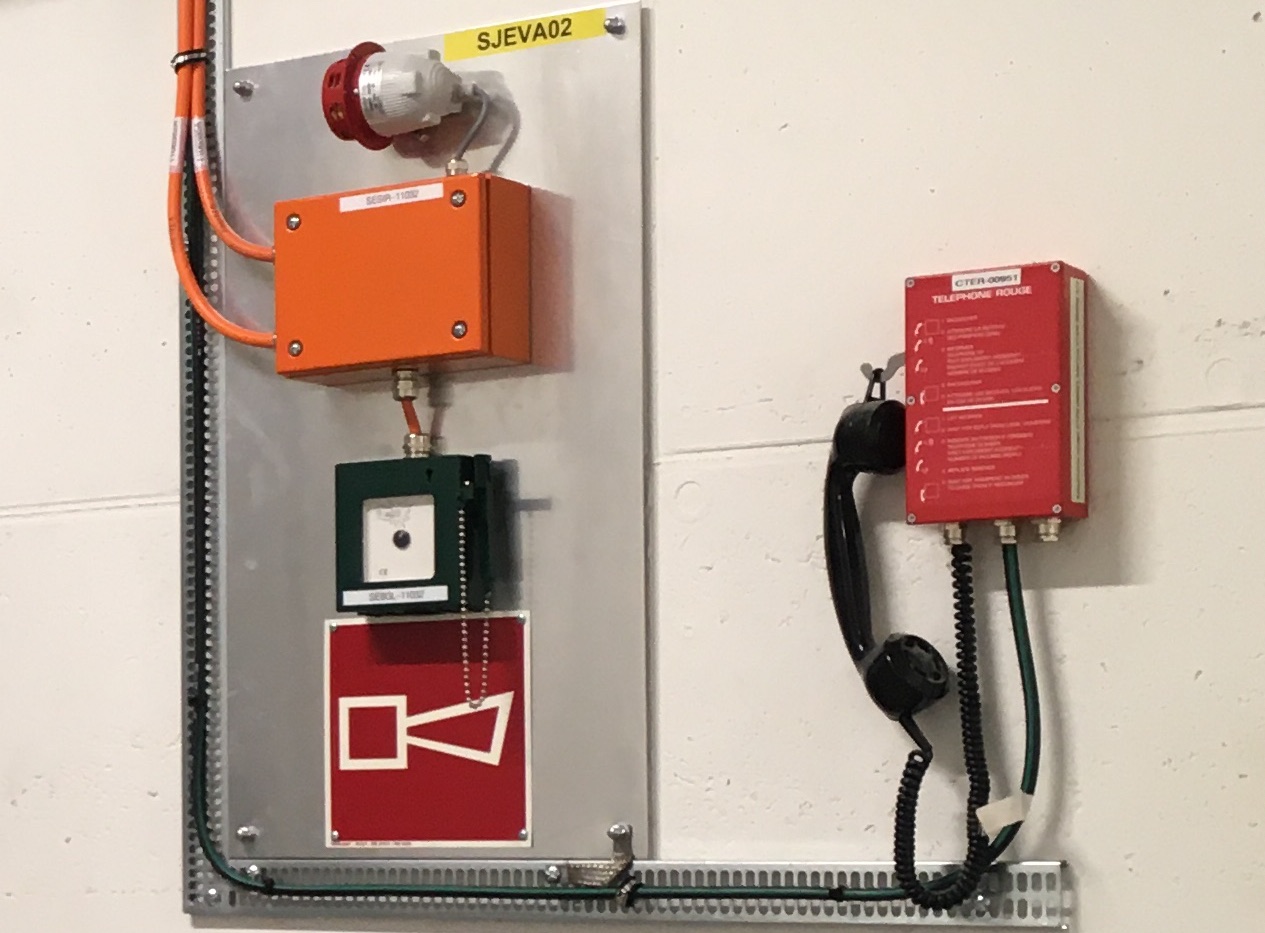}
     \caption{\label{fig:AC:evacandrt} Manual evacuation call point (``break-the-glass''), siren, and red emergency telephone.}
\end{figure}

Alarm systems for fire and oxygen deficiency hazards and evacuation as well as emergency communication will also be necessary. For the installation phase, basic services are to be installed during LS3, which can then be extended to cover any possible risks of the experimental phase. The alarm systems in question are:

\begin{itemize}

\item {\bf Fire detection:}
Smoke detection by aspiration tubes is currently the most effective method of early detection of fire. It is proposed to install a detection unit at the top of the shaft, from which detection tubes can be installed reaching the bottom of the shaft (see the left panel of Fig.~\ref{fig:AC:firedet}). Cabling for the fire detection system is pulled from the main Fire Detection System rack of Point 4 in building SR4 (see the right panel of Fig.~\ref{fig:AC:firedet}).

\item {\bf ODH detection:} 
Based on previous experience it is assessed that the probability of a large He release (the most credible incident, MCI) in PX46 is small, as demonstrated by earlier real He releases in the RF area, where the helium gas has been largely contained in the LHC tunnel and driven away from UX45 towards other LHC sites, due to the effect of the regular ventilation~\cite{Hakulinen:2022}. Therefore, it is estimated at this time that the existing ODH detection already present at the top of the PX46 shaft and the SX4 building will be sufficient.

\item {\bf Evacuation:}
Any hazard detection will need to be communicated to the personnel present on site by the evacuation system. Currently, simple sirens exist at the LHC for this purpose. However, the LHC evacuation system is being consolidated during the LS3 to a modern voice alarm system, which will likely be the type of solution finally foreseen to be installed at the top of the shaft, and which will give a general evacuation signal in case of a fire or ODH detection. In any case, it will be the responsibility of the personnel to evaluate the situation and to decide the best evacuation option based on the location of the hazard, either via the elevator or through UX45. In addition, two manual call points, i.e., ``break-the-glass devices'' (see Fig.~\ref{fig:AC:evacandrt}), are to be installed in the shaft, one at the top and one at the bottom. Activating the call point by pushing the button will sound the evacuation alarm and alert the CERN Fire Brigade just as a fire or an ODH alarm would.

\item {\bf Emergency communication:}
A standard emergency communication device used in the LHC, a Red Telephone (see  Fig.~\ref{fig:AC:evacandrt}), is to be installed at the bottom of the shaft. Lifting the receiver will dispatch a Level 3 alarm to CERN Fire Brigade and open direct voice communication with CERN Safety Control Room (SCR).

\end{itemize}

Although not strictly part of the alarm and emergency communication systems, it is worth mentioning here that it is foreseen to include the provision for a GSM repeater at the top of the shaft, which has been demonstrated to provide coverage for the entire 143\ m depth, ensuring personnel can communicate via mobile or TETRA radio even at the base of the pit.

\begin{table}
\begin{center}
\caption{Summary of the foreseen access safety and alarm systems and their implementation periods.}
\begin{tabular}{lr}
\hline
{\bf Safety System} & {\bf Implementation Period} \\
\hline
LASS modifications & LS3 \\
Fire detection and manual evacuation call points & LS3 \\
Voice alarm evacuation system & LS3 \\
Emergency communications (Red Telephones) & LS3 \\
SUSI access control and video surveillance & LS3 / after LS3 \\
LACS access control (TBD) & LS3 / after LS3 \\
Laser room Gas / ODH detection (TBD) & after LS3  \\
Laser safety interlock system (TBD) & after LS3 \\
\hline
\end{tabular}
\end{center}
\end{table}

\begin{table}[htbp]
\small
\centering
\caption{\label{tab:tech-reqs} Preliminary technical and infrastructure requirements. It is anticipated that 5 to 10 side-arms will ultimately be required.}
\vspace{3mm}
%\smallskip
\bgroup
\def\arraystretch{1.25}
 \begin{tabular}{|>{\centering\arraybackslash}m{2.5cm}|>{\centering\arraybackslash}m{4cm}|>{\centering\arraybackslash}m{4cm}|>{\centering\arraybackslash}m{4cm}|}
 \hline
 Requirement & Laser Lab & Interferometry region & Side-arm (per side-arm)\\ \hline
 \hline
  Volume & Floor area $\sim$ \SI{100}{m^2} & \SI{1}{m^2} cross-sectional area &  1~m $\times$ 1~m $\times$ 2~m \\ \hline
  Mains power & $\sim$ \SI{35}{kW} (three- and single-phase outlets) & $\mathcal{O}(\SI{100}{W})$ diagnostic and monitoring electronics & $\mathcal{O}(\SI{10}{kW})$\\ \hline
  Control cables & Ethernet, fibre, coaxial & Magnetic coils, diagnostic and monitoring electronics & optical fibres, coaxial, high-power steel-clad fibers\\ \hline
  Temperature stability & \SI{22}{\degreeCelsius} w/ $\pm$ \SI{1}{\degreeCelsius} pk-pk & $<\SI{1}{\degreeCelsius\per\hour}$ & Temperature controlled, NEMA rated enclosure, $<\SI{0.5}{\degreeCelsius}$ pk-pk\\ \hline
  Water cooling & \SI{30}{kW} cooling capacity & n/a & \SI{5}{kW} cooling capacity, $<\pm \SI{1}{\degreeCelsius}$ stability\\ \hline
   Laser safety & Engineering (enclosures, interlocks); admin (training); PPE (glasses) & Already safe (enclosed) & Engineering (enclosures); admin (training); PPE (glasses) \\ \hline
  Gases & Helium, compressed air, Argon & n/a & Helium for commissioning\\ \hline
  Cryogenics & n/a & n/a & n/a \\ \hline
  Ventilation & Air-handling unit capable of temp. spec. & Air-flow to maintain temp. spec. & Air-flow to move \SI{5}{kW} of heat\\ \hline
  Access & Year-round (24 hrs/day) & Access for maintenance (more access during calibration and commissioning) & Year-round 24 hrs/day (more R\&D for fully autonomous atom sources)\\ \hline
  Smoke detector & Yes & Yes & Yes\\ \hline
  Oxygen depletion monitor & Yes & During maintenance & n/a\\ \hline
  Hoisting \newline equipment & n/a & Modular sections $<$ \SI{907}{kg} & n/a \\
 \hline
 \end{tabular}
% \label{tab:tech-reqs}
\egroup
\end{table}

\subsection{Electrical infrastructure} 
\label{sec:site:electrical}

%\textcolor{red}{Charline M.}

\subsubsection{Power requirements and loads}
New electrical loads are expected, for both the AICE experiment itself and for the associated infrastructure. Those related to the infrastructure will be present from LS3 onwards, and those related to the experiment and its laboratory will be installed at a later stage.
The current estimates for electrical requirements are summarized in Tables~\ref{tab:EL_infrastructure} and~\ref{tab:EL_experiment}. All loads are supplied at LV (400 or 230 VAC), but from several different networks: normal, secured, and UPS networks.

\begin{table}
    \centering
%    \begin{tabular}{ccccc}\toprule
    \begin{tabularx}{\linewidth}{X X X X}\toprule
         Load type&  Location of the load&  Electrical network& Maximum load\\\midrule
         \hline
         Lifting platform&  SX4&  Secured network& 22 kW\\
         Shielding door&  Underground&  Normal network& 15 kW\\ \hline
%         Cooling cabinet&  SX4&  Normal network& 87 kW\\
         Ventilation cabinet&  SX4&  Normal network& 56 kW\\ \hline
         Access point PAD and rack&  SX4&  Secured \& UPS networks& 1 kW\\
%         &  Alarm systems&  Shaft&  \multicolumn{2}{X}{Negligible (from existing surface racks)}\\
%         &  LASS&  Underground&  \multicolumn{2}{X}{Negligible (from existing underground installations)}\\ \hline
         Alarm systems&  PX46 Shaft&  Negligible (*)\\
         LASS&  Underground&  Negligible (**)\\ \hline
         AMF&  Underground&  Normal network& 0.4 kW\\ \bottomrule
%    \end{tabular}
    \end{tabularx}
    \caption{Infrastructure-related power loads. (*) From existing surface racks; (**) from existing underground installations. }
    \label{tab:EL_infrastructure}
\end{table}

\begin{table}
    \centering
%     \begin{tabular}{ccccc}\toprule
    \begin{tabularx}{\linewidth}{X X X X}\toprule
         Load type&  Location of the load&  Electrical network& Maximum load\\\midrule
         \hline
         Vacuum equipment&  Powered from SX4&  Normal network& 112 kW (during commissioning)\\
 & & &24 kW (during operation)\\ \hline
         Laser room equipment&  Laser laboratory&  UPS network& 27 kW\\
         &  &  Normal network& 46 kW\\ \hline
         Atom sources&  PX46 Shaft&  Normal network& 100 kW\\ \hline
         Cooling cabinet&  SX4&  Normal network& 87 kW\\ \hline
%         Ventilation cabinet&  SX4&  Normal network& 56 kW\\ \hline         
         General infrastructure&  Laser laboratory&  Normal network& 35 kW\\ \bottomrule
%    \end{tabular}
    \end{tabularx}
    \caption{Experiment-related power loads}
    \label{tab:EL_experiment}
\end{table}

The infrastructure-related loads result in a total of 158 kW on normal network, 22 kW on secured network and 1 kW on UPS-secured network.

The experiment-related loads result in a total of 293 kW on normal network, and 27 kW on UPS network.

\subsubsection{Powering solution and associated infrastructure works}
To accommodate the new loads introduced by the experiment and its associated infrastructure, a new distribution switchboard is required. This switchboard would supply the loads connected to the normal network. It must be designed to supply both the loads commissioned during LS3 and those of the experiment, and therefore be capable of handling a total load of 451 kW (maximum consumption of all loads). A minimum 650 A LV switchboard, referred to here as EXD1/4X, should therefore be installed near the shaft and future laser laboratory.

The power supply could be provided from the main switchboard of building EBD1/4X, whose current consumption (maximum 450 A, measured during YETS~25-26) is well below its maximum capacity (2000 A). However, there are not enough available spare outgoing feeders to supply a 650 A feeder from the switchboard. As the switchboard is quite old, it is not possible to modify it directly to add new feeders. The replacement of EBD1/4X would therefore be required under the   distribution scheme proposed currently.

The power supply for the lifting platform, which requires a secured power supply, could be provided directly from the secured network switchboard located in the building, ESD1/4X. The lifting platform would require a 32 A feeder, which is currently available as an equipped spare feeder on the ESD1/4X. 
%The platform's power supply cabinet could be directly connected to this feeder.

The existing UPS located in SX4 does not have sufficient capacity to supply the remaining loads needing UPS protection, which would require 28 kW. The proposed solution is to install a new UPS with a minimum capacity of 30 kVA for 10 minutes, dedicated to the AICE infrastructure and experiment. This UPS would be supplied from EXD1/4X and would supply a new UPS distribution cabinet, referred to here as EOJ3/4X. The loads could then be supplied from this cabinet.
% \textcolor{red}{OB: Sergio, perhaps you can look at this: does a minimum capacity of 30~kVA provide sufficient margin for a 28~kW load and the required 10-minute autonomy?}

\begin{figure}
    \centering
    \includegraphics[width=0.8\linewidth]{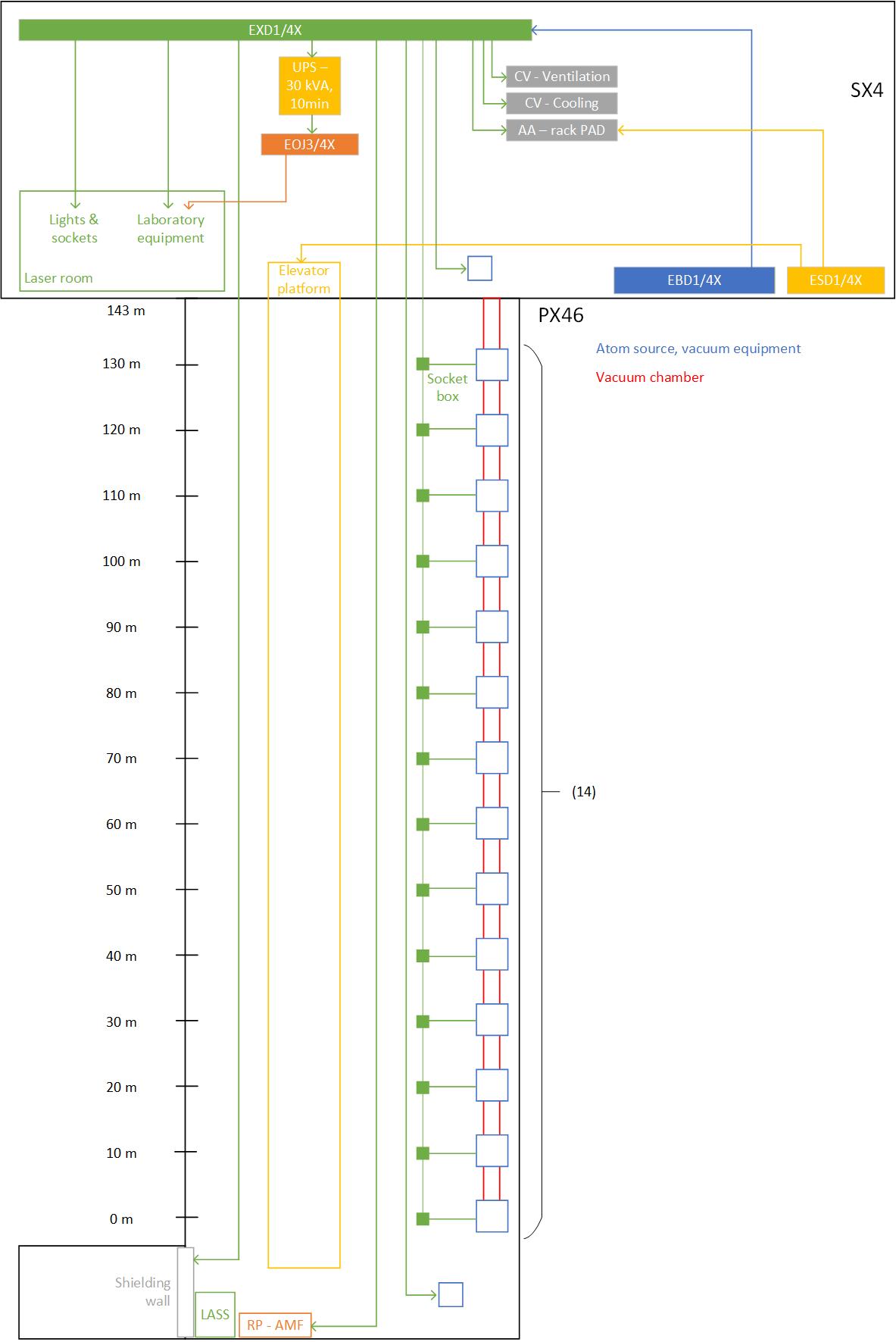}
    \caption{Layout of the proposed electrical distribution network for AICE and the related infrastructure.}
    \label{fig:elec_distrib}
\end{figure}

The access-point PAD rack, that requires UPS and secured networks for 1 kW, would be powered from the newly installed UPS, through EOJ3/4X, and from an existing spare feeder of ESD1/4X.
Figure~\ref{fig:elec_distrib} shows the proposed electrical distribution network, including the various loads and equipment to be installed.

\subsubsection{Signal and fibre optic cabling}
The underground LASS installation, the safety and alarms systems will require signalling cables to be installed during LS3 as described below. 

A detailed assessment of the signal and optical fibre requirements required along the shaft and on the surface laboratory for the later installation of the AICE experiment is not yet available.  Therefore, the impact on the existing infrastructure, the new required infrastructure (installation of cable trays, consolidation of the fibre distribution point in SX4 and UX45) and its related planning cannot yet be established.

\subsubsection{Schedule and constraints}
The first step of the schedule will be to power the lifting platform, at the top of the shaft in SX4. After the lifting platform has been installed and commissioned, the necessary work for the infrastructure in the shaft and underground to be completed during LS3 will be performed. During this period, all the powering of the new loads related to the infrastructure will have to be done, as well as the signal and fibre optic cabling.

At a later stage, the powering for the experiment and its laboratory, as well as related signal and fibre optic cabling, will be done in parallel with their installation.

The main constraints on schedule for the electrical and cable installations are related to the modifications of the EBD1/4X switchboard: its possible replacement should be decided at an early stage, due to long ordering lead-times, and to allow the organization of this work, which would also cause impacts in the area. It is also important to consider the cable trays integration into the infrastructure: for example, if cables are required to cross the shielding door, their quantity and types should be evaluated to foresee necessary cable trays for them and include them in the design of the door, and assess any RP impact of possible openings. Finally, owing to the heavy workloads during the LS3 phase, the installation and cabling activities would need to be planned at an early stage so that they can be incorporated into the LS3 schedule. It is understood that activities pertinent to the experiment may be delayed until after LS3. However, if consolidation of the fibre optic network is required, these works will need to be carried out only during technical stops of the LHC, owing to access constraints to the related buildings.

\subsection{Cooling and ventilation}
\label{sec:site:CV}
% \textcolor{red}{Olivier C.-L., Roberto. A. B.}

Existing Heating, Ventilation and Air Conditioning (HVAC) equipment for the LHC UX45 is not expected to impact AICE (the average air velocity in PX46 is \SI{0.16}{\metre\per\second}, as reported in the conceptual feasibility study~\cite{Arduini:2851946}, resulting in a turbulent air flow with a Reynolds number of the order of 100,000), nor would AICE construction impact HVAC operation, provided the existing top cover of the PX46 shaft is retained. 

Existing chilled water production in the neighbouring surface building SU4 does not have sufficient extra capacity for adding AICE as new user, and its annual shutdowns are not compatible with AICE operation. A new station would have to be installed on the surface, which would allow AICE to operate independently from the LHC and its technical stops. Its installation could be done in parallel with construction of the experiment.

The option of local chillers to cool the AICE sidearms has been assessed, but has not been retained, for the following main reasons:
\begin{itemize}
    \item Safety: the most commonly used environmentally friendly refrigerants are either flammable or slightly flammable, introducing some risk in the event of a leak or release into the shaft;
    \item Maintenance: the maintenance requirements of a chiller are much higher than more standard and less complex hydraulic equipment such as pumps, valves, heat exchangers etc.;
    \item Vibration: a local chiller generates unwanted additional noise and vibrations, whose impact should be carefully assessed (depending on the selected machine and its working conditions). 
\end{itemize}

The proposed solution for cooling the AICE sidearms is instead a compact cooling skid for each sidearm, supplied by the aforementioned new dedicated chilled water production unit installed on the surface.
New pipework will be installed in the shaft to connect each skid, to be accessed using elevator platform. Each skid will be supported by a dedicated steel frame anchored on the shaft wall, separated from the supports of the platform and interferometer to avoid transmitting any vibration. 

\begin{figure}
    \centering
    \includegraphics[width=0.75\linewidth, angle=90]{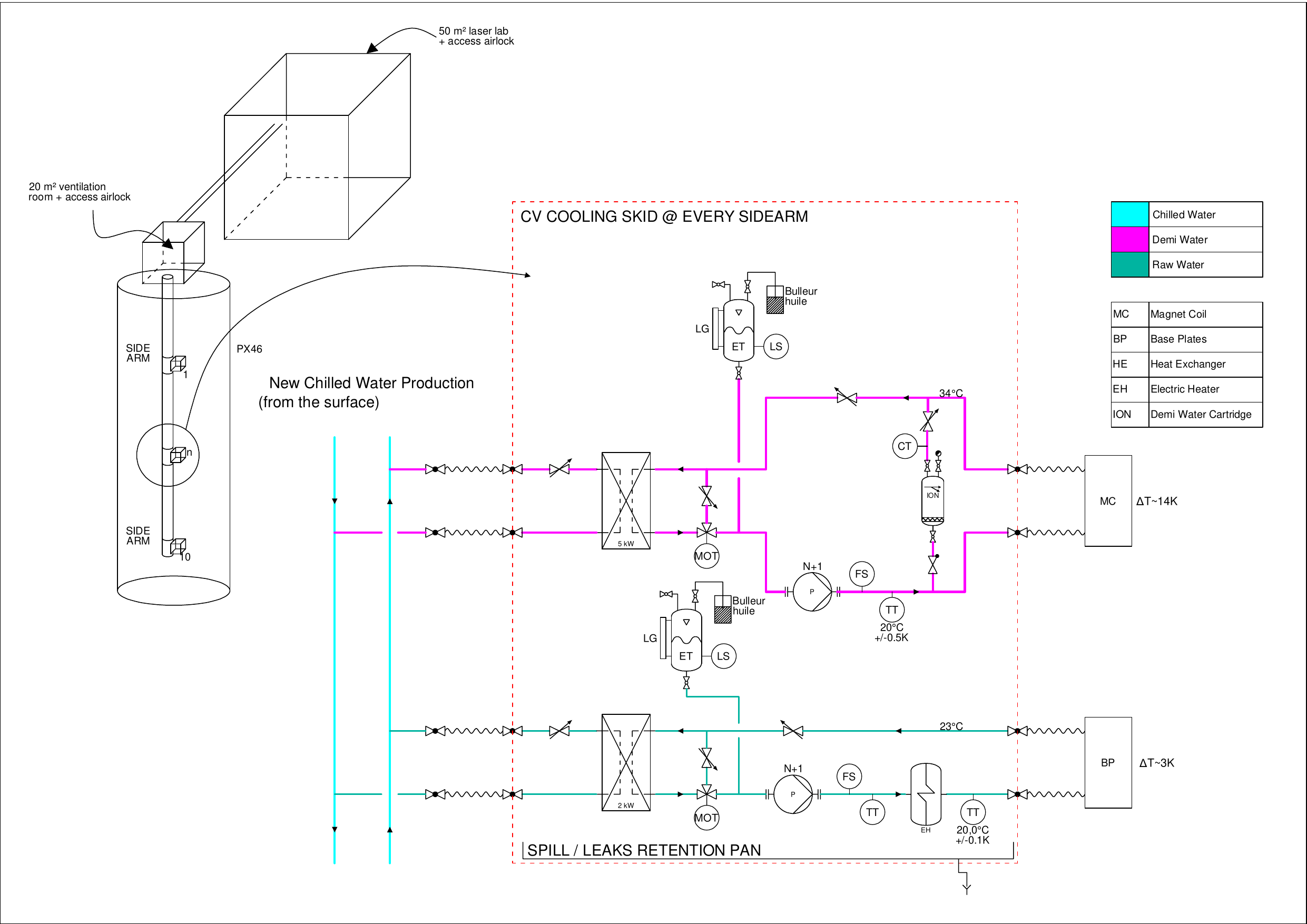}
    \caption{Schematic illustration and Piping and Instrumentation Diagram (PID) of the proposed cooling scheme of the AICE side-arms, including the new chilled water distribution circuit}
    \label{fig:chilled_water}
\end{figure}

The proposed cooling solution is illustrated in Fig.~\ref{fig:chilled_water}. Two cooling circuits are foreseen:
\begin{itemize}
    \item One for the magnet coils (approximate cooling capacity of up to \SI{5}{\kilo\watt}, for compatibility with future upgrades), supplying demineralized water at \SI{20}{\degreeCelsius} with a temperature spread of approximately \SI{14}{\degreeCelsius}  (return temperature of \SI{34}{\degreeCelsius}) at the magnets with 100\% duty cycle: the guaranteed temperature stability is $\pm$ \SI{1.0}{\degreeCelsius};
    \item One for the baseplates (approximate cooling capacity of up to \SI{2}{\kilo\watt}, for compatibility with future upgrades), supplying water at \SI{20}{\degreeCelsius} with a temperature spread of approximately \SI{3}{\degreeCelsius}  (return temperature of \SI{23}{\degreeCelsius}), an electric heater is foreseen to ensure the required $\pm$ \SI{0.1}{\degreeCelsius} stability.
\end{itemize}

The pumps' redundancy is N+1 to increase the cooling systems' availability, while heat exchangers are not redundant: suitable downtime shall be taken into account for periods of heat exchanger cleaning or replacement and equipment maintenance activities.

Two new HVAC systems will be installed. One will be dedicated to the approximately  \SI{20}{\square\metre} ventilation room constructed in building SX4 on top of the shaft, guaranteeing unchanged ventilation management of the underground areas. The ventilation room will be provided with an access airlock to keep the existing pressure cascade between PX46 and SX4.
The other will be dedicated to the $\sim$\SI{100}{\square\metre} laser laboratory constructed in building SX4: as for the ventilation room, the laser laboratory will be provided with an access airlock to ensure a proper dynamic confinement.
The CV systems' power and control cubicle will be installed within the SX4 building.

The access doors illustrated in Figs.~\ref{fig:CE:chicane} and \ref{fig:CE:door} have a meshed insert. This stems from the requirement of having some airflow at the bottom of PX46 when the shielding wall is constructed in order to avoid stagnation and accumulation of CO2 creating a potential hazard for AICE personnel. Meshed doors comply with this requirement, while not adding any extra impact in case of fire or of a helium release accident, since PX46 and UX45 are already communicating via TU46.

 As mentioned in Section~\ref{sec:civil} no major HVAC activities will be required during  LS3, since most of the work can be delayed until the construction phase of AICE.

\subsection{Laser laboratory and beam delivery} 
\label{sec:laser-lab-infrastructure}

% \textcolor{red}{Eduardo G., Olivier C. - with input from Richard H.}

Four possible locations for the laser laboratory have been considered: one at the base of PX46 and three on the surface. The full laser laboratory space requirements could not be accommodated at the base of PX46, and installation, service access and maintenance would be more complicated than in a surface location, so this option is not favoured. Two of the surface options involved constructing extensions of the existing SX4 building on either its northern or southern side. These options have been discarded provisionally in favour of locating the laser laboratory within the existing SX4 structure. The possibility of locating the laser laboratory in the existing rooms has been considered, but the preferred option that we present here is a new room in the open SX4 area, as seen in Fig.~\ref{fig:CE:ventilation}. In this case the laser beams would be transported to PX46 via either a trench in the floor of SX4 or via pipes above ground level, so as not to interfere with transportation requirements. 

%{\color{red}Richard H comment: Need to explain these 4 options here, in brief - Eduardo? Also, I edited this section to refer more to section 4 on laser systems, and to explain which beams are phase stabilised and power amplified etc.}

The detector laser systems are described in Section~\ref{sec:laser}; these laser systems determine the requirements for the laser laboratory and beam delivery infrastructure. To maximise operational stability and simplify maintenance, the majority of the laser systems will be installed in a dedicated laser laboratory located outside the PX46 shaft. Housing the optical infrastructure in a controlled environment significantly reduces thermal fluctuations, vibration levels and access constraints compared with installation directly on the experimental platform.

\begin{table}[htbp]
\centering
\caption{Preliminary requirements for the AICE laser laboratory.}%\textcolor{red}{Check with Olivier}. {\color{red} Richard H comment: Eduardo please check the Electrical power, maybe underestimated (see our emails). Also can we have 24/7 access if not too difficult? Daytime would be sufficient for most things but fewer restrictions would be better.} \textcolor{blue}{Sergio: access 24/7 YES; Updated power from 20-30 kW to 75 kW from info collected by EL services, see table 13} }
\vspace{3mm}
\begin{tabular}{ll}
\hline
Parameter & Requirement \\
\hline
Floor area & $\sim 100$ m$^2$ \\
Temperature stability & $\pm1^\circ$C peak-to-peak \\
Relative humidity & 40--60\% \\
Electrical power & 75 kW  \\
Cooling water & CERN standard service \\
Network connectivity & CERN controls network \\
Optical table area & $\sim 25$ m$^2$ \\
24/7 access requirement & Yes  \\
Radiation environment & No \\
\hline
\end{tabular}
\label{tab:laser_lab_requirements}
\end{table}

The laser laboratory accommodates the strontium cooling and interferometry systems, optical frequency references, frequency-comb infrastructure, diagnostics and control electronics. Sufficient space is reserved for future ytterbium systems and additional optical benches associated with advanced atom-optics development.
Laser light is transferred from the laboratory to the interferometer through optical fibres, but the exceptional case of the clock interferometry beam is delivered close to the shaft by free space optics, with only a short length of fibre to define the spatial mode of the beam while avoiding stimulated Brillouin scattering~\cite{kobyakov_stimulated_2010}. For the clock interferometry beam, active cancellation of fibre-induced phase fluctuations is essential for long-baseline single-photon interferometry, as uncompensated optical path variations would otherwise contribute directly to the measured phase. For the cooling and trapping lasers, it is sufficient to mitigate fibre-induced phase fluctuations passively by routing the fibres through insulated ducting, since the target linewidths of fibre-delivered light are in the kilohertz range or above (see Table~\ref{tab:laser_overview}).

% Richard H: I commented out the table below, because (i) I'm not sure what details are needed here which aren't described already in section 4; (ii) other than the clock interferometry beam, the fibres don't need to be phase stabilised, and (iii) there's no clear need for staged specifications Baseline vs Stretch.

% \begin{table}[htbp]
% \centering
% \caption{Target performance of phase-stabilised optical links.}
% \vspace{3mm}
% \begin{tabular}{lll}
% \hline
% Parameter & Baseline & Stretch \\
% \hline
% Maximum fibre length & 300 m & 300 m \\
% Residual phase noise & $<1$ mrad & $<0.3$ mrad \\
% Fractional frequency transfer error &
% $<10^{-18}$ &
% $<10^{-19}$ \\
% Availability & $>95\%$ & $>98\%$ \\
% \hline
% \end{tabular}
% \label{tab:fibre_links}
% \end{table}

At the experimental shaft, local optical-distribution modules near each atom source provide beam conditioning, power monitoring, beam shaping and alignment diagnostics. In cases where $\gtrsim \qty{100}{\milli\watt}$ is required for the atom sources, low laser power is delivered by fibre near to the atom source, to act as a frequency reference while avoiding stimulated Brillouin scattering~\cite{kobyakov_stimulated_2010}, and local high-power lasers or power amplifiers are installed to provide sufficient light at the atoms. These stations contain only the components required for local beam delivery, and laser amplifiers only where necessary, thereby minimising the amount of sensitive equipment installed underground. This modular architecture also facilitates future upgrades and maintenance interventions.

\begin{table}[htbp]
\centering
\caption{Optical links between the laser laboratory and the interferometer. See Table~\ref{tab:laser_overview} for the full list of related laser specifications.}
\vspace{3mm}
\begin{tabular}{lll}
\hline
Beam & Wavelength(s) & Transport \\
\hline
Sr clock beam & 698 nm & Free-space with short optical fibre, phase-stabilised \\
Sr low-power & 679/707/487 nm & Optical fibre to each atom source, power-stabilised \\
Sr high-power & 461/689 nm & Optical fibre to each atom source, power-amplified \\
Yb clock beam & 578 nm & Free-space with short optical fibre, phase-stabilised \\
Yb low-power & 649/770 nm & Optical fibre to each atom source, power-stabilised \\
Yb high-power & 399/556 nm & Optical fibre to each atom source, to be power-amplified
\\ \hline
\end{tabular}
\label{tab:beam_delivery}
\end{table}

The beam-delivery network is designed to be compatible with both the baseline strontium programme and future dual-species operation. In particular, the fibre routing, support structures and optical interfaces are specified to accommodate additional wavelengths without major modifications to the installed infrastructure.
Laser safety will follow CERN standards for high-power visible and near-infrared laser systems. Beam paths will be enclosed wherever practicable, access interlocks will be implemented, and all operational procedures will be incorporated into the experiment safety documentation.

\subsection{Compatibility with LHC and HL-LHC operations}

Comprehensive technical evaluations, including the Conceptual Feasibility Study \cite{Arduini:2851946} and the subsequent Implementation Study \cite{Arduini:2025jhe}, have established that AICE is fully compatible with existing and future accelerator operations. A central finding of these assessments is that the two facilities can operate concurrently without mutual interference, allowing AICE to pursue its fundamental physics goals while the High-Luminosity LHC (HL-LHC) carries out its primary scientific mission.

\textbf{AICE's Impact on LHC Operations.} AICE is designed to be a ``passive" installation that does not constrain the exploitation of the LHC. A primary consideration is the management of the PX46 access shaft, which is essential for the transport of technical equipment such as superconducting magnets and RF cryomodules. AICE’s footprint occupies only a small fraction of the shaft’s cross-section. This leaves a substantial transport area of nearly 60\,m$^2$ entirely free for accelerator maintenance, which has been shown in detailed transport studies to be sufficient. Furthermore, the radiation shielding wall in the TX46 gallery includes a motorized frame and a large opening designed to facilitate the frequent movement of heavy LHC equipment during technical stops.

The experiment’s technical requirements are also well within existing infrastructure margins. The total electrical load of AICE is negligible compared to the 1.25 MVA transformer capacity at Point 4, requiring no major upstream interventions. To ensure climate control is not disrupted, AICE includes a purpose-built ventilation room at the top of the shaft that acts as an airlock. This maintains the pressure gradient necessary for consistent ventilation management of the underground caverns, ensuring AICE does not interfere with the accelerator’s air renewal systems.

\textbf{LHC's Impact on AICE Operations.} Conversely, our studies have shown that the proximity of the LHC and its high-power systems does not impede the sensitive quantum measurements of the atom interferometer~\cite{Arduini:2851946,Arduini:2025jhe}, see Section~\ref{sec:noise}. Measurement campaigns have found that electromagnetic and radiofrequency noise from the LHC RF power plant are not a concern for the experiment. While ambient magnetic fields vary during LHC magnet ramps, these changes are slow and can be compensated by the experiment’s dedicated magnetic shielding. We note also that the seismic environment at Point 4 is well-characterised; cultural noise is important only above the frequency spectrum relevant for AICE.

Personnel safety and operational access for AICE are guaranteed through specific civil engineering interventions. The construction of an 0.8\,m thick concrete shielding wall in the TX46 gallery reduces radiation levels at the base of the shaft to those of a Supervised Radiation Area, permitting full access even during active HL-LHC beam operations, see Subsection~\ref{sec:site:safety:RP}. Furthermore, the installation of a high-speed elevator platform ensures that emergency evacuations for AICE personnel can be completed within the mandatory two-minute window, effectively managing fire or cryogenic helium hazards without introducing new risks to the wider LHC facility.

By completing these essential infrastructure modifications during the LHC LS3 window, the preparatory works will be successfully decoupled from the experiment’s later construction phase. This plan allows AICE to be installed and commissioned during LHC Run 4, leveraging CERN’s unique infrastructure without impacting the accelerator's exploitation

\subsection{Sustainability considerations}
AICE is an experiment of moderate scale, siting within existing CERN infrastructure at the LHC point 4. No major infrastructure works will thus be required, in contrast to a greenfield experiment. Complements to the existing CERN infrastructure, such as the elevator platform, or the new closed-cycle chilled water supply system, would be required also at different sites. The only significant modification to the existing CERN infrastructure is the shielding wall required for radiological protection (see Section~\ref{sec:civil:shielding}). It is foreseen that {about a half} of the volume of the wall will be in the form of concrete blocks, and it is already planned to reuse blocks from the CERN inventory, thereby cutting approximately in half the quantity of concrete to be poured.

AICE will also necessarily follow the environmental sustainability practices defined by the CERN management \cite{Sustainability-CERN}, in particular for any component that will be procured through CERN administrative processes or fabricated at CERN, in alignment with the ESPP 2026 recommendations on sustainability.

A full Life Cycle Assessment (LCA) will be conducted at the TDR phase.

%\newpage

\section{Infrastructure Preparation and Civil Engineering}
\label{sec:civil}

In this Section we describe the major construction works that need to be performed during LS3, in order to make the PX46 shaft available for the assembly of the AICE experiment. To define the scope boundary of these enabling works and manage project risks, Table~\ref{tab:aice_scope} categorizes the required interventions as mandatory or desirable, identifying the technical and programmatic consequences should delays occur. The major activities include notably the civil engineering works for the shielding wall, the construction of the concrete sliding door allowing access to the tunnel for equipment transport, the construction of the elevator platform and of the surface ventilation room. The required services that need to be provided during LS3 for these infrastructures to be realized and operated have been discussed in the previous Section, and their prioritization is also included in Table~\ref{tab:aice_scope}. 

We finally present in this Section the foreseen planning for the works, the current budget estimate for their realization, and preliminary ideas for the organization required for overseeing these works.

\begin{table}[htbp]
\centering
\caption{AICE infrastructure and experiment scope boundary and consequences of delay.} 
\label{tab:aice_scope}
\small % Slightly smaller font to ensure fit
\begin{tabularx}{\textwidth}{ l >{\raggedright\arraybackslash}p{3cm} >{\raggedright\arraybackslash}p{2.2cm} X }
\hline
\textbf{Category} & \textbf{Work Item} & \textbf{System / Package} & \textbf{Consequence of Delay} \\ \hline
\textbf{Mandatory (LS3)} & \textbf{Radiation Shielding Wall \& Doors} & Civil Engineering & Protects personnel in the PX46 shaft base during HL-LHC beam operations. \\
\textbf{Mandatory (LS3)} & \textbf{Elevator Access Platform} & Handling Engineering & Essential for safe routine access and the mandatory 2-minute emergency evacuation window. \\ 
\textbf{Mandatory (LS3)} & \textbf{LASS Modifications \& Safety Doors} & Access \& Alarms & Required to decouple PX46 from the interlocked LHC safety zone for independent experiment access. \\ 
\textbf{Mandatory (LS3)} & \textbf{Fire Detection \& Red Telephones} & Access \& Alarms & Basic regulatory safety requirement for any continued occupancy of the underground shaft. \\ 
\textbf{Mandatory (LS3)} & \textbf{Radiation Monitor AFM} & Safety & Prerequisite for classifying the shaft as a Supervised Radiation Area. \\ 
\textbf{Mandatory (LS3)} & \textbf{Ventilation Room (Airlock)} & Civil Engineering & Protects airflow stability. \\ 
\textbf{Desirable (LS3)} & \textbf{Electrical Switchboard Upgrades} & Electrical Engineering & Necessary to power shielding doors, elevator, and safety systems installed during LS3; delay results in major power cuts if upgraded in Run 4. \\ 
\textbf{Desirable (LS3)} & \textbf{Additional SX4 Hoist} & Handling Engineering & Essential for precise handling in narrow shaft spaces; delay slows platform construction and detector installation. \\ 
\textbf{Desirable (LS3)} & \textbf{Laser beam transport} & Civil Engineering & Demolition works required for a trench in SX4 floor; final choice in TDR; delay may impact ventilation room structure. \\ 
\textbf{Post-LS3} & \textbf{Vacuum System \& Telescope} & Experiment Infrastructure & Primary scientific infrastructure in PX46; installation must follow LS3 site readiness. \\ 
\textbf{Post-LS3} & \textbf{Atom Sources \& Laser Systems} & Experiment Scientific Package  & Core of the scientific instrument; deployment is staged following infrastructure completion. \\ 
\textbf{Post-LS3} & \textbf{Laser Laboratory} & Experiment Infrastructure & Reserved space in SX4 for master lasers and frequency references. \\ 
\textbf{Post-LS3} & \textbf{Chilled Water Unit \& Cooling Skids} & Cooling Engineering  & Required for magnet and source cooling; not required for initial site preparation. \\ 
\textbf{Post-LS3} & \textbf{Dedicated UPS \& Lab Powering} & Electrical Engineering & Provides stable power for scientific equipment; can be installed during detector assembly. \\ \hline
\end{tabularx}
\end{table}

\subsection{Shielding wall in TX46}
\label{sec:civil:shielding}
% \textcolor{red}{Responsible: Tamara Bud}

\subsubsection{Civil Works Overview}
\begin{figure}[h!]
    \centering
    \hspace{1cm}
    \includegraphics[width=0.45\textwidth]{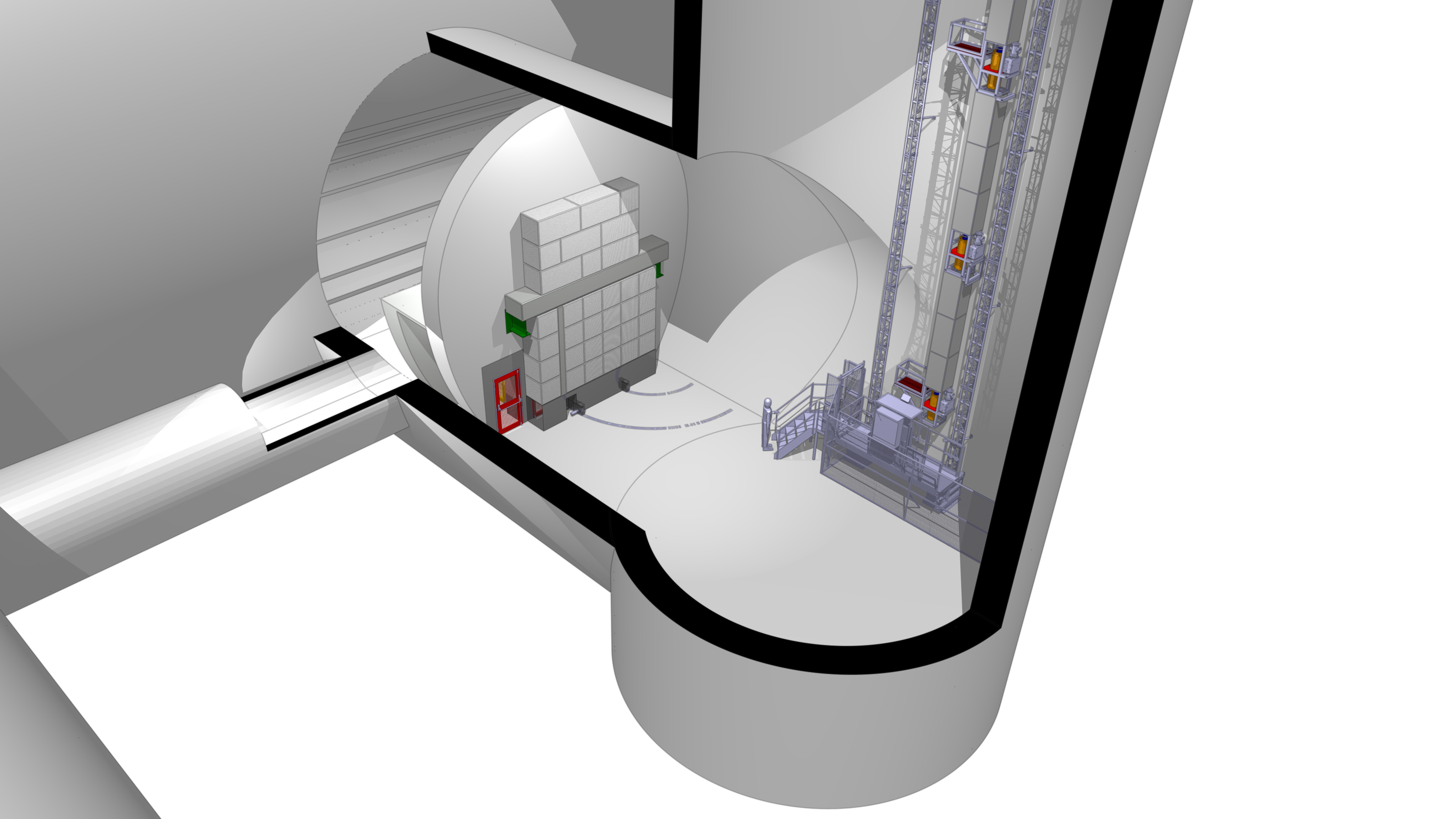}
    \includegraphics[width=0.45\textwidth]{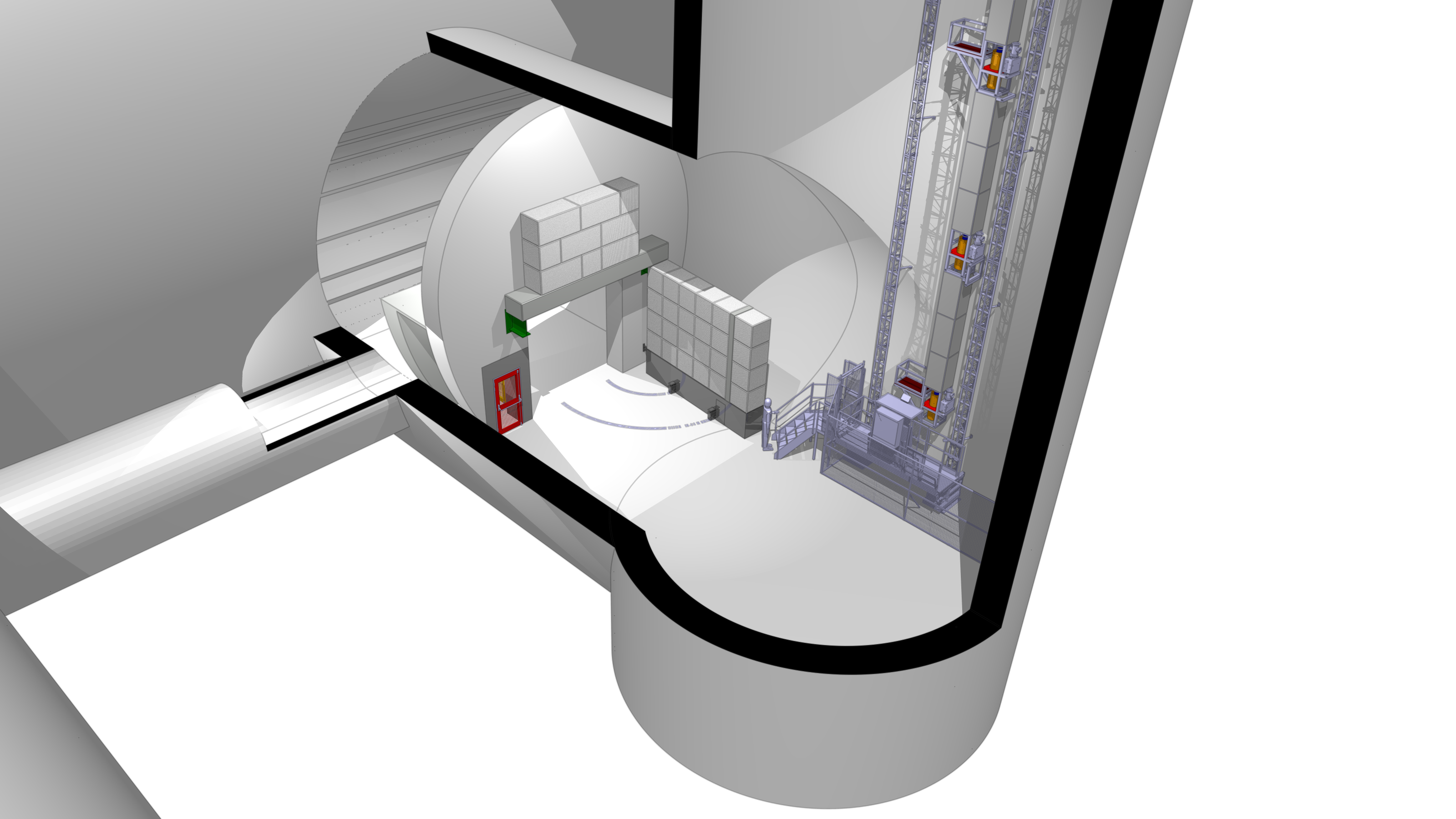}
    \caption{\label{fig:CE:chicane}Shielding wall with the proposed equipment access doors in closed ({\it left panel}) and open ({\it right panel}) configuration. The bottom landing of the elevator platform is also visible.}
\end{figure}

\begin{figure}[h!]
    \centering
    \includegraphics[width=0.9\textwidth]{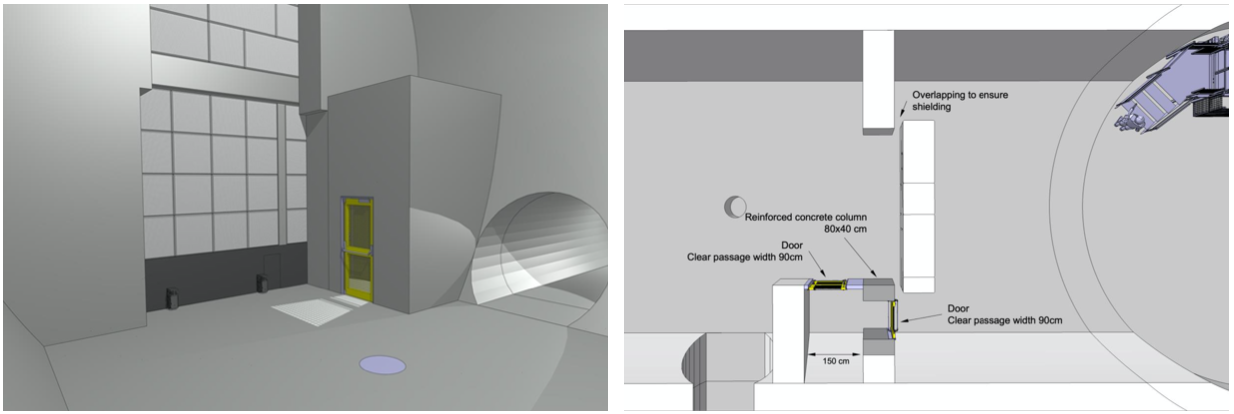}
    \caption{\label{fig:CE:door} {\it Left panel}: Detail of the shielding wall, showing the door for personnel between the UX45 cavern and the PX46 shaft. {\it Right panel}: Two-dimensional schematic view of the shielding wall, showing the layout of the access doors for equipment and personnel.}
\end{figure}

{As part of the enabling works for AICE during LS3, several civil engineering interventions will be needed. This Section provides an overview of these works and outlines the process to implement them.
The largest portion of these interventions will be in the UX45 cavern, where a shielding wall separation is required to allow regular access to the bottom of the PX46 shaft, including during LHC beam operations. The requirement for a shielding door with a 4\ m$\times$4\ m opening comes from the need to replace failing equipment in the UX45 cavern within a day, including during LHC operations. 
This system would be completely shielded, including inside the frame. A preliminary design of such a system has been proposed that demonstrates the feasibility, based on the design of similar equipment implemented recently for HIE-ISOLDE. Images of these proposed designs can be seen in Figs.~\ref{fig:CE:chicane} and \ref{fig:CE:door}.}

{In addition to the 16m$^2$ lower portion of the shielding wall that will be designed so that it can be moved relatively frequently, there will also be an upper portion of 12m$^2$ of shielding blocks. These will be designed to be moved only once, during the LHC decommissioning stage. Supporting this upper portion of shielding blocks there will be a reinforced concrete lintel or beam located between the two shielding block sections.}

{The civil engineering works involved in the construction of the shielding wall are:
\begin{itemize}
    \item Casting of a circular concrete wall in the UX45 cavern with an opening for shielding blocks;
    \item Construction of a reinforced concrete beam between the top and bottom portions of the shielding blocks;
    \item Construction of an access room in UX45 between the LHC and PX46;
    \item Coordination with the transport team for construction and installation of movable shielding blocks.
\end{itemize}}

\begin{figure}[h!]
    \centering
    \includegraphics[width=0.9\textwidth]{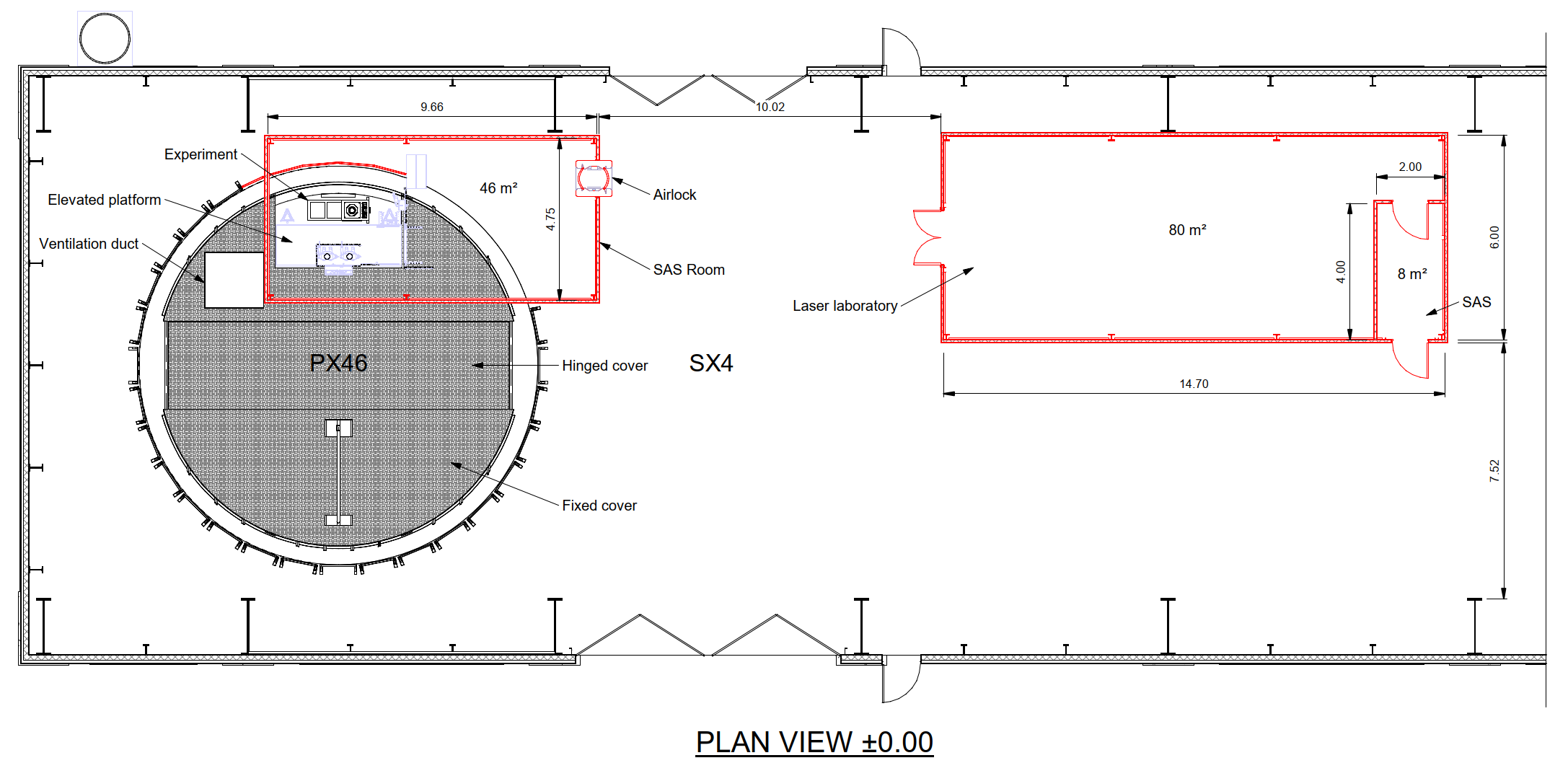}
    \includegraphics[width=0.9\textwidth]{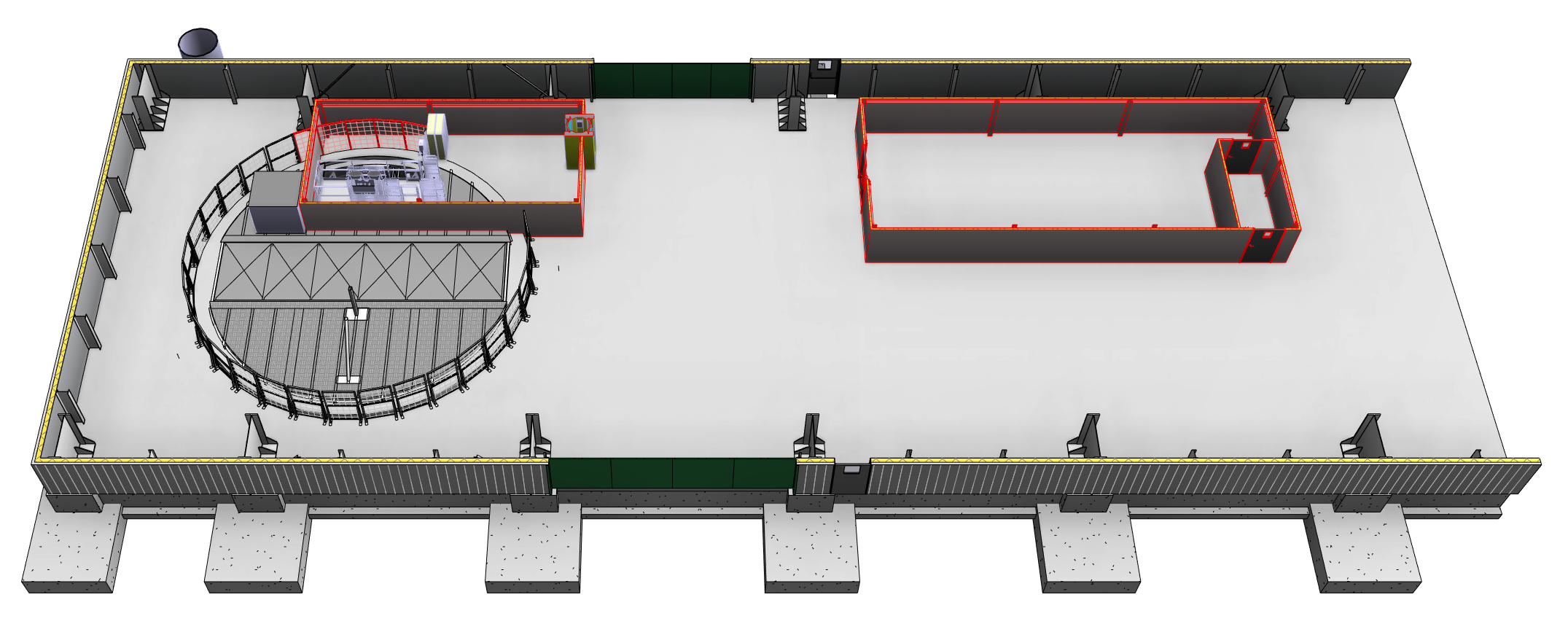}
    \includegraphics[width=0.85\textwidth]{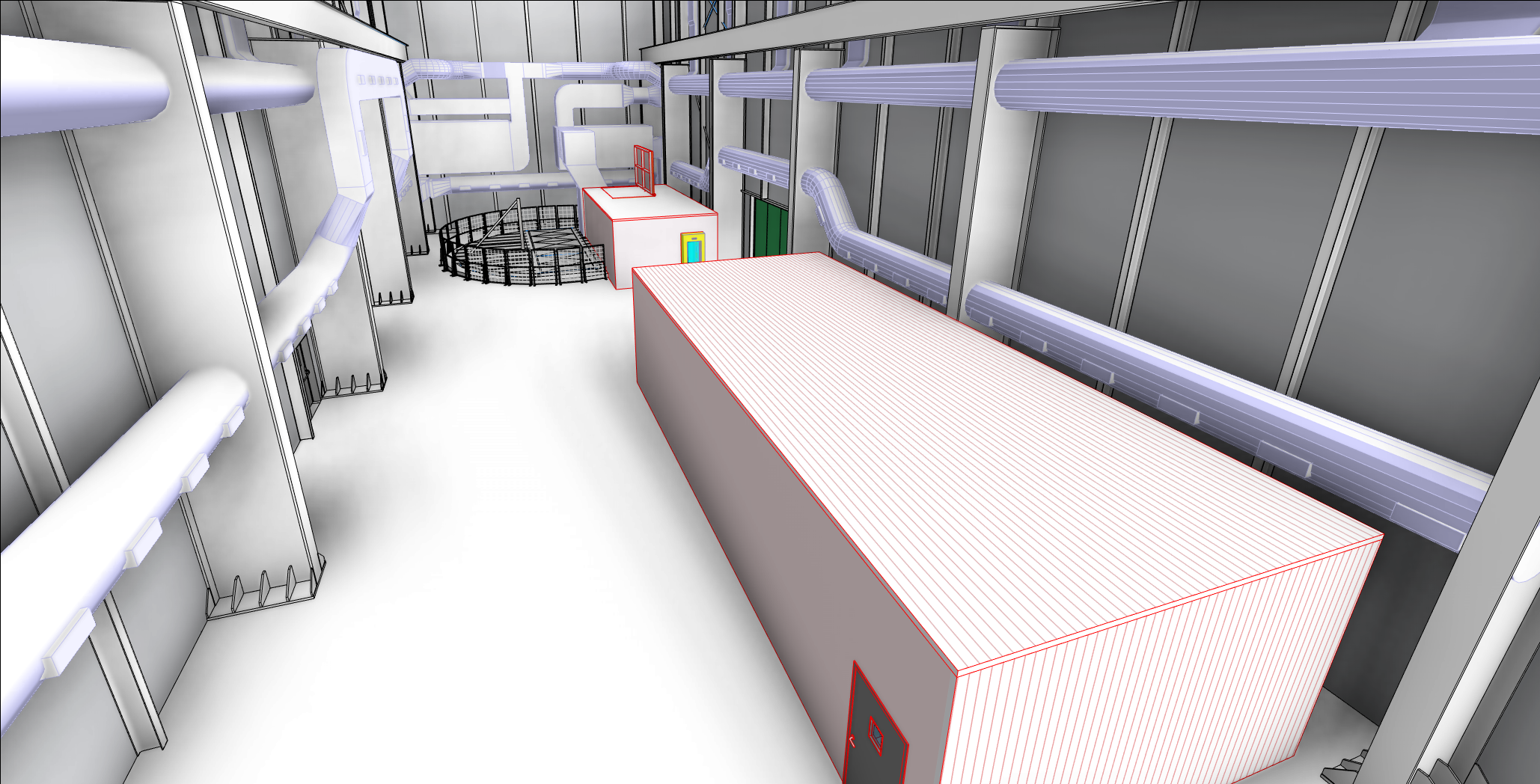}
    \caption{\label{fig:CE:ventilation} Two- and three-dimensional visualisations of the steel-framed ventilation surface building and of the preferred laser laboratory location in SX4.}
\end{figure}

{The second portion of civil engineering works required in LS3 will be on the surface, where a new ventilation room will be constructed at the top of the PX46 shaft in SX4. This is currently being considered as a steel-framed building of dimensions 4\ m$\times$5\ m$\times$3.5\ m. This will be complemented by the laser laboratory, which at present level of maturity of the technical studies is foreseen at the other end of the surface building SX4. The laser beam will be transported from the laboratory to the experiment in a beampipe located either in a trench to be excavated in the concrete floor of the building (which is the current preferred solution), or running along the side wall of the building. The final choice will be made in time for a possible realization of the trench during LS3. Three-dimensional visualisations of these buildings can be seen in Fig.~\ref{fig:CE:ventilation}.}

\subsubsection{Civil Engineering Design Studies}
{In order to facilitate the civil engineering works described above, certain civil engineering design studies will need to be undertaken before construction can begin.
For the shielding wall in UX45, a structural design study will be needed from external engineering consultants. Based on their input, it may be necessary to also carry out geotechnical and site investigation works in the location of the proposed shielding wall. This would be to establish the structural capacity of the existing tunnel lining in the cavern, as well as any additional information required about the ground in that location.}

{It will also be necessary to ask an external company to undertake a structural study for the ventilation room on the surface. This is 
because it is intended to rest partially on the lid of the PX46 shaft, so the loading on this lid will have to be checked and the structure of the ventilation room adjusted if necessary.}

{Once these studies have been completed and detailed drawings for the works have been provided by the engineering consultants, construction studies can begin. These will involve a detailed bill of quantities and programmes from the contractor selected for the works, as well as integration and logistics coordination with other CERN teams, e.g., Transport and Radiation Protection. This process will be greatly accelerated if it is possible to undertake the construction works using the CERN Framework Contract. This would mean that a tender process could be avoided and instead the contractor company that is already used for other LS3 CERN activities could be deployed for the task.}

\subsection{Elevator platform and handling equipment}
\label{sec:civil:platform}

% \textcolor{red}{Responsible: Damien Lafarge}

{The equipment required in order to implement AICE in the PX46 shaft consists of:
\begin{itemize}
\item An elevator platform providing access to the experiment at all relevant heights, and
\item Specific handling equipment for lowering and raising components of the experiment adjacent to the walls of the PX46 shaft.
\end{itemize}}

{A suitable elevator design has been proposed by a consultancy firm~\cite{XLIndustries:2023}, which provides a normal operation mode and an evacuation mode complying with the safety requirements outlined in Section~\ref{sec:site:safety}. The main characteristics of this platform are the following:
\begin{itemize}
\item  Total maximum useful load: 500 kg (with the possibility of redesign for up to 1000~kg);
\item  Space for two operators in a protective booth, in order to avoid any risk of personal injuries during fast descent;
\item  Electrical power from a secure network with backup batteries in case of power outage, and capacity of operating emergency services (lighting and communications) for at least two hours;
\item  Nominal speed in automatic mode 40 m/min;
\item  Nominal speed in manual mode 12 m/min;
\item  Controlled descent in case of evacuation at a speed of 70~m/min to comply with the requirement of a 2-minute maximum descent time for evacuation, as set out in the Feasibility Study Report~\cite{Arduini:2851946};
\item  Compliance with the EN1495 standard.
\end{itemize}}

\begin{figure}[h!]
	\centering
    \includegraphics[width=0.45\textwidth]{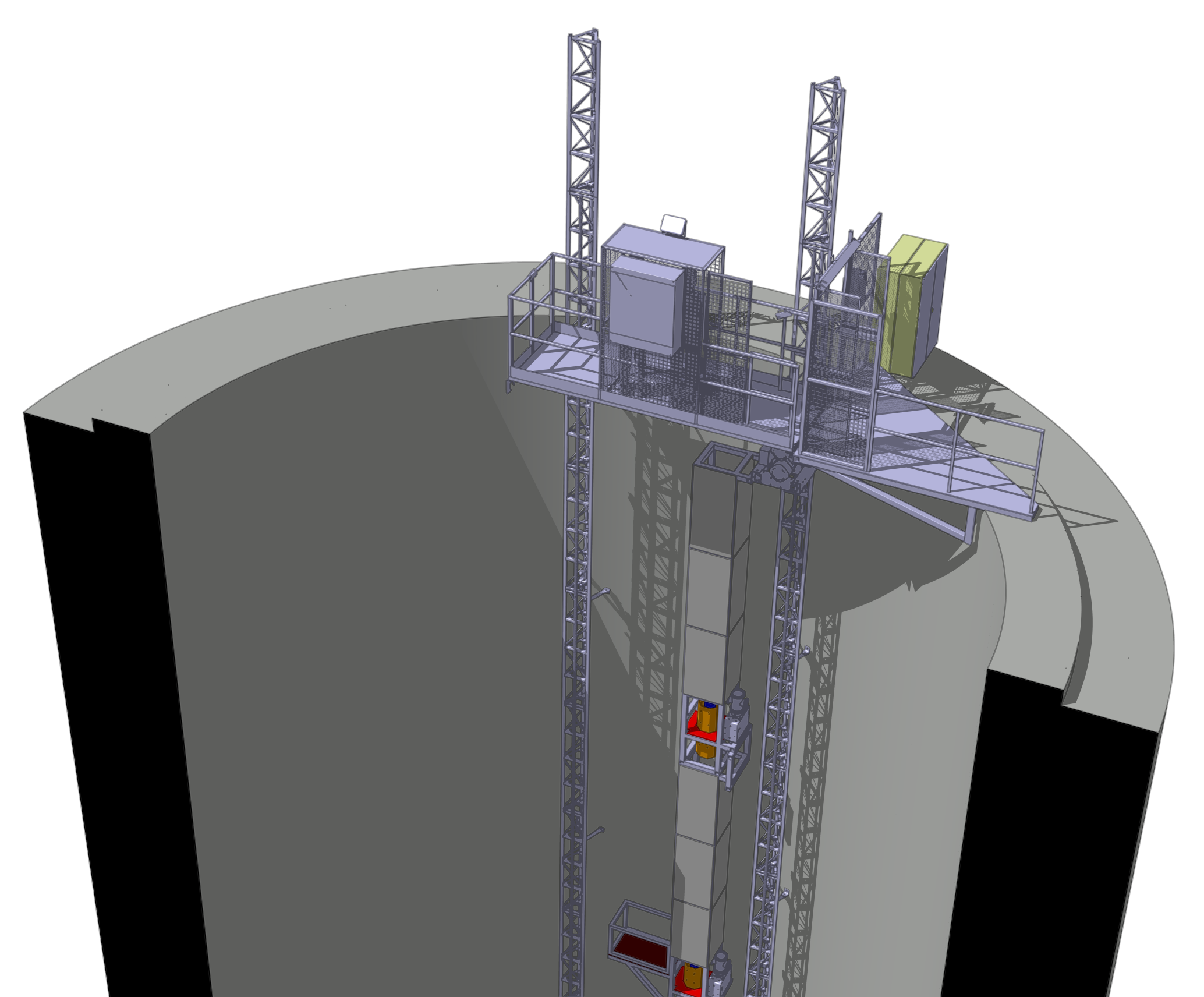}
    \includegraphics[width=0.45\textwidth]{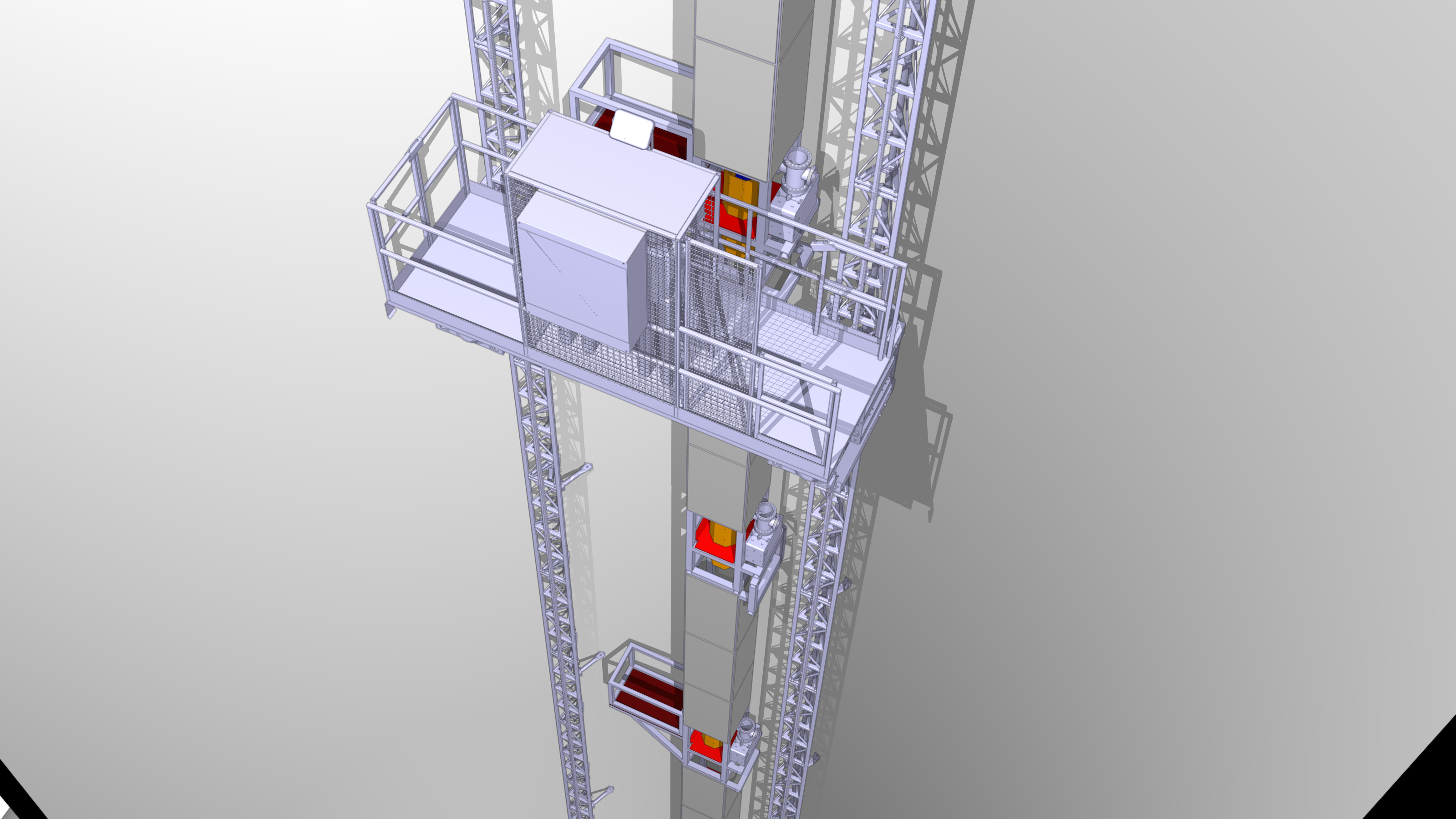}
	\caption{\label{fig:Platform1} {\it Left panel}: 3D view of the top landing of the elevator platform in SX4. {\it Right panel}: 3D view of the elevator platform during operation in PX46~\cite{XLIndustries:2023}.}
\end{figure}

{The proposed custom-made elevator platform is illustrated in Fig.~\ref{fig:Platform1}. It allows routine access to the experiment, stopping at the locations of atom sources in the side arms of the experiment, via the automatic operation mode like that of a normal lift. In the event of an emergency evacuation, it could reach the surface sufficiently rapidly for the operators to escape through the SX4 building and, should evacuation to the surface not be possible, it could descend to the bottom of PX46 with the required speed as specified.}

{The platform would also be used during the installation of AICE. In this case, a manual operation mode has been envisaged, where the operators could start and stop the platform manually (moving at reduced speed) and thus be able to work at any point along the height of PX46. Horizontal forces during experimental work will be taken into account for the structural design of the mast to ensure stability. Provisions for avoiding vertigo, such as opaque side panels or possibly a fully enclosed platform, and extensible flooring to allow operators to comfortably reach the wall or working on the side arms and the atom sources, will be included in the final design. After the installation of the platform, the lid of the PX46 shaft will be adapted to host the surface landing of the platform and support the ventilation room.}

{An anchor point installed above the platform in the surface building will provide the possibility of rescuing the operators in the event of mechanical failure of the platform, although the probability of this happening concurrently with a fire or helium release event is considered negligible. In such an event an alternative option could be to suspend a rescue nacelle from the lifting equipment described below, installed on the overhead crane in the SX4 surface building. It could be used autonomously by a Fire Brigade rescue team. The design of the platform should allow enough room and capacity to support the two operators in normal operation plus two rescuers in case of an intervention.}

\begin{figure}[h!]
	\centering
	\includegraphics[width=.46\textwidth]{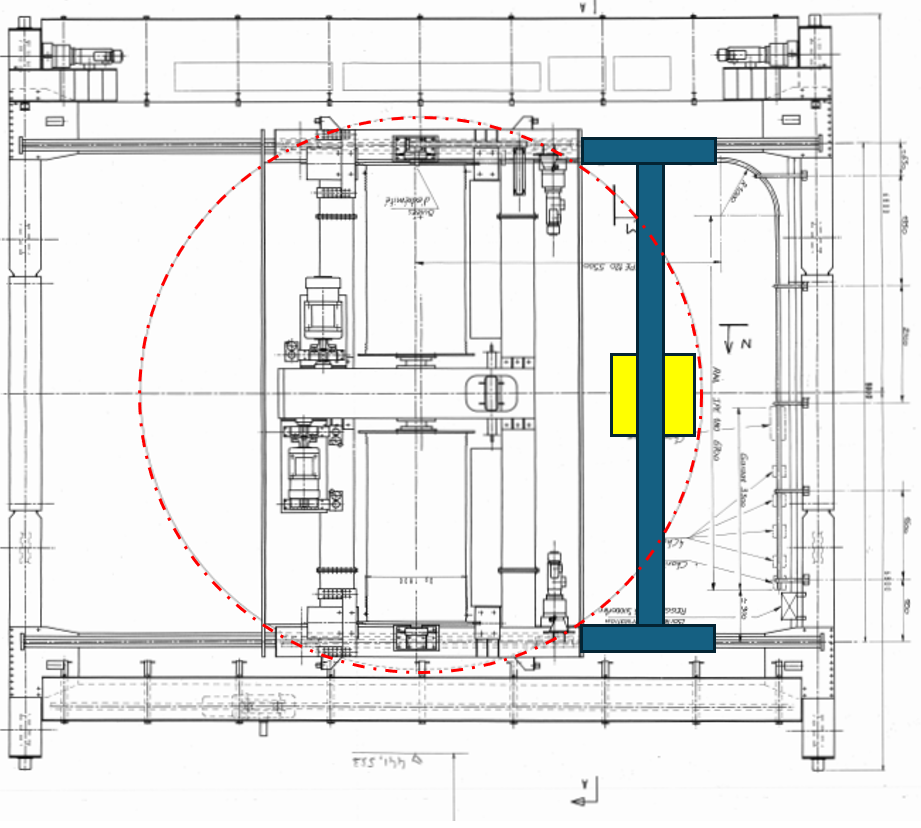}
	\includegraphics[width=.52\textwidth]{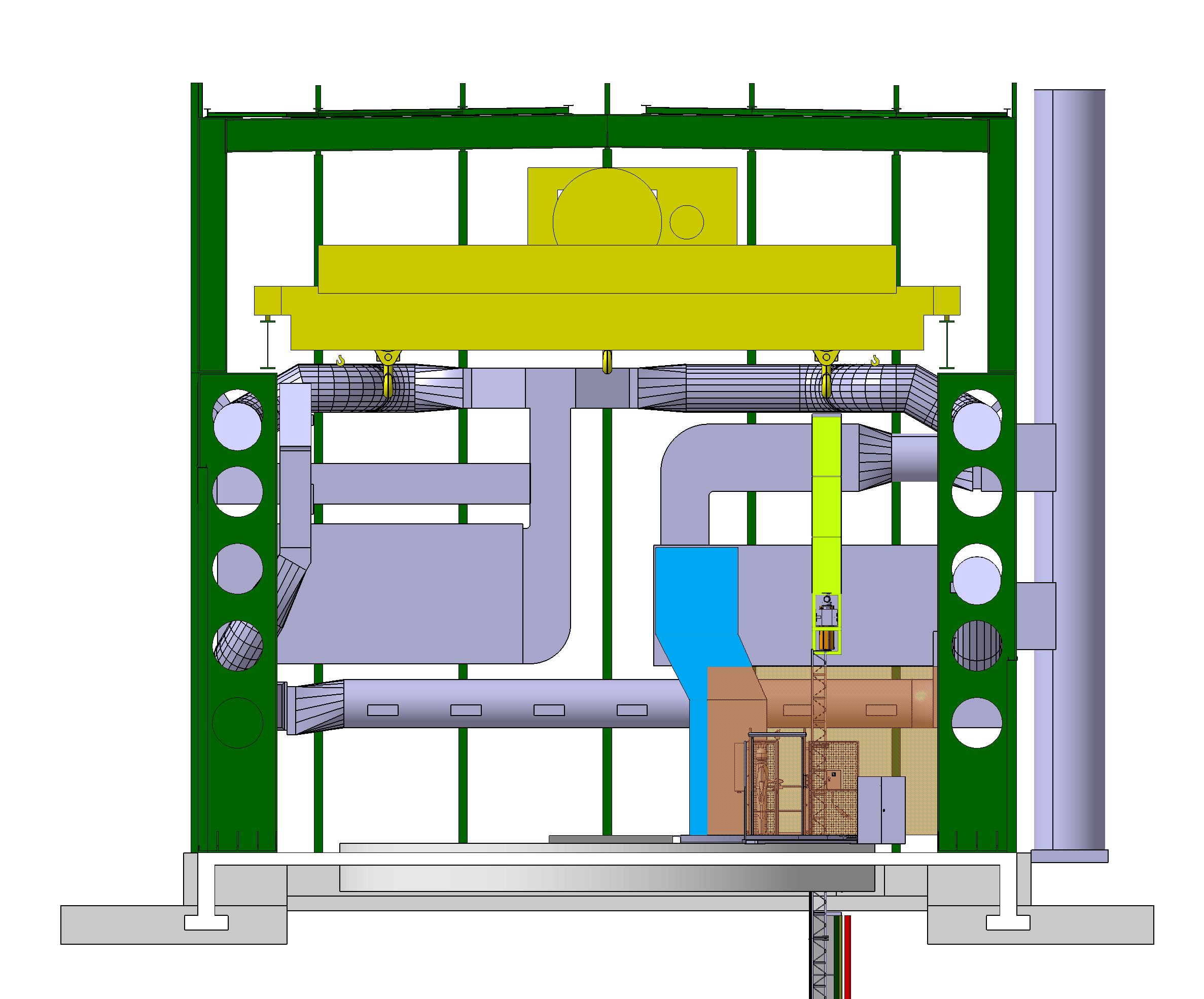}
    \caption{\label{fig:lifting} {\it Left panel}: Top view of the SX4 overhead crane equipped with new handling equipment dedicated to AICE. {\it Right panel}: Side view of the SX4 surface building during handling operations: the overhead crane is in yellow, ventilation pipes are in blue, the AICE component is in light green, the ventilation room is in orange.}
\end{figure}

{The handling of the different elements of AICE during the installation phase will be possible thanks to additional equipment fixed to the main hoist of the SX4 overhead crane (see the left panel of Fig.~\ref{fig:lifting}).}

{This solution allows for precise handling in the narrow space between the wall and the access platform over the entire height of the shaft, the cables of the main overhead crane hoist having too wide a span to fit into that volume. (See the right panel of Fig.~\ref{fig:lifting}).}

% \begin{figure}[h!]
%    \centering
%    \includegraphics[width=.49\textwidth]{figures/Laser_Lab_Plan.png}
%    \includegraphics[width=.48\textwidth]{figures/Component_Handling.png}

%    \caption{
%        \label{fig:laser-lab-handling}
%        \textit{Left panel}: Overhead view of the laser laboratory and the new handling equipment in SX4 dedicated to AICE.
%        \textit{Right panel}: \textcolor{red}{Repeated figure!} Rendering of the procedure for lowering detector components from SX4 into the PX46 shaft.
%        \textcolor{red}{Eduardo thinks the laser laboratory should be larger. Also there is concern about vibration and noise from the ventilation system. The laser laboratory should perhaps be located at the far end of SX4, with laser beams transported to PX46 via a trench or a ``periscope''. Need cost estimates.
%        }
%    }
% \end{figure}

\subsection{Schedule}
\label{sec:civil:schedule}

% \textcolor{red}{Sergio}

\begin{figure}[ht]
  \centering
  \includegraphics[width=1.0\linewidth]{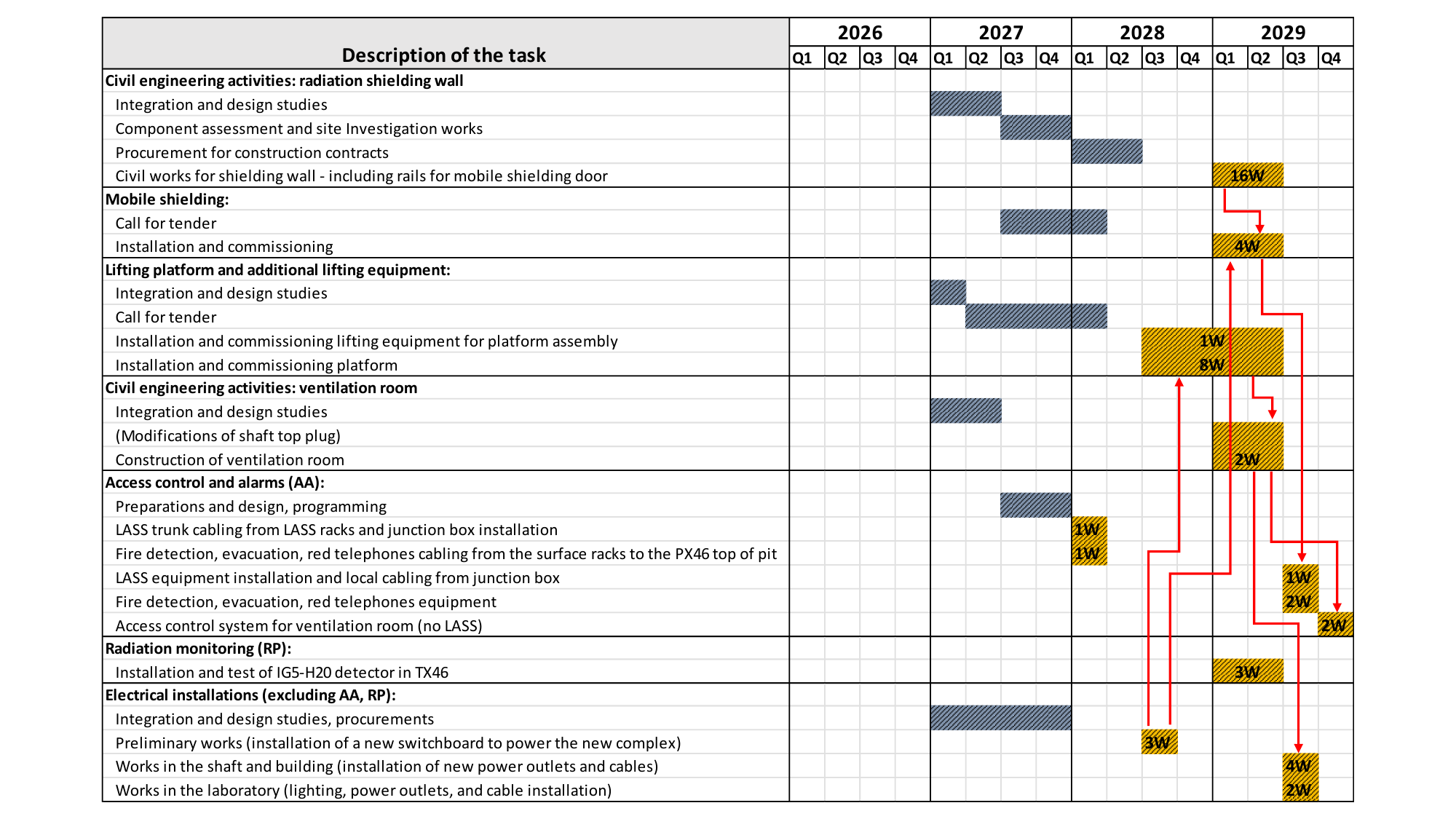}
  \caption{Gantt chart outlining the timeline of the major work items foreseen during LS3. Preparatory activities are indicated in grey and field work is indicated in orange. Arrows indicate dependencies. (Updated from the version in~\cite{Arduini:2025jhe}, as presented at~\cite{LS3C.2026.05.20}.) }
  \label{fig:Gantt}
\end{figure}

{A preliminary schedule for the realisation of the site preparation works described in this section has been developed, and is illustrated in Gantt form in Fig.~\ref{fig:Gantt}. This schedule assumes timely approval of the project and its budget, ideally by end of 2026, but is subject to several constraints. Civil engineering works must be completed before the installation of the movable shielding door and of the personnel access doors, and before instrumenting them and interfacing with the LASS. The crane in SX4 will require additional lifting equipment for the assembly of the lifting platform. The tendering phase of all this equipment has a significant lead time. Finally, it is not feasible to install fire detection in the shaft before the lifting platform is in place. These constraints, which are indicated by red arrows in Fig.~\ref{fig:Gantt}  are mostly logical and do not introduce any significant delays compared to a technically-limited schedule.}

{The schedule illustrated in Fig.~\ref{fig:Gantt} is matched to the present draft LS3 schedule. A potential window of opportunity is between January 2028 and June 2029 during which LHC-related activities in Point 4 are limited, allowing the AICE-related works to be realised~\cite{LS3S:LHC-PM-MS-0022}. This will need to be reviewed regularly in collaboration with the CERN group in charge of LS3 planning. At present, the relevant technical services have specific time frames for HL-LHC installation work, which result in a preferred window for the AICE works in the first half of 2029. An earlier implementation of all or part of the works relevant to this project might also be technically feasible, depending on the approval time and on the available funding and resources.}

{It should  be noted that the duration of many tasks in Fig.~\ref{fig:Gantt}, indicated by their duration in weeks, is substantially shorter than the time slot allocated to them. This means that the allocation of these tasks is flexible within their time slot, provided that the dependencies of works between different services are accommodated, as indicated by the arrows. Moreover, there is a substantial time gap between the end of the engineering studies of each major activity, and its execution. If the LS3 master planning would change, resulting in an earlier access to the UX45/TX46/PX46 complex, some of our major activities, and in particular the construction of the shielding wall, could be scheduled at a earlier date, which would help reduce potential risks arising from any delays in construction works. }

{We emphasise that the availability of the resources needed to perform the described activities lies beyond the scope of this document, and would be decided by CERN management upon acceptance of this Technical Proposal. Most of the activities can be outsourced to contractors. Preliminary consultations with the concerned services and their line management showed a strong degree of support for AICE and indicated that the required coordination effort by CERN staff is within acceptable margins.}

\subsection{Infrastructure cost estimate}
\label{sec:civil:cost}

%\textcolor{red}{Sergio, Oliver}

{The main infrastructure cost drivers, namely the construction of the shielding wall and the large movable shielding door, the elevator platform including modifications of the lid of the PX46 shaft with the ventilation room, and all the necessary modifications to the access control system, the fire detection and evacuation alarms, and the emergency communication devices, have been reviewed above. Cost estimates were made for all these systems in~\cite{Arduini:2025jhe}, updating the less detailed estimates given in the initial feasibility study~\cite{Arduini:2851946}: the figures are summarised in Table~\ref{tab:LS3_costs}, together with an indication of the prospective expenditure profile. These estimates can be considered Class 4~\cite{bib:DOEGuide}, with a lower range of uncertainty between -15\% and -30\% and an upper range between +20\% and +50\%. Also included in Table~\ref{tab:LS3_costs} are an allowance for preparatory works and studies during 2026 and a modest contingency allowance.
We note that the present cost estimate does not include infrastructure that is pertinent only to the atom interferometer equipment, such as a new chilled water production and distribution system and electrical and service cabling systems for both the atom interferometer and its laser laboratory, which are not needed for preparing the site during LS3.}

\begin{table}[htbp]
\centering
\caption{Class 4 budget estimates for civil engineering, access and alarm system modifications, lifting platform and other heavy handling and transport equipment to be installed during LS3, with indications of the prospective expenditure profile and a contingency allowance. Some expenses are advanced to 2028 at contract award stage, even if construction works is performed in 2029. }
\bigskip
\begin{tabular}{|l|c|}
\hline
{\bf Description}                                    & {\bf Cost (CHF)} \\
\hline
2026 & \\
Integration and installation studies & 40,000 \\
\hline
2027 & \\
Civil engineering design services                    & {130,000} \\
Handling engineering design services                    & {50,000} \\
\hline
2028 & \\
Civil engineering works (including door rails)  & {260,000} \\
Additional hoist in SX4                              & {130,000} \\
Lifting platform                                     & {350,000}  \\
Movable shielding door (structure and concrete blocks)                   & {100,000}  \\
\hline
2029 & \\
LHC Access Safety System (LASS)                      & {65,000} \\
Fire detection, alarms, emergency communications     & {30,000}  \\
RP monitor                                           & {20,000}  \\
Modifications of shaft lid (including ventilation room)                   & 75,000 \\

\hline
Experiment integration studies                       & 100,000 \\
\hline
Contingency & 150,000 \\
\hline
{\bf Grand total}                                    & {\bf 1,500,000} \\
\hline
\end{tabular}
\label{tab:LS3_costs}
\end{table}

The LS3 estimate covers only the civil-engineering enabling works listed above, which are provided by CERN as host-laboratory preparation. The LS3-phase electrical works specific to the experiment (replacement of the EBD1/4X switchboard, installation of EXD1/4X and the powering of loads commissioned during LS3, estimated at 0.20--0.25~MCHF) are not part of this civil budget and are shown as a separate host-preparation line in Table~\ref{tab:cost}, their funding attribution between CERN and the collaboration remaining to be discussed. The complete AICE capital-cost estimate, including post-LS3 experiment infrastructure and the scientific instrument, is summarised in Table~\ref{tab:cost}.

\subsection{Organization and supervision of the works during LS3}
\label{sec:civil:LS3}

\begin{figure}[h!]
\centering
	\includegraphics[width=.95\textwidth]{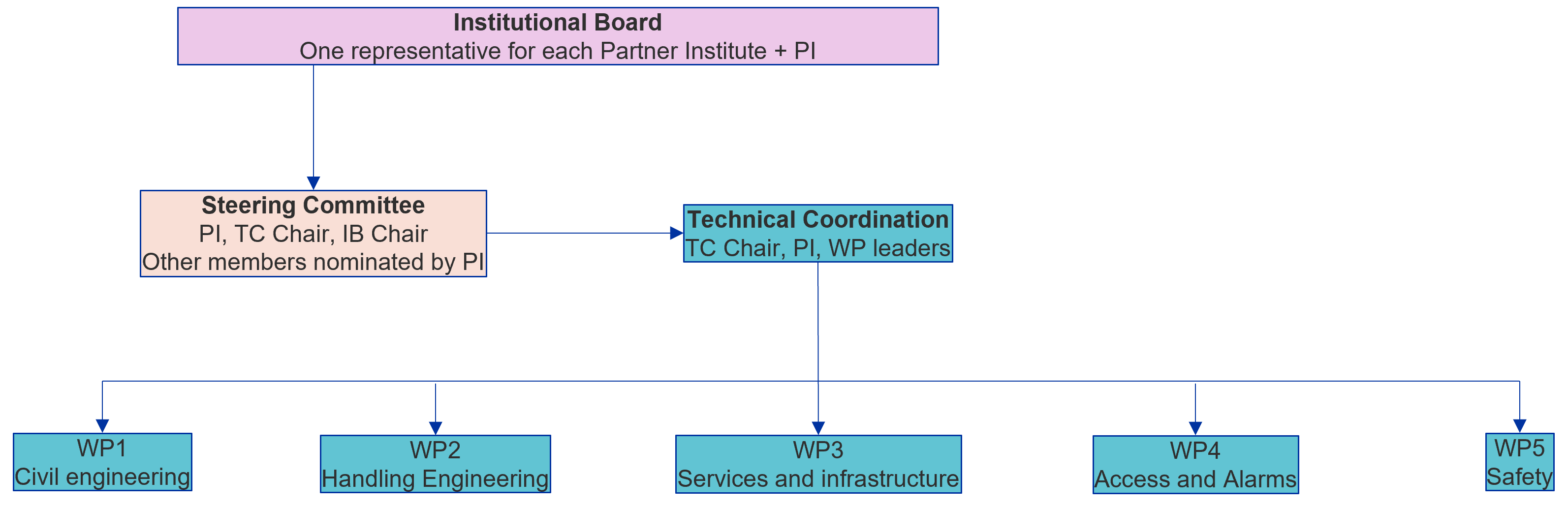}
    \caption{\label{fig:LS3_organigram} Organigram for the oversight of the works to be completed during LS3.}
\end{figure}

{Figure~\ref{fig:LS3_organigram} shows the organigram proposed for overseeing the works required during LS3. These are divided into 5 work packages that will each be led by a CERN staff member. The work packages will be coordinated by the Technical Coordinator, also a CERN staff member. The TC will be a member of the AICE Steering Committee, which will be chaired by the PI. The Collaboration will be led by the Institutional Board, which will be composed of representatives of each of the Partner Institutes and the members of the Steering Committee.}

%===========================================================================

% Float barrier: prevents floats from the preceding chapter appearing inside
% this Section's opening. Uses placeins if loaded, otherwise a page break.
\providecommand{\FloatBarrier}{\clearpage}
\FloatBarrier

\section{Noise and Systematics}
\label{sec:noise}

%\textcolor{red}{Samuel and Leonardo, with support from Michael and Daniel}

\subsection{Noise budget and methodology}
\label{sec:noise:budget}

%\textcolor{red}{To be checked by Samuel and Leonardo}

{The sensitivity ultimately achieved by AICE will depend on the control of technical and environmental disturbances across its measurement band. This Section presents the current noise and systematics assessment, bringing together measurements at PX46, subsystem studies and model estimates. It identifies the mechanisms by which the different noise sources couple to the instrument, summarises the evidence currently available, and defines the measurements and mitigation work required as the facility design and site characterisation progress. The noise budget is therefore an evolving part of the experimental programme rather than a completed description of the operational instrument.}

{The sensitivity projections in Section~\ref{sec:physics} principally describe the intrinsic reach of AICE at the atomic shot-noise level. They do not constitute a complete prediction of the sensitivity that will be achieved in operation, and, with the exception of the gravitational-wave strain figure in Section~\ref{sec:gw}, noise contributions are not superposed on the physics figures. Their status and potential impact are assessed collectively in this Section. Where sufficiently developed models and site inputs are available, the corresponding degradation is propagated into representative physics observables. Where a quantitative estimate is not yet available, the present evidence, the relevant design requirement and the required validation programme are stated explicitly.}

{Most of the technical and environmental contributions considered here are not presently identified as exceeding the atomic shot-noise target, although several still require validation with the final AICE systems. Gravity gradient noise (GGN) warrants a more detailed treatment because it acts directly on the atoms and may dominate at the low-frequency end of the measurement band. We therefore evaluate its possible impact on the gravitational-wave and scalar ULDM sensitivities and set out the associated programme of site measurements, modelling and mitigation. The unmitigated GGN estimates provide reference environmental-impact scenarios before these measures are applied, and they should not be interpreted as the expected operational noise floor. A consolidated overview of all contributions is given in Section~\ref{sec:noise:summary}.}

{Three distinct effects are referred to by similar names in the atom-interferometry literature, and they are treated separately throughout this proposal.}
\begin{itemize}
\item {\it Mechanical seismic vibration} is the physical motion of the ground, of the support structures and of the optical elements. It enters the measurement principally through the laser phase and through the motion of the retroreflecting optics. Because a common laser addresses all atom ensembles, this contribution is largely common mode in a gradiometer and is strongly suppressed in the differential observable (Section~\ref{sec:noise:detector}).
\item {\it Fluctuating gravity gradient noise}, also called Newtonian noise, is the time-dependent gravitational acceleration produced by moving mass density in the vicinity of the instrument. It acts directly on the atoms in free fall, and it is not removed by mechanical isolation. A spatially uniform component of the fluctuating field cancels in the differential observable, but the component associated with spatial variation of the field across the baseline survives, because the two ends of the gradiometer sample different local mass distributions. This is the subject of Section~\ref{sec:noise:GGN}.
\item {\it Static gravity gradients} are the time-independent or slowly varying gradients of the local gravitational field. They do not limit the broadband noise floor, and they are treated as biases and systematic effects in the equivalence-principle and fine-structure-constant measurements described in Sections~\ref{sec:ep} and~\ref{sec:alpha}.
\end{itemize}

%\subsection{Laser, electromagnetic and seismic noise}

\subsection{Detector technical noise}
\label{sec:noise:detector}

%\textcolor{red}{Eduardo, Richard, Samuel - does this look OK? Links to Secs. 4.3-4.5?}

\subsubsection{Laser phase noise}

As the laser phase is imprinted onto the atomic wavefunction with each light pulse, technical laser noise can result in noise on the interferometric phase. In the AICE facility, this effect will be largely suppressed by the use of a gradiometer configuration, where several atom ensembles are addressed by a common laser so that laser noise is common-mode and suppressed in the differential measurement. In particular, single-photon atom optics, as proposed in clock atom interferometers including AICE, generally allow for greater cancellation of laser phase noise that is well-suited to very-long-baseline experiments~\cite{Yu:2011GRG,Graham:2012sy,Dimopoulos2008}. Common-mode noise rejection in differential interferometers has been tested and demonstrated in table-top experiments and deployed systems~\cite{Stray2022,baynham_prototype_2026}, including in single-photon clock interferometry, demonstrating operation at the Standard Quantum Limit.

Several studies~\cite{MAGIS-100:2021etm,Graham:2012sy,Hogan2011AGIS-LEO} have evaluated the residual laser noise, which generally arises from velocity differences between the two atomic clouds, and indicate that it can be kept below the atom-shot-noise target for the parameter ranges considered, subject to validation at the AICE laser specification, see Tables~\ref{tab:laser_overview} and~\ref{tab:clock_laser_requirements}.
%\textcolor{red}{[link to laser specifications in Sec. 4.3]}.

In addition to imprinting phase noise, laser frequency noise can also reduce the transfer efficiency of atom-optics pulses, leading to a loss of contrast and through that a degradation of phase sensitivity. While these effects are the subject of ongoing international work~\cite{Chiarotti2022,Jiang2026}, notably how they scale up as a function of momentum separation, they can generally be reduced through the use of tailored pulse-shaping techniques~\cite{Dunning2014,Lellouch2023,LMTClockHogan2022}. Synergies with the TVLBAI programme, which has a dedicated work package in that area, will enable investigation and implementation of such techniques as AICE develops, offering alternatives that could allow the laser stability requirements of the experiment to be relaxed.

\subsubsection{Wavefront aberration}
%% NEW (30/7/26 from CAM)

Wavefront distortions in light-pulse atom interferometers have been studied and modelled extensively~\cite{Wicht2005,Fils2005,Louchet-Chauvet2011, Hogan_2011,Karcher2018,Abe_2021,Pagot2024,Pagot2026,Mouelle2026}.
Aberrations in the atom-optics laser beam, arising from finite optical quality and propagation through imperfect elements, couple to the interferometer through the spatially dependent phase of
the laser field sampled by the atoms at each atom--light interaction. The effect grows with the order of LMT and can result in both a differential interferometer phase bias, affecting mainly time-integrated
measurements such as fine-structure-constant measurements and equivalence-principle tests, and shot-to-shot phase noise arising from fluctuations in the atomic motion or optical phase profile, affecting all programs, including GW and ULDM detection.

Estimating and characterizing these effects is highly dependent on the detailed phase map of the optical field and will be addressed as part of the AICE instrument development and characterization program. In general, wavefront aberrations are conveniently described as a spatial Fourier (or Zernike) spectrum. The resulting interferometer response depends on both the spatial-frequency content of the aberration spectrum and the transfer function of the interferometer to each transverse mode \(k_t\).

\textit{Aberration-mode filtering.}
Similarly to AION and MAGIS, the AICE experiment benefits from long-baseline propagation of the interferometer beam, which provides intrinsic spatial filtering of high-spatial-frequency aberrations. During
free-space propagation, Fourier components with large \(k_t\) diffract out of the main beam. A subsequent magnifying telescope further rescales the surviving wavefront structure while preserving the filtering of
high-\(k_t\) components, so the effective aberration spectrum at the interferometer location is dominated by smooth, low-order distortions with characteristic transverse scales on the order of centimeters or larger.
%The characteristic diffraction length is \(z_{\mathrm{diff}}\sim\ell^2/\lambda\); propagation over \(100\text{--}140~\mathrm{m}\) at \(698~\mathrm{nm}\) gives a crossover scale \(\ell\sim1~\mathrm{cm}\).

The use of high-quality large-aperture optics ensures that no significant additional distortions are created during these filtering stages.
%\textcolor{red}{[required wavefront-flatness specification to be provided]}.
In the limit of ideal optical components, the remaining transverse phase structure is given by the intrinsic Gaussian wavefront curvature of the beam.

\textit{Interferometer response.}
The interferometer phase response to a given transverse aberration mode \(k_t\), with amplitude \(\epsilon(k_t)\), arises from the discrete sampling of the laser phase at the interferometer pulse locations. The response is largest for modes whose phase varies appreciably between the positions sampled over the LMT sequence while remaining spatially coherent across the atomic ensemble.

For large transverse wave numbers \(k_t\), the wavefront varies on length scales shorter than the atomic motion and cloud size, resulting in reduced phase coherence between pulse samples and an effective averaging out of the contribution over the ensemble, leading to suppression of their contributions to both the differential bias and shot-to-shot phase noise.

For small \(k_t\), the wavefront varies slowly over the transverse extent of the atomic trajectories, and these low-spatial-frequency modes are strongly suppressed in the mean differential bias by common-mode rejection. However, this rejection remains finite because the beam propagation modifies the aberrated field between the interferometer locations and differences between the atomic sources cause the two ensembles to sample it differently~\cite{Pagot2026}. The residual bias can be evaluated by combining independent measurements of the optical phase map and atomic phase-space distribution with numerical simulation of the complete pulse sequence, as demonstrated in precision measurements of the fine-structure constant~\cite{Bouchendira2011, parker2018, morel2020}.
Spatially resolved in-situ measurements provide a complementary method for characterizing and correcting the effective wavefront bias sampled by the atoms~\cite{Mouelle2026}.

These modes contribute to shot-to-shot phase noise through fluctuations in the initial transverse positions, velocities, sizes, and expansion rates of the clouds, as well as through beam-pointing fluctuations and temporal changes in the optical phase profile, all of which lead to imperfect common-mode cancellation. These effects are relevant to all AICE operating modes and are of particular importance for the time-resolved ULDM and gravitational-wave programs, where fluctuations within the target frequency band constitute a direct noise background~\cite{Hogan_2011,Abe_2021,Mouelle2026}.

The contribution of these modes to shot-to-shot noise can be evaluated in the Gaussian-curvature limit~\cite{Hogan_2011, Mouelle2026}. For the AICE baseline configuration, \(L=140~\mathrm{m}\), \(T_{\mathrm{grad}}=2.5~\mathrm{s}\), and \(n\simeq10^3\), it is expected that the contribution of these dominant modes can be reduced below the Baseline goal of \(\delta\phi\sim10^{-4}~\mathrm{rad}/\sqrt{\mathrm{Hz}}\), provided that the beam geometry and LMT efficiency are jointly optimized and the atom-source fluctuations are controlled at the \(\mathcal{O}(10~\mu\mathrm{m})\) level in position and the \(\mathcal{O}(10~\mu\mathrm{m\,s^{-1}})\) level in velocity~\cite{Mouelle2026}. These requirements are comparable to current state-of-the-art performance~\cite{PhysRevA.101.033606,Asenbaum_2020}.

%\textcolor{red}{[Input from Sections~4.3--4.5 and the final optical design is required. Comparison with the AICE \(\mathrm{rad}/\sqrt{\mathrm{Hz}}\) target also requires the final effective independent measurement rate.]}

Further suppression toward
\(10^{-5}~\mathrm{rad}/\sqrt{\mathrm{Hz}}\) will rely on precise stabilization or monitoring of the relevant atomic and optical parameters. Measurements of the final atom-cloud positions, together with prior
characterization of the source dynamics, can constrain the relevant initial positions and velocities. These parameters may be combined with the characterized wavefront and interferometer model to estimate correlated phase contributions. Spatially-resolved measurements provide an additional route by allowing correlations between the measured phase and transverse atomic motion to be learned and subtracted from the data, with simulations indicating that the residual contribution can be reduced to the atom-shot-noise level~\cite{Mouelle2026}.

Additional recoil-induced corrections arising from wavefront gradients have been analyzed in Ref.~\cite{Abe_2021}. For the smooth, low-spatial-frequency distortions expected to dominate after long-baseline propagation, the smallness of the recoil provides suppression of this contribution.

In summary, wavefront-induced phase shifts have been modelled sufficiently precisely and will be suppressed through long-baseline spatial filtering, gradiometric common-mode rejection, atom-source stabilization and characterization, and spatially-resolved detection. The required optical flatness and corresponding source and imaging specifications will be established through the complete AICE wavefront noise model.

\subsection{Environmental noise}
\label{sec:noise:environment}

{This Section presents the environmental measurements performed at the PX46 shaft and in the neighbouring SX4 building. These measurements characterise the site directly, and provide input to the gravity-gradient-noise model of Section~\ref{sec:noise:GGN:model}.}

\subsubsection{Electromagnetic noise}
\label{sec:electromagnetic}

\begin{figure}[h!]
    \centering
\includegraphics[width=0.9\textwidth]{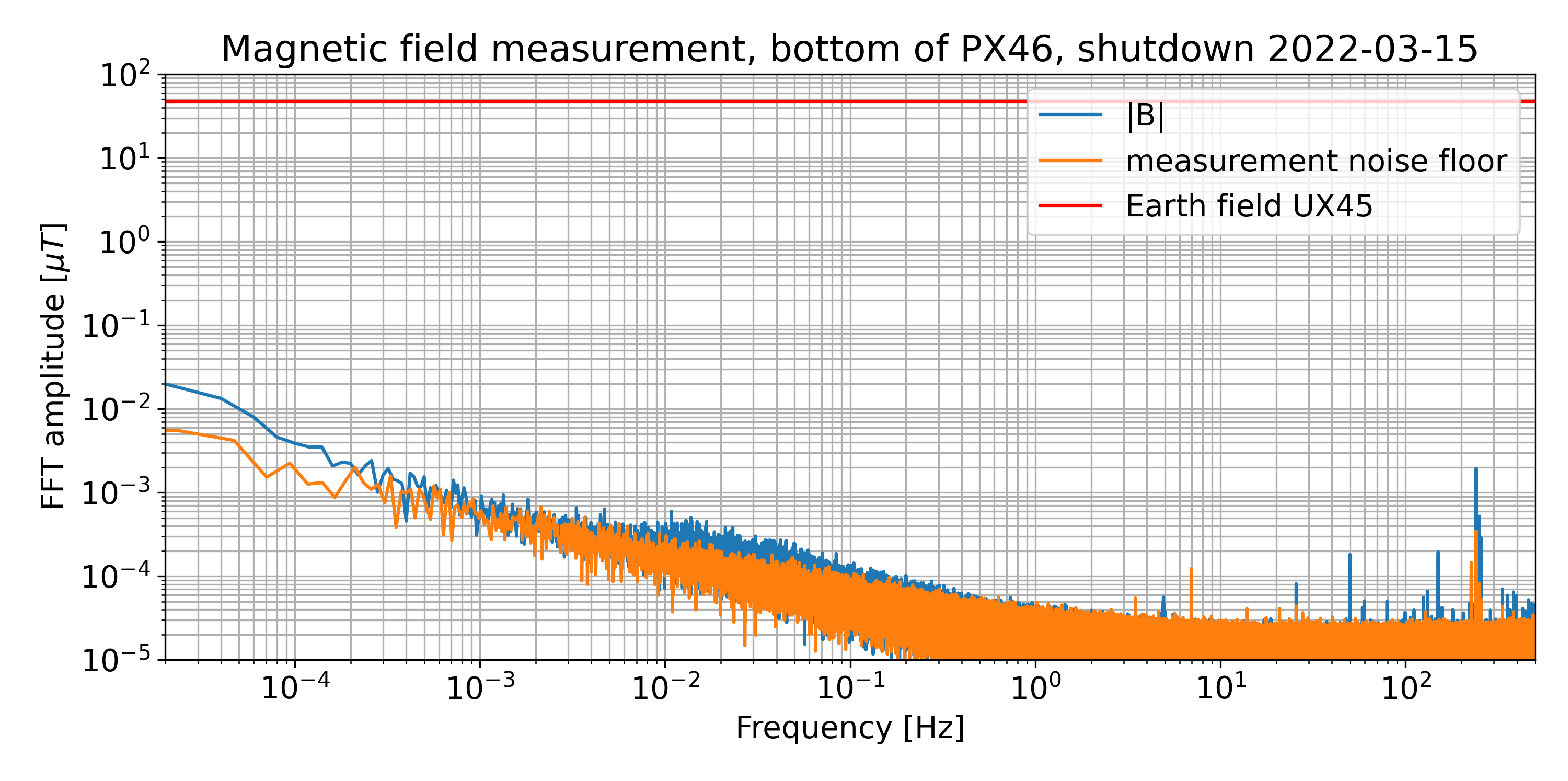}
    \caption{Fast Fourier transform of magnetic field measurements made at the base of the PX46 shaft during the LHC shutdown, showing for illustration the ambient magnetic field of the Earth as a red line~\cite{Arduini:2851946}.}
    \label{fig:MagField_cycle_freq}
\end{figure}

Figure~\ref{fig:MagField_cycle_freq} shows a Fast Fourier transform of magnetic field measurements made at the base of the PX46 shaft during an LHC shutdown. We see that the noise level is far below the magnitude of the Earth's ambient magnetic field, shown for reference as a red line. This is also the case at the top of the PX46 shaft. Measurements show a few discrete spectral lines at the mains frequency and its harmonics. However, no significant `broadband' magnetic field perturbation was identified within the dynamic range and the noise floor of our measurement set-up.  We conclude from these studies that the EM noise generated by the LHC systems is not expected to limit the operation of AICE in the PX46 shaft.

\subsubsection{The LHC machine cycle}
\label{sec:machine_cycle}

\begin{figure}[h!]
    \centering
    \includegraphics[width=0.8\textwidth]{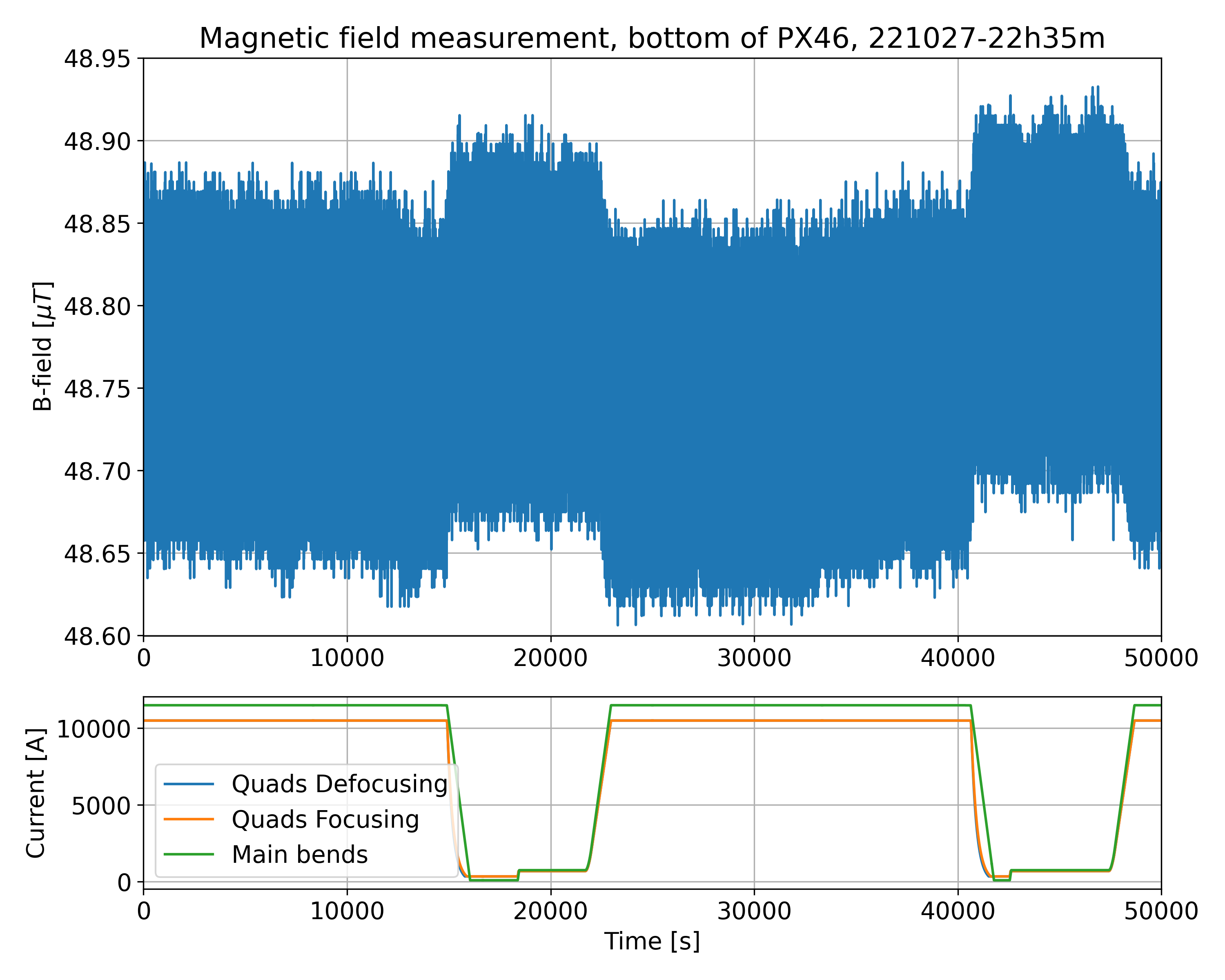}
    \caption{Low-frequency magnetic field measurements at the bottom of PX46 during the LHC machine cycle~\cite{Arduini:2851946}.}
    \label{fig:MagField_cycle_time}
\end{figure}

Figure~\ref{fig:MagField_cycle_time} shows the results of measurements at the bottom of PX46 of the magnetic field over two typical LHC machine cycles with  flat-top plateaux and ramps up and down, including also the pre-cycles. 

{The double-layer mu-metal magnetic shielding foreseen for AICE described in Section~\ref{sec:shielding} reduces the field produced by the LHC machine, and its variations, to an acceptable level. The magnet system foreseen for AICE can, moreover, be synchronized with the LHC machine cycle and used further to compensate the low-frequency magnetic field variations during the  cycle that are shown in Figure~\ref{fig:MagField_cycle_time}.
The several-minute timescale of the LHC field variations is compatible with compensation by the AICE magnet system.}

\subsubsection{Seismic ground motion}
\label{sec:seismic_noise}

\begin{figure}[!h]
\centering % \begin{center}/\end{center} takes some additional vertical space
\includegraphics[width=0.9\textwidth]{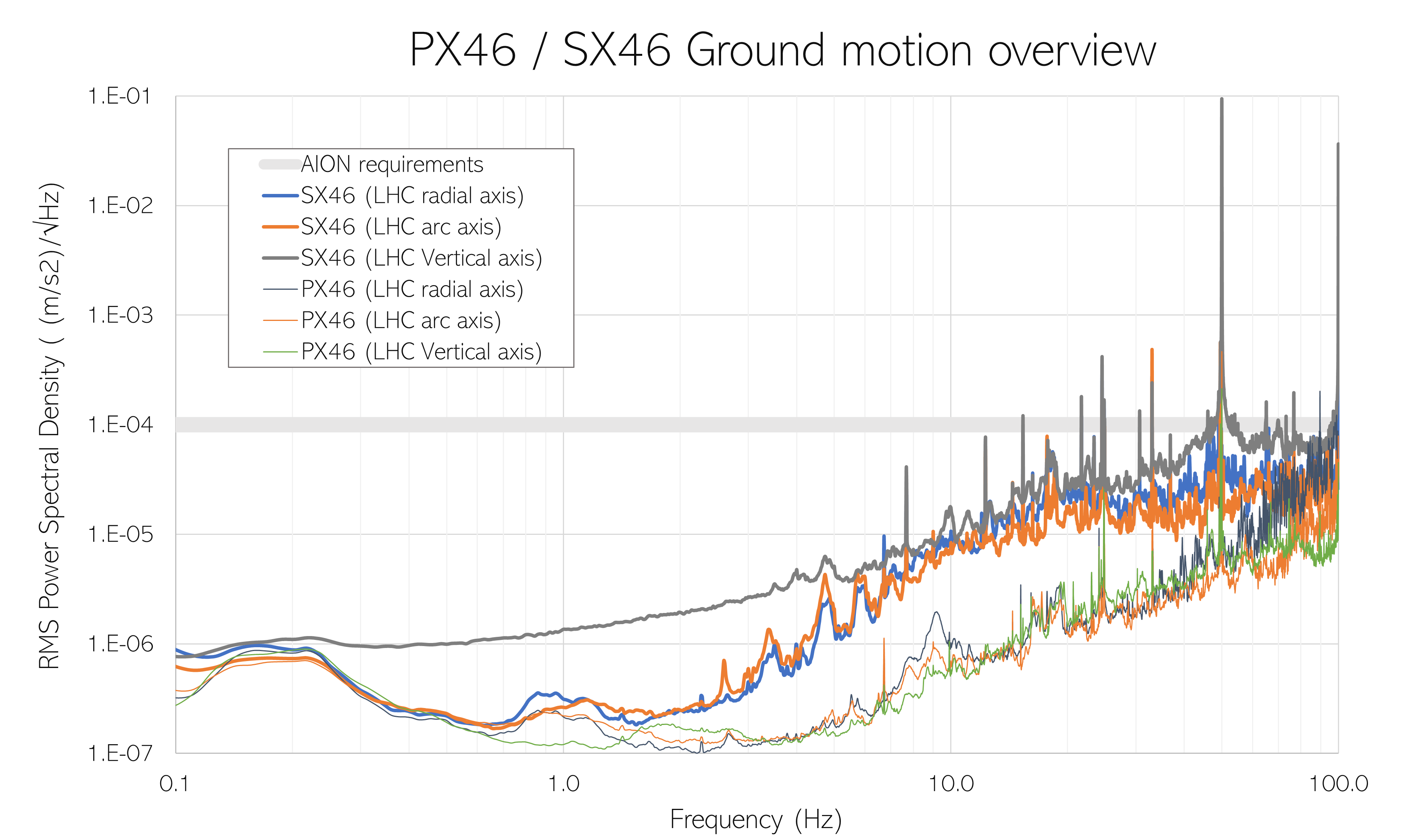}
\caption{\label{fig:RMS powerSpectral Density worst Case} The RMS power spectral density of ground motion along the 3 axes measured in SX4 and at the base of PX46. The data shown are for 64s time blocks, averaged over the worst of the 6 days of measurement.}
\end{figure}

{The RMS power spectral density of seismic activity in both SX4 and at the base of the PX46 shaft was measured along the three orthogonal axes for 6 days~\cite{Arduini:2851946}. The vertical ground motion in SX4 varied during this campaign, despite the fixed positions of the sensors. This variation is probably mainly due to the local seismicity. Similar behaviour was observed at the base of PX46, but with less variability. The averages of the spectra estimated from 64~s-long time intervals for the worst day for which measurements were made are shown in Figure~\ref{fig:RMS powerSpectral Density worst Case}. Below 1 Hz, and except for the SX4 vertical direction, all the curves are quite similar, as expected. The peak at 0.2 Hz corresponds to the well-known microseism peak that is related to the coupling of ocean waves with the solid Earth \cite{Bertoldi2024}.}

%\begin{figure}[!h]
%\centering % \begin{center}/\end{center} takes some additional vertical space
%\includegraphics[width=0.8\textwidth]{figures/vertical_dispalcement_surface_maxmin.pdf}
%\qquad
%\caption{\label{fig:noisemodels} The RMS spectral density of surface  vertical displacement measurements, compared with the New High and Low Noise Models (NHNM and NLNM)~\cite{peterson1993observations}. The shaded band corresponds to the difference between the minimum and maximum daily measurements. \textcolor{red}{We need a new version of this plot that shows the displacement PSDs at the top and and the bottom of the shaft.}}
%\end{figure}

{Above 1 Hz, the power spectral density varies by a factor of up to 10 because of cultural noise. These variations can originate either from local excitation sources, such as the cooling and ventilation (CV) systems, or from changes in the seismic amplitude with time. The importance of CV equipment was apparent in the measurements in SX4, leading to a strong peak in the spectrum at a frequency of 50 Hz.
Except for the 50 Hz disturbances coming from the CV equipment, which will require mitigation, all the ground motion measurements performed in SX4 and PX46 are below the AICE requirements for the stability of the laser system.} 
{The surface  vertical displacement measurements at frequencies $< 1$~Hz generally lie below the Peterson NHNM but above the NLNM~\cite{peterson1993observations}, albeit with fluctuations.} %The shaded regions in Fig.~\ref{fig:noisemodels} correspond to the difference between the measured minimum and maximum power spectral density levels reported in Fig.~\ref{fig:RMS powerSpectral Density worst Case}.}

{Measurements are available at more than one level, and the vertical spectra differ appreciably between them above the microseism peak, the higher location being the noisier of the two, which is consistent with its proximity to surface activity and to the cooling and ventilation plant. This non-uniformity along the shaft is an essential input to the gravity-gradient-noise estimate, and it is the reason why the model of Section~\ref{sec:noise:GGN:model} treats the two ends of the gradiometer separately rather than assuming a single representative spectrum. The spectra used as input to that model span approximately $0.1$--$100$~Hz, and the behaviour below this range, which matters for the longest baselines, is presently obtained by extrapolation.}

\subsubsection{Air flow and local environmental disturbances}
\label{sec:noise:airflow}

{Air movement in the shaft, driven principally by the cooling and ventilation system described in Section~\ref{sec:site:CV}, is a potential source of local disturbance through more than one mechanism. Turbulent flow produces mechanical and acoustic excitation of the vacuum vessel and of the support structure, which couples to the measurement in the same way as mechanical seismic vibration and is suppressed by the same common-mode mechanisms to the extent that the excitation is shared by the interferometers being differenced. It also produces local fluctuations of the air density inside and around the shaft, which act gravitationally on the atoms and therefore contribute to the near-field gravity gradient noise discussed in Section~\ref{sec:noise:GGN}. Finally, it imposes operational constraints on the temperature stability of the optical elements. Quantitative estimates for the PX46 flow conditions are under evaluation as part of the site characterisation programme, together with the measurements listed in Section~\ref{sec:noise:GGN:mitigation}, and the scope for reducing or modulating the flow in the shaft during data-taking periods will be assessed against the ventilation requirements set out in Section~\ref{sec:site:CV}.}

\subsection{Gravity Gradient Noise}
\label{sec:noise:GGN}

\subsubsection{Physical origin and scope}
\label{sec:noise:GGN:origin}

{Any time-dependent redistribution of mass in the vicinity of the instrument produces a fluctuating gravitational field, and hence a fluctuating acceleration of the freely falling atoms. This gravity gradient noise, also termed Newtonian noise, is a well-studied limitation for terrestrial gravitational-wave detectors~\cite{Harms:2019dqi}, and it has two properties that distinguish it from the other environmental noise sources considered above.}

{First, it acts on the atoms directly rather than through the apparatus. Mechanical isolation of the optical elements, however good, does not attenuate it, because the perturbation is a gravitational field rather than a physical contact force. Secondly, only part of it is common-mode. A perturbation that produces the same acceleration at both ends of the gradiometer cancels in the differential observable, in the same way as laser phase noise and platform vibration. The two ends of a long gradiometer are, however, separated by a distance comparable to the distance to the dominant nearby mass fluctuations, so they sample different local gravitational environments, and the differential part of the perturbation survives. The effect grows in relative importance towards low frequency, both because the ground motion driving it rises steeply and because the conversion from acceleration to displacement scales as $\omega^{-2}$. It is therefore the noise source most likely to limit AICE at the low-frequency end of its measurement band, which is precisely the region that carries the lowest-mass ULDM reach and the gravitational-wave reach.}

{Two physically distinct sources contribute. Seismic gravity gradient noise arises from density perturbations in the ground, driven by the seismic field measured in Section~\ref{sec:seismic_noise}, and is treated in Section~\ref{sec:noise:GGN:model}. Atmospheric gravity gradient noise arises from pressure, temperature and wind-driven density fluctuations in the air, and is estimated from meteorological data using a different model, described in Section~\ref{sec:atmospheric_noise}. The two are computed from different inputs and are reported separately in what follows.}

\subsubsection{From seismic ground motion to interferometer phase noise}
\label{sec:noise:GGN:model}

{The estimate proceeds in four steps, from the measured seismic acceleration spectra to an equivalent phase-noise amplitude spectral density at the gradiometer output. Because the endpoint is a phase noise, the same model serves every observable, and the propagation into gravitational-wave strain and into ULDM coupling reach is deferred to Section~\ref{sec:noise:GGN:impact}. The treatment follows Refs.~\cite{Badurina:2021lna,Badurina:2025lin}. The environmental-displacement model supplies a common input to the different observables. The interrogation time $T_{\rm grad}$, the LMT order $n$ and the shot-noise target $\delta\phi_{\rm noise}$ are taken from Table~\ref{tab:roadmap}, while the separation $\Delta z$ between the two ends of the gradiometer and the detailed response function are specified for each operating configuration in Section~\ref{sec:noise:GGN:impact}.}

{{\it Step 1: ground displacement.} The measured vertical acceleration spectra $S_a^{X}(f)$ at the upper and lower locations $X = {\rm upper},\,{\rm lower}$ are converted to ground displacement spectra,
\begin{equation}
    S_\xi^{X}(f) = \frac{S_a^{X}(f)}{\omega^{4}}\,, \qquad \omega = 2\pi f \,.
    \label{eq:ggn_disp}
\end{equation}
Each spectrum is described by a continuous broken power law fitted to the data over the measured range, approximately $0.1$--$100$~Hz, and extended to $10^{-3}$~Hz by power-law extrapolation of the lowest measured segment.}

{{\it Step 2: Newtonian coupling.} The ground displacement is converted to the effective displacement of the atoms produced by the associated fluctuating gravitational acceleration. For a perturbation of a half-space of local density $\rho_X$ this gives
\begin{equation}
    S_x^{X}(f) = \left( \frac{2\pi G \rho_X}{\omega^{2}} \right)^{2} S_\xi^{X}(f)
               = \left( \frac{2\pi G \rho_X}{\omega^{4}} \right)^{2} S_{a}^{X}(f) \,.
    \label{eq:ggn_newton}
\end{equation}
As modelling assumptions, local densities of $2400$ and $1800\,{\rm kg\,m^{-3}}$ are taken at the upper and lower locations respectively, consistent with the geology of the CERN site (Fig.~\ref{fig:CE:Stratigraphy}), and $2700\,{\rm kg\,m^{-3}}$ is taken for the kilometre-scale configuration, consistent with the geology around the ventilation shaft at the Sedrun site~\cite{Guinchard:2026cen}. The geometrical prefactor $2\pi$ follows from the simplified half-space treatment. Geological layering, the composition of the seismic wave field, the geometry of the sources and their depth all modify it, and the resulting order-unity modelling uncertainty is carried into the uncertainty band shown in Section~\ref{sec:noise:GGN:assumptions}.}

{{\it Step 3: interferometer response.} The effective displacement at each end is imprinted on the interferometer phase through the effective wavevector of the atom optics, $\keff = n k_0$. Writing $H_{\rm upper}(f)$ and $H_{\rm lower}(f)$ for the complex transfer functions from the effective displacement at each location to the gradiometer phase, the phase-noise power spectral density due to GGN is
\begin{equation}
    S_{\phi,{\rm GGN}}(f) = \left|H_{\rm upper}\right|^{2} S_x^{\rm upper}(f)
                            + \left|H_{\rm lower}\right|^{2} S_x^{\rm lower}(f)
                            - 2\,{\rm Re}\!\left[ H_{\rm upper} H_{\rm lower}^{*}\,
                              S_x^{\rm upper,\,lower}(f) \right] \,,
    \label{eq:ggn_cross}
\end{equation}
where $S_x^{\rm upper,\,lower}$ is the cross spectrum between the two locations. The transfer functions follow from the Doppler term of Eq.~(31) of Ref.~\cite{Badurina:2025lin}, which for a Mach-Zehnder sequence of interrogation time $T_{\rm grad}$ takes the form
\begin{equation}
    H_{X}(f) = 4\, \keff \, \sin^{2}\!\left(\pi f T_{\rm grad}\right) \, \mathcal{K}_{X}(f) \,,
    \qquad X = {\rm upper},\,{\rm lower} \,,
    \label{eq:ggn_phase}
\end{equation}
with $\mathcal{K}_{X}(f)$ collecting the large-momentum-transfer and light-propagation kernels at location $X$, normalised so that $\mathcal{K}_{X} \to 1$ in the limit in which the light travel time along the baseline is short compared with the signal period. The Shapiro and Einstein contributions identified in Ref.~\cite{Badurina:2025lin} are subleading here and are not included.}

{{\it Step 4: present benchmark.} The cross spectrum appearing in Eq.~(\ref{eq:ggn_cross}) has not been incorporated into the present estimate, which sets it to zero and neglects the difference between the endpoint kernels, using a common effective kernel $\mathcal{K}_{\rm upper} = \mathcal{K}_{\rm lower} \equiv \mathcal{K}$ that retains the large-momentum-transfer and light-propagation dependence used in the corresponding impact curves. The GGN phase-noise amplitude spectral density used for the curves shown below is then
\begin{equation}
    \delta\phi_{\rm GGN}(f) = 4\, \keff \, \sin^{2}\!\left(\pi f T_{\rm grad}\right)
                              \, \mathcal{K}(f) \, \sqrt{ S_x^{\rm upper}(f) + S_x^{\rm lower}(f) } \,,
    \label{eq:ggn_reduced}
\end{equation}
in ${\rm rad}/\sqrt{\rm Hz}$, so that the contributions of the two ends are added in quadrature. This is the reduced response used to produce the impact estimates of Section~\ref{sec:noise:GGN:impact}. The consequences of setting the cross term to zero, which is not a bound in either direction, are discussed in Section~\ref{sec:noise:GGN:assumptions}, and its measurement is a priority of the characterisation programme of Section~\ref{sec:noise:GGN:mitigation}.}

{The response $\sin^{2}(\pi f T_{\rm grad})$ vanishes at a discrete set of frequencies determined by the interrogation time. These zeros are physical for a fixed pulse sequence. For broadband reach plots a smooth response is therefore shown in place of the oscillatory one. Two conventions are used below, and they are not equivalent. An operating envelope takes, at each frequency, the response of the best interrogation configuration that can be operated there, on the grounds that the facility is not restricted to a single fixed sequence. A root-mean-square substitution instead replaces the oscillatory factor by its RMS value over a cycle, which is the appropriate choice when the signal frequency is not known in advance. The convention used is stated with each of the impact estimates in Section~\ref{sec:noise:GGN:impact}.}

{Combining Eq.~(\ref{eq:ggn_reduced}) in quadrature with the atom shot noise $\delta\phi_{\rm noise}$ of Table~\ref{tab:roadmap} gives the total phase-noise amplitude spectral density used for the impact estimates,
\begin{equation}
    \delta\phi_{\rm tot}(f) = \sqrt{\delta\phi_{\rm noise}^{2} + \delta\phi_{\rm GGN}^{2}(f)} \,.
    \label{eq:ggn_total}
\end{equation}
The order-of-magnitude modelling uncertainty is applied to the GGN term alone, before this combination, so that the resulting band closes onto the shot-noise curve in the frequency range where shot noise dominates.}

\subsubsection{Atmospheric gravity gradient noise}
\label{sec:atmospheric_noise}
\label{sec:noise:GGN:atmos}

\begin{figure}[h!]
    \centering
    \includegraphics[width=0.325\textwidth]{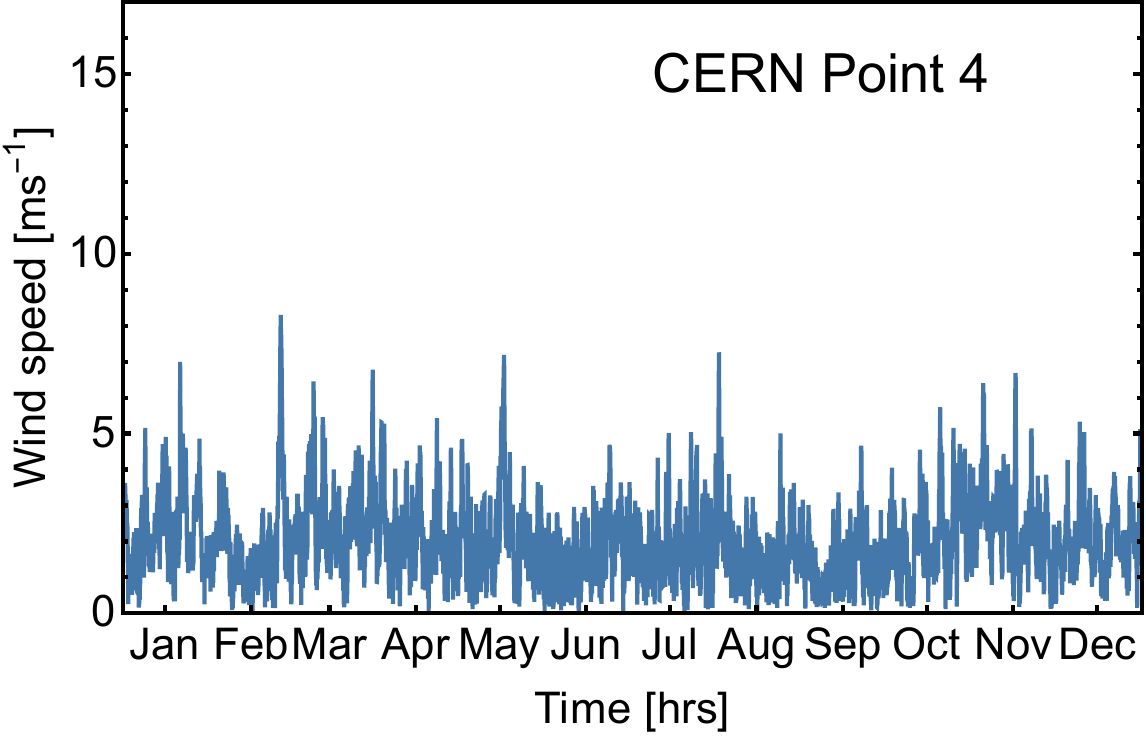}
    \includegraphics[width=0.325\textwidth]{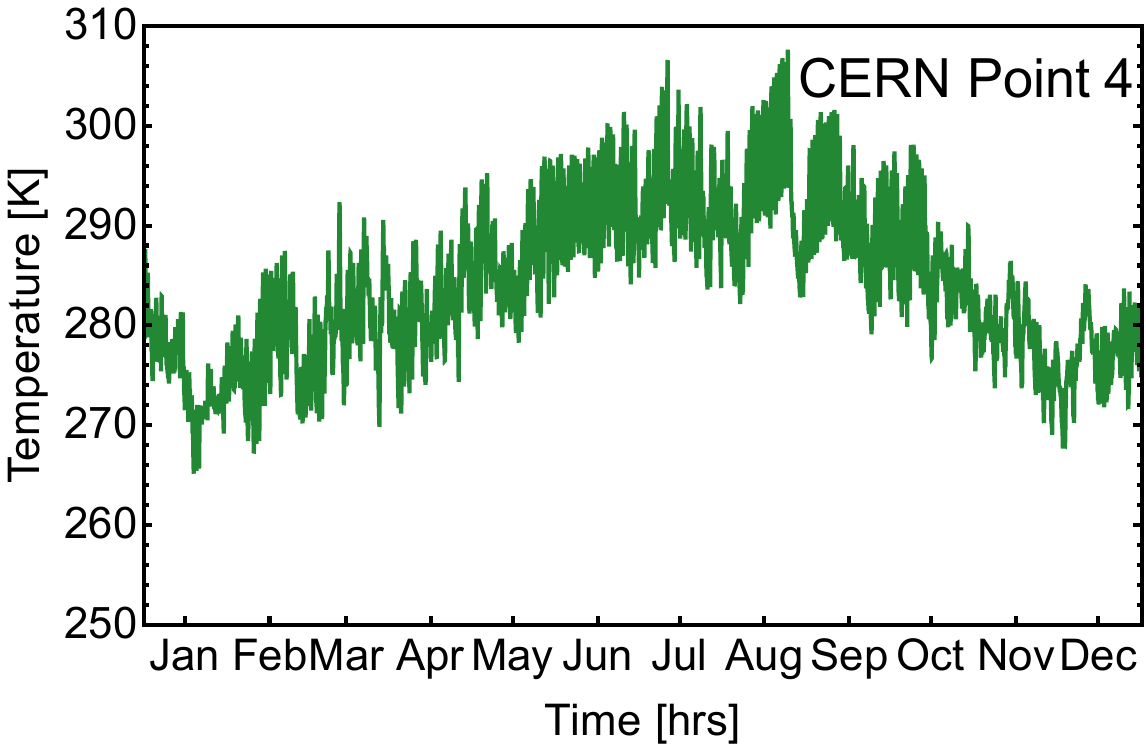}
    \includegraphics[width=0.325\textwidth]{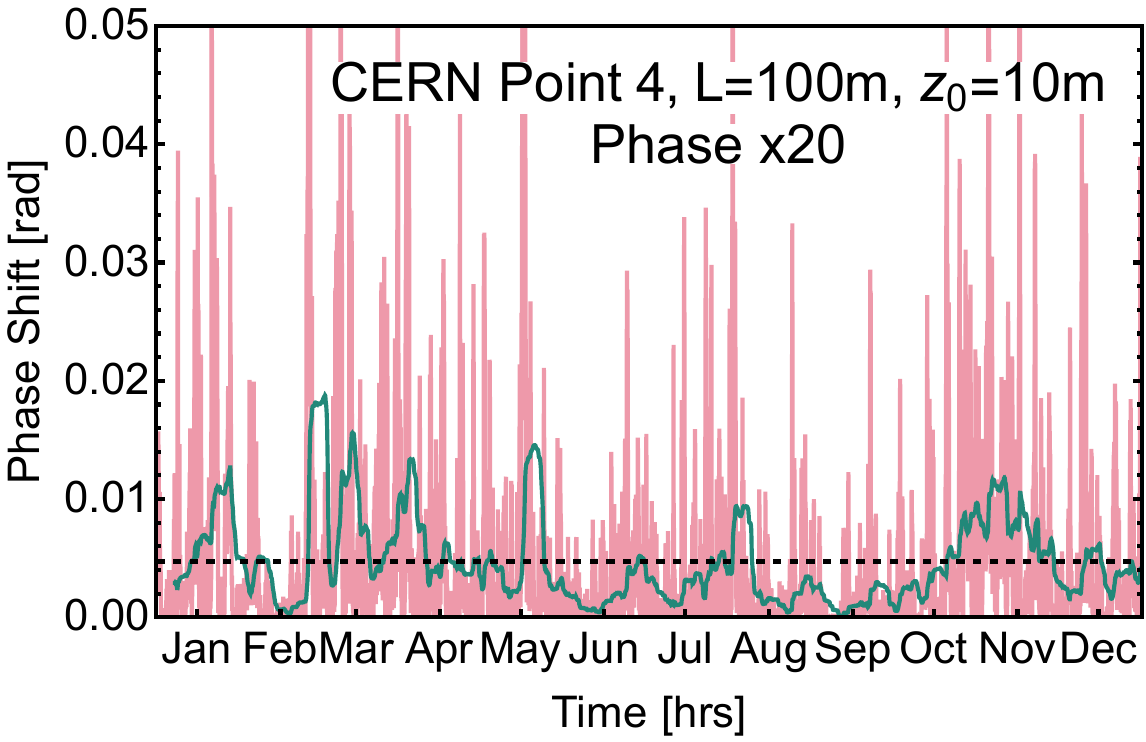}
    \caption{ERA5-derived time series of wind speed and ambient temperature at CERN Point~4, together with the corresponding modelled atmospheric-GGN phase shift~\cite{Carlton:2024lqy}. The atmospheric quantities are data from the ERA5 database~\cite{ERA5} rather than local measurements, and the phase series is modelled from them. In the phase-shift panel the trace is scaled for visibility by the factor indicated on the panel, the black dashed line shows the mean across the whole year and the green line a weekly moving average.}
    \label{fig:Atmospheric_GGN}
\end{figure}

{In addition to the seismic activity discussed above, another source of GGN is the atmosphere, through fluctuations in the pressure, wind speed and temperature. The noise due to these sources at the CERN site has been estimated in~\cite{Carlton:2024lqy} using meteorological data from the ERA5 database~\cite{ERA5}, as seen in Fig.~\ref{fig:Atmospheric_GGN}. (We note that atmospheric conditions at CERN are relatively favourable compared to other sites studied in~\cite{Carlton:2024lqy}.)}\vspace{3mm}

\begin{figure}[h!]
    \centering
\includegraphics[width=0.9\textwidth]{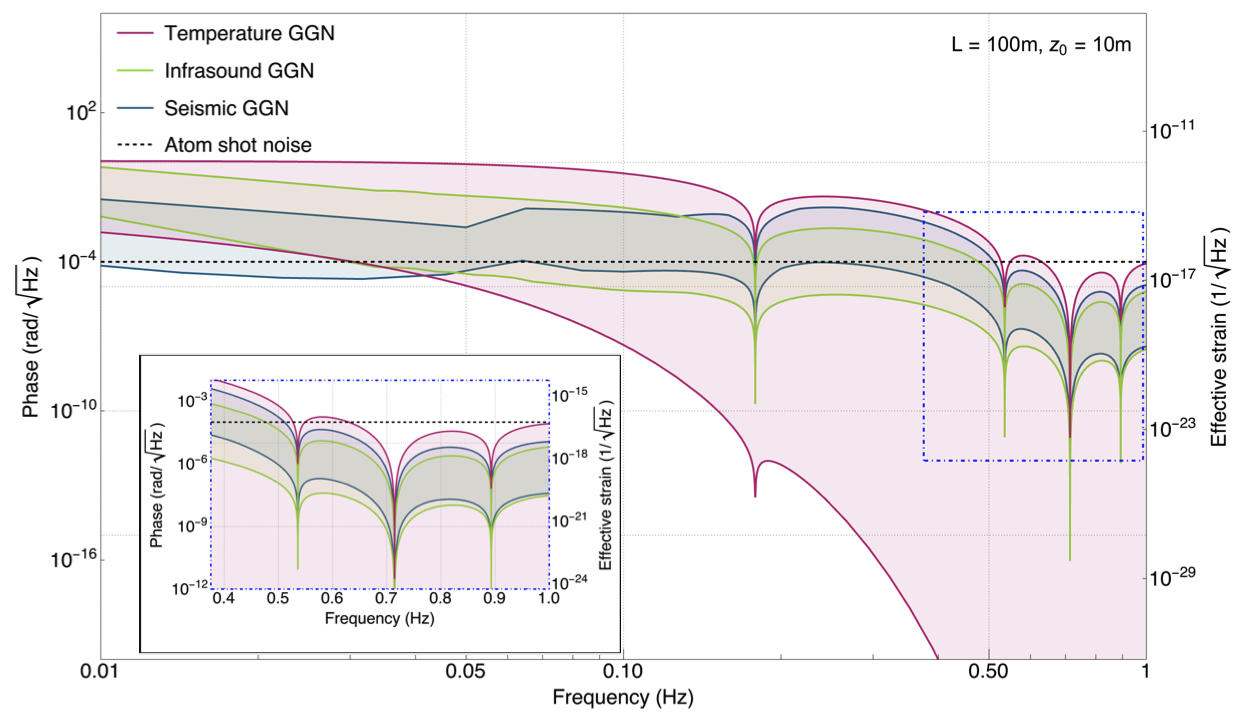}
    \caption{Global upper and lower bounds for seismic GGN, and estimates of the atmospheric pressure and temperature contributions, reproduced from~\cite{Carlton:2024lqy}. These are evaluated for a representative $100$~m baseline with the lower interferometer at a depth $z_0 = 10$~m, and are therefore not the site-specific PX46 estimate of Section~\ref{sec:noise:GGN:model}, which uses the AICE geometry of Table~\ref{tab:roadmap} and the measured upper and lower spectra. The black dotted line shows the expected level of atom shot noise in the AICE Baseline configuration of Table~\ref{tab:roadmap}. The blue dot-dashed box shows a reduced region plotted as a zoom.}
    \label{fig:Atmospheric_Phase_Noise}
\end{figure}

{The resulting atmospheric pressure and temperature contributions to the AICE GGN are shown in Fig.~\ref{fig:Atmospheric_Phase_Noise}~\cite{Carlton:2024lqy},\footnote{This reference also contains an indicative analysis of the atmospheric noise for a 1\ km atom interferometer.} where they are compared with global upper and lower bounds on the seismic noise and with the Baseline atom-shot-noise amplitude spectral density from Table~\ref{tab:roadmap}. {The atmospheric noise could be better modelled, and potentially mitigated, by real-time monitoring of atmospheric conditions.}}

{Two qualifications apply when reading Fig.~\ref{fig:Atmospheric_Phase_Noise} alongside the seismic estimates of this Section. The seismic band shown there represents global upper and lower bounds rather than the site-specific model of Section~\ref{sec:noise:GGN:model}, and the atmospheric and seismic contributions are obtained from different inputs using different models and have not been combined. The impact estimates of Section~\ref{sec:noise:GGN:impact} therefore contain the seismic contribution only. The atmospheric term is of comparable interest at the lowest frequencies and is reported separately.}

\subsubsection{Modelling assumptions and uncertainties}
\label{sec:noise:GGN:assumptions}

{The seismic-GGN estimate rests on a number of modelling choices, which are collected in Table~\ref{tab:ggn_assumptions} together with their effect and their interpretation.}

{Some of the choices are conservative, in the sense that a more complete treatment would be expected to reduce the estimated impact, the clearest case being that no subtraction or vetoing of any kind is applied. Others are uncertain in both directions, in particular the geometrical prefactor of the Newtonian coupling, the extrapolation of the fitted spectra below the measured range, and the correlation between the two ends of the gradiometer. For this reason the unmitigated curve should be understood as the largest seismic impact within the present model, and not as a rigorous upper limit on the effect of GGN.}

{The treatment of the correlation between the two ends is a major source of uncertainty in the result. Setting the cross term in Eq.~(\ref{eq:ggn_cross}) to zero is conservative for a positively correlated far-field disturbance, since the correlated part of such a perturbation cancels in the differential observable and quadrature addition discards that cancellation. It is not conservative in general. A density perturbation located between the two atom clouds attracts them towards each other, so its contributions to the two ends have opposite sign and add constructively in the difference, and for such near-field sources, of which the cooling and ventilation plant is a potential candidate, quadrature addition underestimates the differential noise. The sign and magnitude of the correlation are therefore a measurement question rather than a modelling choice, and they are addressed directly in Section~\ref{sec:noise:GGN:mitigation}.}

\begin{table}[h!]
\centering
\caption{{Modelling assumptions entering the seismic gravity-gradient-noise estimate of Section~\ref{sec:noise:GGN:model}, with their effect on the estimate and their interpretation.}}
\label{tab:ggn_assumptions}
\vspace{2mm}
\small
\begin{tabularx}{\textwidth}{p{0.26\textwidth} p{0.29\textwidth} X}
\hline
Assumption & Effect & Interpretation \\
\hline
Upper and lower spectra treated separately
  & Captures the measured non-uniformity of the seismic field along the shaft
  & More realistic than a uniform-shaft model; supported by the PX46 measurements \\
Cross spectrum between the two ends not incorporated, and set to zero
  & Contributions of the two ends added in quadrature
  & Conservative for positively correlated far-field motion; potentially optimistic for near-field perturbations between the atom clouds; to be measured \\
Half-space Newtonian coupling, prefactor $2\pi$
  & Converts ground displacement into gravitational acceleration of the atoms
  & Order-unity uncertainty in either direction from geological layering, wave-field composition, source geometry and depth \\
No subtraction or vetoing applied
  & Gives the unmitigated impact
  & Conservative relative to a successful mitigation programme \\
Power-law extrapolation below the measured range
  & Extends the input spectra from $0.1$~Hz down to $10^{-3}$~Hz
  & Model dependent in either direction; affects the lowest-frequency reach only \\
Smooth response substituted for the response zeros
  & Removes the narrow zeros of $\sin^{2}(\pi f T_{\rm grad})$ from the plotted curves
  & Two conventions are used, an operating envelope over interrogation configurations for the GW estimate and an RMS substitution for the scalar estimate, and neither is the literal fixed-$T$ response \\
Assumed densities of the surrounding material
  & Sets the amplitude of the Newtonian coupling, Eq.~(\ref{eq:ggn_newton})
  & Model assumptions representative of the formations traversed; to be refined by measurements at the geological interface \\
Amplitude varied by $\sqrt{10}$ either side of the central estimate
  & Produces the shaded band, spanning a factor of ten between its edges
  & Overall modelling uncertainty of the present estimate; applied to the GGN term before combination with shot noise \\
\hline
\end{tabularx}
\end{table}

\subsubsection{Impact on the AICE sensitivity}
\label{sec:noise:GGN:impact}

{The phase-noise spectrum of Eq.~(\ref{eq:ggn_total}) is propagated into each observable using the response function appropriate to that measurement. The signal response functions themselves are derived in Section~\ref{sec:physics} and are not repeated here. Two representative cases are presented, the gravitational-wave strain and the scalar ULDM couplings, which between them cover the two frequency regions of interest and the two classes of interferometer response used by AICE.}

{\it Gravitational-wave strain.} {The effect on the GW strain sensitivity is shown by the dashed lines and shaded bands of Fig.~\ref{fig:GWs} in Section~\ref{sec:gw}, alongside the shot-noise-limited curves. That estimate is evaluated for the AICE Stretch configuration operated as a single two-source gradiometer, with one source near the top of the shaft and one near the bottom. Because the atoms are launched upwards, the upper source sits approximately $30.7$~m below the top of the shaft, so that the gradiometer baseline is $\Delta z \approx 109$~m rather than the full shaft length. The noise-equivalent characteristic strain is
\begin{equation}
	h_n(f)  =  \sqrt{f \left[ \frac{\delta\phi_{\rm noise}^2}{R(f)^2} + \frac{S_x^{\rm upper}(f) + S_x^{\rm lower}(f)}{\Delta z^2 \, {\rm sinc}^2(\pi f n \tau_{\Delta z})} \right] } \, ,
    \label{eq:ggn_strain}
\end{equation}
where the two terms are the atom shot noise and the endpoint GGN displacement spectra of Eq.~(\ref{eq:ggn_newton}), each referred to strain, and
\begin{equation}
    R(f) = 2 \keff \Delta z \sin(\pi f T) \, \mathrm{sinc}\!\left( \pi f n \tau_{\Delta z} \right) \sin\!\left( \pi f T - \pi f (n-1) \tau_{\Delta z} \right)
    \label{eq:ggn_response}
\end{equation}
is the dimensionless strain response function evaluated at the source orientation that maximises it, with $\keff = n k_0$ and $\tau_{\Delta z} = \Delta z / c$ the light travel time along the gradiometer baseline. For this figure the envelope prescription of Section~\ref{sec:noise:GGN:model} is implemented by scanning $T$ over values below the maximum $T_{\rm grad}$ and taking, at each frequency, $\max_{T \le T_{\rm grad}} |R(f,T)|$, which is the envelope of the resulting family of curves.}

{The estimated seismic GGN lies below the atom shot noise above approximately $1$~Hz and rises above it at lower frequencies, so the degradation affects the low-frequency end of the mid-band. Below approximately $0.1$~Hz the estimate relies on the extrapolation of the input spectra rather than on measured data.}

\begin{figure}[h!]
    \centering
\includegraphics[width=0.48\linewidth]{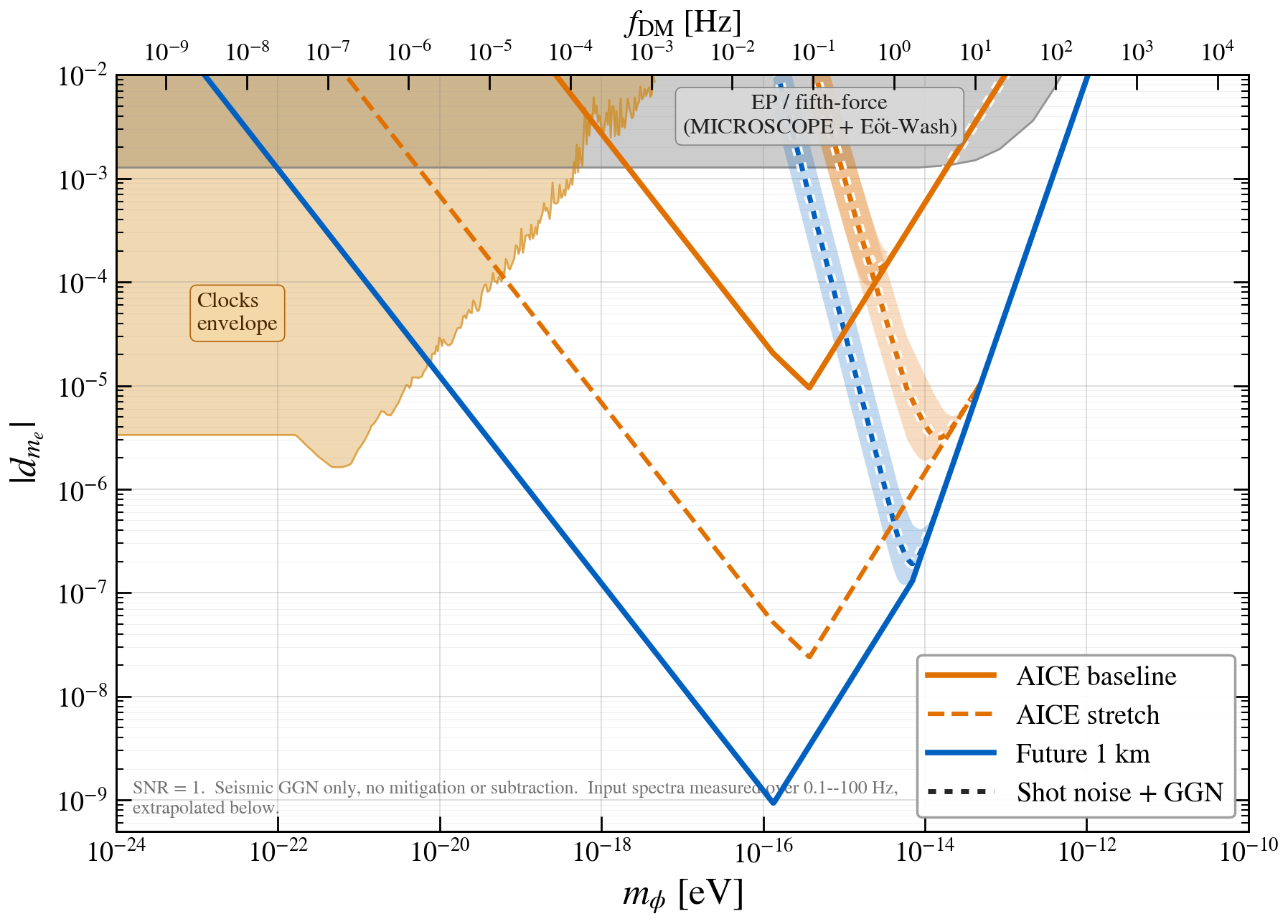}
    \hfill
\includegraphics[width=0.48\linewidth]{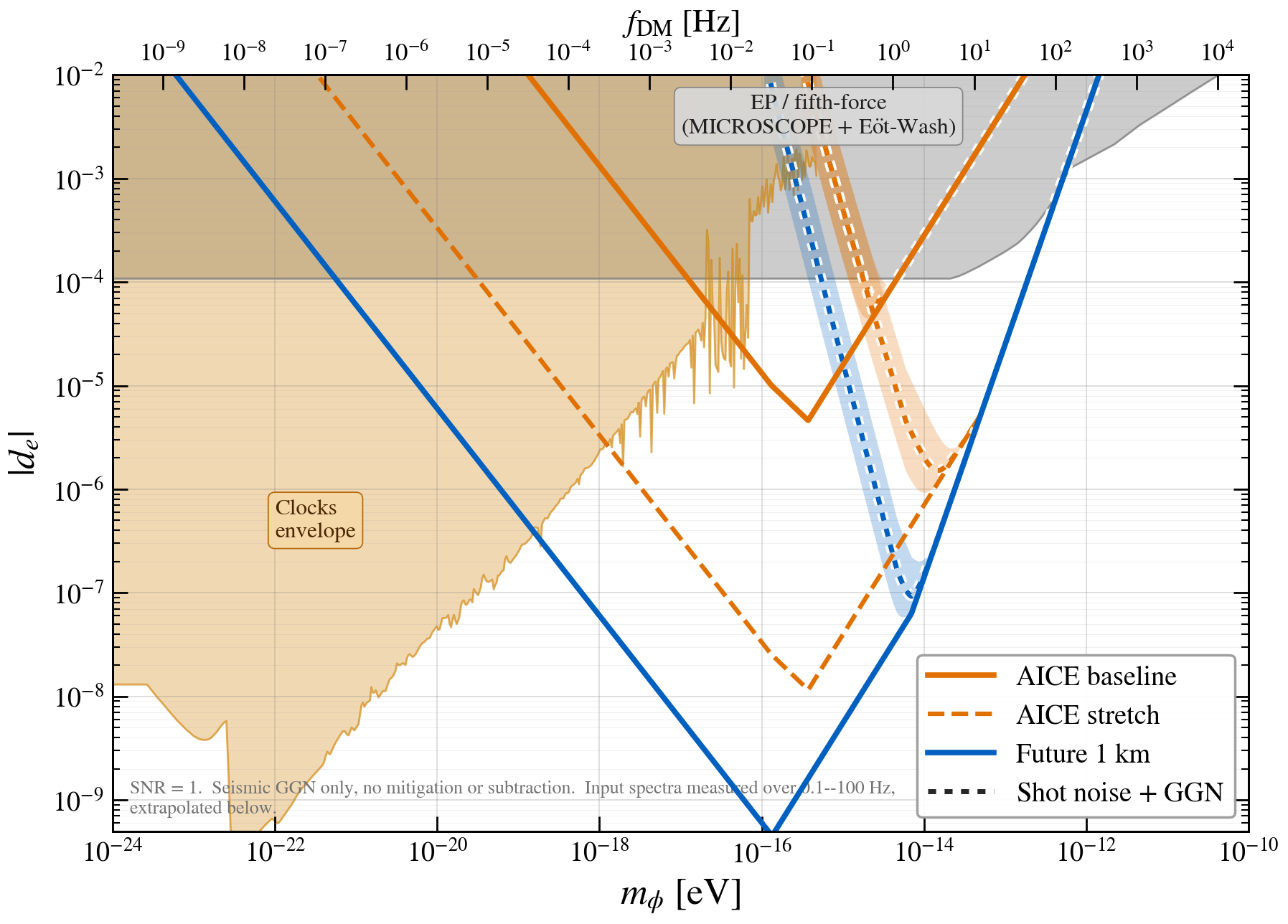}
    \caption{{Impact of the estimated unmitigated seismic gravity gradient noise on the AICE sensitivity to the scalar ULDM electron-mass coupling $|d_{m_e}|$ (left panel) and photon coupling $|d_e|$ (right panel), for the AICE Baseline and Stretch configurations and a possible future 1\ km atom interferometer. The shot-noise-limited curves retain the configuration styling of Fig.~\ref{fig:ULDM_scalar} and are identical to the projections shown in Section~\ref{sec:uldm_scalar}. The corresponding curves including the estimated unmitigated seismic GGN of Section~\ref{sec:noise:GGN:model}, combined in quadrature with that shot noise following Eq.~(\ref{eq:ggn_total}), are dotted. The shaded band of the same colour shows the order-of-magnitude modelling uncertainty on the GGN estimate, applied to the GGN term before combination so that it closes onto the shot-noise curve where shot noise dominates. The Baseline and Stretch configurations share a colour, following Fig.~\ref{fig:ULDM_scalar}. Their dotted curves coincide at low mass, where the GGN-limited reach is the same for the two, and where they separate at higher mass the upper curve is Baseline and the lower is Stretch. No mitigation or subtraction is included, and the atmospheric contribution of Section~\ref{sec:atmospheric_noise} is not included. Both panels are produced by the same code as the shot-noise-limited projections of Section~\ref{sec:uldm_scalar}, from the parameters of Table~\ref{tab:roadmap}.}}
    \label{fig:ULDM_GGN}
\end{figure}

{\it Scalar ULDM couplings.} {The effect on the reach in the scalar ULDM couplings is shown in Fig.~\ref{fig:ULDM_GGN}, for the electron-mass coupling $|d_{m_e}|$ in the left panel and the photon coupling $|d_e|$ in the right panel. The curves are evaluated with the parameters of Table~\ref{tab:roadmap} and the model of Section~\ref{sec:noise:GGN:model}, at fixed $T_{\rm grad}$, with the oscillatory kernels of Eq.~(\ref{eq:ggn_reduced}) replaced by a smooth RMS envelope through the substitution $|\sin x| \to \min\!\left(|x|,\,1/\sqrt{2}\right)$. This preserves the low-frequency scaling while replacing the oscillatory high-frequency response, including its nulls, by the RMS envelope. The corresponding estimate for the kilometre-scale configuration uses the seismic measurements at the Sedrun access shaft to the Gotthard Base Tunnel reported in~\cite{Guinchard:2026cen}. The estimated seismic GGN rises above the atom shot noise below approximately $0.8$, $2.5$ and $5$~Hz for the Baseline, Stretch and kilometre-scale configurations respectively, so the degradation is concentrated at the low-mass end of the accessible range. At the optimum it costs roughly a factor of $10$ in reach for Baseline and of $10^{2}$ for the Stretch and kilometre-scale configurations, and it moves the mass of best reach upwards by about a factor of seven for Baseline and by more than an order of magnitude for the other two. The Stretch degradation is the larger of the two 140~m cases even though its GGN-limited reach is the same as that of Baseline, because its shot-noise target is a factor of $400$ lower and the crossover therefore occurs at a higher frequency. This is directly visible in Fig.~\ref{fig:ULDM_GGN}, where the two dotted curves of the 140~m configurations converge at low mass. In the unmitigated scenario the Baseline reach is degraded to approximately the level of the existing equivalence-principle bound over most of the affected mass range, so that its GGN curve in Fig.~\ref{fig:ULDM_GGN} lies largely within the already-excluded region. On this estimate the Baseline configuration would add little scalar reach in the GGN-dominated band without mitigation. This is the principal motivation for the characterisation programme of Section~\ref{sec:noise:GGN:mitigation}.}

{The GGN-limited reach is, to leading order, independent of the LMT order. In the low-frequency region where seismic GGN dominates, the dependence on the effective wavevector is common to the scalar signal and to the GGN-induced phase, both of which are imprinted on the atoms through the same atom-optics pulses and therefore carry the same factor $\keff = n k_0$, which cancels in the ratio that determines the coupling reach. Increasing the LMT order improves the shot-noise-limited sensitivity, as reflected in the ladder of Table~\ref{tab:roadmap}, but it does not produce the corresponding improvement once GGN dominates. Residual configuration dependence remains through the finite-frequency response kernels of Eq.~(\ref{eq:ggn_phase}). Realising the full Stretch and kilometre-scale improvement in the GGN-dominated part of the band therefore requires corresponding progress in characterisation and mitigation.}

{The scalar channel is presented as the representative quantitative example, being the primary gradiometer mode of the facility and the mode for which the GGN treatment is most fully developed. The same environmental inputs and the same phase-noise model can be propagated through the other operating modes, but the response to GGN is configuration-dependent and requires mode-specific treatment. In the dual-species vector and equivalence-principle mode, a gravitational acceleration common to both species cancels at leading order in the differential observable, with the residual sensitivity set by scale-factor matching and by the spatial separation of the species. In the spin-dependent axion mode the signal does not scale with the LMT order, whereas the motion-induced phase noise of Eq.~(\ref{eq:ggn_phase}) does, so the cancellation described above does not apply and the GGN scaling with $n$ differs from the scalar case. These modes will be evaluated with the same framework as the site model is refined.}

\subsubsection{Characterisation and mitigation}
\label{sec:noise:GGN:mitigation}

{The characterisation and mitigation of gravity gradient noise is an active area of research in both atom interferometry and gravitational-wave detection, and the estimates above describe a scenario in which none of the available measures is applied. The assumptions of Table~\ref{tab:ggn_assumptions} identify the measurements needed to replace them with site-specific inputs. AICE foresees the following programme, whose early elements are compatible with the site preparation works described in Section~\ref{sec:civil}.}

\begin{itemize}
\item {\it Synchronised measurements at the two ends of the shaft.} Time-synchronised seismometers at the top and bottom of PX46 allow the cross spectrum of Eq.~(\ref{eq:ggn_cross}) to be measured directly as a function of frequency, replacing the zero-correlation benchmark by a measured coherence. This is a high-priority measurement.
\item {\it Measurements at intermediate levels along the shaft.} Sensors at the planned positions of the atom sources characterise the seismic field at the relevant heights rather than only at the extremes of the shaft, and measurements near the moraine/molasse interface (see Fig.~\ref{fig:CE:Stratigraphy}) constrain the densities and the coupling model of Eq.~(\ref{eq:ggn_newton}).
\item {\it Near-field monitoring inside the shaft.} Local sensors near the cooling and ventilation plant and other identified equipment address the near-field perturbations for which quadrature addition is not conservative, and support the identification and, where possible, the removal or relocation of anthropogenic sources.
\item {\it Measurements in neighbouring shafts and galleries.} The PZ45 and PM45 shafts, and the caverns and horizontal galleries at Point~4 (see Fig.~\ref{fig:iso-view-PX46}), make possible additional sampling of the local seismic field and a lever arm on its spatial correlation length.
\item {\it Atmospheric monitoring.} Real-time monitoring of pressure, temperature and wind at the site constrains the atmospheric contribution of Section~\ref{sec:atmospheric_noise} and would support its modelling and potential subtraction.
\item {\it Witness-sensor subtraction.} With a measured correlation between the seismic field and the interferometer phase, the GGN contribution can be estimated from an array of witness sensors and subtracted from the data. This is a standard technique in gravitational-wave detectors~\cite{Harms:2019dqi} and is a mitigation candidate rather than a demonstrated capability at the required level for AICE.
\item {\it Multi-source and multi-gradiometer combinations.} A facility with several atom sources along the baseline provides more than one gradiometer pair, and the different pairs respond differently to the signal and to a local gravitational perturbation. Combinations of pairs can therefore be constructed that suppress the GGN contribution relative to the signal, as studied for ULDM searches in~\cite{Badurina:2022ngn}. The multi-source architecture described in Section~\ref{sec:concept:facility} retains this option.
\item {\it Resonant and multi-loop operation.} Operating in a resonant or multi-loop mode concentrates the response in a narrow frequency band, which changes the weighting of the noise spectrum and allows the noisiest parts of the band to be avoided. This is a complementary frequency-selection strategy rather than a removal of GGN, and its interaction with the mitigation measures above will be studied as part of the operating-mode optimisation of Section~\ref{sec:concept:mode:readout}.
\end{itemize}

{Validation of these measures with commissioning data is included in the commissioning programme of Section~\ref{sec:ops}, and the resulting characterisation of the site is one of the deliverables that a facility at PX46 would provide for future longer-baseline detectors.}

\subsection{Summary and experimental implications}
\label{sec:noise:summary}

{The present assessment has not identified any technical or environmental contribution that has been shown to impose an irreducible noise floor above the atomic shot-noise target. This does not mean that all contributions have already been demonstrated to lie below that target. Several technical noise sources are expected to be controllable through the gradiometer configuration, the instrument design and established engineering measures (Table~\ref{tab:noise_summary}), but their residual levels must be validated with the final AICE systems. Airflow and turbulence in the shaft remain to be quantified, while the present unmitigated model predicts that seismic gravity gradient noise exceeds atom shot noise at the low-frequency end of the measurement band.}

{For each contribution identified in this Section, the AICE programme includes a corresponding route to characterisation, control or mitigation. The most demanding element of this programme is likely to be the mitigation of gravity gradient noise at low frequencies. Several complementary pathways are available, including direct measurements of seismic coherence, monitoring and control of local sources, witness-sensor subtraction, combinations of multiple gradiometers and optimisation of the interrogation sequence. These methods provide a credible route towards reducing the GGN contribution~\cite{Harms:2019dqi,Badurina:2022ngn}, but they must be developed and demonstrated at the scale and sensitivity required for AICE. Establishing their performance, and determining the frequency range over which operation can approach the atom-shot-noise reference, are therefore central objectives of the AICE characterisation and commissioning programme.}

{The present seismic estimate quantifies the scale of this challenge. Above approximately $5$~Hz the shot-noise-limited projections of Section~\ref{sec:physics} are not expected to be materially affected for the configurations considered, subject to the validations noted above. At lower frequencies the unmitigated estimate exceeds the atom shot noise below approximately $0.8$, $2.5$ and $5$~Hz for the Baseline, Stretch and kilometre-scale configurations, and the two reference scenarios of Section~\ref{sec:noise:GGN:impact} differ at the sensitivity optimum by roughly one order of magnitude in coupling reach for Baseline and by roughly two for the Stretch and kilometre-scale configurations. Increased momentum transfer does not by itself close this gap, and the estimate rests on site inputs and correlation assumptions that the measurements of Section~\ref{sec:noise:GGN:mitigation} are designed to replace.}

{Table~\ref{tab:noise_summary} documents this assessment contribution by contribution, stating for each the conclusion relative to the atom-shot-noise target, the basis on which it rests, its treatment in the sensitivity projections, and the action required.}

\begin{table}[p]
\centering
\caption{{Summary of the assessment of this Section. For each contribution the Table gives the present assessment relative to the atom-shot-noise target, the basis on which that assessment rests, the treatment adopted in the sensitivity projections of Section~\ref{sec:physics}, and the action required.}}
\label{tab:noise_summary}
\vspace{2mm}
\scriptsize
\renewcommand{\arraystretch}{1.15}
\begin{tabularx}{\textwidth}{p{0.135\textwidth} p{0.26\textwidth} p{0.175\textwidth} p{0.155\textwidth} X}
\hline
Contribution & Present assessment & Basis & Treatment in projections & Required action \\
\hline
Atom shot noise
  & Intrinsic statistical reference defining the sensitivity target
  & Prototype operation at the standard quantum limit~\cite{AION:2025igp}
  & Defines the projections (Table~\ref{tab:roadmap})
  & Higher-flux sources for Baseline and Stretch (Sec.~\ref{sec:atom-sources}) \\
Laser phase and frequency noise
  & Expected below the target after common-mode rejection; residual not yet validated
  & Prototype demonstrations~\cite{Stray2022,baynham_prototype_2026}; analyses~\cite{MAGIS-100:2021etm,Graham:2012sy}
  & Not included
  & Validation at the final AICE laser specification \\
Wavefront aberration
  & Expected below the target, conditional on spatial filtering, $\lambda/100$ optics and $\mathcal{O}(10^6)$ atoms
  & Modelling of aberration filtering and response~\cite{MAGIS-100:2021etm}
  & Not included
  & AICE-specific optical-system analysis and wavefront specification \\
Electromagnetic noise
  & No limiting broadband disturbance identified within the measurement sensitivity
  & PX46 measurements~\cite{Arduini:2851946}
  & Not included
  & Commissioning validation, with the shielding of Sec.~\ref{sec:detector} \\
LHC machine cycle
  & Several-minute timescale compatible with planned compensation
  & PX46 full-cycle measurements~\cite{Arduini:2851946}
  & Not included
  & Demonstration of synchronised compensation \\
Mechanical seismic vibration
  & Below the laser-system requirement, except the 50~Hz CV line
  & 6-day campaigns in SX4 and PX46~\cite{Arduini:2851946}
  & Not included
  & 50~Hz mitigation; validation of common-mode suppression \\
Seismic GGN
  & Expected to exceed shot noise below $\sim 0.8$--$5$~Hz (configuration-dependent), unmitigated
  & PX46 upper and lower spectra; Sedrun for the km scale~\cite{Guinchard:2026cen}
  & Excluded from ULDM curves; GW-figure overlay; benchmark in Sec.~\ref{sec:noise:GGN:impact}
  & Characterisation and mitigation programme (Sec.~\ref{sec:noise:GGN:mitigation}) \\
Atmospheric GGN
  & Comparable importance at the lowest frequencies; not a local measurement
  & Analysis of ERA5 data~\cite{Carlton:2024lqy,ERA5}
  & Not included; reported separately
  & Local monitoring for modelling and potential subtraction \\
Air flow and turbulence
  & No quantitative conclusion presently available
  & Under study for PX46
  & Not included
  & Characterisation; flow reduction assessed against ventilation requirements \\
Static gravity gradients
  & Systematic bias, not a broadband noise contribution
  & Standard treatment in precision interferometry
  & Systematic in the EP and $\alpha$ analyses
  & Addressed in Sections~\ref{sec:ep} and~\ref{sec:alpha} \\
\hline
\end{tabularx}
\end{table}

% Keep the summary table within this Section: barrier before the next chapter.
\FloatBarrier
\section{Commissioning and Operations}
\label{sec:ops}

%\textcolor{red}{John, Oliver, Sergio. Following text copy/pasted form the Answer to the LHCC Referees}

The AICE operational model is organised in three phases, with progressively decreasing requirements for shaft access and on-site personnel. A central design principle is that routine operation is concentrated in the surface laser laboratory, while shaft access is limited to defined and planned interventions.

\subsection{Construction and Installation}

During construction, regular access to the PX46 shaft will be required for installation of the vacuum system, atom sources, optical infrastructure and associated services. Shaft access will be carried out by trained personnel operating in pairs, in accordance with CERN safety rules and the elevator two-person requirement. All personnel will hold the relevant CERN safety certifications, including work-at-height and other applicable qualifications as discussed in Section~\ref{sec:ops:risk}.

Activities will be coordinated through standard CERN planning procedures. This phase requires concentrated person power during active installation periods but does not entail continuous presence beyond scheduled work.

\subsection{Commissioning}

The commissioning phase will involve structured and recurring shaft access as atom sources, laser systems and interferometric sequences are brought into operation and characterised. Access remains pair-based via the certified elevator platform, with active commissioning campaigns typically involving up to six specialists on site.

The required expertise in cold-atom operation, laser stabilisation and alignment, and vacuum diagnostics is already well established within the AION and TVLBAI collaborations. As detector performance is established and systems stabilise, the frequency of shaft access decreases.

\subsection{Standard Operation}

In steady-state operation, shaft interventions are expected to be infrequent. The operational model is based on the following elements.

\subsubsection{Surface-based operation}
The large majority of routine operational activity takes place in the surface laser laboratory in the SX4 building, where laser tuning, stabilisation and system monitoring are performed. Access requires standard laser safety certification, which is routinely held by collaboration members operating cold-atom systems. No additional shaft-specific provisions are required for this work.

\subsubsection{Remote monitoring}
Extended data-taking campaigns will be supervised remotely from participating institutes using established run-control and data-quality monitoring systems, minimising the requirement for on-site staffing.

\subsubsection{Shaft access by exception}
In the event of a hardware fault within the shaft, data-taking will be paused until trained personnel are available to carry out the repair. The physics programme does not require an immediate intervention capability. Repairs will be scheduled and executed by qualified personnel operating in pairs via the elevator platform.

\subsubsection{Planned maintenance and upgrades}
Major modifications such as installation of additional atom sources, changes of atomic species, or vacuum interventions will be planned well in advance and carried out during dedicated intervention periods coordinated with CERN.

\subsubsection{Person-power in steady state}
Standard physics running is expected to require only a small surface-based presence, typically one to two persons during active data-taking campaigns, with most monitoring and run control performed remotely across the collaboration. Continuous 24-hour on-site staffing is not foreseen as a technical requirement of the experiment. Instead, a structured on-call model will be implemented, whereby qualified personnel can attend the site if adjustments in the surface laser laboratory are required.

Should CERN safety regulations impose additional attendance requirements, these will be fully respected and integrated into the operational model. Based on the present design, however, no permanent round-the-clock shift-based presence is required for safe or stable operation.

\subsection{Operational model}

AICE is conceived as a facility supporting multiple physics programmes. Long data-taking campaigns are envisaged, interspersed with planned breaks for reconfiguration, maintenance or upgrades. This operational rhythm naturally accommodates scheduled shaft interventions without impacting the LHC physics programme.

The required skills in cold-atom experimental work, laser safety certification, and standard CERN safety qualifications are well established within the collaboration and fully consistent with comparable precision-physics facilities. The operational concept therefore relies on structured, safety-compliant and scheduled access procedures rather than continuous on-site rapid intervention capability. The issue of operating safely on the platform is discussed in more detail in Sections~\ref{sec:civil:platform} and~\ref{sec:ops:risk}.
%in the parallel submission addressing infrastructure aspects of AICE.

We emphasise that PX46 is currently the only vertical ${\cal O}(100)$~m site in Europe for which the full technical preparation strategy, infrastructure modifications, cost evaluation and installation pathway have been developed to an implementation level compatible with the required project schedule. This places CERN in a unique position to realise a 100~m class vertical atom interferometer in the near term.

%Yes. After checking the broader literature, the overall model is appropriate, but the previous version was too specific for the present design maturity.

%The evidence supports three general requirements:

%- MAGIS-100 establishes cyclic, camera-based readout, synchronised diagnostics and per-shot monitoring, but not a detailed online architecture. [MAGIS-100](https://arxiv.org/pdf/2104.02835)
%- MIGA demonstrates that remote-controlled, continuously monitored subsystems are realistic for a large underground atom interferometer. [MIGA laser system](https://arxiv.org/abs/1911.12209)
%- ELGAR distinguishes science data from ancillary environmental data, while explicitly leaving the detailed storage and computing infrastructure for later study. [ELGAR, Section 8.3](https://arxiv.org/pdf/1911.03701)

%I therefore recommend specifying functions, interfaces and safety boundaries without committing to particular monitored quantities, processing algorithms, alarm logic or access-control implementation.

\subsection{Controls and data acquisition}

The AICE controls and data-acquisition system will coordinate the experimental sequence and record the fluorescence images from each detector cycle together with the timing, configuration and auxiliary monitoring information required for their interpretation. The architecture will support synchronous operation of the atom sources and detection stations and preserve the association between the science data and the relevant laser, magnetic, vacuum and environmental conditions.

An online data-quality monitoring (DQM) system will provide prompt indicators of detector and subsystem performance, allowing operators to identify loss of signal quality, drifts or hardware faults during extended data-taking campaigns. The MAGIS~\cite{MAGIS-100:2021etm} and MIGA~\cite{Sabulsky:2019rll} collaborations have each established the feasibility of suitable remote DQM systems for large subsurface experiments. In the case of AICE, the quantities to be monitored, alarm thresholds and degree of automated processing will be defined during detailed design and commissioning, when the final detector configuration and operational experience are available.

Run status and DQM information will be available for remote supervision by authorised collaborators. CERN access and safety interlocks will remain independent of the scientific controls and DAQ. Data will be buffered and stored locally before transfer to the host-provided long-term archive. This functionality is covered by the experimental-controls, imaging and diagnostics, and DAQ and online-computing provisions in the cost estimate; no separate DQM cost item is required.

\subsection{Operational risks}
\label{sec:ops:risk}

This Section summarises the preliminary risk assessment and safety implementation strategy for AICE in the PX46 shaft at LHC Point 4. The following matrix and safety provisions are derived from the conceptual feasibility and implementation studies conducted between 2023 \cite{Arduini:2851946} and 2025 \cite{Arduini:2025jhe}. It is understood that a formal, detailed risk matrix for the final experiment will be defined by the competent safety officers through a comprehensive risk analysis when drafting the final Technical Design Report.

The risk matrix shown in Table~\ref{Personnel risk} identifies primary hazards for the AICE personnel, their qualitative probability, potential impact, and the mitigation strategies required to ensure concurrent operation with the High-Luminosity LHC (HL-LHC).
We note that laser exposure hazard is present only during commissioning and operation of AICE.
We also note that the occurrence of a mechanical failure of the platform coincident with a fire or helium release accident would be unlikely.

\begin{table}[h!]
\small
\centering
%\begin{tabularx}{\textwidth}{{2cm}p{2.2cm}p{1.8cm}p{2cm}X}
\begin{tabularx}{\textwidth}{
>{\raggedright\arraybackslash}p{2.1cm}p{2.3cm}p{1.8cm}p{2.5cm}X}
\toprule
\textbf{Hazard} & \textbf{Cause} & \textbf{Probability} & \textbf{Impact}  & \textbf{Mitigation Measures} \\ 
& & & \textbf{(Unmitigated)} & \\ \midrule
\textbf{Ionizing Radiation} & Accidental LHC beam loss near~PX46 base. & Small. & \textbf{High}: Dose $>$ 20 mSv/year (safety limit). & 80-cm concrete shielding wall in TX46; interlocked LASS doors; RP monitoring \cite{rp-study}. \\ \addlinespace
\textbf{Fire and Smoke} & Fire in UX45 or adjacent areas, smoke propagation via PX46. & Possible. & \textbf{High}: Suffocation and reduced visibility. Burns. & High-speed elevator platform (evacuation $<$ 2 min either via SX4 or TX46/UX45); smoke detection; visual and audible alarms; AUG-proof bidirectional communication means; mandatory self-rescue masks \cite{fire-safety}. \\ \addlinespace
\textbf{Fire and Smoke} & Fire in AICE and related equipment. & Possible. & \textbf{High}: Suffocation and reduced visibility. Burns. & Same as above and: choice of materials; smoke detection in close proximity; selective depowering of the electronics  \cite{fire-safety}. \\ \addlinespace
\textbf{Fire and Smoke} & Fire on the AICE mobile platform. & Possible. & \textbf{High}: Suffocation and reduced visibility. Burns. Evacuation impeded & Same as above and: extinguishing equipment on the platform; mechanical failsafe to bring the platform to the bottom. \\ \addlinespace
\textbf{Oxygen Deficiency (ODH)} & Helium release from LHC magnets or RF cryomodules. & Small. & \textbf{High}: Asphyxiation  due to helium displacement. & Pressure-relief doors divert the He flow from LHC tunnel to PM shaft; ventilation design in UX45/TU46; High-speed elevator platform; ODH detection; mandatory self-rescue masks \cite{helium-release,Hakulinen:2022}. \\ \addlinespace
\textbf{Mechanical Failure} & Elevator platform malfunction. & Small. & \textbf{Moderate}: Personnel stranded on platform or at bottom of PX46. & Mains power + battery backup for the platform; subsidiary evacuation via rescue nacelle or harness with firemen \cite{elevator}. \\ \addlinespace
\textbf{Class 4 laser exposure} & Laser malfunction. & Possible. & \textbf{High}: Personal injury, blindness. & Laser laboratory interlocked enclosures; protective eyewear; controlled access. \\ \bottomrule
\end{tabularx}
\caption{Personnel risk matrix for AICE at PX46.}
\label{Personnel risk}
\vspace{5mm}
\end{table}

Table~\ref{LHC risk} identifies the principal risks for the continuing operation of the LHC machine itself, due to the presence of the AICE experiment and human activities therein.

Table~\ref{AICE risk} identifies the principal risks for the AICE experiment due to LHC-related activities in PX46.

% ~~\\
% ~~\\

\begin{table}[h!]
\small
\centering
%\begin{tabularx}{\textwidth}{{2cm}p{2.2cm}p{1.8cm}p{2cm}X}
\begin{tabularx}{\textwidth}{
>{\raggedright\arraybackslash}p{2.3cm}p{3.0cm}p{1.8cm}p{2.5cm}X}
\toprule
\textbf{Hazard} & \textbf{Cause} & \textbf{Probability} & \textbf{Impact}  & \textbf{Mitigation Measures} \\ 
& & & \textbf{(Unmitigated)} & \\ \midrule
\textbf{LHC beam dump} & Elevator platform malfunction and personnel at the bottom of PX46 evacuating through LASS doors. & Small. & \textbf{High}: Loss of patrol and beam restart taking several hours. & Mains power + battery backup for the platform; subsidiary evacuation via rescue nacelle or harness with firemen \cite{elevator}. Operators mandatory training. \\ \bottomrule
\end{tabularx}
\caption{Principal HL-LHC risk due to AICE at PX46.}
\label{LHC risk}
%\vspace{5mm}
\end{table}

%~~\\
%~~\\

\begin{table}[h!]
\small
\centering
%\begin{tabularx}{\textwidth}{{2cm}p{2.2cm}p{1.8cm}p{2cm}X}
\begin{tabularx}{\textwidth}{
>{\raggedright\arraybackslash}p{2.3cm}p{3.0cm}p{1.8cm}p{2.5cm}X}
\toprule
\textbf{Hazard} & \textbf{Cause} & \textbf{Probability} & \textbf{Impact}  & \textbf{Mitigation Measures} \\ 
& & & \textbf{(Unmitigated)} & \\ \midrule
\textbf{Damage to the AICE experiment} & Loss of control of material being lifted with the main crane in PX46, hitting the experiment. & Small. & \textbf{High}: Damage to main vacuum chamber or experimental platforms. & Only authorized operators for the crane. Implement an anti-rotation system for suspended loads. \\ \bottomrule
\end{tabularx}
\caption{Principal AICE material risk.}
\label{AICE risk}
%\vspace{5mm}
\end{table}

Key provisions to be implemented in order to mitigate risks may be summarised as follows.
\begin{itemize}
    \item \textbf{Decoupling:} Implementation of a shielding wall in the TX46 gallery allows the PX46 shaft to be classified as a Supervised Radiation Area. This enables personnel to access and maintain the experiment during LHC beam operation~\cite{Devienne:2026};
    \item \textbf{Emergency Evacuation:} The custom lifting platform is a primary safety device. It must reach the surface or bottom in less than 2 minutes and is equipped with redundant power supplies and a mechanical brake for controlled descent even in a total power and battery failure scenario. A rescue nacelle can be deployed via the overhead crane as a subsidiary means of evacuation, upon intervention of the CERN fire brigade;
    \item \textbf{Access Control:} Integration into the LHC Access Safety System (LASS) and LHC Access Control System (LACS) ensures that opening the end-of-zone doors in TX46 automatically dumps the LHC beam \cite{Arduini:2025jhe}, and guarantees access only of properly trained operators;
    % \item \textbf{Air Quality:} A 20 m$^2$ ventilation room at the shaft top maintains a 20--40 Pa overpressure to ensure air tightness and unchanged ventilation management of underground areas.
    \item \textbf{Training:} Operators of AICE working in the PX46 shaft will have to undertake the mandatory safety training appropriate for supervised underground areas (electrical safety, radiation protection, emergency evacuation, self-rescue masks, etc.), and specific training such as work at heights, hands-on training on the specific elevator platform, and laser safety training.
\end{itemize}

% This Section has not been checked by the co-authors of the 2023 \cite{Arduini:2851946} and 2025 \cite{Arduini:2025jhe} reports, and serves uniquely the purpose of summarizing in table form the content therein, with the reserve of possible mistakes.

\clearpage % flush operational-risk floats before the cost chapter
%\input{sections/sec_org} old changed to cost 
% v15: EN-EL LS3/post-LS3 split (Change 26) implemented in lockstep with AICE_cost_matrix_v15.xlsx. Supersedes v14.
\section{Cost, Schedule and Risks}
\label{sec:org}

%\textcolor{red}{John, Oliver, Sergio}

\subsection{Project organisation}

AICE is being put forward by an international group of institutes and is embedded in the TVLBAI Proto-Collaboration of 57 institutions in 22 countries, whose endorsement is documented in Appendix~\ref{app:tvlbai-endorsement}. The required expertise in long-baseline atom interferometry has been established through the AION, MAGIS-100 and VLBAI programmes. The construction of the facility will be organised through work packages mapping onto the technical systems described in the preceding chapters. These work packages will provide the framework for the future assignment of institutional responsibilities and for the project-execution model. The formal governance of the collaboration, including institutional commitments, work-package leadership and the allocation of deliverables, will be established in a Collaboration Agreement to be concluded before completion of the Technical Design Report.

\subsection{Experiment cost}
\label{sec:cost}

The cost of the Initial AICE configuration is estimated at conceptual-design level, and is classified as a Class~4 estimate according to Ref.~\cite{bib:DOEGuide}. Values are included where supported by procurement experience, engineering estimates or subsystem costings; preliminary provisions awaiting confirmation are identified explicitly. All currently identified experiment cost lines are quantified. The estimate is intended to establish the scale of the project and support discussions of funding responsibilities, while procurement-level costing will be developed for the Technical Design Report. The estimate covers the full construction of a ready-to-be-commissioned instrument: the figures represent capital costs, and commissioning consumables, recurring operation, maintenance and steady-state personnel costs are not included. General operational spares are excluded; essential installed redundancy and component-level spares are included where they are explicitly part of a subsystem estimate. Source quotations will be normalised to a common price year before the estimate is frozen for submission.

The cost of the enabling infrastructure foreseen for completion during LS3 is taken from the implementation study described in Section~\ref{sec:civil:cost} and is shown separately from the experiment cost, since it constitutes the host-laboratory preparation to be provided by CERN during LS3, ahead of the construction of the experiment proper. The experiment-specific infrastructure and scientific-instrument costs are based on bottom-up engineering estimates, dedicated estimates by the relevant CERN technical groups, and procurement experience from AION-10 and related long-baseline atom-interferometer developments. The Initial configuration costed here comprises three strontium interferometer sources in the shaft and one surface reference source; the latter is not included in the source count $N_\mathrm{source}$ used in the performance roadmap. The fixed infrastructure is dimensioned for the planned facility upgrade path, whereas the scientific-instrument subtotal covers only the Initial strontium configuration. The ytterbium system and additional atom sources required for later Baseline and Stretch operation and other campaign-specific upgrades are not included in the present subtotal.

Subsystem-level engineering, integration, installation and implementation effort is included within the relevant work-package estimates; no separate central project-execution line is carried. Before the estimate is frozen for the Technical Design Report, each work-package basis will explicitly identify the delivery and implementation effort included. A uniform 20\% contingency is adopted for this conceptual estimate, applied to all quantified experiment direct-cost items and shown as a single explicit line; the line ranges themselves additionally reflect the estimate spread of each item. No additional experiment contingency is applied to the LS3 amount, which is transferred from the implementation study with its own Class-4 treatment. Totals are calculated from the unrounded line values, so the displayed rounded values may differ by 0.1~MCHF when added directly. MAGIS-100 costs are used as consistency checks for host-laboratory infrastructure alone, since the Fermilab estimate did not include the scientific instrument, which was delivered through in-kind contributions.

The resulting cost breakdown is summarised in Table~\ref{tab:cost}. The experiment cost totals 9.7--14.8~MCHF, or 11.6--17.8~MCHF including contingency, with the LS3 host preparation (1.5~MCHF civil works and 0.20--0.25~MCHF experiment-related electrical works) shown separately.

\begin{table}[htbp]
\centering
\caption{\label{tab:cost} Conceptual-design (Class~4) cost estimate for the AICE facility, for the Initial single-species strontium configuration with three interferometer atom sources and one surface reference source. Figures are in MCHF, at the indicative conversion rates USD$\to$CHF 0.80, EUR$\to$CHF 0.95 and GBP$\to$CHF 1.10, excluding recoverable VAT. Individual lines are rounded to 0.05~MCHF and subtotals to 0.1~MCHF; totals are calculated from the unrounded line values and may differ by 0.1~MCHF when the displayed values are added directly. All currently identified experiment lines are quantified. The LS3 host preparation, provided by CERN, is shown separately from the experiment cost.}
\smallskip
\small
\begin{tabular}{@{}>{\raggedright\arraybackslash}p{11.6cm}r@{}}
\toprule
{\bf Item} & {\bf Cost (MCHF)} \\
\midrule
\multicolumn{2}{@{}l}{\emph{Experiment infrastructure and installation (post-LS3)}} \\
\addlinespace[2pt]
Long-baseline vacuum system including telescope vessel & 2.0--4.0 \\
Magnetic shielding and field control (baseline double-layer architecture) & 1.4--2.2 \\
Mechanical support, shaft attachment and installation logistics & 0.45 \\
Experiment-facility controls, interlocks and facility networking (excluding LS3 systems) & 0.30--0.65 \\
Cooling and ventilation systems & 0.80--0.90 \\
Electrical distribution and supply (post-LS3 phase) & 0.10--0.15 \\
Seismic and survey network & $\sim$0.10 \\
SX4 fit-out, furniture, local services and terminal conditioning (central CV plant/distribution and upstream electrical supply excluded; Section~\ref{sec:laser-lab-infrastructure}) & 0.50--0.60 \\
\addlinespace[2pt]
{\bf Infrastructure subtotal} & {\bf 5.7--9.1} \\
\midrule
\multicolumn{2}{@{}l}{\emph{Scientific instrument, controls and DAQ}} \\
\addlinespace[2pt]
Atom-source packages (three interferometer + one surface reference) & 0.60--0.70 \\
Laser systems incl.\ local high-power amplification per source & 0.90--1.30 \\
Frequency reference: comb, reference cavity, stabilisation branches & 0.45--0.55 \\
Laser-laboratory optics and beam distribution & 0.50--0.60 \\
In-vacuum optics and positioning & 0.30--0.45 \\
Experimental controls and environmental monitoring & 0.20--0.30 \\
Imaging and diagnostics & 0.05--0.10 \\
AICE-specific scientific-system integration & 0.20--0.40 \\
Data acquisition, storage and online computing, incl.\ the dedicated DAQ network & 0.35--0.70 \\
Clock-laser transport and delivered-beam conditioning (laser laboratory to shaft) & 0.40--0.55 \\
\addlinespace[2pt]
{\bf Scientific instrument subtotal} & {\bf 4.0--5.7} \\
\midrule
{\bf Experiment direct total} & {\bf 9.7--14.8} \\
{\bf Contingency (20\% of quantified experiment direct cost)} & {\bf 1.9--3.0} \\
{\bf Experiment total including contingency} & {\bf 11.6--17.8} \\
\midrule
\multicolumn{2}{@{}l}{\emph{Host-laboratory preparation (provided by CERN)}} \\
\addlinespace[2pt]
LS3 infrastructure preparation (Section~\ref{sec:civil:cost}) & 1.5 \\
LS3-phase experiment electrical works (EBD1/4X replacement, EXD1/4X, powering of LS3-installed loads; Section~\ref{sec:civil:cost}) & 0.20--0.25 \\
\addlinespace[2pt]
{\bf Combined programme total: experiment plus CERN LS3 preparation} & {\bf 13.3--19.6} \\
\bottomrule
\end{tabular}
\end{table}
\clearpage % keep the summary cost table with the cost chapter

\paragraph{Installation and infrastructure.} The long-baseline vacuum line includes the main vacuum envelope and its supporting structure, transfer beam pipe, telescope vacuum vessel, interconnecting chambers, pumping and instrumentation, assembly and leak testing. For costing purposes the telescope vacuum vessel is included within this envelope; its detailed configuration remains to be confirmed, and no separate cost line is assumed. The line excludes the atom-source vacuum packages, optical elements and actuators, magnetic system, and permanent shaft support structure, which are costed separately. The magnetic-shielding estimate corresponds to the double-layer architecture identified as suitable in Section~\ref{sec:shielding}. The 1.4~MCHF lower scenario combines the approximately 15\% reduction indicated by the latest quotation for the dominant material component with further series-production, assembly and logistics economies; it is not a single-layer baseline and will be confirmed with the shielding engineering lead before the estimate is frozen. The experiment-facility controls line excludes the LASS, fire, evacuation, emergency-communication and radiation monitoring systems already included in the LS3 estimate. The cooling and ventilation line, comprising a dedicated chiller, cooling skids for the atom-source sidearms, the central ventilation plant and its primary distribution serving the ventilation room and laser laboratory, is based on a dedicated estimate at current CERN cooling-and-ventilation contract prices. The electrical scope is split between phases following the installation schedule: the LS3-phase works (replacement of the EBD1/4X switchboard, installation of EXD1/4X and the powering of loads commissioned during LS3) are carried in the host-laboratory preparation, while the post-LS3 electrical distribution and supply for the experiment and laboratory form the experiment infrastructure line; both are based on a dedicated first estimate by the CERN electrical group. These estimates exclude signal and fibre-optic cabling, their cable trays, and internal consolidations. 
% \textcolor{red}{[OB: Sergio, please check this EN-EL split (LS3-phase CHF 0.20--0.25\,M; post-LS3 CHF 0.10--0.15\,M; total CHF 0.30--0.40\,M) is compatible with what we have received.]} 
The SX4 fit-out line covers the room preparation, furniture, bare optical tables and supports, local services and safety systems, and terminal environmental conditioning of the surface laser laboratory, including final filtration, local sensing and control and room-side distribution interfaces. The central cooling and ventilation plant and primary distribution and the upstream electrical supply are carried in the corresponding infrastructure lines, the scientific optical equipment in the instrument block, and the delivery of light to the shaft in the clock-laser transport line. Of the 0.45~MCHF mechanical-support and installation logistics provision, 0.20~MCHF is assigned to mechanical support and shaft attachment and 0.25~MCHF to residual installation logistics. 
%These precautionary figures include a margin of approximately 50\% above the initial engineering communication and are adopted pending confirmation by the responsible group leader. 
% Vacuum-system assembly, installation and leak testing remain within the vacuum line and are not duplicated in this provision.

\paragraph{Scientific instrument.} The cold-atom equipment is costed for the Initial strontium configuration from itemised, procurement-based costings and scales with the number of atom sources and optical stations rather than with baseline length. The principal AICE-specific cost relative to a laboratory-scale instrument is local high-power laser amplification at each atom source, and a dedicated frequency reference is included since no in-kind infrastructure is assumed; the technical rationale is given in Section~\ref{sec:laser}. The laser-system line is supported by a quote-based itemisation, and the AICE-specific integration line covers the extended beam distribution, source-local interfaces and CERN-compatible experimental safety and diagnostics. The data-acquisition line is a bottom-up estimate of the readout, timing, online-processing, storage and network systems, and has been confirmed as a reasonable envelope against the MAGIS-100 experience. The transport of the clock laser from the surface laboratory to the shaft, comprising a dedicated relay beamline and a delivered-beam conditioning and Coriolis-compensation station, is costed from a scope-based engineering estimate. A quantified descoping option, drop-mode operation of the atom sources, would reduce the estimate by approximately 0.35~MCHF and is retained as a cost sensitivity rather than adopted in the Initial baseline.

\paragraph{Open items.} 
%No cost line remains marked TBD. The 0.45~MCHF mechanical-support and installation provision remains preliminary pending group-leader confirmation.
The remaining technical confirmations concerning the detailed telescope configuration and the magnetic-shielding lower scenario are retained within the existing Class~4 envelopes and do not create additional cost lines. The division of financing between CERN and the collaborating funding agencies remains to be discussed.

\subsection{Schedule}
\label{sec:schedule}

The overall project timeline is presented in Section~\ref{sec:staged} and Fig.~\ref{fig:timeline}. The LS3 enabling works and their detailed Gantt chart are described in Section~\ref{sec:civil:schedule}. Post-LS3 detector installation and the subsequent 12--18 month commissioning programme are described in Section~\ref{sec:ops}. The absolute dates will be updated once the duration and access assumptions for the post-LS3 installation campaign have been finalised.

\subsection{Project risks}
The preliminary project risk assessment for AICE in the PX46 shaft at LHC Point~4 draws on the conceptual feasibility and implementation studies conducted between 2023~\cite{Arduini:2851946} and 2025~\cite{Arduini:2025jhe}. The identified operational hazards, their qualitative probabilities and the corresponding mitigation measures, covering personnel, LHC operation and the experiment itself, are presented in Section~\ref{sec:ops:risk}, with the underlying safety and hazard-mitigation framework described in Section~\ref{sec:site:safety}. A formal risk register and detailed safety analysis will be developed for the Technical Design Report by the project management team together with the relevant CERN technical and safety experts.
\section{Conclusions}
\label{sec:conclusions}

AICE is proposed as a staged facility for long-baseline atom interferometry. Its shared infrastructure is designed to support several complementary measurements over the lifetime of the experiment, rather than a single fixed programme. The initial scientific focus is on searches for scalar, vector $B{-}L$ and pseudoscalar ultralight dark matter, using strontium and ytterbium sources in several interferometer geometries. Measurements of the fine-structure constant and the universality of free fall are also foreseen, alongside further tests of gravity and quantum mechanics. At later stages, the same infrastructure can be used for a pathfinding study of gravitational waves in the mid-frequency band between terrestrial laser interferometers and LISA. A further aim is to obtain operating experience relevant to the design of future kilometre-scale detectors.

The technical basis for AICE draws on more than a decade of work in the UK, the US, Germany, France and China. AION, MAGIS-100, VLBAI, MIGA and ZAIGA have demonstrated or advanced the main elements required for long-baseline atom interferometry: single-photon clock-transition interferometry, large momentum transfers, differential gradiometry and dual-species operation. This work includes 10\,m-scale prototypes as well as the design and construction of instruments at the 100\,m scale. Of particular relevance to AICE is the demonstration of differential interferometry on the 698\,nm strontium clock transition, which validates its core measurement principle. Moving to $\mathcal{O}(100)$\,m remains a significant integration and engineering task, but it does not require an untested experimental concept.

At CERN, PX46 is the candidate site for combining these developments in a single instrument. The feasibility study completed in 2023~\cite{Arduini:2851946} and the implementation study completed in 2025~\cite{Arduini:2025jhe} found the shaft suitable, developed a site-specific integration concept and identified no fundamental technical obstacles. The proposed arrangement preserves the use of PX46 for LHC transport, is compatible with concurrent HL-LHC operation and benefits from existing CERN infrastructure and expertise.

%The current schedule places the enabling works in LS3, preferably in the first half of 2029. 
The current schedule aligns with the LS3 schedule, allowing the enabling works in PX46 to be carried out during the first half of 2029.
Site preparation could then be completed separately from detector installation, allowing installation and commissioning during Run~4 from 2030 and staged scientific operation thereafter, extending to about 2040. PX46 is currently the only vertical $\mathcal{O}(100)$\,m site in Europe for which infrastructure modifications, costs and an installation sequence have been studied to this level on the required schedule. A facility there would complement MAGIS-100 in the United States and make possible coordinated measurements with an international network of long-baseline instruments.

The staging plan separates permanent infrastructure from components expected to evolve. The shaft infrastructure, vacuum envelope, support structure and main services would be installed once and sized for the full programme. Atom sources, laser systems and measurement configurations could then be added or upgraded separately. The Initial strontium instrument defines the route from commissioning to the first physics measurements. Further sources, ytterbium capability and improvements to the laser and atom-source systems would support the Baseline and Stretch programmes without requiring reconstruction of the facility. This sequence defines the scope of the first stage while allowing later choices to be informed by operating experience.

The projected sensitivities depend on the benchmark performance assumptions used in this proposal. For the Baseline configuration, the reach for ultralight-dark-matter couplings is competitive with, and in some cases exceeds, current laboratory bounds. Under the Stretch assumptions, the projected sensitivity to axion--nucleon couplings improves by about one order of magnitude on the SN\,1987A astrophysical bound, whose interpretation depends on supernova modelling. The precision-measurement programme aims at an independent single-photon determination of the fine-structure constant with $\sigma_\alpha/\alpha \lesssim 10^{-11}$, which would bear on the discrepancy between the caesium and rubidium recoil measurements. The corresponding target for tests of the universality of free fall is $\eta \lesssim \mathrm{few}\times 10^{-16}$.

No contribution considered in the present noise and systematics assessment has been shown to impose an irreducible floor above the atomic shot-noise target. This conclusion is provisional: several contributions must still be tested with the final AICE systems. In particular, the unmitigated estimate of seismic gravity gradient noise lies above the shot-noise target at the low-frequency end of the measurement band. Site measurements and tests of the available mitigation methods are therefore required during commissioning and early operation.

At the present conceptual design stage, the Class~4 cost estimate for the full construction of the Initial AICE experiment, corresponding to a ready-to-be-commissioned instrument, is 11.6 to 17.8~MCHF; this figure includes a uniform 20\% contingency applied to the quantified direct cost of 9.7 to 14.8~MCHF. In addition, LS3-phase preparatory works in the PX46 shaft are foreseen: 1.5~MCHF of civil-engineering works provided by CERN as host-laboratory preparation, together with 0.20 to 0.25~MCHF of experiment-related electrical works whose funding attribution remains to be discussed. These preparatory works make the shaft ready to host the experiment, rather than forming part of the experiment cost itself. The combined programme total is 13.3 to 19.6~MCHF (Section~\ref{sec:cost}). Costs for the subsequent Baseline and Stretch upgrades are not included; these are expected to proceed through incremental investments, each currently anticipated to be of the order of 5~MCHF, and each therefore only a fraction of the Initial construction cost.

The allocation of costs between CERN and the participating collaborating institutes is under discussion. A clear structure is beginning to emerge, with responsibilities for several elements already being identified and acknowledged by prospective contributors. This division of responsibilities will be developed further at the forthcoming workshop and then finalised for inclusion in the Technical Design Report.

Several tasks must be completed before construction can begin. The next design phase should resolve the remaining subsystem interfaces and engineering choices, complete the laser-laboratory and mechanical-integration designs, and produce the risk register, safety case, procurement plan and project-execution model. Further site measurements are needed both to constrain the design and to specify a programme for characterising and mitigating low-frequency gravity gradient noise. The outstanding costs and the allocation of funding and in-kind contributions must also be settled. These tasks define the scope of a Technical Design Report; none has so far revealed a fundamental obstacle to the proposed facility.

The next phase can draw on a site-specific implementation concept, demonstrated core technologies and a modular detector design. The associated scientific programme has the endorsement of the TVLBAI Proto-Collaboration, which comprises over 55 institutions in more than 20 countries. Its contribution to the 2026 update of the European Strategy for Particle Physics identifies AICE as the key European step towards kilometre-scale detectors. The AICE community workshop at CERN on 1--3 September 2026 is intended to refine the technical baseline, assign work-package responsibilities, develop the funding strategy and establish the route to the Technical Design Report.

On this basis, we ask CERN to endorse the preparation of an AICE Technical Design Report and to enable the proposed LS3 preparatory works on the schedule described above.

\section*{Acknowledgements}

We gratefully acknowledge valuable technical advice and information from the following CERN colleagues: Kincso Balazs, Maria Barberan Marin, Oliver Boettcher, Nicolas Broca, Marco Buzio, Paolo Chiggiato, Nicholas Chritin, Jean-Pierre Corso, Olivier Crespo-Lopez, Pierre Durand, Lucie Elie, Christelle Gaignant, Grégory Godineau, Mike Lamont, Francesca Luoni, Raphael Langlois, Maddalena Maietta, Marija Majstorovic, Alexander Pierre Marion, Rui Nunes, Mario Parodi, Mariano Pentella, Vitor Rios, Melvyn Rouchouse, Marc Timmins, Jean-Philippe Tock, Katarzyna Turaj and Heinz Vincke.

\bibliographystyle{JHEP}
\bibliography{CERNAI,AICE_ULDM,ALPHA,WAVEFRONT}

% ---------------------------------------------------------------
% Appendices
% ---------------------------------------------------------------
\newpage
\appendix

\section{TVLBAI Support and Endorsement}
\label{app:tvlbai-endorsement}

The AICE Facility is an integral part of the Terrestrial Very Long
Baseline Atom Interferometer (TVLBAI) Proto-Collaboration activities
and enjoys the full support of the TVLBAI community. The TVLBAI
Proto-Collaboration was established through a Memorandum of
Understanding and signed by 57 institutions in 22 countries with 4
additional observer institutions~\cite{TVLBAIMOU,TVLBAISIG} at the date of this Report,
representing a coordinated international effort to develop atom
interferometers beyond the 10\,m scale. The scientific objectives and
coordinated approach of the TVLBAI community have been developed through
three major international workshops, each with over 200 participants
from the particle physics, atomic physics, astrophysics and cosmology
communities~\cite{TVLBAISummary,abdalla_terrestrial_2025,TVLBAI3},
and the community has provided input to the 2026 update of the European
Strategy for Particle Physics~\cite{TVLBAIESPP}.

The proposed experiment at CERN has been identified by the TVLBAI Study
Group as a priority demonstrator within their comprehensive roadmap for
long-baseline atom interferometry. The TVLBAI activity fully endorses
this initiative as a cornerstone activity that directly advances the
TVLBAI scientific and technical objectives. As outlined in the TVLBAI
roadmap, this experiment represents a key component of the envisioned
global network of detectors with baselines $\mathcal{O}(100)$\,m that
will provide unique sensitivity to bosonic ultra-light dark matter
couplings and enable first explorations of gravitational wave signals
with frequencies $\mathcal{O}(1)$\,Hz.

As a TVLBAI-supported activity, the AICE Facility benefits from:
\begin{itemize}
\item Coordinated international expertise from the 57 participating
  TVLBAI institutions across 22 countries;
\item Shared technological development across the global TVLBAI
  network, including exchange of approaches using various atomic species
  (rubidium, strontium, ytterbium) and geometries;
\item Integrated scientific planning as part of the broader TVLBAI
  demonstrator programme leading towards km-scale detectors in the
  mid-2030s;
\item Collaborative resource mobilisation leveraging the established
  international framework for information exchange and coordination;
\item Access to the TVLBAI international topical working groups and
  regular coordination meetings.
\end{itemize}

The TVLBAI Proto-Collaboration recognises CERN's PX46 site as uniquely
promising for demonstrating the feasibility and scientific potential of
$\mathcal{O}(100)$\,m scale atom interferometry. The AICE Facility will
serve as a flagship TVLBAI demonstrator, providing crucial validation of
technologies and methodologies that will inform the development of the
broader TVLBAI network, including future km-scale detectors.

We foresee that a significant fraction of the TVLBAI Proto-Collaboration
will actively contribute to and participate in the AICE Facility,
bringing together the collective expertise and resources of this
established international community to ensure its success.

\section{Definitions of acronyms}
\label{app:Acronyms}

\setlength{\parindent}{0pt}

\noindent\textbf{AEDGE}: Atomic Experiment for Dark Matter and Gravity Exploration in Space

\noindent\textbf{AI}: Atom Interferometer/Interferometry

\noindent\textbf{AICE}: Atom Interferometer CERN Experiment

\noindent\textbf{AION}: Atom Interferometer Observatory and Network

\noindent\textbf{ALP}: Axion-Like Particle

\noindent\textbf{AMF}: Area Mixed Field (Radiation Monitor)

\noindent\textbf{ASN}: Atom Shot Noise

\noindent\textbf{AUG}: Arrêt Général d'Urgence~\footnote{General Emergency Stop}

\noindent\textbf{AURIGA}: Antenna Ultracriogenica Risonante per l'Indagine Gravitazionale Astronomica~\footnote{Ultracryogenic Resonant Bar Gravitational Wave Detector}

\noindent\textbf{BEC}: Bose-Einstein Condensate

\noindent\textbf{BH}: Black Hole

\noindent\textbf{BRM}: Bottom Retroreflecting Mirror

\noindent\textbf{BSM}: Beyond the Standard Model

\noindent\textbf{CF}: ConFlat

\noindent\textbf{CV}: Cooling and Ventilation

\noindent\textbf{DFG}: Degenerate Fermi Gas

\noindent\textbf{DM}: Dark Matter

\noindent\textbf{ELGAR}: European Laboratory for Gravitation and Atom-interferometric Research

\noindent\textbf{EM}: ElectroMagnetic

\noindent\textbf{EP}: Equivalence Principle

\noindent\textbf{EMC}: ElectroMagnetic Compatibility

\noindent\textbf{ESPP}: European Strategy for Particle Physics

\noindent\textbf{GGN}: Gravity Gradient Noise

\noindent\textbf{GW}: Gravitational Wave

\noindent\textbf{HL-LHC}: High-Luminosity LHC

\noindent\textbf{HVAC}: Heating, Ventilation and Air Conditioning

\noindent\textbf{ICC}: Interconnecting Chamber

\noindent\textbf{IMBH}: Intermediate-Mass Black Hole

\noindent\textbf{KAGRA}: KAmioka GRAvitational wave experiment

\noindent\textbf{LACS}: LHC Access Control System

\noindent\textbf{LASS}: LHC Access Safety System

\noindent\textbf{LHC}: Large Hadron Collider

\noindent\textbf{LIGO}: Laser Interferometer Gravitational Observatory experiment

\noindent\textbf{LISA}: Laser Interferometer Space Antenna experiment

\noindent\textbf{LMT}: Large Momentum Transfer

\noindent\textbf{LS}: Long Shutdown

\noindent\textbf{LS3}: Long Shutdown Three

\noindent\textbf{LSBB}: Laboratoire Souterrain {\` a} Bas Bruit~\footnote{Low-Noise Underground Laboratory}

\noindent\textbf{LV}: Low Voltage

\noindent\textbf{LVK}: LIGO, Virgo and KAGRA

\noindent\textbf{MAD}: Material Access Device

\noindent\textbf{MAGIS}: Matter-wave Atomic Gradiometer Interferometric Sensor

\noindent\textbf{MICROSCOPE}: Micro-Satellite {\` a} tra{\^ i}n{\' e}e Compens{\' e}e pour l’Observation du Principe d’Equivalence~\footnote{Micro-Satellite with Compensated Drag for Observing the Principle of Equivalence}

\noindent\textbf{MCI}: Maximum Credible Incident

\noindent\textbf{MIGA}: Matter wave-laser based Interferometer Gravitation Antenna

\noindent\textbf{MOT}: Magneto-Optic Trap

\noindent\textbf{NEG}: Non-Evaporable Getter

\noindent\textbf{NEMA}: National Electrical Manufacturers Association 

\noindent\textbf{NHNM}: New High-Noise Model

\noindent\textbf{NLNM}: New Low-Noise Model

\noindent\textbf{ODH}: Oxygen Deficiency Hazard

\noindent\textbf{PAD}: Personal Access Device

\noindent\textbf{PBC}: Physics Beyond Colliders

\noindent\textbf{PI}: Principal Investigator

\noindent\textbf{PM}: Puits Mat{\' e}riel~\footnote{Access shaft with stairs and lift used for the transfer of equipment}

\noindent\textbf{PPE}: Personal Protection Equipment

\noindent\textbf{PPTA}: Parkes Pulsar Timing Array

\noindent\textbf{PX}: Puit eXp{\' e}rience~\footnote{Access shaft to experimental cavern for (formerly) LEP or (currently) LHC detectors}

\noindent\textbf{PX46}: Access shaft at LHC Point 4

\noindent\textbf{RF}: Radio Frequency

\noindent\textbf{RP}: Radiation Protection

\noindent\textbf{SMBH}: Super Massive Black Hole

\noindent\textbf{SM}: Standard Model

\noindent\textbf{SNR}: Signal-to-Noise Ratio

\noindent\textbf{SRF}: Superconducting RF

\noindent\textbf{SUSI}: Syst{\`e}me de SUrveillance des SItes~\footnote{Site surveillance system}

\noindent\textbf{SU4}: Surface building dedicated to cooling and ventilation at Point 4

\noindent\textbf{SX4}: Surface building on top of the PX46 shaft

\noindent\textbf{TC}: Technical Coordinator

\noindent\textbf{TDR}: Technical Design Report

\noindent\textbf{TETRA}: Terrestrial Trunked Radio, formerly known as Trans-European Trunked Radio

\noindent\textbf{TVLBAI}: Terrestrial Very-Long-Baseline Atom Interferometer

\noindent\textbf{TX46}: Access gallery at LHC Point 4

\noindent\textbf{UFF}: Universality of Free Fall

\noindent\textbf{ULDM}: Ultra-Light Dark Matter

\noindent\textbf{UPS}: Uninterruptible Power Supply

\noindent\textbf{UX45}: Experimental cavern at LHC Point 4

\noindent\textbf{VLBAI}: Very-Long-Baseline Atom Interferometer

\noindent\textbf{WEP}: Weak Equivalence Principle

\noindent\textbf{WIMP}: Weakly Interacting Massive Particle

\noindent\textbf{YETS}: Year-End Technical Stop

\noindent\textbf{ZAIGA}: Zhaoshan Long-baseline Atom Interferometer Gravitation Antenna experiment

\end{document}